\documentclass[a4paper,11pt]{article}
\pdfoutput=1
\usepackage{amsmath,bm,amssymb,color,mathtools}

\usepackage{booktabs}
\usepackage{multirow}
\usepackage{hhline}

\usepackage{graphicx}
\usepackage{jheppub} 
\usepackage{xspace}

\renewcommand{\tabcolsep}{2em}

\DeclareRobustCommand{\NNLOJET}{{\normalfont\textsc{NNLOjet}}\xspace}

\def\Q2{\left(Q^{2}\right)}
\def\e{\epsilon}

\def\d{{\rm d}}

\def\l({\left(}
\def\r){\right)}

\def\nf{N_{F}}

\def\e{\epsilon}

\def\d{\hbox{d}}

\def\nf{N_{F}}
\def\nn{\nonumber}

\def\ba{\begin{eqnarray}}
\def\ea{\end{eqnarray}}
\def\wt{\widetilde}

\def\ds{\d\hat{\sigma}}

\def\ph#1{\phantom{#1}} 
\def\snip{\nonumber\\} 

\def\l{(}
\def\r{)}

\title{NNLO QCD corrections to jet production in deep inelastic scattering}
\author{James Currie$^a$, Thomas Gehrmann$^b$, Alexander Huss$^c$, Jan Niehues$^b$}
\affiliation{$^a$Institute for Particle Physics Phenomenology, Durham University, South Road, Durham, DH1 3LE, UK\\
$^b$ Department of Physics, Universit\"at Z\"urich,
Winterthurerstrasse 190,\\CH-8057 Z\"urich, Switzerland\\
$^c$  Institute for Theoretical Physics, ETH,
Wolfgang-Pauli-Strasse 27,\\CH-8093 Z\"urich, Switzerland}
\emailAdd{james.currie@durham.ac.uk}
\emailAdd{thomas.gehrmann@uzh.ch}
\emailAdd{ahuss@itp.phys.ethz.ch}
\emailAdd{jan@physik.uzh.ch}
\abstract{Hadronic jets in deeply inelastic electron-proton collisions  are produced by the scattering of a parton
from the proton with the virtual gauge boson mediating the interaction. The HERA experiments have performed
precision measurements of inclusive single jet production and di-jet production in the Breit frame, which provide
important constraints on the strong coupling constant and on parton distributions in the proton. We describe
the calculation of the next-to-next-to-leading order (NNLO) QCD corrections to these processes, and assess their
size and impact. A detailed comparison with data from the H1 and ZEUS experiments highlights that
inclusive single jet production displays a better perturbative convergence than di-jet production. We also observe
that the event selection cuts in some of the di-jet measurements of both H1 and ZEUS induce
an infrared sensitivity that destabilises the perturbative stability of the predictions.
Our results open up new opportunities for QCD precision studies with jet production observables in deep inelastic
scattering.
 }
\preprint{{IPPP/17/19, ZU-TH 03/17}}

\begin{document}
\maketitle
\allowdisplaybreaks

\section{Introduction}

The HERA electron-proton collider produced a very precise data set on deeply inelastic scattering (DIS) processes,
both fully inclusively as well as for specific hadronic final states~\cite{disbook,Newman:2013ada}. Especially
jet final states
in deep inelastic scattering contain important information on the distribution of partons in the
proton, which are a crucial ingredient for the prediction of any cross section at the LHC and
at other future hadron colliders. Single jet inclusive and di-jet cross sections in deep inelastic
scattering~\cite{lo} are among the few precisely measured
processes~\cite{h1a12,h1b1,h1c2,h1d12,h1e12,h1highq2,h1lowq2,zeusa2,zeusb1,zeusc12,zeusd1,zeus2j}
 that provide sensitivity to the strong coupling constant and
the  gluon distribution already at tree level.
The importance of these cross sections for
precision QCD studies has long been known. However, having the relevant hard sub-process
coefficient functions only at next-to-leading order (NLO) in
perturbative QCD~\cite{graudenz,mirkes,jetvip,nagy}, the HERA
deep inelastic jet data
could not be confronted with a sufficiently precise theory description to
fully exploit their physics potential.

Owing to methodological advances in the calculation of two-loop amplitudes~\cite{laporta,gr,henn} and in the treatment of
infrared singular real radiation~\cite{secdec,qtsub,ourant,currie,stripper,njettiness,trocsanyi}, an increasing number of collider processes have been computed
in fully differential form to
next-to-next-to-leading order (NNLO) QCD accuracy. Following earlier results on the Drell-Yan process~\cite{dynnlo,babisdy},
on Higgs production~\cite{babishiggs,hnnlo} and on $e^+e^-\to 3j$~\cite{our3j,weinzierl3j}, NNLO results have been
obtained in recent years for $pp\to\gamma\gamma$~\cite{twogamma},  $pp\to VH$~\cite{vh}, $pp\to V\gamma$~\cite{vgamma},
$pp\to t\bar t$~\cite{czakon1,czakon2}, $pp\to H+j$~\cite{hjet,ourhj},
$pp\to W+j$~\cite{wjet},  $pp\to Z+j$~\cite{ourzj,zjet}, $pp\to \gamma+X$~\cite{mcfmgam},
$pp\to ZZ$~\cite{zz}, $pp\to WW$~\cite{ww},  $pp\to ZW$~\cite{zw}, $ep\to 1j$~\cite{abelof} and $pp\to 2j$~\cite{2jnew}.
All these calculations were implemented in the form of parton-level event generators (some of
the lower-multiplicity processes have also become part of the latest version of the MCFM code~\cite{mcfmnnlo}),
which provide
full kinematical information on all final state particles, and consequently allow to account for the precise
definition (jet algorithm, kinematical acceptance cuts) of observables used in the experimental analyses.

Calculations of collider observables to NNLO require to combine three types of parton-level contributions, which are
individually infrared divergent: double-real radiation, single-real radiation at one loop and two-loop virtual contributions.
To implement these contributions into a parton-level event generator, a method to extract and recombine
their infrared (IR) singular parts is required. Our group has developed the
antenna subtraction method~\cite{ourant,currie} for this purpose. This method forms the basis of the \NNLOJET code,
which provides the necessary infrastructure and building blocks to implement NNLO corrections to
different collider processes. Up to now,
$pp\to Z+j$~\cite{ourzj}, $pp\to H+j$~\cite{ourhj} and $pp\to 2j$~\cite{2jnew} have been implemented in \NNLOJET.

The same framework is used in the calculation of NNLO corrections to jet production in DIS, and first results on
di-jet final states were reported in a short letter~\cite{disprl}. This paper extends this calculation to single-jet
inclusive production in DIS, and provides a detailed documentation of its implementation in the \NNLOJET framework.
Section~\ref{sec:kin} establishes the notation and describes the kinematical
situation of jet production in DIS. The calculation of the NNLO corrections
is described in Sec.~\ref{sec:nnlo}, which also addresses in detail the handling of initial-state partons in the
antenna subtraction method. Our results for the NNLO corrections
to single jet inclusive production and di-jet production in DIS
are discussed in detail in Secs.~\ref{sec:1j} and~\ref{sec:2j} where we also compare to
the available data from the HERA experiments. We conclude with Sec.~\ref{sec:conc}.

\section{Kinematics of jet production in deep inelastic scattering}
\label{sec:kin}
The basic interaction in deep inelastic lepton-proton scattering is mediated by a virtual gauge boson. The kinematics of the
fully inclusive process can be inferred from the momenta of the incoming particles and  of the outgoing lepton:
$$l(k)+p(P) \to l'(k') + X(p_X),$$
such that a four-momentum $q=k-k'$ is transferred to the proton.
Measurements are carried out in terms of the following variables (neglecting the proton mass):
\begin{equation}
s_{lp} = (k+P)^2\,,\quad Q^2 = -q^2\,, \quad x=\frac{Q^2}{2q\cdot P}\,, \quad y = \frac{q\cdot P}{k\cdot P} = \frac{Q^2}{xs_{lp}}.
\end{equation}
The underlying parton-level process is the scattering of a quark from the proton with the virtual boson. At leading order, this quark carries a
momentum fraction $x$ of the proton momentum and scatters by an angle $\gamma_h$, defined as

\begin{equation}
\cos\gamma_h=\frac{(1-y)xE_{p}-yE_l}{(1-y)xE_{p}+yE_l},
\end{equation}
with $E$ denoting the energy.

\begin{figure}[t]
\centering
\includegraphics[width=8cm]{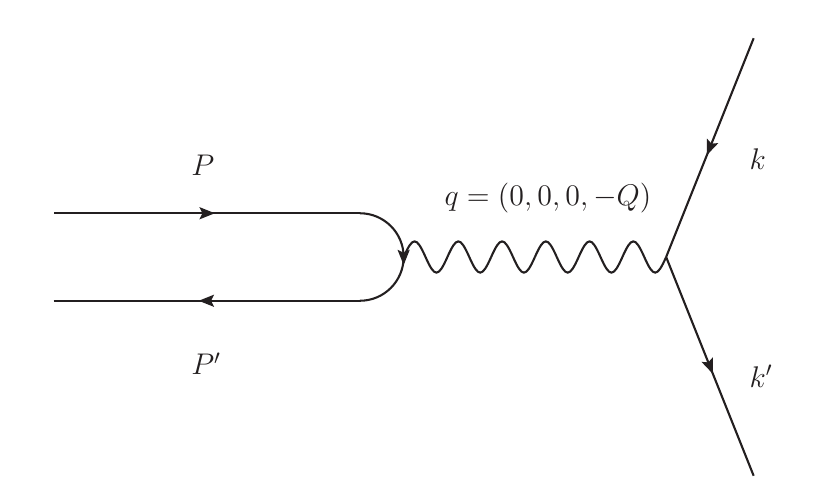}
\caption{An illustration of the basic hard scattering process in the Breit frame with incoming proton momentum $P$, incoming lepton momentum $k$,
virtual boson exchange momentum $q$, outgoing proton momentum $P'$ and outgoing lepton momentum $k'$. }
\label{fig:breit}
\end{figure}

More detailed information on the underlying parton-level dynamics can be gained by examining the hadronic final state $X$. This is often
analysed in the Breit frame of reference, Figure~\ref{fig:breit}, which is defined by requiring proton and gauge boson momenta to
take the form
\begin{equation}
P_B=(Q/(2x),0,0,Q/(2x))\,, \quad q_B=(0,0,0,-Q) \,,
\end{equation}
with $Q=\sqrt{Q^2}$. Momenta in the Breit frame are indicated by a subscript $B$. The Lorentz transformation from the laboratory frame
to the Breit frame can be determined from the measured lepton kinematics.
In this frame, the leading order DIS process results in an outgoing quark with vanishing transverse momentum, such that higher order
contributions to DIS can be resolved by looking at final state objects with non-vanishing transverse momentum $p_{T,B}$ in the Breit frame.

Of particular interest is jet production in the Breit frame, which has a large cross section and allows the measurement of a variety of different distributions that can provide constraints on the parton content of the proton and on the strong coupling constant.
Inclusive DIS at leading order is at ${\cal{O}}(\alpha^2\alpha_s^0)$ and only contains quarks in the initial-state; sensitivity to the strong coupling,
$\alpha_{s}$ and initial-state gluons only arises at NLO. In contrast, jet production is sensitive to $\alpha_{s}$ and both initial-state quarks and gluons
already at LO.

Jets are reconstructed in the Breit frame based on the transverse momenta $p_{T,B}$ (or equivalently transverse energies $E_{T,B}$),
pseudorapidity $\eta_B = -\ln \tan(\Theta_B)$ (with $\eta_B>0$ in the incoming proton direction) and
azimuthal angle $\phi_B$ of individual particles. Jets are formed from the particles by using an iterated clustering algorithm based on
a distance measure~\cite{gavin}:
$$
d_{ij} = \min(p_{T,B,i}^{2\delta},p_{T,B,j}^{2\delta}) R_{ij}^2\;,\quad d_{iB} = p_{T,B,i}^{2\delta} R_0^2\;,
$$
where $R_0$ is a resolution parameter and $\delta=1$ for the $k_T$ algorithm and  $\delta=-1$ for the anti-$k_T$ algorithm. If the minimal
distance measure is $d_{ij}$, then objects $i$ and $j$ are recombined, if it is $d_{iB}$, then object $i$ is called a jet and removed from
the list.
The recombination of
particle clusters is done in the $E_T$ scheme, as
$$ p_{T,B} = \sum_i p_{T,B,i}\,, \quad \eta_B = \sum_i \frac{p_{T,B,i} \eta_{B,i}}{p_{T,B}}\,, \quad \phi_B = \sum_i \frac{p_{T,B,i} \phi_{B,i}}{p_{T,B}}\;.$$
This scheme results in massless clusters and consequently massless jets, and $p_{T,B} =E_{T,B}$. The jets are accepted
if they are within rapidity cuts and above a minimum transverse momentum. They are ordered in decreasing transverse momentum, and
denoted as $j1$, $j2$ and so forth.

The HERA electron-proton collider operated in its run II with beam energies $E_e = 27.5$~GeV and $E_p=920$~GeV, resulting in
$\sqrt{s_{ep}}=318$~GeV. The HERA experiments H1 and ZEUS measured jet production cross sections~\cite{zeus2j,h1highq2,h1lowq2}
using the $k_T$ algorithm with $R_0=1$ as
function of the Breit frame jet variables
\begin{eqnarray}
M_{jj} = M_{12} & =& \sqrt{(p_{j1}+p_{j2})^2}\,, \nonumber \\
\bar{E}_{T,B} = \langle p_T\rangle_2 & =& \frac{1}{2} \left(p_{T,B,j1} + p_{T,B,j2} \right)\,, \nonumber \\
\xi = \xi_2 &=& x\left(1+\frac{M_{12}^2}{Q^2}\right)\,, \nonumber \\
\eta^* &=& \frac{1}{2} \left| \eta_{B,j1}-\eta_{B,j2} \right|\,. \label{eq:breitvar}
\end{eqnarray}
For di-jet production at leading order, $\xi_2$ can be identified with the momentum fraction of the incoming parton relative to the proton momentum.

\section{Calculation of NNLO corrections}
\label{sec:nnlo}

To calculate jet production at NNLO, all matrix elements which contribute to the process at the desired order in $\alpha_{s}$ must be
considered. Beyond leading order, radiative corrections can proceed via virtual loops or additional on-shell final-state particles,
both of which can modify the cross section for an observable in non-trivial ways. We use dimensional regularisation with $d=4-2\epsilon$
space-time dimensions as regulator of both ultraviolet and infrared divergences in these contributions.
Inclusive and semi-inclusive jet cross sections involve final-states containing up to one additional jet at NLO and
up to two additional jets at NNLO, relative to the LO final-state.

The theoretical prediction for a given $n$-jet final state at NNLO
receives double-real~(RR) corrections from all relevant tree-level $(n+2)$-parton matrix elements, real-virtual~(RV) corrections which involve
the interference between one-loop and tree-level $(n+1)$-parton amplitudes, and double-virtual~(VV) corrections involving the interference
between two-loop and tree-level $n$-parton amplitudes as well as the one-loop $n$-parton contribution interfered with itself.
For di-jet production in deep inelastic scattering, the relevant matrix elements have been known for a long time~\cite{meRR,meRV,meVV}. The RR
and RV contributions are also part of the NLO corrections to tri-jet production in DIS~\cite{nagy}, and  the relevant matrix elements
can now be generated automatically using standard tools~\cite{openloops,blackhat,gosam,amcnlo}.

\subsection{Structure of the NNLO cross section}
\label{sec:structure}

The colourless ingredients of the matrix element as well as the QCD coupling and colour factors can be pulled into an overall factor for each sub-process; this leaves the colour-stripped matrix element as the basic object on which to perform the subtraction, which is appropriate as this is the function which contains all the IR singularities. The overall factor at leading-order for initial-state parton flavour $i$, is given by
\ba
{\cal{N}}_{i}^{LO}&=&4(4\pi\alpha)^{2}V{\cal{S}}_{i}{\cal{C}}_{i}\bigg(\frac{\alpha_{s}}{2\pi}\bigg)\frac{\bar{C}(\e)}{C(\e)} ,
\ea
where $V=N^2-1$, $N=3$ the number of colours, $\alpha_{s}$ is the strong coupling and $\alpha$ is the fine structure constant. The initial-state colour averaging factor for the incoming parton is given by
\ba
{\cal{C}}_{q}=\frac{1}{N}&&{\cal{C}}_{g}=\frac{1}{V}.
\ea
The initial-state spin averaging factors are given by
\ba
{\cal{S}}_{q}=\frac{1}{s_{q}s_{e}}&&{\cal{S}}_{g}=\frac{1}{s_{g}s_{e}},
\ea
where the number of spin states for the various initial-state particles are given by $s_{e}=s_{q}=2,\ s_{g}=2(1-\e)$, and
the factors
\ba
{C}(\e)=\frac{(4\pi)^\e}{8\pi^2}e^{-\e\gamma}&,&\bar{C}(\e)=8\pi^2 {C}(\e)
\ea
which are arising from dimensional regularisation
are introduced.

The nomenclature for squared matrix elements we adopt in this paper is to denote the various types of colour stripped
squared matrix elements with the shorthand:
\ba
M_{n}^{\gamma,\ell}
\ea
where $M$ can stand for $A$ (all gluons), $B$ (one quark-pair), $C$ (two quark pairs of distinct flavour), $D$ (interference terms for two quark pairs of identical flavour), $n$ denotes the number of gluons in the process and $\ell$ the number of loops. The case of two quark pairs of the same flavour can be decomposed into distinct-flavour matrix elements ($C$-type) and interference-only contributions ($D$-type). In the list of momentum arguments, the two outermost momenta denote the primary quark pair.
A secondary quark pair is enclosed by semi-colons in the momentum list.
In the case of matrix elements containing two quark pairs it is also necessary to distinguish which quark line the exchanged electroweak boson couples to. Sub-leading orders in colour are indicated by $\tilde{M}$, and
closed fermion loop contributions by $\hat{M}$.

The different contributions to the cross section at
leading order are given in Tab.~\ref{tab:LO}. The antiquark-initiated processes are implemented explicitly in the \NNLOJET program but considered only implicitly in this
paper as they are closely related to the quark-initiated processes by relabelling.
 \begin{table}
\centering
\begin{tabular}{|c|c|c|c|}
\hline
 level & sub-process & factor & notation \\
\hline
 \multirow{2}{*}{LO} & \multirow{1}{*}{$q\gamma\to qg$} &$ {\cal{N}}_{q}^{LO}$ & $B_{1}^{\gamma,0}(\hat{1},3,2)$\\
 \cline{2-4}
 & \multirow{1}{*}{$g\gamma\to q\bar{q}$} &$ {\cal{N}}_{g}^{LO}$ &$B_{1}^{\gamma,0}(2,\hat{1},3)$ \\
 \cline{1-4}
\hline
\end{tabular}
\caption{Summary of the scattering channels and matrix elements contributing at leading order. The argument with a caret denotes
the initial-state parton, and the lepton momenta are omitted from the argument list.}
\label{tab:LO}
\end{table}

To simplify the equations in this paper we absorb all additional powers of the coupling relative to leading-order into overall factors suitable for the correction being considered. We purposefully do not absorb additional colour factors into these factors so it is always clear which colour factor we are considering. To this end we define the following NLO factors for the real~(R) and virtual~(V) contributions,
\ba
{\cal{N}}_{i}^{R}&=&{\cal{N}}_{i}^{LO}\bigg(\frac{\alpha_{s}}{2\pi}\bigg)\frac{\bar{C}(\e)}{C(\e)},\\
{\cal{N}}_{i}^{V}&=&{\cal{N}}_{i}^{LO}\bigg(\frac{\alpha_{s}}{2\pi}\bigg)\bar{C}(\e)=C(\e)~{\cal{N}}_{i}^{R} ,
\ea
which are used to define the contributions to the cross section at NLO summarised in Tabs.~\ref{tab:NLOq} and~\ref{tab:NLOg}.
 \begin{table}
\centering
\begin{tabular}{|c|c|c|c|}
\hline
 level & sub-process & factor & notation \\
\hline
 \multirow{4}{*}{R} & \multirow{2}{*}{$q\gamma\to qgg$} &$ {\cal{N}}_{q}^{R}\cdot{N}/{2!}$ & $B_{2}^{\gamma,0}(\hat{1},i,j,2)$\\
 \cline{3-4}
 & $ $ & $ {\cal{N}}_{q}^{R}\cdot (-1)/(2!N)$ & $\tilde{B}_{2}^{\gamma,0}(\hat{1},3,4,2)$ \\
\cline{2-4}
 & \multirow{2}{*}{$q\gamma\to qq\bar{q}$} &$ {\cal{N}}_{q}^{R}$ & $C_{0}^{\gamma,0}(\hat{1};3,4;2)$ \\
 \cline{3-4}
 &  & $  {\cal{N}}_{q}^{R}\cdot (-1)/N$ & $D_{0}^{\gamma,0}(\hat{1},2;4,3)$ \\
 \cline{1-4}
 \multirow{3}{*}{V} & \multirow{2}{*}{$q\gamma\to qg$} & $ {\cal{N}}_{q}^{V}\cdot{N}$ & $B_{1}^{\gamma,1}(\hat{1},3,2)$ \\
   \cline{3-4}
 &  & $ {\cal{N}}_{q}^{V}\cdot(-1)/{N}$ & $\tilde{B}_{1}^{\gamma,1}(\hat{1},3,2)$  \\
     \cline{2-4}
 &  \multirow{1}{*}{$q\gamma\to qg^{\circ}$} & ${\cal{N}}_{q}^{V}\cdot\nf$ & $\hat{B}_{1}^{\gamma,1}(\hat{1},3,2)$ \\
   \cline{1-4}
\hline
\end{tabular}
\caption{Summary of the quark-channel partonic sub-processes and colour factors for the NLO calculation. Non-numeric arguments
denote summation of gluon permutations. e.g. the matrix element $B_{2}^{\gamma,0}(\hat{1},i,j,2)$ includes a summation $\{i,j\}\in P(3,4)$
where $P$ denotes the set of permutations. Each $^{\circ}$ denotes a closed quark loop.}
\label{tab:NLOq}
\end{table}

 \begin{table}
\centering
\begin{tabular}{|c|c|c|c|}
\hline
 level & sub-process & factor & notation \\
\hline
 \multirow{2}{*}{R} & \multirow{2}{*}{$g\gamma\to q\bar{q}g$} &$ {\cal{N}}_{g}^{R}\cdot{N}$ & $B_{2}^{\gamma,0}(3,{i},j,4)$\\
 \cline{3-4}
 & $ $ & $ {\cal{N}}_{g}^{R}\cdot (-1)/N$ & $\tilde{B}_{2}^{\gamma,0}({3},\hat{1},2,4)$ \\
\cline{1-4}
 \multirow{3}{*}{V} & \multirow{2}{*}{$g\gamma\to q\bar{q}$} & $ {\cal{N}}_{g}^{V}\cdot{N}$ & $B_{1}^{\gamma,1}({2},\hat{1},3)$ \\
   \cline{3-4}
 &  & $ {\cal{N}}_{g}^{V}\cdot(-1)/{N}$ & $\tilde{B}_{1}^{\gamma,1}({2},\hat{1},3)$  \\
     \cline{2-4}
 &  \multirow{1}{*}{$g\gamma\to q\bar{q}^{\circ}$} & ${\cal{N}}_{g}^{V}\cdot\nf$ & $\hat{B}_{1}^{\gamma,1}({2},\hat{1},3)$ \\
   \cline{1-4}
\hline
\end{tabular}
\caption{Summary of the gluon-channel partonic sub-processes and colour factors for the NLO calculation.
The matrix element $B_{2}^{\gamma,0}(3,i,j,4)$ includes a summation $\{i,j\}\in P(\hat{1},2)$, where $\hat{1}$ is the initial-state gluon.
Each $^{\circ}$ denotes a closed quark loop.}
\label{tab:NLOg}
\end{table}

At NNLO we define the set of overall factors,
\ba
{\cal{N}}_{i}^{RR}&=&{\cal{N}}_{i}^{LO}\bigg(\frac{\alpha_{s}}{2\pi}\bigg)^{2}\frac{\bar{C}(\e)^{2}}{C(\e)^{2}},\\
{\cal{N}}_{i}^{RV}&=&{\cal{N}}_{i}^{LO}\bigg(\frac{\alpha_{s}}{2\pi}\bigg)^{2}\frac{\bar{C}(\e)^{2}}{C(\e)}=C(\e)~{\cal{N}}_{i}^{RR},\\
{\cal{N}}_{i}^{VV}&=&{\cal{N}}_{i}^{LO}\bigg(\frac{\alpha_{s}}{2\pi}\bigg)^{2}\bar{C}(\e)^{2}=C(\e)~{\cal{N}}_{i}^{RV}=C(\e)^{2}~{\cal{N}}_{i}^{RR},
\ea
which are used to consistently define the various contributions to the NNLO cross section in Tabs.~\ref{tab:channelsq} and~\ref{tab:channelsg}.
\begin{table}
\centering
\begin{tabular}{|c|c|c|c|}
\hline
 level & sub-process & factor & notation \\
\hline
 \multirow{7}{*}{RR} & \multirow{3}{*}{$q\gamma\to qggg$} &$ {\cal{N}}_{q}^{RR}\cdot{N^{2}}/{3!}$ & $B_{3}^{\gamma,0}(\hat{1},i,j,k,2)$\\
 \cline{3-4}
 & $ $ & $ {\cal{N}}_{q}^{RR}\cdot (-1)/3!$ & $\tilde{B}_{3}^{\gamma,0}(\hat{1},i,j,k,2)$ \\
\cline{3-4}
 &  & $ {\cal{N}}_{q}^{RR}\cdot 1/3!\cdot\big(\frac{N^2+1}{N^2}\big)$ &$ \tilde{\tilde{B}}_{3}^{\gamma,0}(\hat{1},3,4,5,2)$ \\
  \cline{2-4}
 & \multirow{4}{*}{$q\gamma\to qq\bar{q}g$} &$ {\cal{N}}_{q}^{RR}\cdot{N}$ & $C_{1}^{\gamma,0}(\hat{1},5;4,3;2)$ \\
 \cline{3-4}
 & &$ {\cal{N}}_{q}^{RR}\cdot{(-1)/N}$ & $\tilde{C}_{1}^{\gamma,0}(\hat{1},5;3,4;2)$ \\
 \cline{3-4}
 &  & $  {\cal{N}}_{q}^{RR}\cdot (-1)/2$ & $D_{1}^{\gamma,0}(\hat{1},5,2;4,3)$ \\
 \cline{3-4}
 &  & $  {\cal{N}}_{q}^{RR}\cdot 1/(2N^2)$ & $\tilde{D}_{1}^{\gamma,0}(\hat{1},5,2;4,3)$ \\
  \cline{1-4}
 \multirow{10}{*}{RV} & \multirow{3}{*}{$q\gamma\to qgg$} & $ {\cal{N}}_{q}^{RV}\cdot{N^{2}}/{2!}$ & $B_{2}^{\gamma,1}(\hat{1},i,j,2)$ \\
   \cline{3-4}
 &  & $ {\cal{N}}_{q}^{RV}\cdot(-1)/{2!}$ & $\tilde{B}_{2}^{\gamma,1}(\hat{1},i,j,2)$  \\
    \cline{3-4}
  &  & $ {\cal{N}}_{q}^{RV}\cdot1/(2!N^2)$ & $\tilde{\tilde{B}}_{2}^{\gamma,1}(\hat{1},3,4,2)$ \\
    \cline{2-4}
 &  \multirow{2}{*}{$q\gamma\to qgg^{\circ}$} & ${\cal{N}}_{q}^{RV}\cdot N\nf/2!$ & $\hat{B}_{2}^{\gamma,1}(\hat{1},i,j,2)$ \\
  \cline{3-4}
  &  & $ {\cal{N}}_{q}^{RV}\cdot(-\nf)/(2!N)$ & $\hat{\tilde{B}}_{2}^{\gamma,1}(\hat{1},3,4,2)$ \\
   \cline{2-4}
 &  \multirow{4}{*}{$q\gamma\to qq\bar{q}$} & ${\cal{N}}_{q}^{RV}\cdot N$ & $C_{0}^{\gamma,1}(\hat{1};3,4;2)$ \\
  \cline{3-4}
  &  & $ {\cal{N}}_{q}^{RV}\cdot(-1)/N$ & ${\tilde{C}}_{0}^{\gamma,1}(\hat{1};3,4;2)$  \\
  \cline{3-4}
  &  & $ {\cal{N}}_{q}^{RV}\cdot 1/2$ & $D_{0}^{\gamma,1}(\hat{1},2;4,3)$ \\
  \cline{3-4}
  &  & $ {\cal{N}}_{q}^{RV}\cdot(1)/(2N^2)$ & $\tilde{D}_{0}^{\gamma,1}(\hat{1},2;4,3)$  \\
  \cline{2-4}
 &  $q\gamma\to qq\bar{q}^{\circ}$& ${\cal{N}}_{q}^{RV}\cdot \nf$ & $\hat{C}_{0}^{\gamma,1}(\hat{1};3,4;2)$ \\
  \cline{3-4}
  &  & $ {\cal{N}}_{q}^{RV}\cdot(\nf)/(2N)$ & $\hat{D}_{0}^{\gamma,1}(\hat{1},2;4,3)$  \\
   \cline{1-4}
 \multirow{6}{*}{VV} & \multirow{3}{*}{$q\gamma\to qg$} & ${\cal{N}}_{q}^{VV}\cdot N^2$ & $B_{1}^{\gamma,2}(\hat{1},3,2)$ \\
  \cline{3-4}
 &  & ${\cal{N}}_{q}^{VV}\cdot 1$ & $\tilde{B}_{1}^{\gamma,2}(\hat{1},3,2)$  \\
  \cline{3-4}
 &  & ${\cal{N}}_{q}^{VV}\cdot 1/N^2$ & $\tilde{\tilde{B}}_{1}^{\gamma,2}(\hat{1},3,2)$ \\
  \cline{2-4}
 &  \multirow{2}{*}{$q\gamma\to qg^{\circ}$} & ${\cal{N}}_{q}^{VV}\cdot N\nf$ & $\hat{B}_{1}^{\gamma,2}(\hat{1},3,2)$ \\
  \cline{3-4}
 & & ${\cal{N}}_{q}^{VV}\cdot \nf/N$ & $\hat{\tilde{B}}_{1}^{\gamma,2}(\hat{1},3,2)$  \\
  \cline{2-4}
 &  $q\gamma\to qg^{\circ\circ}$ & ${\cal{N}}_{q}^{VV}\cdot \nf^2$ & $\hat{\hat{B}}_{1}^{\gamma,2}(\hat{1},3,2)$ \\
\hline
\end{tabular}
\caption{Summary of the quark-channel partonic sub-processes and colour factors for the NNLO calculation and the notation used for the colour-stripped squared matrix elements. Non-numeric arguments denote summation over permutations of partons, e.g. $B_{3}^{\gamma,0}(\hat{1},i,j,k,2)$ is summed over the six permutations of the labels $\{i,j,k\}\in P(3,4,5)$. The lepton momentum labels are suppressed. Each $^{\circ}$ denotes a closed quark loop.}
\label{tab:channelsq}
\end{table}
\begin{table}[!t]
\begin{tabular}{|c|c|c|c|}
\hline
 level & sub-process & factor & notation\\
\hline
 \multirow{7}{*}{RR} & \multirow{3}{*}{$g\gamma\to q\bar{q}gg$} &$ {\cal{N}}_{g}^{RR}\cdot{N^{2}}/{2!}$ & $B_{3}^{\gamma,0}(4,i,j,k,5)$\\
 \cline{3-4}
 & $ $ & $ {\cal{N}}_{g}^{RR}\cdot (-1)/2!$ & $\tilde{B}_{3}^{\gamma,0}(4,i,j,k,5)$\\
\cline{3-4}
 &  & $ {\cal{N}}_{g}^{RR}\cdot 1/2!\cdot\big(\frac{N^2+1}{N^2}\big)$ &$ \tilde{\tilde{B}}_{3}^{\gamma,0}(4,\hat{1},2,3,5)$\\
  \cline{2-4}
 & \multirow{4}{*}{$g\gamma\to q\bar{q}q\bar{q}$} &$ {\cal{N}}_{g}^{RR}\cdot{2N}$ & $C_{1}^{\gamma,0}(3,\hat{1};5,4;2)$\\
 \cline{3-4}
 & &$ {\cal{N}}_{g}^{RR}\cdot{(-1)/2N}$ & $\tilde{C}_{1}^{\gamma,0}(3,\hat{1};5,4;2)$\\
 \cline{3-4}
 &  & $  {\cal{N}}_{g}^{RR}\cdot (-1)/4$ & $D_{1}^{\gamma,0}(3,\hat{1},2;4,5)$\\
 \cline{3-4}
 &  & $  {\cal{N}}_{g}^{RR}\cdot 1/(4N^2)$ & $\tilde{D}_{1}^{\gamma,0}(3,\hat{1},2;4,5)$\\
  \cline{1-4}
 \multirow{5}{*}{RV} & \multirow{3}{*}{$g\gamma\to q\bar{q}g$} & $ {\cal{N}}_{g}^{RV}\cdot N^{2}$ & $B_{2}^{\gamma,1}(3,i,j,4)$\\
   \cline{3-4}
 &  & $ {\cal{N}}_{g}^{RV}\cdot(-1)$ & $\tilde{B}_{2}^{\gamma,1}(3,i,j,4)$ \\
    \cline{3-4}
  &  & $ {\cal{N}}_{g}^{RV}\cdot1/N^2$ & $\tilde{\tilde{B}}_{2}^{\gamma,1}(3,\hat{1},2,4)$ \\
    \cline{2-4}
 & \multirow{2}{*}{$g\gamma\to q\bar{q}g^{\circ}$} & ${\cal{N}}_{g}^{RV}\cdot N\nf$ & $\hat{B}_{2}^{\gamma,1}(3,\hat{1},2,4)$\\
  \cline{3-4}
  &  & $ {\cal{N}}_{g}^{RV}\cdot(-\nf)/N$ & $\hat{\tilde{B}}_{2}^{\gamma,1}(3,\hat{1},2,4)$ \\
   \cline{1-4}
 \multirow{5}{*}{VV} & \multirow{3}{*}{$g\gamma\to q\bar{q}$} & ${\cal{N}}_{g}^{VV}\cdot N^2$ & $B_{1}^{\gamma,2}(2,\hat{1},3)$ \\
  \cline{3-4}
 &  & ${\cal{N}}_{g}^{VV}\cdot 1$ & $\tilde{B}_{1}^{\gamma,2}(2,\hat{1},3)$ \\
  \cline{3-4}
 &  & ${\cal{N}}_{g}^{VV}\cdot 1/N^2$ & $\tilde{\tilde{B}}_{1}^{\gamma,2}(2,\hat{1},3)$ \\
  \cline{2-4}
 &  \multirow{2}{*}{$g\gamma\to q\bar{q}^{\circ}$} & ${\cal{N}}_{g}^{VV}\cdot N\nf$ & $\hat{B}_{1}^{\gamma,2}(2,\hat{1},3)$ \\
  \cline{3-4}
 & & ${\cal{N}}_{g}^{VV}\cdot \nf/N$ & $\hat{\tilde{B}}_{1}^{\gamma,2}(2,\hat{1},3)$ \\
  \cline{2-4}
 &  $g\gamma\to q\bar{q}^{\circ\circ}$ & ${\cal{N}}_{g}^{VV}\cdot \nf^2$ & $\hat{\hat{B}}_{1}^{\gamma,2}(2,\hat{1},3)$ \\
\hline
\end{tabular}
\caption{Summary of the gluon-channel partonic sub-processes and colour factors for the NNLO calculation. Non-numeric arguments
denote summation over the set of gluons, which may include the initial-state gluon. e.g. the five-parton matrix element  $B_{3}^{\gamma,0}(4,i,j,k,5)$
represents a sum over the permutations $\{i,j,k\}\in P(\hat{1},2,3)$. The four-parton matrix element $B_{2}^{\gamma,1}(3,i,j,4)$
represents a sum over $\{i,j\}\in P(\hat{1},2)$. Each $^{\circ}$ denotes a closed quark loop.}
\label{tab:channelsg}
\end{table}

Occasionally, in the subtraction terms listed in App.~\ref{app:nlosub} and App.~\ref{app:nnlosub}, we retain some information about the coupling of the
vector boson to the quark line. In the case of a matrix element containing two distinct flavour quark lines (of flavours $q$ and $Q$) in an unresolved limit where two of the quarks
become collinear to form a gluon, it is often necessary to retain the information about which quark line remains. The reduced two-quark matrix element in the subtraction term
is then denoted by either of the two terms,
\ba
B_{n,q}^{\gamma,\ell},&&B_{n,Q}^{\gamma,\ell}
\ea
to distinguish the different flavoured quark lines. Similarly it is sometimes necessary to use a matrix element which has been symmetrised over the final-state
quark pair's momenta; we denote these matrix elements like so,
\ba
\bar{B}_{n}^{\gamma,\ell}(1_{q},i_{g},\cdots,j_{g},2_{\bar{q}})&=&\sum_{P(1_{q},2_{\bar{q}})}{B}_{n}^{\gamma,\ell}(1_{q},i_{g},\cdots,j_{g},2_{\bar{q}}).
\ea

Aside from the colour factors given in Tabs.~\ref{tab:LO}--\ref{tab:channelsg} there are also contributions carrying the charge-weighted sum over flavours
\ba
N_{F,\gamma}=\frac{(\sum_{q}e_{q})^2}{\sum_{q}e_{q}^{2}}  .
\ea
These occur in multi-quark interferences between amplitudes where the vector boson couples to lines of different quark flavour or 1-loop
matrix elements where the vector boson couples to the closed quark loop. The contributions coming from this overall factor are finite
and only amount to a negligible correction to the cross section and are not included in our calculation.

\subsection{Application of the antenna subtraction method}

With the notation defined in Sec.~\ref{sec:structure} we can write the NNLO correction to the cross section, with incoming parton species $a$, in the form,
\ba
\ds_{a}^{NNLO}&=&\int_{\Phi_{n+2}}\ds_{a}^{RR}\nn\\
&+&\int_{\Phi_{n+1}}\bigg(\ds_{a}^{RV}+\ds_{a}^{MF1}\bigg)\nn\\
&+&\int_{\Phi_{n\phantom{+1}}}\bigg(\ds_{a}^{VV}+\ds_{a}^{MF2}\bigg) ,
\ea
where $\ds_{a}^{RR},\ \ds_{a}^{RV}\ \ds_{a}^{VV}$ are the contributions from
tree-level, one- and two-loop squared matrix elements. They are decomposed into individual quark-initiated and gluon-initiated
sub-processes in Tabs.~\ref{tab:channelsq} and~\ref{tab:channelsg}.
The $\ds_{a}^{MF1,2}$ are the
 counter-terms arising from the mass-factorisation of the physical PDFs.

It is well known that each of these contributions contains singularities which render them individually ill-defined: $\ds_{a}^{RR}$ contains single- and double-unresolved divergences when integrated over the $(n+2)$-parton phase space. $\ds_{a}^{RV}+\ds_{a}^{MF1}$ contains explicit poles as deep as ${\cal{O}}(\e^{-2})$
 in the dimensional regularisation parameter $\e$ as well as single-unresolved divergences when integrated over the $(n+1)$-parton phase space. $\ds_{a}^{VV}+\ds_{a}^{MF2}$ contains explicit poles as deep as ${\cal{O}}(\e^{-4})$.

To ensure the cancellation of all explicit poles in $\e$ and render each phase space integral finite, one constructs three local subtraction terms and recasts the cross section
in the form,
\ba
\ds_{a}^{NNLO}&=&\int_{\Phi_{n+2}}\bigg(\ds_{a}^{RR}-\ds_{a}^{S}\bigg)\nn\\
&+&\int_{\Phi_{n+1}}\bigg(\ds_{a}^{RV}-\ds_{a}^{T}\bigg)\nn\\
&+&\int_{\Phi_{n\phantom{+1}}}\bigg(\ds_{a}^{VV}-\ds_{a}^{U}\bigg).
\label{eq:antsub}
\ea
Various methods for the construction of these subtraction terms at NNLO have been put forward in the
literature~\cite{secdec,qtsub,ourant,currie,stripper,njettiness,trocsanyi}. We use the antenna subtraction method at
NNLO~\cite{ourant,currie}. In this method, antenna functions are introduced to account for all unresolved partonic
radiation off a pair of two radiator partons~\cite{ant}. The antenna subtraction term
for each colour-ordered squared matrix element is then constructed as a sum
(over all pairs of radiators) of products of antenna functions with
reduced squared matrix elements of lower multiplicity.
 Each antenna function is integrated analytically over the sub-space of
 unresolved parton momenta in $d=4-2\e$ dimensions and added back into
a lower multiplicity final-state contribution as a Laurent series in $\e$, thus ensuring the analytic cancellation of all explicit poles.
The following subsections provide a detailed description of the construction of the antenna
subtraction terms in Eq.~\eqref{eq:antsub}.

\subsubsection{Phase-space mappings}

To ensure that each antenna subtraction term converges to the unresolved limits of the full
squared matrix element that it is intended for, the momenta in its reduced squared matrix element must be
appropriate to this limit. However, by its construction, the antenna subtraction term is defined over the full phase space.
To ensure the correct factorisation behaviour in all unresolved limits consequently requires a
factorisation of the full phase space into a reduced phase space and a so-called antenna phase space
corresponding to the unresolved radiation in each antenna subtraction term. The momenta
of the reduced phase space are constructed from the original momenta by a phase space mapping, which is typically a
non-linear transformation.
 This ensures the correct factorisation behaviour of the subtraction term in the unresolved limits, but also
the factorisation of the phase space itself into disjoint spaces so that the analytic integration
can be achieved over each antenna sub-space.

The form of the mapping depends on the kinematic configuration of the two hard radiators involved in the
antenna function, final-final~(FF)~\cite{ourant}, initial-final~(IF)~\cite{nloant}
or initial-initial~(II)~\cite{nloant}. Since DIS processes have only one parton in the initial state, only the
FF and IF mappings can occur.
In each configuration there exists a mapping appropriate
to single-unresolved ($n+1\to n$ particles) and double-unresolved ($n+2\to n$ particles) configurations.
The subtraction terms are then constructed such that the antenna function depends on potentially unresolved unmapped momenta, whereas the reduced-multiplicity squared
matrix element (and any event selection criteria) depends only on the mapped momenta.

\subsubsection{Subtraction for double-real~(RR) contributions}
\label{sec:rrsub}

The double-real matrix elements contain both single- and double-unresolved divergences and the subtraction term therefore has to be able to regulate both types of limits.
The single-unresolved limits of the matrix element, $M_{n+2}^{\gamma,0}$, when considering at least $n$ jets, are dealt with by subtraction terms of the general form,
\ba
X_{3}^{0}(\{p_{n+2}\})M_{n+1}^{\gamma,0}(\{\tilde{p}_{n+1}\})J_{n}^{(n+1)}(\{\tilde{p}_{n+1}\})  , \label{eq:dsiga}
\ea
where $X_{3}^{0}$ denotes an antenna function, which may in general be in the FF, IF or II configuration and is a function of three of the momenta from
the ($n+2$)-parton momentum set, $\{p_{n+2}\}$. $J_n^{(n+1)}$ represents the jet function which builds at least $n$ jets from $n+1$ partons. The
reduced matrix element and the jet algorithm depend only on the ($n+1$)-parton mapped momentum set, $\{\tilde{p}_{n+1}\}$.

Colour-connected double-unresolved limits in the matrix elements are regulated using a four-parton antenna function and a double-unresolved
momentum mapping. The subtraction terms for these limits take the form,
\ba
X_{4}^{0}(\{p_{n+2}\})M_{n}^{\gamma,0}(\{\tilde{p}_{n}\})J_{n}^{(n)}(\{\tilde{p}_{n}\}).\label{eq:dsigb}
\ea
Colour-disconnected and almost-colour-connected double-unresolved divergences (where unresolved partons share at most one hard neighbour in the colour ordering) are
removed using subtraction terms with iterated single-unresolved mappings,
\ba
X_{3}^{0}(\{p_{n+2}\})X_{3}^{0}(\{\tilde{p}_{n+1}\})M_{n}^{\gamma,0}(\{\tilde{\tilde{p}}_{n}\})J_{n}^{(n)}(\{\tilde{\tilde{p}}_{n}\}).\label{eq:dsigc}
\ea

These three types of subtraction term are sufficient to remove all potential divergences in the ($n+2$)-parton phase-space integral and allow it
to be computed in $d=4$ using numerical techniques.
The explicit forms of the RR subtraction terms for quark- and gluon-initiated
channels are given in App.~\ref{app:nnlosubRRq} and App.~\ref{app:nnlosubRRg}, respectively.

\subsubsection{Subtraction for real-virtual~(RV) contributions}
\label{sec:rvsub}

The real-virtual matrix elements contain only single-unresolved divergences but unlike the double-real matrix elements, they also contain explicit
poles in $\e$ which have to be cancelled. The explicit poles of the ($n+1$)-parton real-virtual matrix elements, $M_{n+1}^{\gamma,1}$, are cancelled analytically by a subtraction term
of the form,
\ba
{J}_{2,ij}^{1,kl}(\{p_{n+1}\},\e)M_{n+1}^{\gamma,0}(\{p_{n+1}\})J_{n}^{(n+1)}(\{p_{n+1}\}),\label{eq:dsigta}
\ea
where the integrated real radiation
function ${J}_{2,ij}^{1,kl}$ is constructed~\cite{currie} from the integrated form of the antenna function introduced in Eq.~(\ref{eq:dsiga}) and possibly also the kernels
coming from the mass factorisation counterterms. The labels $i$ and $j$ denote the flavours of the hard partons and belong to the set $\{QQ,QG,GQ,GG\}$ for quark-antiquark,
quark-gluon, gluon-quark and gluon-gluon integrated real radiation functions. The labels $k$ and $l$ denote the kinematical configuration and belong to the set $\{FF, IF, II\}$ for final-final, initial-final and initial-initial configurations respectively. In DIS kinematics only FF and IF configurations are present.

The single-unresolved divergences of the real-virtual matrix elements are regulated using two distinct subtraction terms, built to reflect the factorisation behaviour of
one-loop amplitudes. The first term reflects the divergence associated with a tree-like singular function and a one-loop reduced matrix element and has the form,
\ba
X_{3}^{0}(\{p_{n+1}\})M_{n}^{\gamma,1}(\{\tilde{p}_{n}\})J_{n}^{(n)}(\{\tilde{p}_{n}\})\label{eq:dsigtb1}.
\ea
The second type of subtraction term reflects the divergence associated with a one-loop singular function and tree-like reduced matrix element and so involves
a one-loop antenna function,
\ba
X_{3}^{1}(\{p_{n+1}\})M_{n}^{\gamma,0}(\{\tilde{p}_{n}\})J_{n}^{(n)}(\{\tilde{p}_{n}\})\label{eq:dsigtb2}.
\ea

The subtraction terms in Eq.~\eqref{eq:dsigta} remove all explicit poles from the matrix elements yet contain their own single unresolved divergences, whereas the subtraction terms in
Eqs.~\eqref{eq:dsigtb1}, \eqref{eq:dsigtb2} remove all single-unresolved divergences from the matrix elements yet contain their own explicit poles in $\e$. To cancel these remaining
poles and divergences we introduce subtraction terms of the form,
\ba
{J}_{2,ij}^{1,kl}(\{p_{n+1}\},\e)X_{3}^{0}(\{p_{n+1}\})M_{n}^{\gamma,0}(\{\tilde{p}_{n}\})J_{n}^{(n)}(\{\tilde{p}_{n}\})
\ea
and
\ba
X_{3}^{0}(\{p_{n+1}\}){J}_{2,ij}^{1,kl}(\{\tilde{p}_{n}\},\e)M_{n}^{\gamma,0}(\{\tilde{p}_{n}\})J_{n}^{(n)}(\{\tilde{p}_{n}\}) ,
\ea
which render the RV contribution finite and integrable using numerical techniques.
 It should be noted that the reduced
matrix elements are required only through to their finite terms in the $\epsilon$-expansion.
 The explicit form for the RV subtraction terms in quark- and gluon-initiated channels
are given in App.~\ref{app:nnlosubRVq} and App.~\ref{app:nnlosubRVg}, respectively.

\subsubsection{Subtraction for double-virtual~(VV) contributions}

The double virtual matrix elements, $M_{n}^{\gamma,2}$, are integrated over the $n$-parton phase space and exhibit no IR divergences; however they do contain explicit poles
in $\e$ which are cancelled analytically using subtraction terms of the form,
\ba
{J}_{2,ij}^{1,kl}(\{p_{n}\},\e)M_{n}^{\gamma,1}(\{p_{n}\})J_{n}^{(n)}(\{p_{n}\}),\label{eq:dsigua}
\ea
\ba
{J}_{2,ij}^{1,kl}(\{p_{n}\},\e)\otimes{J}_{2,i'j'}^{1,k'l'}(\{p_{n}\},\e)M_{n}^{\gamma,0}(\{p_{n}\})J_{n}^{(n)}(\{p_{n}\}),\label{eq:dsigub}
\ea
and
\ba
{J}_{2,ij}^{2,kl}(\{p_{n}\},\e)M_{n}^{\gamma,0}(\{p_{n}\})J_{n}^{(n)}(\{p_{n}\}).\label{eq:dsiguc}
\ea
The function ${J}_{2,ij}^{2,kl}$ contains the integrated form of the four-parton antenna from Eq.~(\ref{eq:dsigb}), the integrated one-loop antenna from
Eq.~(\ref{eq:dsigtb2}) and several other elements of lower complexity
(essentially products of three-parton antenna functions) which conspire to cancel the poles of the two-loop matrix elements.
The integrated antenna functions up to NNLO were derived in Ref.~\cite{ourant} for FF kinematics, in Ref.~\cite{daleo} for
IF kinematics and in Ref.~\cite{monni} for II kinematics. In DIS processes, only FF and IF kinematics contribute to the
subtraction terms.

Once the RR and RV subtraction terms have been defined, the form of the VV subtraction terms is completely fixed and so for brevity we do not present these
explicitly in an appendix.

\subsection{Initial-state identity changing collinear limits}
\label{sec:IClimits}

\begin{figure}[t]
\centering
\includegraphics[width=8cm]{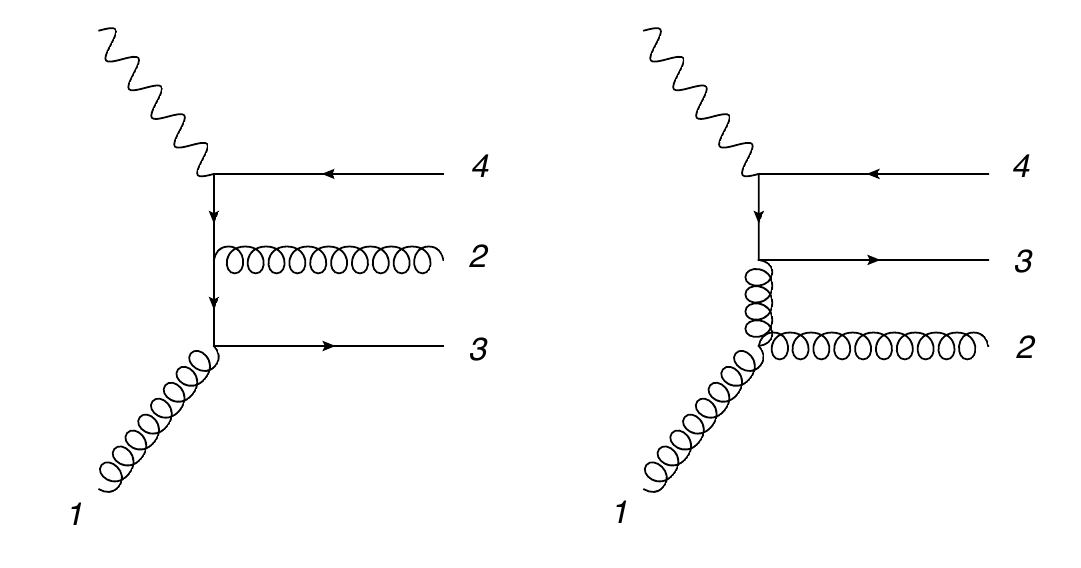}
\caption{A subset of Feynman diagrams contributing to the matrix element $B_{2}^{\gamma,0}(3_{q},\hat{1}_{g},2_{g},4_{\bar{q}})$ with indicative momentum labels for the external legs. }
\label{fig:feyn1}
\end{figure}

In scattering processes with initial-state partons, there is an additional complication that arises when attempting to capture the singularity structure of the matrix element where the identity of the initial-state parton changes due to a collinear limit with a final-state parton: for example an initial-state gluon becoming collinear with a final-state quark, combining to form an initial-state anti-quark. In this situation the matrix element will factorise, as usual, into a product of a splitting function and a reduced multiplicity matrix element, e.g.\ consider the gluon-initiated two-quark--two-gluon DIS sub-process (of which some indicative Feynman diagrams are shown in Fig.~\ref{fig:feyn1}) in a collinear limit between the initial-state gluon and the final-state quark,
\ba
B_{2}^{\gamma,0}(3_{q},\hat{1}_{g},2_{g},4_{\bar{q}})\stackrel{3||\hat{1}}{\longrightarrow}\frac{1}{s_{13}}P_{q\bar{q}\leftarrow G}~B_{1}^{\gamma,0}(\hat{\bar{1}}_{q},2_{g},4_{\bar{q}})  , \label{eq:me1}
\ea
In this limit, the parton entering the reduced matrix element is an anti-quark (denoted by $q$ with an initial-state momentum), and its
momentum is given by:
\ba
p_{\hat{\bar{1}}}&=&p_{\hat{1}}-p_{3}.
\ea

We would like to regulate this kind of divergent limit using an appropriate subtraction term constructed from antennae and reduced matrix elements. A candidate subtraction term would be,
\ba
D_{3}^{0}(3,\hat{1},2)~B_{1}^{\gamma,0}(\hat{\bar{1}}_{q},\wt{(23)}_{g},4_{\bar{q}}).\label{eq:ex1}
\ea
This subtraction term does indeed mimic the matrix element faithfully in the $1_{g}||3_{q}$ limit, which can be seen by considering:
\ba
D_{3}^{0}(3,\hat{1},2)&\stackrel{3||\hat{1}}{\longrightarrow}&\frac{1}{s_{13}}P_{q\bar{q}\leftarrow G} ,\\
p_{\wt{(23)}}&\stackrel{3||\hat{1}}{\longrightarrow}&p_{2} ,\\
p_{\hat{\bar{1}}}&\stackrel{3||\hat{1}}{\longrightarrow}&p_{\hat{1}}-p_{3}.
\ea
An attractive feature of the antenna subtraction method is that one antenna function regulates several unresolved limits. This is possible because of two facts:
\begin{enumerate}
\item the antenna function tends to the appropriate universal function in \emph{each} unresolved limit.
\item the composite momenta for the reduced matrix element tend to the appropriate resolved momenta in \emph{each} unresolved limit.
\end{enumerate}
The former is ensured by a proper definition of the antenna functions, the latter achieved by an appropriate phase-space mapping which interpolates between several unresolved limits.

For soft or collinear limits between final-state partons this procedure is unproblematic, as it is for collinear limits between initial-state partons and final-state partons where the species of the initial-state parton is unchanged (for example, a final-state gluon becoming collinear with an initial-state quark); we refer to these limits as \emph{identity-preserving}~(IP) limits. We refer to the initial-final collinear limits that change the species of the initial-state parton as \emph{identity-changing}~(IC) limits.
When a subtraction term, such as that in Eq.~(\ref{eq:ex1}), attempts to regulate both IP and IC limits then an
inconsistency arises due to the fact that we have to make a choice about the momentum
assignment in the reduced matrix element. This determines in turn the momentum crossing in which
the reduced matrix element is evaluated.
 For example, consider the matrix element in Eq.~(\ref{eq:me1}) and the subtraction term in Eq.~(\ref{eq:ex1}).
 The matrix element contains, among others, divergences in the $\hat{1}_{g}||3_{q}$ and $\hat{1}_{g}||2_{g}$ initial-final collinear limits. As we have seen, the subtraction term in Eq.~(\ref{eq:ex1}) successfully regulates the $\hat{1}_{g}||3_{q}$ limit. In the $\hat{1}_{g}||2_{g}$ the matrix element factorises according to,
\ba
B_{2}^{\gamma,0}(3_{q},\hat{1}_{g},2_{g},4_{\bar{q}})\stackrel{\hat{1}||2}{\longrightarrow}\frac{1}{s_{12}}P_{gg\leftarrow G}~B_{1}^{\gamma,0}(3_{q},\hat{\bar{1}}_{g},4_{\bar{q}})\label{eq:me2},
\ea
where now,
\ba
p_{\hat{\bar{1}}}&=&p_{\hat{1}}-p_{2}.
\ea
The subtraction term in Eq.~(\ref{eq:ex1}) also contains a divergence in the $\hat{1}_{g}||2_{g}$ limit,
\ba
D_{3}^{0}(3,\hat{1},2)~B_{1}^{\gamma,0}(\hat{\bar{1}}_{q},\wt{(23)}_{g},4_{\bar{q}})\stackrel{\hat{1}||2}{\longrightarrow}\frac{1}{s_{12}}P_{gg\leftarrow G}~B_{1}^{\gamma,0}(\hat{\bar{1}}_{q},3_{g},4_{\bar{q}}) .
\ea
This clearly does not regulate the matrix element in this limit because the reduced matrix element in the subtraction term has the quark as its initial-state parton. This choice is appropriate for the $\hat{1}_{g}||3_{q}$ limit, but inappropriate for the $\hat{1}_{g}||2_{g}$ which requires an initial-state gluon in the reduced matrix element. The phase-space mapping can interpolate between resolved momenta in different unresolved limits but cannot interpolate between different crossings of the reduced matrix element.

There are several solutions to this problem of mixed IP and IC limits. A simple solution is to partial fraction the antenna such that it only contains either the IP or IC singularities, thus fixing the choice of crossing for the reduced matrix element. Following this approach we would divide the subtraction term into two terms, one handling each unresolved limit, e.g.
\ba
&+&d_{3,g\to q}^{0}(3,\hat{1},2)~B_{1}^{\gamma,0}(\hat{\bar{1}}_{q},\wt{(23)}_{g},4_{\bar{q}})\nn\\
&+&d_{3,g\to g}^{0}(3,\hat{1},2)~B_{1}^{\gamma,0}(\wt{(23)}_{q},\hat{\bar{1}}_{g},4_{\bar{q}}),
\ea
where the first line only contains the $\hat{1}_{g}||3_{q}$ limit and the second line only contains the $\hat{1}_{g}||2_{g}$ limit and they carry different reduced matrix elements. The disadvantage of this approach is that it relies on our ability to partial fraction the antenna function \emph{and} successfully integrate the sub-antenna analytically. The three-parton antennae relevant for NLO calculations have been partial fractioned in this way and successfully integrated~\cite{nloant} so all ingredients are available to apply this method at NLO. However, the problem of mixed IP and IC limits is also present in three-parton one-loop and four-parton antennae used for NNLO subtraction terms. Partial fractioning these more complicated antennae is a non-trivial task and integrating the resulting sub-antennae analytically poses a substantial technical challenge.

An alternative to this method, which requires no additional analytic integrations, is to construct subtraction terms which remove all IP limits from the matrix element. The antennae in these subtraction terms will generally contain IP and IC collinear limits, but the reduced matrix element is chosen to be appropriate for IP limits only. The strategy is then to use quark-antiquark antennae to remove the spurious IC limits from these subtraction terms. Once this is achieved, the genuine IC limits of the matrix element can be removed using the same set of quark-antiquark antennae but now carrying the crossing of the reduced matrix element appropriate to the genuine IC limit. For example, consider the combination of matrix elements,
\ba
B_{2}^{\gamma,0}(3_{q},\hat{1}_{g},2_{g},4_{\bar{q}})+B_{2}^{\gamma,0}(3_{q},2_{g},\hat{1}_{g},4_{\bar{q}}).\label{eq:me3}
\ea
These contain the initial-final IP collinear limit $\hat{1}_{g}||2_{g}$, initial-final IC limits $\hat{1}_{g}||3_{q}$, $\hat{1}_{g}||4_{\bar{q}}$, as well as several final-state soft and collinear limits which are irrelevant to this discussion. We can regulate these matrix elements with the subtraction term,
\ba
+D_{3}^{0}(3,\hat{1},2)~B_{1}^{\gamma,0}(\wt{(23)}_{q},\hat{\bar{1}}_{g},4_{\bar{q}})\nn\\
+D_{3}^{0}(4,\hat{1},2)~B_{1}^{\gamma,0}(3_{q},\hat{\bar{1}}_{g},\wt{(24)}_{\bar{q}})\nn\\
-A_{3}^{0}(3,\hat{1},4)~B_{1}^{\gamma,0}(2_{q},\hat{\bar{1}}_{g},\wt{(34)}_{\bar{q}})\nn\\
+A_{3}^{0}(3,\hat{1},4)~B_{1}^{\gamma,0}(\hat{\bar{1}}_{q},2_{g},\wt{(34)}_{\bar{q}}).\label{eq:ex3}
\ea
The first three lines carry crossings of the reduced matrix element appropriate for IP limits (gluon initiated). The fourth line carries the crossing of the reduced matrix element relevant to the IC limits (quark initiated). The first two lines regulate the $\hat{1}_{g}||2_{g}$ IP limits but contain spurious IC $\hat{1}_{g}||3_{q}$ and $\hat{1}_{g}||4_{\bar{q}}$ limits. These spurious singularities are then removed by the quark-antiquark antenna in the third line, which by necessity also carries a gluon-initiated matrix element. The fourth line is similar to the third insofar as it captures IC limits, but now uses a quark-initiated reduced matrix element to regulate the genuine limit. Taken together, the block of terms in Eq.~(\ref{eq:ex3}) successfully disentangles and removes all IP and IC divergences for the block of matrix elements in Eq.~(\ref{eq:me3}).

The advantage of this method is that although it is more cumbersome at NLO, as in the example considered here, the same method can be applied to combinations of three-parton one-loop antennae and four-parton antennae to disentangle the NNLO IP and IC limits. This method requires no new integrals and all ingredients are readily available such that the problem is reduced to constructing an appropriate subtraction term. For example, the subset of RR matrix elements for gluon-initiated DIS,
\ba
\sum_{P(i,j)}\bigg[B_{3}^{\gamma,0}(3_{q},\hat{1}_{g},i_{g},j_{g},4_{\bar{q}})+B_{3}^{\gamma,0}(3_{q},i_{g},\hat{1}_{g},j_{g},4_{\bar{q}})+B_{3}^{\gamma,0}(3_{q},i_{g},j_{g},\hat{1}_{g},4_{\bar{q}})\bigg]  , \label{eq:ICNNLO}
\ea
contain the IC triple-collinear limits $(\hat{1}_{g}||3_{q}||i_{g})$, $(\hat{1}_{g}||3_{q}||j_{g})$, $(\hat{1}_{g}||4_{\bar{q}}||i_{g})$, $(\hat{1}_{g}||4_{\bar{q}}||j_{g})$, as well as the
IP triple-collinear limits $(\hat{1}_{g}||i_{g}||j_{g})$, $(3_{q}||i_{g}||j_{g})$, $(4_{\bar{q}}||i_{g}||j_{g})$. Following the strategy outlined above at NLO, we construct a subtraction term that regulates the IP limits using full
antennae,
\ba
&&+D_{4}^{0}(3,\hat{1},i,j)~B_{1}^{\gamma,0}(\wt{(3ij)}_{q},\hat{\bar{1}}_{g},4_{\bar{q}})\nn\\
&&+D_{4}^{0}(3,i,\hat{1},j)~B_{1}^{\gamma,0}(\wt{(3ij)}_{q},\hat{\bar{1}}_{g},4_{\bar{q}})\nn\\
&&+D_{4}^{0}(3,i,j,\hat{1})~B_{1}^{\gamma,0}(\wt{(3ij)}_{q},\hat{\bar{1}}_{g},4_{\bar{q}})\nn\\\nn\\
&&+D_{4}^{0}(4,\hat{1},i,j)~B_{1}^{\gamma,0}(3_{q},\hat{\bar{1}}_{g},\wt{(4ij)}_{\bar{q}})\nn\\
&&+D_{4}^{0}(4,i,\hat{1},j)~B_{1}^{\gamma,0}(3_{q},\hat{\bar{1}}_{g},\wt{(4ij)}_{\bar{q}})\nn\\
&&+D_{4}^{0}(4,i,j,\hat{1})~B_{1}^{\gamma,0}(3_{q},\hat{\bar{1}}_{g},\wt{(4ij)}_{\bar{q}}).
\ea
Of course, these subtraction terms contain spurious IC limits which can be removed using quark-antiquark four parton antennae,
\ba
&&-A_{4}^{0}(3,\hat{1},i,4)B_{1}^{\gamma,0}(\wt{(3i4)}_{q},\hat{\bar{1}}_{g},j_{\bar{q}})\nn\\
&&-A_{4}^{0}(3,i,\hat{1},4)B_{1}^{\gamma,0}(\wt{(3i4)}_{q},\hat{\bar{1}}_{g},j_{\bar{q}})\nn\\
&&-A_{4}^{0}(3,\hat{1},j,4)B_{1}^{\gamma,0}(\wt{(3j4)}_{q},\hat{\bar{1}}_{g},i_{\bar{q}})\nn\\
&&-A_{4}^{0}(3,j,\hat{1},4)B_{1}^{\gamma,0}(\wt{(3j4)}_{q},\hat{\bar{1}}_{g},i_{\bar{q}}).
\ea
The genuine IC triple-collinear limits of the matrix elements in Eq.~\eqref{eq:ICNNLO} can then be regulated by the same set of quark-antiquark antenna, but now
carrying a quark- or antiquark-initiated reduced matrix element,%
\footnote{In the case of the axial coupling of the vector boson to the quark line ($Z$-exchange contribution), the line reversal symmetry of
the partial amplitude (implicitly used in this example) is broken and so appropriate symmetrization over the final-state quarks should be employed to reinstate it.}
\ba
&&+A_{4}^{0}(3,\hat{1},i,4)B_{1}^{\gamma,0}(\hat{\bar{1}}_{q},j_{g},\wt{(3i4)}_{\bar{q}})\nn\\
&&+A_{4}^{0}(3,i,\hat{1},4)B_{1}^{\gamma,0}(\hat{\bar{1}}_{q},j_{g},\wt{(3i4)}_{\bar{q}})\nn\\
&&+A_{4}^{0}(3,\hat{1},j,4)B_{1}^{\gamma,0}(\hat{\bar{1}}_{q},i_{g},\wt{(3j4)}_{\bar{q}})\nn\\
&&+A_{4}^{0}(3,j,\hat{1},4)B_{1}^{\gamma,0}(\hat{\bar{1}}_{q},i_{g},\wt{(3j4)}_{\bar{q}}).
\ea
The alternative would be to partial fraction the $D_{4}^{0}$ antenna into sub-antennae which is significantly more difficult than the partial fractioning of the $D_{3}^{0}$ antenna
due to the presence of overlapping single- and double-unresolved limits. The integration of such a sub-antenna would also present significant technical challenge and additional
master integrals.

In practice, we employ a mixture of approaches to the subtraction terms constructed for this calculation. We use the known NLO partial fractioned antennae where possible to simplify the subtraction term and use a carefully constructed combination of antennae for the genuine NNLO IC limits such as the triple collinear or one-loop single collinear IC limits. In principle all partial fractioned antennae could be eliminated from the subtraction terms, but this would only serve to make the construction more baroque so we take an optimal approach and use partial fractioned NLO antennae where possible.

\subsection{Implementation into \NNLOJET and validation}

The \NNLOJET code is a parton-level event generator that provides the framework for the implementation
of jet production processes to NNLO accuracy, using the antenna subtraction method. Besides
containing the event generator infrastructure (phase-space integration, event handling and analysis routines), it
supplies the unintegrated and integrated antenna functions and the phase-space mappings relevant to all kinematical
situations. The multi-dimensional phase space integration is performed using the adaptive Monte Carlo
integrator VEGAS~\cite{vegas}. To avoid numerical instabilities in the matrix elements, all parton-parton
Mandelstam invariants are required to beeither $y_0=10^{-7}$ or $y_0=3\cdot10^{-7}$ times the electron-parton centre-of-mass energy squared. We
have validated that our results are insensitive to variations of $y_0$ by one order of magnitude in each direction.
The implementation of processes in the \NNLOJET framework requires the availability of the matrix
elements for all RR, RV and VV processes, and the construction of the antenna subtraction terms. \NNLOJET provides
testing routines to verify the point-wise convergence of the subtraction,
as documented for example in Ref.~\cite{joao}. Processes included in \NNLOJET  up to now
are $Z$ and $Z+j$ production~\cite{ourzj}, $H$ and $H+j$ production~\cite{ourhj} as well as
 di-jet production in hadron-hadron collisions~\cite{2jnew}.

 Our \NNLOJET implementation~\cite{disprl} of jet production in DIS uses the same matrix elements~\cite{meRR,meRV,meVV}
 as were used in $Z+j$ production~\cite{ourzj}, however in different kinematical crossings.
While the phase space for $Z+j$ production corresponds to a single crossing region of the matrix elements,
three different crossing regions are required~\cite{graudenz,disme} to describe each DIS process, depending on the
relative size of $Q^2$ compared to the parton-parton invariants. The matrix element contributions are summarised in
Tabs.~\ref{tab:channelsq} and~\ref{tab:channelsg}. The antenna subtraction terms relevant to each of these contributions
are collected in the Appendix.

 To validate our
 implementation of the tree-level and one-loop matrix elements,
 we compared the NLO predictions for di-jet and tri-jet production against SHERPA~\cite{sherpa} (in
 DIS kinematics~\cite{hoeche}), which
uses OpenLoops~\cite{openloops} to automatically generate the  one-loop contributions at NLO. The antenna
subtraction is then verified by testing the convergence of subtraction terms and matrix elements in all unresolved limits
and by the infrared pole cancellation between the integrated subtraction terms
and the two-loop matrix elements.

\subsection{Scale dependence of the NNLO cross section}
\label{sec:scales}
The coupling constant renormalisation and mass factorisation procedures are performed at
a renormalisation scale $\mu_r$ and factorisation scale $\mu_f$, where the
strong coupling constant  $\alpha_s(\mu_r)$ and parton
distribution functions $f_i(x,\mu_f)$ are evaluated, respectively.

Beyond leading order, the fixed-order contributions to the
 parton-level cross sections contain an explicit dependence on $\mu_r$ and $\mu_f$,
which compensates the dominant scale dependence from the coupling constant and parton distributions at the
previous orders. These scale-dependent terms can be predicted from the renormalisation group equations
(our normalisation conventions are summarised in the appendix of Ref.~\cite{ourhj}).
Starting from the evaluation of the
DIS di-jet
cross section at fixed default values $\mu_f=\mu_r=\mu_0$ (which can be chosen dynamically event-by-event):
\begin{eqnarray}
\sigma(\mu_0,\mu_0,\alpha_s(\mu_0))
&=& \left(\frac{\alpha_s(\mu_0)}{2\pi}\right) \hat{\sigma}_{i}^{(0)}\otimes f_i(\mu_0)
+\left(\frac{\alpha_s(\mu_0)}{2\pi}\right)^{2} \hat{\sigma}_{i}^{(1)}\otimes f_i(\mu_0)  \nonumber \\ &&
+\left(\frac{\alpha_s(\mu_0)}{2\pi}\right)^{3} \hat{\sigma}_{i}^{(2)}\otimes f_i(\mu_0)
+{\cal O} (\alpha_s^{4}) \,,
\end{eqnarray}
the full scale dependence of the cross section can be predicted in terms of
\begin{equation}
L_r = \log \left(\frac{\mu_r^2}{\mu_0^2}\right) ,\qquad
L_f = \log \left(\frac{\mu_f^2}{\mu_0^2}\right).
\end{equation}

It takes the form:
\ba
\lefteqn{\sigma(\mu_r,\mu_f,\alpha_s(\mu_r),L_r,L_f)
=}\nonumber \\ &&
\left(\frac{\alpha_s(\mu_r)}{2\pi}\right) \hat{\sigma}_{i}^{(0)}\otimes f_i(\mu_f)
+\left(\frac{\alpha_s(\mu_r)}{2\pi}\right)^{2} \hat{\sigma}_{i}^{(1)}\otimes f_i(\mu_f)
  \nonumber \\ &&
 +L_r \, \left(\frac{\alpha_s(\mu_r)}{2\pi}\right)^{2} \, \beta_0 \,
 \hat{\sigma}_{i}^{(0)}\otimes f_i(\mu_f)
 +L_f\, \left(\frac{\alpha_s(\mu_r)}{2\pi}\right)^{2}
\Big[
- \hat{\sigma}_{i}^{(0)}\otimes  \left( P_{ik}^{(0)}\otimes f_k(\mu_f)\right)
\Big] \nonumber \\ &&
+\left(\frac{\alpha_s(\mu_r)}{2\pi}\right)^{3} \hat{\sigma}_{i}^{(2)}\otimes f_i(\mu_f)
+L_r \, \left(\frac{\alpha_s(\mu_r)}{2\pi}\right)^{3} \left(  2 \, \beta_0 \,
 \hat{\sigma}_{i}^{(1)} + \, \beta_1 \, \hat{\sigma}_{i}^{(0)}\right)
 \otimes f_i(\mu_f)  \nonumber \\ &&
 +L_r^2 \, \left(\frac{\alpha_s(\mu_r)}{2\pi}\right)^{3}\, \beta^2_0 \,
 \hat{\sigma}_{i}^{(0)}\otimes f_i(\mu_f)
 \nonumber \\ &&
+L_f\, \left(\frac{\alpha_s(\mu_r)}{2\pi}\right)^{3}
\Big[
-  \hat{\sigma}_{i}^{(1)}\otimes  \left( P_{ik}^{(0)}\otimes f_k(\mu_f)\right)
-  \hat{\sigma}_{i}^{(0)}\otimes  \left( P_{ik}^{(1)}\otimes f_k(\mu_f)\right)
\Big] \nonumber \\ &&
 +L_f^2\, \left(\frac{\alpha_s(\mu_r)}{2\pi}\right)^{3}
\left[
 \frac{1}{2} \hat{\sigma}_{i}^{(0)}\otimes \left(
P_{ik}^{(0)}\otimes P_{kl}^{(0)}\otimes f_l(\mu_f)\right)
+\frac{1}{2} \beta_0 \, \hat{\sigma}_{i}^{(0)}\otimes  \left( P_{ik}^{(0)}\otimes f_k(\mu_f)\right)
  \right] \nonumber \\ &&
+L_f L_r\, \left(\frac{\alpha_s(\mu_r)}{2\pi}\right)^{3}
\Big[
- 2\, \beta_0\, \hat{\sigma}_{i}^{(0)}\otimes  \left( P_{ik}^{(0)}\otimes f_k(\mu_f)\right) \Big]
+{\cal O} (\alpha_s^{4}) \,.
\label{eq:fullscale}
\ea
In the above expressions, summation over the parton indices is implicit.
Equation (\ref{eq:fullscale}) can be used to compute the cross section at multiples of an initially chosen scale, and was
also employed to perform detailed validations of our implementation of the different contributions to the NNLO corrections
to di-jet production.

\section{Inclusive jet production}
\label{sec:1j}

Inclusive jet production in deep inelastic scattering has been widely studied by the
H1~\cite{h1a12,h1b1,h1d12,h1e12,h1highq2,h1lowq2}
and ZEUS~\cite{zeusb1,zeusc12,zeusd1} experiments at DESY HERA.
The jet measurements are preformed in the Breit frame, where the transverse momentum requirement on the jet ensures
the existence of a partonic recoil, even if only a single jet is reconstructed in the kinematical acceptance.
Jets are reconstructed using the $k_T$-algorithm with $R_0=1$.

In this section, we use the kinematic criteria  (see Tab.~\ref{tab:incjkin} below)
used in the final H1 measurements~\cite{h1highq2,h1lowq2} to discuss several generic features of
the NNLO corrections to inclusive jet production, followed by an in-depth comparison of the newly derived
predictions to the H1 data~\cite{h1highq2,h1lowq2}.

\subsection{Structure of the inclusive jet production cross section at NNLO}

\begin{figure}
  \centering
  \includegraphics[width=0.4\linewidth]{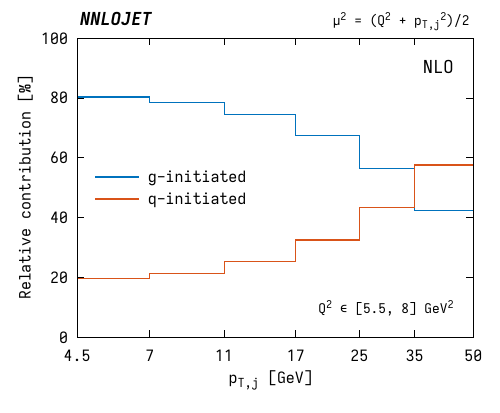}
  \includegraphics[width=0.4\linewidth]{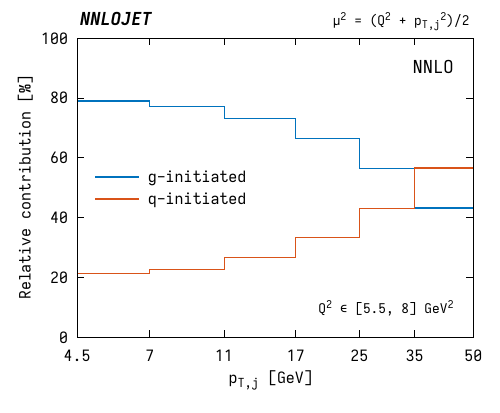}
  \caption{
    Quark- and gluon-initiated contributions to the inclusive jet transverse momentum distribution at NLO and NNLO.
  }
  \label{fig:qginit}
\end{figure}

Inclusive jet production in the Breit frame receives leading order contributions both from quark-initiated and
gluon-initiated processes. The relative magnitude of these depends
on the final-state kinematics.
Figure~\ref{fig:qginit} shows the relative contributions from quark
and gluon initial states to inclusive jet production, evaluated at NLO and NNLO for a representative
range in $Q^2$, using the cuts from the H1 low-$Q^2$ analysis (see below). We observe that at low transverse momentum, $p_T\leq 10$~GeV, inclusive jet production is mainly
gluon-initiated (to almost 80\%). With increasing $p_T$, the quark-initiated contribution becomes more and more important, reaching 50\% at around
$p_T\approx 30$~GeV, and further increasing towards higher $p_T$.
NNLO corrections affect the relative importance of the different initial states only marginally.
The overall behaviour of the parton fractions remains largely unchanged at higher $Q^2$.

\begin{figure}
  \centering
  \includegraphics[width=0.55\linewidth]{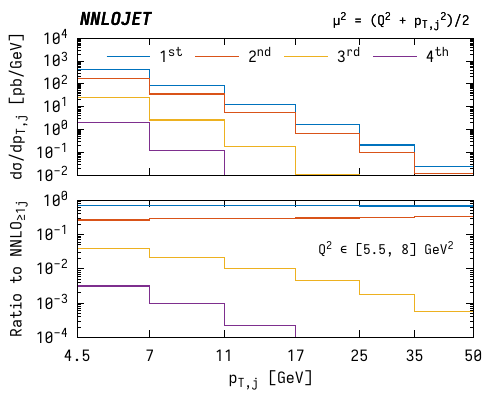}
  \caption{
    Contributions to inclusive jet production ($k_T$-algorithm with $R_0=1$) from first, second, third, and fourth jet.
  }
  \label{fig:jetmult}
\end{figure}

The inclusive jet production cross section receives contributions from different jet multiplicities. At NNLO, final states with up to four identified jets contribute.
The jets are ordered in transverse momentum. Figure~\ref{fig:jetmult} displays the contribution of the first, second, third and fourth jet
to the inclusive jet distribution at NNLO for a given bin in $Q^2$. It can be seen that the leading jet and subleading jet contribute about 70\% and
25\% to
the distribution over the full $p_T$ range. This behaviour can be understood from the fact that the jet production is measured in the Breit frame,
where the leading order process will always yield a $p_T$-symmetric two-jet final state. The jets are not necessarily both found inside the
rapidity cut, such that in some fraction of the events, only the leading jet is observed. Furthermore, real radiation from higher order corrections
can lower the transverse momentum of the second jet compared to the first one, such that the same event will enter the $p_T$
distribution at a larger value with the first jet than with the second jet. Due to the sharp decrease of the distribution with increasing $p_T$, the relative
importance of the second jet is lower than of the first jet.
The contributions from the third and fourth jet are at the level of a few per-cent and a few per-mille respectively at
low $p_T$. Their importance
decreases to higher $p_T$, which can be understood from the limited final-state phase space that is available for multi-jet production
at large transverse momenta.
\begin{table}
\begin{center}
\begin{tabular}{ccc}
H1 (high-$Q^2$) &  H1 (low-$Q^2$) & ZEUS \\ \hline
 $150<Q^2/\mbox{GeV}^2<15000$ &  $5.5<Q^2/\mbox{GeV}^2<80$ & $125<Q^2/\mbox{GeV}^2$ \\
 $0.2<y<0.7$ & $0.2<y<0.6$  & $-0.65<\cos \, \gamma_{h}<0.65$ \\
 5~GeV$<p^B_T<$50~GeV & 4.5~GeV$<p^B_T<$50~GeV & 8~GeV$<p^B_T$  \\
  $-1.0<\eta^L<2.5$ &  $-1.0<\eta^L<2.5$ & $-1.5<\eta^B<2$ \\
\hline
\end{tabular}
\end{center}
\caption{Kinematical cuts used to define the inclusive jet phase space in the measurements of H1 at high-$Q^2$~\cite{h1highq2} and low-$Q^2$~\cite{h1lowq2}, and ZEUS~\cite{zeusc12} .\label{tab:incjkin}}
\end{table}

\subsection{Comparison to HERA data}

\begin{figure}
  \begin{tabular}{r@{}l}
    \includegraphics[trim={ 0pt 24pt  0pt  0pt},clip]{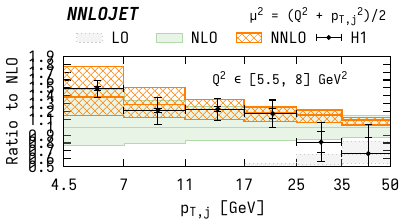} &
    \includegraphics[trim={ 9pt 24pt  0pt 24pt},clip]{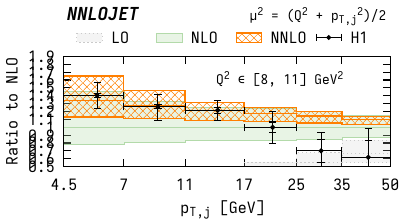} \\[0pt]
    \includegraphics[trim={ 0pt 24pt  0pt 24pt},clip]{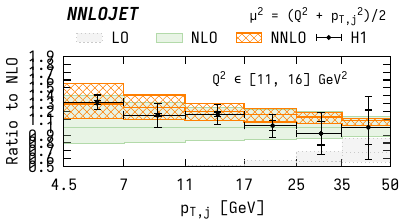} &
    \includegraphics[trim={ 9pt 24pt  0pt 24pt},clip]{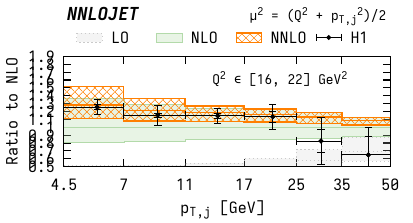} \\[0pt]
    \includegraphics[trim={ 0pt 24pt  0pt 24pt},clip]{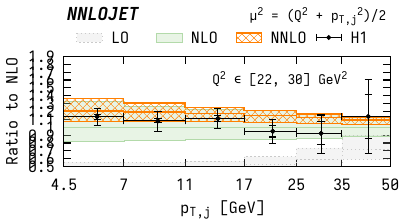} &
    \includegraphics[trim={ 9pt 24pt  0pt 24pt},clip]{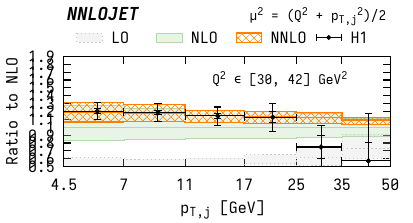} \\[0pt]
    \includegraphics[trim={ 0pt  0pt  0pt 24pt},clip]{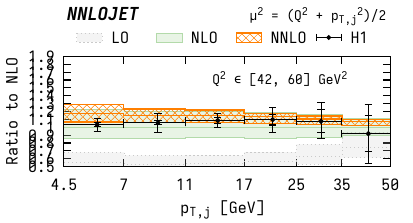} &
    \includegraphics[trim={ 9pt  0pt  0pt 24pt},clip]{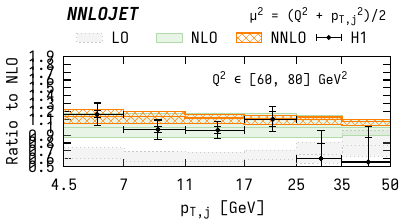}
  \end{tabular}
  \begin{tabular}{r@{}l}
    \includegraphics[trim={ 0pt 24pt  0pt  0pt},clip]{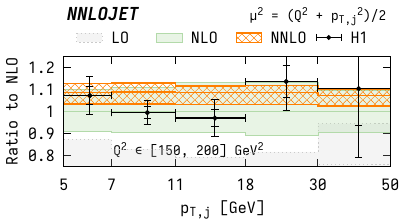} &
    \includegraphics[trim={ 9pt 24pt  0pt 24pt},clip]{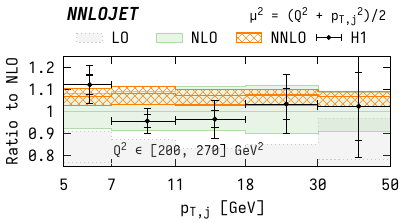} \\[0pt]
    \includegraphics[trim={ 0pt 24pt  0pt 24pt},clip]{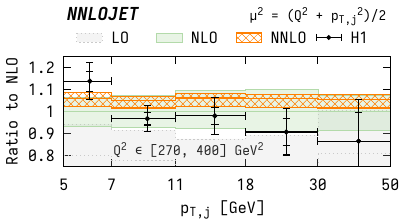} &
    \includegraphics[trim={ 9pt 24pt  0pt 24pt},clip]{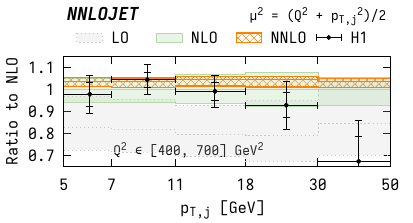} \\[0pt]
    \includegraphics[trim={ 0pt  0pt  0pt 24pt},clip]{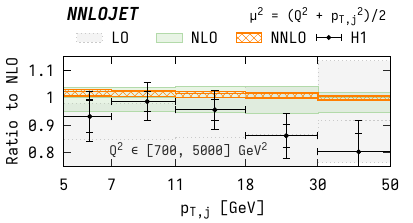} &
    \includegraphics[trim={ 9pt  0pt  0pt 24pt},clip]{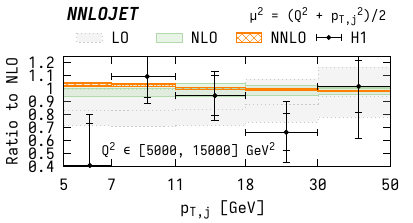}
  \end{tabular}
  \caption{
    Inclusive jet production cross section as a function of the jet transverse momentum $p_{T,B}$ in bins of $Q^2$, compared to H1 data.
  }
  \label{fig:h1inc1}
\end{figure}

\begin{figure}
  \begin{tabular}{r@{}l}
    \includegraphics[trim={ 0pt 24pt  0pt  0pt},clip]{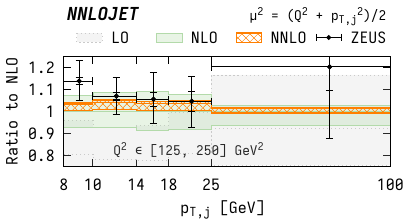} &
    \includegraphics[trim={ 9pt 24pt  0pt 24pt},clip]{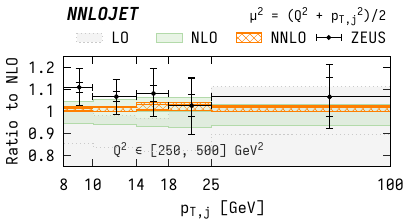} \\[0pt]
    \includegraphics[trim={ 0pt 24pt  0pt 24pt},clip]{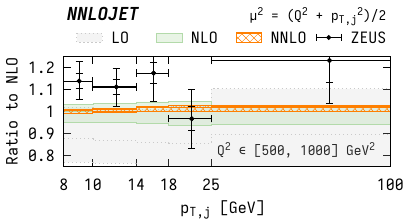} &
    \includegraphics[trim={ 9pt 24pt  0pt 24pt},clip]{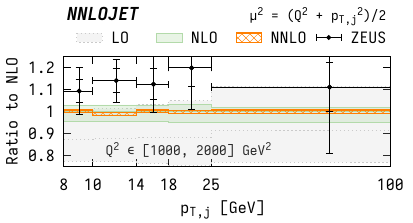} \\[0pt]
    \includegraphics[trim={ 0pt  0pt  0pt 24pt},clip]{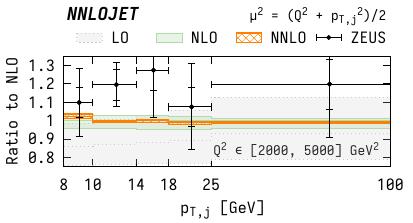} &
    \includegraphics[trim={ 9pt  0pt  0pt 24pt},clip]{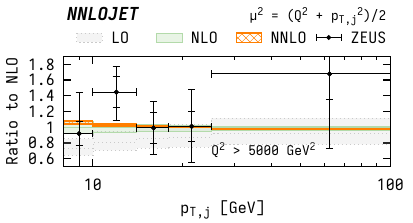}
  \end{tabular}
  \caption{
    Inclusive jet production cross section as a function of the jet transverse momentum $p_{T,B}$ in bins of $Q^2$, compared to ZEUS data.
  }
  \label{fig:ZEUSinc}
\end{figure}

Inclusive jet production in the Breit frame (using the inclusive $k_T$ algorithm with a massless $E_T$ recombination scheme)
was measured
double differentially in $Q^2$ and $p_T^B$ by the H1 experiment, which distinguishes a low-$Q^2$~\cite{h1lowq2} and a high-$Q^2$
sample~\cite{h1highq2} and by the ZEUS experiment~\cite{zeusd1}. The
event selection criteria for all three studies are summarised in Tab.~\ref{tab:incjkin}.
We compute theoretical predictions at LO, NLO and NNLO,
always using the same set of parton distribution functions (NNPDF3.0 NNLO) with $\alpha_s(M_Z)=0.118$. Our predictions use a dynamical
central scale choice $\mu^2_r=\mu^2_f=(Q^2+p_{T,B}^2)/2$, and the theory uncertainty is determined from a seven-point scale variation with rescaling factors $[1/2,2]$.
The theoretical predictions are
multiplicatively corrected for hadronization effects, $Z$-boson exchange, and QED radiative corrections using the correction tables from the experimental papers~\cite{h1highq2,h1lowq2}.
These corrections vary between 0.88 and 1.01 for the H1 low-$Q^2$ data, between 0.9 and 1.10 for the H1 high-$Q^2$ data, and between 0.83 and 1.02 for the ZEUS data.

Figure~\ref{fig:h1inc1} compares our NNLO predictions to the H1 data. We observe that the NNLO corrections are very substantial at low-$Q^2$ and
low-$p_T^B$, with an up to 40\% enhancement with respect to NLO. These large corrections are within the NLO
uncertainty band (close to the upper edge), and
result in a residual theory uncertainty of 10--20\% even at NNLO. Especially at low $Q^2$, the shape and normalisation of the theory prediction
changes significantly going from NLO to NNLO, and results in a considerably improved theoretical description of the data,
as already statistically quantified in the experimental H1 study~\cite{h1lowq2}. A  very similar pattern is also
observed for the ZEUS measurement shown in Fig.~\ref{fig:ZEUSinc}.
With increasing $Q^2$, the
size of the NNLO corrections decreases, accompanied with very small residual theoretical uncertainties (decreasing from 5\% at
$Q^2=150$~GeV$^2$ to less than 2\% at 5000~GeV$^2$). In this region, the combination of precision data with the newly derived NNLO corrections
has clearly the potential to provide important new constraints for precision QCD studies.

\section{Di-jet production}
\label{sec:2j}

In the Breit frame, di-jet production and single jet inclusive production in deep inelastic scattering are
mediated by the same Born level processes and are closely related. In contrast to single jet inclusive production, where
only the rapidity and transverse momentum of the jet can be studied,
di-jet final states allow for more kinematical observables to be reconstructed (see section~\ref{sec:kin} above).
Typically, di-jet cross sections are measured inclusively based on the two leading jets in an event, i.e.\ including events
with more than two reconstructed jets. Inclusive di-jet production was measured by the
H1~\cite{h1a12,h1c2,h1d12,h1e12,h1highq2,h1lowq2} and ZEUS~\cite{zeusa2,zeusc12,zeus2j} experiments at DESY HERA.

In this section, we adapt the event selection (see Tab.~\ref{tab:2jkin} below)
used in the final H1 measurements~\cite{h1highq2,h1lowq2} to discuss several generic features of
the NNLO corrections to inclusive jet production, followed by an in-depth comparison of our NNLO
predictions to the ZEUS~\cite{zeus2j} and H1~\cite{h1highq2,h1lowq2} data.

\begin{figure}
  \centering
  \begin{tabular}{r@{}l}
    \includegraphics[trim={ 0pt 32pt  0pt  0pt},clip,scale=.8]{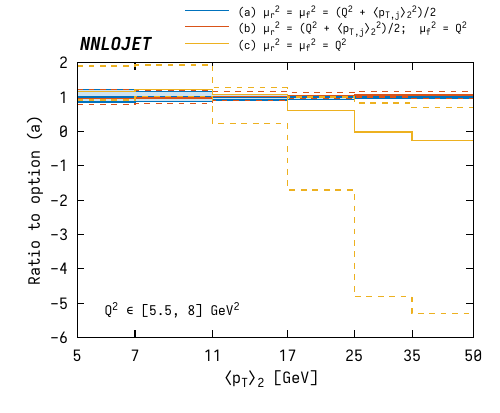} &
    \includegraphics[trim={12pt 32pt  0pt 25pt},clip,scale=.8]{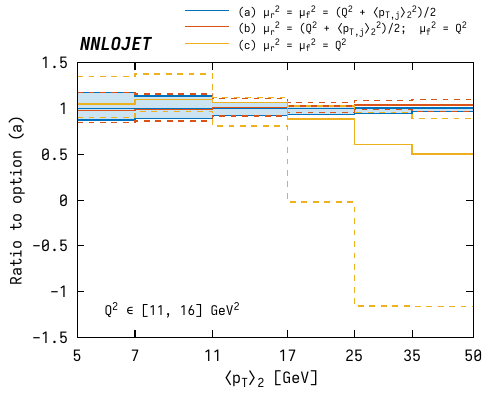} \\[0pt]
    \includegraphics[trim={ 0pt  0pt  0pt 25pt},clip,scale=.8]{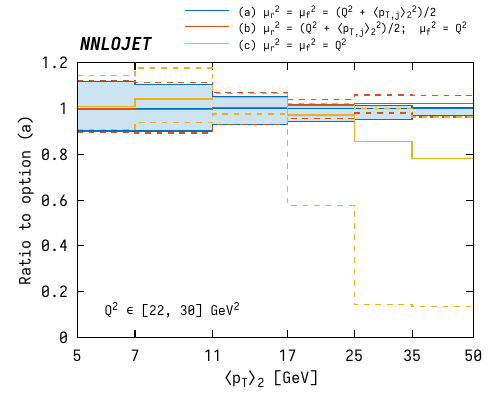} &
    \includegraphics[trim={12pt  0pt  0pt 25pt},clip,scale=.8]{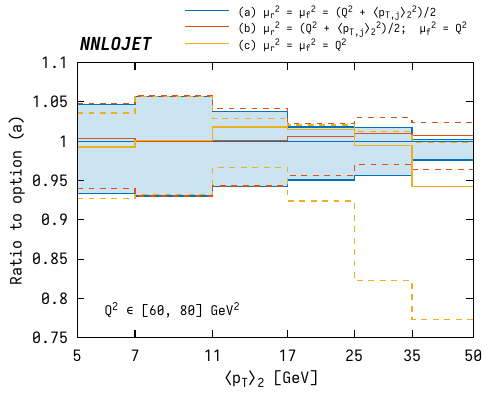}
  \end{tabular}
  \caption{
    Di-jet production cross sections for different scale settings.
  }
  \label{fig:2jscl}
\end{figure}

\subsection{Scale setting in the di-jet production cross section at NNLO}

The dependence of the NNLO cross section on the renormalisation and factorisation scales has been derived in Sec.~\ref{sec:scales}, where it can be seen that each order in
the perturbative series compensates the dominant scale dependent terms from the previous order.
Consequently, the residual scale dependence of a fixed-order prediction is commonly used to estimate the
error on the prediction resulting from missing higher order corrections.
The scale dependence of the theoretical prediction is quantified by choosing central values for $\mu_f$ and $\mu_r$, and
then evaluating the cross section for variations around these central scales, typically by a factor two.

These central scale
values should reflect the dynamics of the process. In processes involving only a single mass scale (such as for instance
inclusive deep inelastic scattering, depending only on the photon virtuality $Q^2$), the central scale choice is unambiguous
(at most up to a constant factor). For processes involving multiple physical scales, several choices are possible (and a priori
equally well motivated). The only guiding principle for choosing the central scales in this case is the occurrence of
large logarithmic terms in each order in perturbation theory, which spoil the convergence of the perturbative expansion
and indicate an inappropriate choice of the central scale.

Di-jet production in DIS depends on two scales: the photon virtuality $Q^2$ and the average transverse momentum of the
two jets $\langle p^B_{T}\rangle_2$. Using the kinematical cuts of some of the bins in the H1 low-$Q^2$ di-jet measurement
(which are discussed in detail in the following subsection) as an example, we study different choices for the central scales in
Fig.~\ref{fig:2jscl}. \begin{minipage}[h]{\textwidth}
 We compare the following three options:
\begin{eqnarray*}
\text{(a)}&& \mu_f^2= \mu_r^2 = \left(Q^2+\langle p^B_{T}\rangle_2^2\right)/2 , \\
\text{(b)}&& \mu_f^2= Q^2 , \quad \mu_r^2 = \left(Q^2+\langle p^B_{T}\rangle_2^2\right)/2 ,  \\
\text{(c)}&& \mu_f^2= \mu_r^2 = Q^2 .
\end{eqnarray*}
\end{minipage}
All three options were considered previously in comparisons of NLO predictions to the H1 and ZEUS
jet data~\cite{h1a12,h1b1,h1c2,h1d12,h1e12,h1highq2,h1lowq2,zeusa2,zeusb1,zeusc12,zeusd1,zeus2j}. Option
(a) represents the average of both hard scales in the jet production process, and is used as default throughout this paper; option (b) is used frequently in the experimental studies, with the argument that the partonic structure
of the proton ($\mu_f$) is resolved by the
virtual photon, while it is hard QCD emission ($\mu_r$) that yields the two-jet final state; finally option (c) is entirely based on
the photon virtuality to describe the hardness of the interaction.

We observe that the scale choices (a) and (b) yield similar predictions, except in the region of large transverse
momentum.
Especially at large $Q^2$, the choice (b) results in slightly higher cross section predictions, accompanied with larger scale uncertainties.
In contrast, scale choice (c) results in unphysical
predictions (negative cross sections) if applied at low $Q^2$ and large transverse momentum. These results confirm
earlier observations made at NLO~\cite{graudenz,mirkes,jetvip,nagy}. By examining the analytical
form of some of the NLO contributions~\cite{mirkes,jetvip} for scale
choice (c), the emergence of large logarithmic corrections in the
jet transverse momenta could be established. These corrections are largely compensated in the
hard coefficient functions for scale choices (a) and (b), which are clearly preferable in terms of reliability and
perturbative stability. It should be pointed out that the large logarithmic terms alone (which can be inferred
from threshold resummation~\cite{klasen}) do not properly account for the
bulk of the NNLO corrections, as observed in Ref.~\cite{h1lowq2}.
\begin{table}
\begin{center}
\begin{tabular}{ccc}
H1 (high-$Q^2$) &  H1 (low-$Q^2$) & ZEUS\\ \hline
 $150<Q^2/\mbox{GeV}^2<15000$ &  $5.5<Q^2/\mbox{GeV}^2<80$ &  $125<Q^2/\mbox{GeV}^2<20000$ \\
 $0.2<y<0.7$ & $0.2<y<0.6$ & $0.2<y<0.6$ \\
 5~GeV$<p^B_T<$50~GeV & 4~GeV$<p^B_T$  &  $E^{B}_{T} > 8$~GeV  \\
  $-1.0<\eta^L<2.5$ &  $-1.0<\eta^L<2.5$&  $-1.0<\eta^L<2.5$ \\
  $M_{12}>16$~GeV & &  $M_{12}>20$~GeV \\
\hline
\end{tabular}
\end{center}
\caption{Kinematical cuts used to define the inclusive di-jet phase space in the measurements of H1 (high-$Q^2$~\cite{h1highq2} and low-$Q^2$~\cite{h1lowq2}) and ZEUS~\cite{zeus2j}.\label{tab:2jkin}}
\end{table}

\subsection{Comparison to HERA data}

\begin{figure}
  \centering
  \includegraphics[width=.45\linewidth]{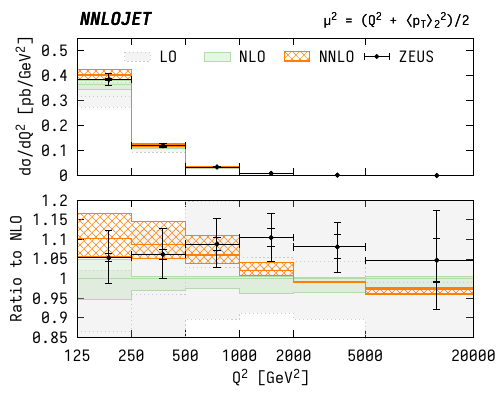}
  \includegraphics[width=.45\linewidth]{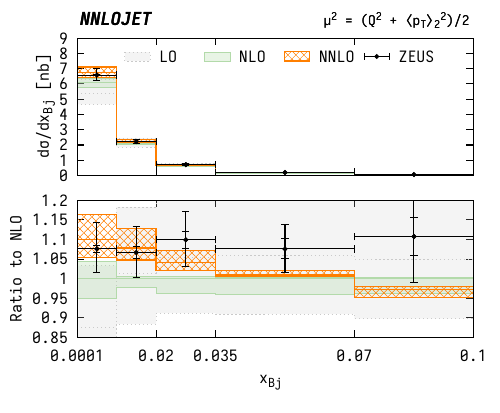}
  \caption{
    Inclusive di-jet production cross section as a function of the electron variables $Q^2$ (left) and $x$ (right), compared to ZEUS data.
  }
  \label{fig:zeusq2x}
\end{figure}

\begin{figure}
  \centering
  \includegraphics[width=.45\linewidth]{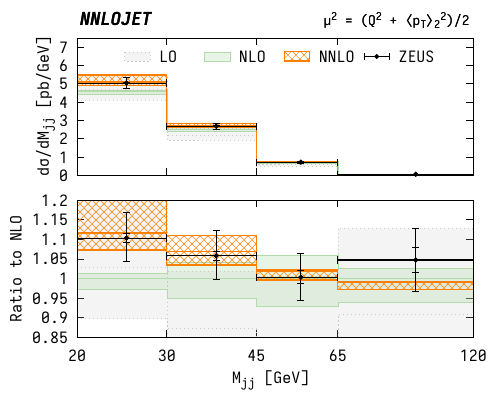}
  \includegraphics[width=.45\linewidth]{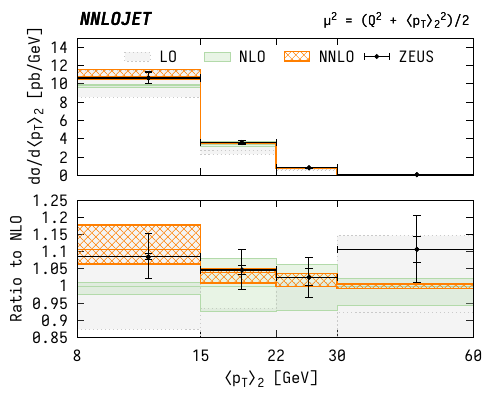}
  \caption{
    Inclusive di-jet production cross section as a function of $\langle p^B_{T}\rangle_2$ (left) and $M_{jj}$ (right), compared to ZEUS data.
  }
  \label{fig:zeusetm}
\end{figure}

Inclusive di-jet production was measured by both HERA experiments: H1~\cite{h1highq2,h1lowq2} provides
double-differential results in
$Q^2$ and $\langle p^B_{T}\rangle_2$ or $Q^2$ and $\xi_2$, using the $k_T$ and anti-$k_T$ jet algorithms in the Breit frame. ZEUS~\cite{zeus2j} uses only the
$k_T$ jet algorithm and provides single-differential results in $\bar{E}_T^B=\langle p^B_{T}\rangle_2$,
$Q^2$, $M_{jj}$, $\eta^*$, $\xi=\xi_2$
as well as double-differential results in $Q^2$ and $\xi$ or $\bar{E}_T^B$.
The kinematical cuts in the measurements are summarised in Tab.~\ref{tab:2jkin}.
We compute theoretical predictions at LO, NLO and NNLO,
always using the same set of parton distribution functions (NNPDF3.0 NNLO) with $\alpha_s(M_Z)=0.118$. Our predictions use the central
scales $\mu^2_r=\mu^2_f=(Q^2+\langle p_{T}\rangle_2^2)/2$ and the theory uncertainty is determined from a seven-point scale variation with
rescaling factors $[1/2,2]$.

\begin{figure}
  \centering
  \begin{tabular}{r@{}l}
    \includegraphics[trim={ 0pt 24pt  0pt  0pt},clip]{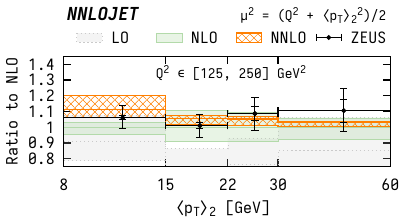} &
    \includegraphics[trim={ 9pt 24pt  0pt 24pt},clip]{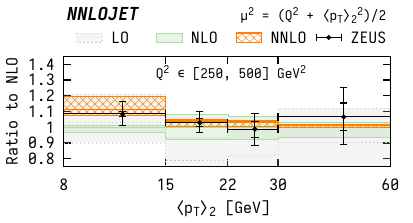} \\[0pt]
    \includegraphics[trim={ 0pt 24pt  0pt 24pt},clip]{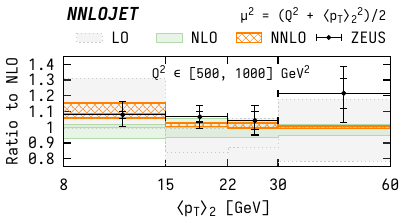} &
    \includegraphics[trim={ 9pt 24pt  0pt 24pt},clip]{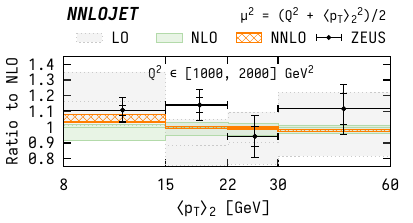} \\[0pt]
    \includegraphics[trim={ 0pt  0pt  0pt 24pt},clip]{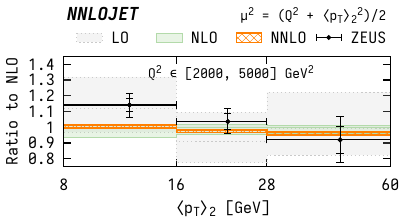} &
    \includegraphics[trim={ 9pt  0pt  0pt 24pt},clip]{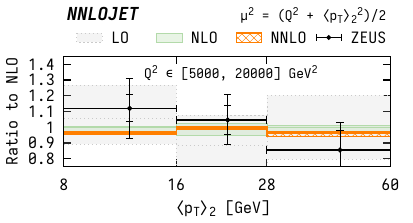}
  \end{tabular}
  \caption{
    Inclusive di-jet production cross section as a function of $\langle p^B_{T}\rangle_2$ in bins of $Q^2$, compared to ZEUS data.
  }
  \label{fig:zeusetq2}
\end{figure}

\begin{figure}
  \centering
  \includegraphics[width=.45\linewidth]{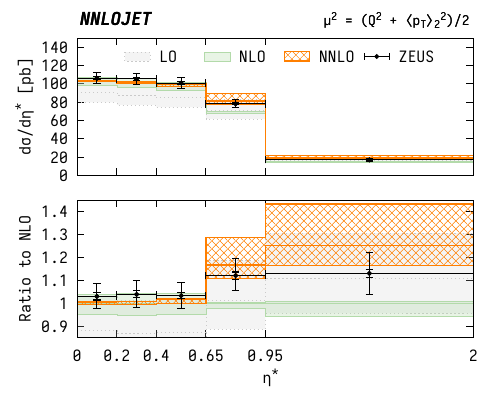}
  \includegraphics[width=.45\linewidth]{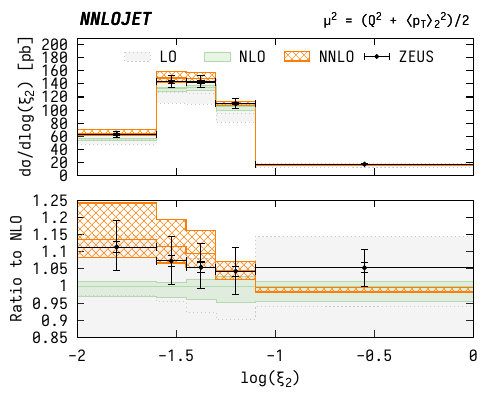}
  \caption{
    Inclusive di-jet production cross section as a function of $\eta^*$ (left) and $\log(\xi_2)$ (right), compared to ZEUS data.
  }
  \label{fig:zeusetaxi}
\end{figure}

The theoretical predictions are
multiplicatively corrected for hadronization effects, the $Z$-boson exchange contribution, and QED radiative effects using the correction tables from the respective experimental papers~\cite{h1highq2,h1lowq2,zeus2j}.
These corrections are very similar in magnitude to those in inclusive jet production.
They vary between 0.88 and 1.01 for the H1 low-$Q^2$ data, between 0.92 and 1.09 for the H1 high-$Q^2$ data, and
between 0.88 and 1.14 for the ZEUS data.

Figure~\ref{fig:zeusq2x} displays the inclusive di-jet cross section
for the ZEUS kinematics as a function of the electron variables $Q^2$ (left) and of $x$ (right). We observe the
NNLO corrections to be sizeable especially at low values of $Q^2$ or $x$, where they enhance the NLO prediction by about 10\%. In this region, the
scale dependence of the NNLO prediction is as large as at NLO, or even larger.
We note that the data is better described by the NNLO prediction than at NLO.
A similar pattern is observed in the distributions in $\langle p^B_{T}\rangle_2$
and $M_{jj}$ shown in Fig.~\ref{fig:zeusetm}, with sizeable NNLO
corrections in the lower range of the distributions. In both these distributions this range clearly correlates with the approach to the infrared limit,
as can be seen even more clearly in the double differential distribution in $\langle p^B_{T}\rangle_2$  and $Q^2$, Fig.~\ref{fig:zeusetq2}.
In this limit,  the QCD coupling constant increases and logarithmically enhanced contributions could deteriorate the convergence of the
perturbative fixed-order expansion. This issue is further aggravated by the interplay of the $M_{jj}$ cut (see Tab.~\ref{tab:2jkin}) with the
transverse momentum requirements on the final state jets, as discussed previously in Ref.~\cite{disprl},
and elaborated upon in more detail below.
The relatively small scale dependence of the NLO predictions in this range is likely an
artefact originating from a cross-over of the upper and lower edges of the scale-band, as investigated in detail for hadronic di-jet production in Ref.~\cite{2jnew}. The scale variation at NNLO therefore provides the more realistic assessment of the theoretical uncertainty.

\begin{figure}
  \centering
  \begin{tabular}{r@{}l}
    \includegraphics[trim={ 0pt 24pt  0pt  0pt},clip]{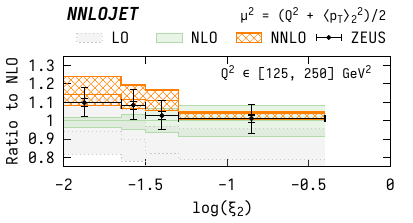} &
    \includegraphics[trim={ 9pt 24pt  0pt 24pt},clip]{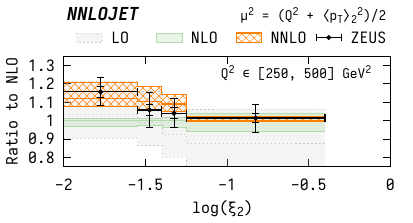} \\[0pt]
    \includegraphics[trim={ 0pt 24pt  0pt 24pt},clip]{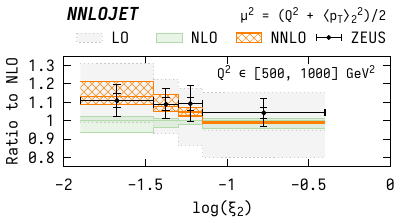} &
    \includegraphics[trim={ 9pt 24pt  0pt 24pt},clip]{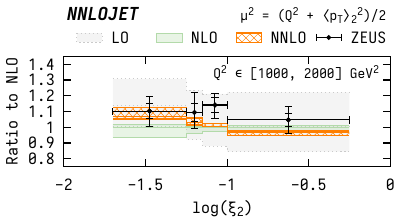} \\[0pt]
    \includegraphics[trim={ 0pt  0pt  0pt 24pt},clip]{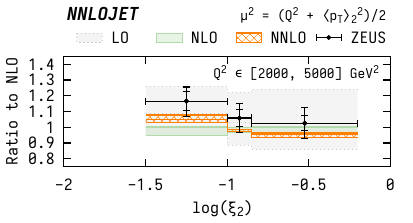} &
    \includegraphics[trim={ 9pt  0pt  0pt 24pt},clip]{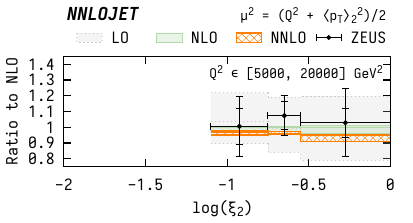}
  \end{tabular}
  \caption{
    Inclusive di-jet production cross section as a function of $\log(\xi)$ in bins of $Q^2$, compared to ZEUS data.
  }
  \label{fig:zeusxiq2}
\end{figure}

The di-jet cross section as function of $\eta^*$ and of $\log(\xi_2)$ is shown in Fig.~\ref{fig:zeusetaxi}. While good perturbative convergence is
observed in the plateau region $\eta^*<0.65$, NNLO corrections turn out to be very sizeable at higher rapidities.
The perturbative instability in this region was already pointed out and explained by the ZEUS collaboration~\cite{zeus2j}.
Even so, going from NLO to NNLO we observe a much improved description of the data in this region.
The $\log(\xi_2)$ correlates most directly with the parton distributions, indicating the importance of NNLO corrections in describing the data at momentum fractions in the medium range $0.01<\xi<0.1$.
The double-differential distribution in $\log(\xi_2)$ and $Q^2$, Fig.~\ref{fig:zeusxiq2},
further illustrates this impact, showing
a coherent pattern of sizeable NNLO corrections that are crucial in describing the data both in normalisation and shape over the full $Q^2$ range.
It will be very interesting to further study the impact
of these data in the determination of parton distributions at NNLO.

\begin{figure}
  \centering
  \begin{tabular}{r@{}l}
    \includegraphics[trim={ 0pt 24pt  0pt  0pt},clip]{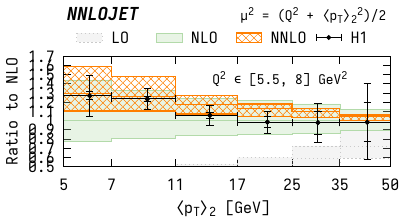} &
    \includegraphics[trim={ 9pt 24pt  0pt 24pt},clip]{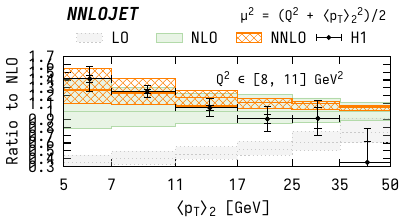} \\[0pt]
    \includegraphics[trim={ 0pt 24pt  0pt 24pt},clip]{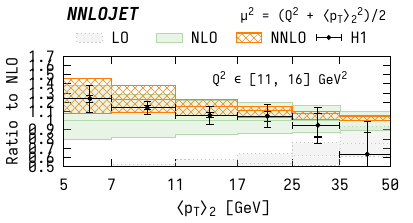} &
    \includegraphics[trim={ 9pt 24pt  0pt 24pt},clip]{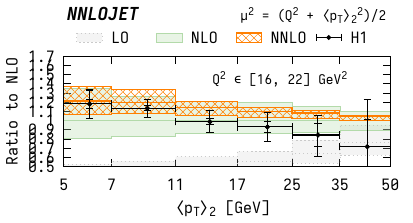} \\[0pt]
    \includegraphics[trim={ 0pt 24pt  0pt 24pt},clip]{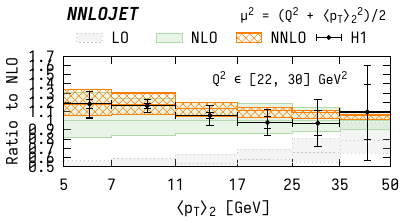} &
    \includegraphics[trim={ 9pt 24pt  0pt 24pt},clip]{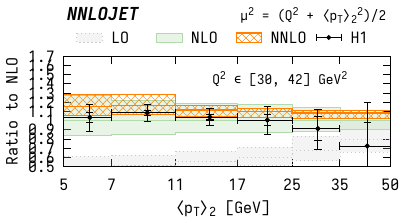} \\[0pt]
    \includegraphics[trim={ 0pt  0pt  0pt 24pt},clip]{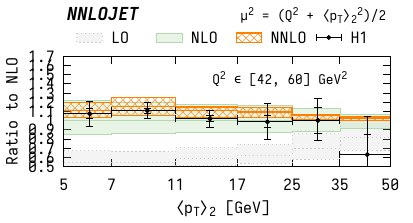} &
    \includegraphics[trim={ 9pt  0pt  0pt 24pt},clip]{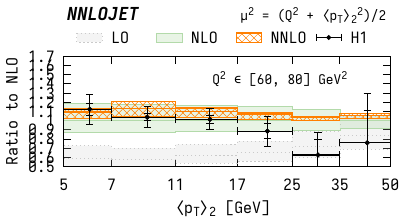}
  \end{tabular}
  \caption{
    Inclusive di-jet production cross section as a function of $\langle p^B_{T}\rangle_2$ in bins of $Q^2$, compared to H1 low-$Q^2$ data.
  }
  \label{fig:h1lowq2}
\end{figure}

\begin{figure}
  \centering
  \begin{tabular}{r@{}l}
    \includegraphics[trim={ 0pt 24pt  0pt  0pt},clip]{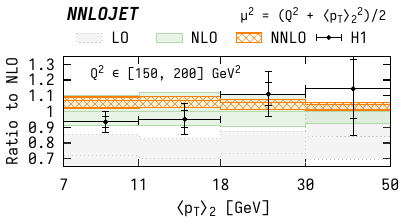} &
    \includegraphics[trim={ 9pt 24pt  0pt 24pt},clip]{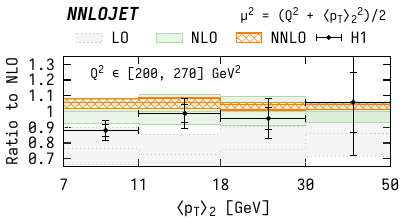} \\[0pt]
    \includegraphics[trim={ 0pt 24pt  0pt 24pt},clip]{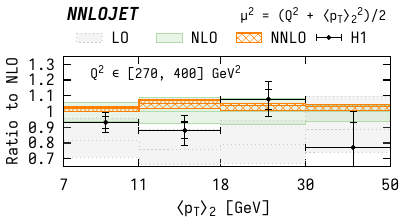} &
    \includegraphics[trim={ 9pt 24pt  0pt 24pt},clip]{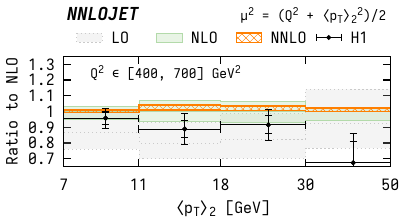} \\[0pt]
    \includegraphics[trim={ 0pt  0pt  0pt 24pt},clip]{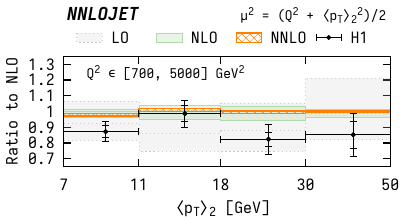} &
    \includegraphics[trim={ 9pt  0pt  0pt 24pt},clip]{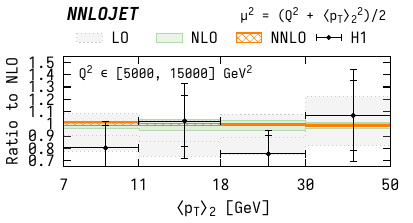}
  \end{tabular}
  \caption{
    Inclusive di-jet production cross section as a function of $p_{T,2}^B$ in bins of $Q^2$, compared to H1 high-$Q^2$ data.
  }
  \label{fig:h1highq2pt}
\end{figure}

The H1 dataset is divided into a high-$Q^2$ and a low-$Q^2$ sample, see Tab.~\ref{tab:2jkin}. For the low-$Q^2$ sample~\cite{h1lowq2},
double-differential distributions were measured in $Q^2$ and $\langle p^B_{T}\rangle_2$, which we compare to our NNLO calculation in Fig.~\ref{fig:h1lowq2}.
Compared to NLO, the NNLO corrections enhance the prediction of the di-jet cross section at lower values of $\langle p^B_{T}\rangle_2$, leading to a considerable
improvement in the description of the H1 data, as already pointed out in Ref.~\cite{h1lowq2}. Moreover, we observe an excellent convergence of the perturbative
series and a considerable reduction of the theory uncertainty in going from NLO to NNLO, which are typically reduced by a factor of two or more.
This highlights the potential of these data for
future precision studies of parton distributions and the strong coupling constant.

\begin{figure}
  \centering
  \begin{tabular}{r@{}l}
    \includegraphics[trim={ 0pt 24pt  0pt  0pt},clip]{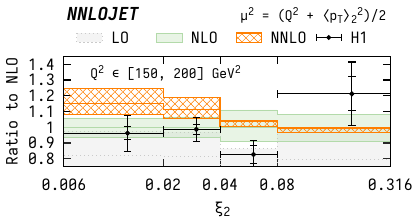} &
    \includegraphics[trim={ 9pt 24pt  0pt 24pt},clip]{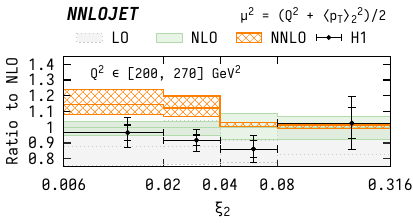} \\[0pt]
    \includegraphics[trim={ 0pt 24pt  0pt 24pt},clip]{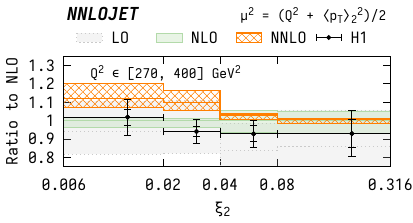} &
    \includegraphics[trim={ 9pt 24pt  0pt 24pt},clip]{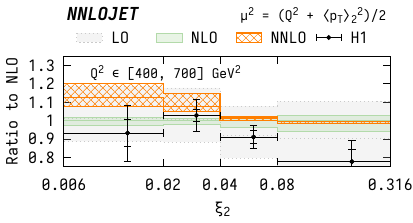} \\[0pt]
    \includegraphics[trim={ 0pt  0pt  0pt 24pt},clip]{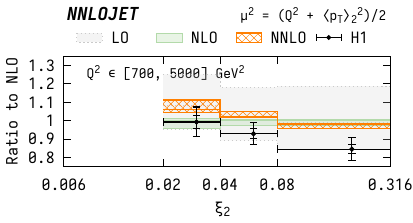} &
    \includegraphics[trim={ 9pt  0pt  0pt 24pt},clip]{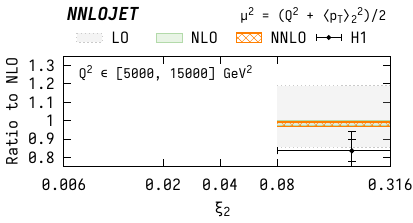}
  \end{tabular}
  \caption{
    Inclusive di-jet production cross section as a function of $\xi_2$ in bins of $Q^2$, compared to H1 high-$Q^2$ data.
  }
  \label{fig:h1highq2xi}
\end{figure}

For the high-$Q^2$ sample, slightly different event selection
criteria are applied: in particular, a minimum value of $M_{jj}$ is imposed. In
a previous work~\cite{disprl}, we have already studied the impact of the NNLO corrections in the  double differential distributions in $Q^2$ and $\langle p^B_{T}\rangle_2$
for the H1 high-$Q^2$ sample. Figure~\ref{fig:h1highq2pt} collects these results.
A similar pattern of improved in the theoretical uncertainty can be seen here with NNLO corrections that are well captured within the NLO uncertainty estimates and residual scale bands that are greatly reduced.

Figure~\ref{fig:h1highq2xi} compares the NNLO predictions for double-differential distributions in $Q^2$ and $\xi_2$ to the H1 high-$Q^2$ measurement (these distributions are not available for the H1 low-$Q^2$ study).
We observe that the quantitative behaviour is very similar to the ZEUS distributions, Fig.~\ref{fig:zeusxiq2}.
At LO, $\xi_2$ is directly related to the momentum fraction carried by the incoming parton, such that Fig.~\ref{fig:h1highq2xi} indicates the
kinematical range where the H1 data can potentially improve the determination of parton distributions. Recalling the definition of $\xi_2$ (\ref{eq:breitvar}),
we moreover observe that the H1 high-$Q^2$ data set typically probes lower values of $\xi_2$
than its low-$Q^2$ counterpart (which is due to the
transverse momentum requirement on the final state jets, and contrasts with the kinematical correlations in inclusive DIS).

\begin{figure}
  \centering
  \begin{tabular}{r@{}l}
    \includegraphics[trim={ 0pt 24pt  0pt  0pt},clip]{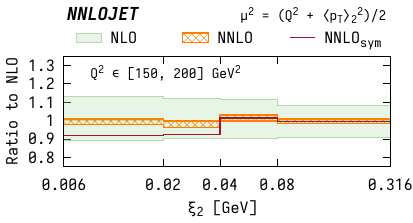} &
    \includegraphics[trim={ 9pt 24pt  0pt 24pt},clip]{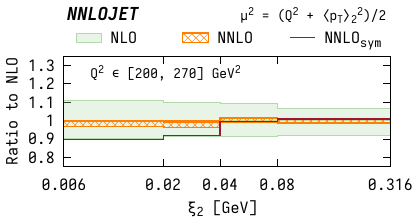} \\[0pt]
    \includegraphics[trim={ 0pt 24pt  0pt 24pt},clip]{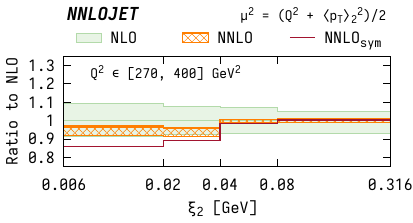} &
    \includegraphics[trim={ 9pt 24pt  0pt 24pt},clip]{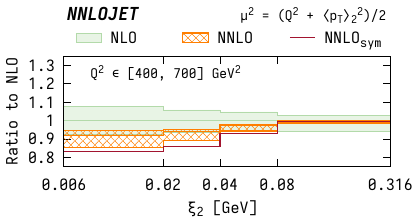} \\[0pt]
    \includegraphics[trim={ 0pt  0pt  0pt 24pt},clip]{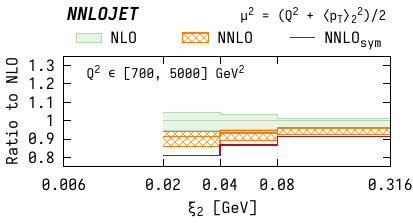} &
    \includegraphics[trim={ 9pt  0pt  0pt 24pt},clip]{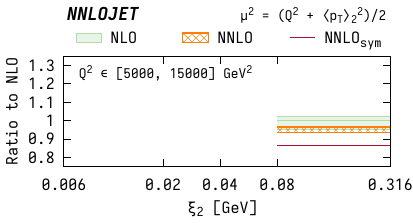}
  \end{tabular}
  \caption{
    Inclusive di-jet production cross section as a function of $\xi_2$ in bins of $Q^2$ with asymmetric cuts on the two jets.
    The red line indicates the NNLO prediction with symmetric cuts of Fig.~\protect{\ref{fig:h1highq2xi}}.
  }
  \label{fig:xiasym}
\end{figure}

Finally, we comment on the potential perturbative instabilities that may arise from the symmetric cuts on $p^B_{T}$ combined with a cut on $M_{jj}$.
To illustrate the impact, we re-evaluated
the double differential distribution in $Q^2$ and $\xi_2$ for a different set of jet cuts:
$p^B_{T,j1}>5$~GeV,  $p^B_{T,j2}>4$~GeV. The result is shown in Fig.~\ref{fig:xiasym}, where
we observe a very substantial improvement in the perturbative convergence, compared to the cuts used in the
H1 analysis~\cite{h1highq2}.

\section{Conclusions and Outlook}
\label{sec:conc}

In this paper, we described the calculation of the second-order (NNLO) QCD corrections to jet production in
deep inelastic scattering. By defining jets in the Breit frame of reference, this process requires at least two partons in the
final state, thereby providing sensitivity on the gluon distribution and the strong coupling constant already at leading order.
We consider inclusive production of single jets and of di-jet systems in the Breit frame, which start both at the same
perturbative order.

Our calculation uses the antenna subtraction method to cancel infrared singularities among parton-level sub-processes of
different multiplicity. The application of this method to processes with one hadron in the initial state requires the
dedicated treatment of infrared-singular splittings that change the parton identity. Our calculation is implemented into
the \NNLOJET parton-level event generator framework, which was also used recently for
the NNLO corrections to $pp\to Z+j$~\cite{ourzj}, $pp\to H+j$~\cite{ourhj} and $pp\to 2j$~\cite{2jnew}.

The HERA experiments H1 and ZEUS have measured inclusive single jet and di-jet production over a broad
kinematical range. We observe that the NNLO corrections to inclusive single jet production modify the shape of the
kinematical distributions, which are now described considerably better than at NLO. The corrections
are moderate in size
(five to ten per cent) except for very low jet transverse momenta or low photon virtuality $Q^2$, and their inclusion
substantially reduces the scale uncertainty on the predictions, typically well below the experimental statistical and
systematical uncertainty.

The NNLO corrections to di-jet production are found to be sizeable, and often well outside the scale uncertainty
band of the NLO predictions over a broad kinematical range for many of the data sets from ZEUS and H1.
However, the inclusion of NNLO corrections substantially improve the agreement with data across the board highlighting their importance for doing precision phenomenology using these measurements.
Potential instabilities due to the interplay between jet $p_T$ and $M_{jj}$ cuts have been investigated by performing dedicated comparisons of the predictions with and without asymmetric $p_T$ cuts, revealing a more stable convergence in the former.

Our NNLO results open up several opportunities for precision phenomenology with jet observables in
deep inelastic scattering.
The determination of parton distributions at NNLO from a global fit is currently dominated by
processes that are only quark-initiated at leading order (inclusive DIS, Drell-Yan processes). Constraints on the gluon distribution come
mainly from indirect effects (scaling violations) or from the inclusion of data from processes (like jet-production) where the full NNLO corrections are
unknown. In these cases, the NNLO corrections to the hard process cross sections are either approximated using some ad-hoc assumptions or
discarded altogether. A recent study~\cite{czakonpdf} on the impact of LHC inclusive top quark cross section data on the
determination of the gluon distribution illustrated the importance of a consistent treatment of NNLO effects (which are known for
top quark production~\cite{czakon1,czakon2}). Our newly derived NNLO corrections to jet production in DIS enable the consistent inclusion of
HERA data on this process into global parton distribution fits at NNLO. The magnitude of the corrections, as well as their kinematical dependence,
makes it likely that their inclusion will lead to modifications of the gluon distribution, also leading to a substantial reduction of
its uncertainty in the crucial region of medium-$x$.

Jet production data from HERA have been used to measure the strong coupling constant
$\alpha_s$~\cite{h1d12,h1highq2,zeusb1}. The error on all these measurements was dominated by the theory
uncertainty inherent to the NLO predictions used in the extraction of $\alpha_s$. This uncertainty was found to be
typically a factor two or more larger than experimental statistical or systematical uncertainties, thereby proving to
be the limiting factor to further improving $\alpha_s$ measurements from jet production in deep inelastic scattering.
Given that the analysis of inclusive deep inelastic scattering
structure function data typically yields values of the $\alpha_s$ at the lower
boundary of the range indicated by other determinations~\cite{pdg},  it will be very interesting
to apply the NNLO corrections derived in this paper to perform an NNLO-accurate determination of the strong
coupling constant from DIS jet production data.

\acknowledgments

We would like to thank Nigel Glover and Thomas Morgan for many
interesting discussions during this project, Stefan H\"oche and
Marek Sch\"onherr for help with the NLO comparisons
against SHERPA, and Daniel Britzger for
many discussions on the H1 jet measurements.
We further thank Robin Sch\"urmann for performing independent re-derivations of integrated initial-final antenna functions, which have led us to uncover an implementation error.
This research was supported in part by the Swiss National Science Foundation (SNF) under contracts 200020-162487 and CRSII2-160814,
 in part by
the UK Science and Technology Facilities Council as well as by the Research Executive Agency (REA) of the European Union under the Grant Agreement PITN-GA-2012-316704  (``HiggsTools'') and  the ERC Advanced Grant MC@NNLO (340983).

\appendix

\section{NLO subtraction terms}
\label{app:nlosub}

In this appendix we include all the subtraction terms for the NLO two-jet calculation. The matrix elements and their associated factors are collected in
Tabs.~\ref{tab:NLOq} and~\ref{tab:NLOg}.

\subsection{Quark-initiated subtraction terms}

The quark channel receives contributions from the processes $q\gamma\to qgg$, $q\gamma\to qq\bar{q}$ at tree level and
$q\gamma\to qg$ at 1-loop.

\subsection*{Real}

Real emission corrections at NLO contain up to three final-state partons and are comprised of
$B_{2}^{\gamma,0}$, $C_{0}^{\gamma,0}$ and $D_{0}^{\gamma,0}$ matrix elements.

\addtocontents{toc}{\protect\setcounter{tocdepth}{2}}
\subsubsection{B-type $\mathcal{O}(N^1)$ contribution}
\addtocontents{toc}{\protect\setcounter{tocdepth}{3}}

The leading-colour contribution is composed of two-quark--two-gluon matrix elements, $B_{2}^{\gamma,0}$, which are
summed over both permutations of the final-state gluons. This matrix element and the corresponding subtraction term are given by
\ba
\sum_{P(i,j)}B_{2}^{\gamma,0}(\hat{1},i,j,2)-\sum_{P(i,j)}B_{2}^{\gamma,0,S}(\hat{1},i,j,2),
\ea
where $P(i,j)$ denotes the permutations of the labels in the set of final-state gluons $\{3,4\}$. The subtraction term
is constructed for a single permutation and given by
\begin{eqnarray}
\lefteqn{{B_{2}^{\gamma,0,S}(\hat{1},i,j,k)=}}\nonumber \\
 \ph{1}&&+d_{3,q}^{0}(1,i,j)\,B_1^{\gamma,0}(\overline{1},(\widetilde{ij}),k)\,J_2^{(2)}(\lbrace p\rbrace_{2}) \nonumber\\
 \ph{2}&&+d_3^{0}(k,j,i)\,B_1^{\gamma,0}(1,(\widetilde{ij}),(\widetilde{kj}))\,J_2^{(2)}(\lbrace p\rbrace_{2}).
\end{eqnarray}

\addtocontents{toc}{\protect\setcounter{tocdepth}{2}}
\subsubsection{B-type $\mathcal{O}(N^{-1})$ contribution}
\addtocontents{toc}{\protect\setcounter{tocdepth}{3}}

\label{sec:nloRBt}

The two-quark--two-gluon matrix element also contributes at sub-leading colour where it is
given by the $\tilde{B}_{2}^{\gamma,0}$ matrix element. The subtracted contribution to
the cross section is given by
\ba
\tilde{B}_{2}^{\gamma,0}(\hat{1},3,4,2)-\tilde{B}_{2}^{\gamma,0,S}(\hat{1},3,4,2).
\ea
The gluons in this function
act as if they are abelian due to the particular combination of interferences used to define it.
This changes the factorization behaviour compared to the leading-colour matrix element and
we construct the subtraction term to reflect this:
\begin{eqnarray}
\lefteqn{{\tilde{B}_{2}^{\gamma,0,S}(\hat{1},i,j,k) =}}\nonumber \\
 \ph{1}&&+A_{3,q}^{0}(1,j,k)\,B_1^{\gamma,0}(\overline{1},i,(\widetilde{jk}))\,J_2^{(2)}(\lbrace p\rbrace_{2}) \nonumber\\
 \ph{2}&&+A_{3,q}^{0}(1,i,k)\,B_1^{\gamma,0}(\overline{1},j,(\widetilde{ik}))\,J_2^{(2)}(\lbrace p\rbrace_{2}) .
\end{eqnarray}

From this subtraction term it is clear that the gluons are colour connected only to the quark and antiquark,
not other gluons; which is what would be expected for an abelian gauge boson.

\addtocontents{toc}{\protect\setcounter{tocdepth}{2}}
\subsubsection{C-type $\mathcal{O}(N_F)$ contribution}
\addtocontents{toc}{\protect\setcounter{tocdepth}{3}}

\label{sec:qnloRC}

With three final-state partons and an initial-state quark we must also evaluate the contribution from
four-quark matrix elements of identical ($q\gamma\to qq\bar{q}$) and non-identical ($q\gamma\to qQ\bar{Q}$) flavours. The full four-quark matrix element,
summed over possible quark flavours, can be written as a term derived from the non-identical
flavour matrix element, $C_{0}^{\gamma,0}$, and an interference term, $D_{0}^{\gamma,0}$.
The $C_{0}^{\gamma,0}$ function contains contributions from diagrams where the vector boson
couples to each quark line and the subtraction term is given by
\ba
\sum_{3,4} \left( C_{0}^{\gamma,0}(\hat{1};4,3;2)-C_{0}^{\gamma,0,S}(\hat{1};4,3;2) \right),
\ea
where
\begin{eqnarray}
\lefteqn{{C_{0}^{\gamma,0,S}(\hat{1};i,j;k) =}} \nonumber \\
 \ph{1}&&+E_3^{0}(k,i,j)\,B_{1,q}^{\gamma,0}(1,(\widetilde{ji}),(\widetilde{ki}))\,J_2^{(2)}(\lbrace p\rbrace_{2}) \nonumber\\
 \ph{2}&&-\frac{1}{2}\,E_{3,q^\prime\to g}^{0}(i,1,k)\,{B}_{1,Q}^{\gamma,0}(j,\overline{1},(\widetilde{ik}))\,J_2^{(2)}(\lbrace p\rbrace_{2}) \nonumber\\
 \ph{3}&&-\frac{1}{2}\,E_{3,q^\prime\to g}^{0}(j,1,k)\,{B}_{1,Q}^{\gamma,0}((\widetilde{jk}),\overline{1},i)\,J_2^{(2)}(\lbrace p\rbrace_{2}) .
\end{eqnarray}

In this subtraction term we note the appearance of the $B_{1,q}^{\gamma,0}$ and $B_{1,Q}^{\gamma,0}$ reduced matrix elements which
were introduced in Sec.~\ref{sec:structure} and which carry over the
information on which quark line the vector boson is coupling to in the reduced
matrix element.

We can see that the first term of the subtraction term regulates the quark-antiquark collinear limit between partons $i$ and $j$
which constitute a quark line of flavour $Q$ and so in the reduced matrix element the vector boson couples to the remaining quark line
of flavour $q$. The remaining subtraction terms, on the other hand, regulate the collinear limit between the quarks $\hat{1}$ and $k$, which are of flavour $q$.
Consequently, the
vector boson in the reduced matrix element couples to the remaining quark line of flavour $Q$.

\addtocontents{toc}{\protect\setcounter{tocdepth}{2}}
\subsubsection{D-type $\mathcal{O}(N^0)$ contribution}
\addtocontents{toc}{\protect\setcounter{tocdepth}{3}}
\label{sec:qnloRD}

As explained above, the four-quark matrix element also contains a term given by the interference of four-quark
amplitudes with different quark assignments, arising from the identical flavour contribution,
\ba
D_{0}^{\gamma,0}(\hat{1},2;3,4).
\ea
We include this contribution but note that it is finite in all unresolved limits and so requires no subtraction term.

\subsection*{Virtual}

The virtual NLO contributions to the cross section arise from the interference of a one-loop amplitude
 with a tree-level amplitude for the process $q\gamma\to qg$. Once all colour factors are stripped out, this
 leaves three separate contributions to the cross section.

\addtocontents{toc}{\protect\setcounter{tocdepth}{2}}
\subsubsection{B-type $\mathcal{O}(N^1)$ contribution}
\addtocontents{toc}{\protect\setcounter{tocdepth}{3}}

The leading-colour contribution and its subtraction term is given by
\ba
B_{1}^{\gamma,1}(\hat{1},3,2)-B_{1}^{\gamma,1,T}(\hat{1},3,2),
\ea
where
\begin{eqnarray}
\lefteqn{{B_{1}^{\gamma,1,T}(\hat{1},i,j) =}} \nonumber \\
  \ph{1}&&-J_{2,QG}^{1,FF}(s_{ji})\,B_1^{\gamma,0}(1,i,j)\,J_2^{(2)}(\lbrace p\rbrace_{2})\nonumber\\
  \ph{2}&&-J_{2,QG}^{1,IF}(s_{1i})\,B_1^{\gamma,0}(1,i,j)\,J_2^{(2)}(\lbrace p\rbrace_{2}).
\end{eqnarray}

All the poles of the one-loop matrix element are cancelled analytically by this subtraction term.

\addtocontents{toc}{\protect\setcounter{tocdepth}{2}}
\subsubsection{B-type $\mathcal{O}(N^{-1})$ contribution}
\addtocontents{toc}{\protect\setcounter{tocdepth}{3}}

The two-quark--one-gluon one-loop matrix element also gives a contribution at subleading colour,
\ba
\tilde{B}_{1}^{\gamma,1}(\hat{1},3,2)-\tilde{B}_{1}^{\gamma,1,T}(\hat{1},3,2) ,
\ea
where
\begin{eqnarray}
\lefteqn{{\tilde{B}_{1}^{\gamma,1,T}(\hat{1},i,j) =}} \nonumber \\
  \ph{1}&&-J_{2,QQ}^{1,IF}(s_{1j})\,B_1^{\gamma,0}(1,i,j)\,J_2^{(2)}(\lbrace p\rbrace_{2}).
\end{eqnarray}

The integrated real radiation function in this subtraction term, which contains all the explicit poles, is a function of the quark and antiquark
momenta but not the gluon momentum. This can be traced back to the colour connection of the virtual gluon in the loop;
at subleading colour the virtual gluon is colour connected only to the quark line.

\addtocontents{toc}{\protect\setcounter{tocdepth}{2}}
\subsubsection{C-type $\mathcal{O}(N_F)$ contribution}
\addtocontents{toc}{\protect\setcounter{tocdepth}{3}}

The one-loop diagrams which contain a closed quark loop form a distinct matrix element, $\hat{B}_{1}^{\gamma,1}$,  carrying a factor of $\nf$.
The poles of this matrix element are cancelled by the subtraction term $\hat{B}_{1}^{\gamma,1,T}$. We must also include the
integrated subtraction terms coming from Sec.~\ref{sec:qnloRC} where the reduced matrix element is gluon-initiated rather than quark-initiated.

This type of subtraction term was discussed in Sec.~\ref{sec:IClimits} and is referred to as an \emph{identity changing} (IC) term. Such a terms
clearly cannot cancel the poles of the one-loop matrix element as it is proportional to the gluon-initiated reduced matrix element. Instead, the relevant integrated antennae
from the real subtraction term are combined with the IC mass factorization kernels to generate a finite virtual subtraction term, $\hat{B}_{1,q\to g}^{\gamma,1,T}$.
\ba
\hat{B}_{1}^{\gamma,1}(\hat{1},3,2)-\hat{B}_{1}^{\gamma,1,T}(\hat{1},3,2)-\hat{B}_{1,q\to g}^{\gamma,1,T}(\hat{1},3,2),
\ea
where
\begin{eqnarray}
\lefteqn{{\hat{B}_{1}^{\gamma,1,T}(\hat{1},i,j) =}} \nonumber \\
  \ph{1}&&-2\hat{J}_{2,QG}^{1,FF}(s_{ij})\,B_1^{\gamma,0}(1,i,j)\,J_2^{(2)}(\lbrace p\rbrace_{2}),
\end{eqnarray}

\begin{eqnarray}
\lefteqn{{\hat{B}_{1,q\to g}^{\gamma,1,T}(\hat{1},i,j) =}} \nonumber\\
  \ph{1}&&-\frac{1}{2}\,J_{2,GQ,q^\prime \to g}^{1,IF}(s_{i1})\,B_1^{\gamma,0}(i,1,j)\,J_2^{(2)}(\lbrace p\rbrace_{2})\nonumber\\
  \ph{2}&&-\frac{1}{2}\,J_{2,GQ,q^\prime \to g}^{1,IF}(s_{j1})\,B_1^{\gamma,0}(i,1,j)\,J_2^{(2)}(\lbrace p\rbrace_{2}).
\end{eqnarray}

\subsection{Gluon-initiated subtraction terms}

The gluon channel is composed of the tree-level process $g\gamma\to q\bar{q}g$ and the one-loop process $g\gamma\to q\bar{q}$.

\subsection*{Real}

Due to the initial-state gluon, no four-quark matrix elements contribute to the real correction, but only two-quark--two-gluon matrix elements
with one of the gluons crossed into the initial state.

\addtocontents{toc}{\protect\setcounter{tocdepth}{2}}
\subsubsection{B-type $\mathcal{O}(N^1)$ contribution}
\addtocontents{toc}{\protect\setcounter{tocdepth}{3}}

The leading-colour contribution is, as in the quark channel, given by the $B_{2}^{\gamma,0}$ function. We sum over
all gluon permutations, where now one of the gluons is in the initial-state. The subtracted contribution to the cross section
is given by
\ba
\sum_{P(\hat{1},2)}B_{2}^{\gamma,0}(3,\hat{1},2,4)-\sum_{P(\hat{1},2)}B_{2}^{\gamma,0,S}(3,\hat{1},2,4),
\ea
where
\begin{eqnarray}
\lefteqn{{B_{2}^{\gamma,0,S}(i,\hat{1},j,k) =}} \nonumber \\
 \ph{1}&&+d_{3,g}^{0}(i,j,1)\,B_1^{\gamma,0}((\widetilde{ij}),\overline{1},k)\,J_2^{(2)}(\lbrace p\rbrace_{2}) \nonumber\\
 \ph{2}&&+d_{3,g}^{0}(k,j,1)\,B_1^{\gamma,0}((\widetilde{jk}),\overline{1},i)\,J_2^{(2)}(\lbrace p\rbrace_{2}) \nonumber\\
 \ph{3}&&-a_{3,g\to q}^{0}(i,1,k)\,B_1^{\gamma,0}(\overline{1},j,(\widetilde{ik}))\,J_2^{(2)}(\lbrace p\rbrace_{2}) \nonumber\\
 \ph{4}&&-a_{3,g\to q}^{0}(k,1,i)\,B_1^{\gamma,0}((\widetilde{ik}),j,\overline{1})\,J_2^{(2)}(\lbrace p\rbrace_{2}) .
\end{eqnarray}

The final two terms in the subtraction term account for IC limits where the initial-state quark becomes collinear with a final-state quark
or antiquark.

\addtocontents{toc}{\protect\setcounter{tocdepth}{2}}
\subsubsection{B-type $\mathcal{O}(N^{-1})$ contribution}
\addtocontents{toc}{\protect\setcounter{tocdepth}{3}}

The subleading-colour contribution to the process $g\gamma\to q\bar{q}g$ is given by the $\tilde{B}_{2}^{\gamma,0}$ function with one of the
gluons crossed into the initial-state,
\ba
\tilde{B}_{2}^{\gamma,0}(3,\hat{1},2,4)-\tilde{B}_{2}^{\gamma,0,S}(3,\hat{1},2,4).
\ea
The initial-state gluon cannot become soft, but
can participate in a collinear limit with
final-state partons. As the gluons in this matrix element behave as if they were
abelian, no gluon-gluon collinear limits exist and all IF collinear limits are therefore IC limits with the final-state quark or antiquark,
\begin{eqnarray}
\lefteqn{{\tilde{B}_{2}^{\gamma,0,S}(i,\hat{1},j,k) =}}\nonumber \\
 \ph{1}&&-a_{3,g\to q}^{0}(i,1,k)\,B_1^{\gamma,0}(\overline{1},j,(\widetilde{ik}))\,J_2^{(2)}(\lbrace p\rbrace_{2}) \nonumber\\
 \ph{2}&&-a_{3,g\to q}^{0}(k,1,i)\,B_1^{\gamma,0}((\widetilde{ik}),j,\overline{1})\,J_2^{(2)}(\lbrace p\rbrace_{2}) \nonumber\\
 \ph{3}&&+A_3^{0}(i,j,k)\,B_1^{\gamma,0}((\widetilde{ij}),1,(\widetilde{kj}))\,J_2^{(2)}(\lbrace p\rbrace_{2}) .
\end{eqnarray}

\subsection*{Virtual}

The virtual NLO contributions arise from the interference of a one-loop amplitude
with a tree-level amplitude for the process $g\gamma\to q\bar{q}$.

\addtocontents{toc}{\protect\setcounter{tocdepth}{2}}
\subsubsection{B-type $\mathcal{O}(N^1)$ contribution}
\addtocontents{toc}{\protect\setcounter{tocdepth}{3}}

The leading-colour contribution contains a subtraction term which cancels the poles of the one-loop matrix element
and an IC subtraction term which is finite,
\ba
B_{1}^{\gamma,1}(2,\hat{1},3)-B_{1}^{\gamma,1,T}(2,\hat{1},3)-B_{1,g\to q}^{\gamma,1,T}(2,\hat{1},3) ,
\ea
where
\begin{eqnarray}
\lefteqn{{B_{1}^{\gamma,1,T}(i,\hat{1},j) =}} \nonumber \\
 \ph{1}&&-\bigg[ 
 +J_{2,GQ}^{1,IF}(s_{1i})
 +J_{2,GQ}^{1,IF}(s_{1j})\bigg]\,B_1^{\gamma,0}(i,1,j)\,J_2^{(2)}(\lbrace p\rbrace_{2}),
\end{eqnarray}

and the finite IC subtraction term is given by
\begin{eqnarray}
\lefteqn{{B_{1,g\to q}^{\gamma,1,T}(i,\hat{1},j) =}} \nonumber \\
  \ph{1}&&-2J_{2,QQ,g \to q}^{1,IF}(s_{1j})\,\bar{B}_1^{\gamma,0}(1,i,j)\,J_2^{(2)}(\lbrace p\rbrace_{2}).
\end{eqnarray}

\addtocontents{toc}{\protect\setcounter{tocdepth}{2}}
\subsubsection{B-type $\mathcal{O}(N^{-1})$ contribution}
\addtocontents{toc}{\protect\setcounter{tocdepth}{3}}

The subleading-colour matrix element is given by the $\tilde{B}_{1}^{\gamma,1}$ function, whose poles are cancelled by
the subtraction term $\tilde{B}_{1}^{\gamma,1,T}$. The IC integrated subtraction terms coming from Sec.~\ref{sec:nloRBt}
are combined with the IC mass factorization kernels to form the finite subtraction term, $\tilde{B}_{1,g\to q}^{\gamma,1,T}$.
The total subtracted contribution to the cross section is given by
\ba
\tilde{B}_{1}^{\gamma,1}(2,\hat{1},3)-\tilde{B}_{1}^{\gamma,1,T}(2,\hat{1},3)-\tilde{B}_{1,g\to q}^{\gamma,1,T}(2,\hat{1},3) ,
\ea
where
\begin{eqnarray}
\lefteqn{{\tilde{B}_{1}^{\gamma,1,T}(i,\hat{1},j) =}} \nonumber \\
  \ph{1}&&-J_{2,QQ}^{1,FF}(s_{ij})\,B_1^{\gamma,0}(i,1,j)\,J_2^{(2)}(\lbrace p\rbrace_{2}),
  \end{eqnarray}

\begin{eqnarray}
\lefteqn{{\tilde{B}_{1,g\to q}^{\gamma,1,T}(i,\hat{1},j) =}} \nonumber \\
  \ph{1}&&-2J_{2,QQ,g \to q}^{1,IF}(s_{1j})\,\bar{B}_1^{\gamma,0}(1,i,j)\,J_2^{(2)}(\lbrace p\rbrace_{2}).
\end{eqnarray}

\addtocontents{toc}{\protect\setcounter{tocdepth}{2}}
\subsubsection{B-type $\mathcal{O}(N_F)$ contribution}
\addtocontents{toc}{\protect\setcounter{tocdepth}{3}}

The one-loop matrix element with a closed quark loop is given by the $\hat{B}_{1}^{\gamma,1}$ function. This function contains
explicit poles, but as is clear from Tab.~\ref{tab:NLOg}, there is no real subtraction term contributing to this colour factor. This
means there are no integrated antenna functions to cancel the explicit poles of the matrix element. However, the NLO mass factorization
term does contribute to this colour factor and cancels the poles of the matrix element. This is reflected in the fact that the integrated real radiation function
for this process is formed entirely from mass factorization kernels, as is noted in Ref.~\cite{currie}:
\ba
\hat{B}_{1}^{\gamma,1}(2,\hat{1},3)-\hat{B}_{1}^{\gamma,1,T}(2,\hat{1},3),
\ea
where
\begin{eqnarray}
\lefteqn{{\hat{B}_{1}^{\gamma,1,T}(i,\hat{1},j) =}} \nonumber \\
  \ph{1}&&-2\hat{J}_{2,GQ}^{1,IF}(s_{i1})\,B_1^{\gamma,0}(i,1,j)\,J_2^{(2)}(\lbrace p\rbrace_{2}).
\end{eqnarray}

\section{NNLO subtraction terms}
\label{app:nnlosub}

In this appendix, we include the subtraction terms that are used to regulate each colour-stripped matrix element introduced in Sec.~\ref{sec:structure}.
The matrix elements associated with the subtraction terms are listed in Tabs.~\ref{tab:channelsq} and~\ref{tab:channelsg}.

\subsection{Quark-initiated subtraction terms}
\label{app:quark}

The quark channel at NNLO receives contributions from the tree-level sub-processes $q\gamma\to qggg$ and $q\gamma\to qq\bar{q}g$,
the one-loop correction to the processes $q\gamma\to qgg$ and $q\gamma\to qq\bar{q}$ and the two-loop corrections to the process $q\gamma\to qg$.

\subsection*{Double real}
\label{app:nnlosubRRq}

The tree-level five-parton contributions are given by the colour decomposition of the two-quark--three-gluon and four-quark--one-gluon
matrix elements with the quark crossed into the initial-state.

\addtocontents{toc}{\protect\setcounter{tocdepth}{2}}
\subsubsection{B-type $\mathcal{O}(N^2)$ contribution}
\addtocontents{toc}{\protect\setcounter{tocdepth}{3}}

The leading-colour matrix element is summed over all six permutations of the final-state gluons. The subtraction term
is summed only over the cyclic permutations of gluons. The cancellation of divergences occurs once all terms in the sum
are evaluated,
\ba
\sum_{P(i,j,k)}B_{3}^{\gamma,0}(\hat{1},i,j,k,2)-\sum_{P_{C}(i,j,k)}B_{3}^{\gamma,0,S}(\hat{1},i,j,k,2),
\ea
where $P_{C}$ denotes the restricted sum over cyclic permutations. The subtraction term is given by
\begin{eqnarray}
\lefteqn{{B_{3}^{\gamma,0,S}(\hat{1},i,j,k,l) =}} \nonumber\\
 \ph{1}&&+f_3^{0}(i,j,k)\,B_2^{\gamma,0}(1,(\widetilde{ij}),(\widetilde{jk}),l)\,J_2^{(3)}(\lbrace p\rbrace_{3}) \nonumber\\
 \ph{2}&&+f_3^{0}(i,j,k)\,B_2^{\gamma,0}(1,(\widetilde{jk}),(\widetilde{ij}),l)\,J_2^{(3)}(\lbrace p\rbrace_{3}) \nonumber\\
 \ph{3}&&+d_3^{0}(l,k,j)\,B_2^{\gamma,0}(1,i,(\widetilde{jk}),(\widetilde{lk}))\,J_2^{(3)}(\lbrace p\rbrace_{3}) \nonumber\\
 \ph{4}&&+d_3^{0}(l,i,j)\,B_2^{\gamma,0}(1,k,(\widetilde{ji}),(\widetilde{li}))\,J_2^{(3)}(\lbrace p\rbrace_{3}) \nonumber\\
 \ph{5}&&+d_3^{0}(1,k,j)\,B_2^{\gamma,0}(\overline{1},(\widetilde{kj}),i,l)\,J_2^{(3)}(\lbrace p\rbrace_{3}) \nonumber\\
 \ph{6}&&+d_3^{0}(1,i,j)\,B_2^{\gamma,0}(\overline{1},(\widetilde{ij}),k,l)\,J_2^{(3)}(\lbrace p\rbrace_{3}) \nonumber\\
\snip \ph{7}&&+\,D_4^{0,a}(l,i,j,k)\,B_1^{\gamma,0}(1,(\widetilde{ijk}),(\widetilde{lij}))\,J_2^{(2)}(\lbrace p\rbrace_{2}) \nonumber\\
 \ph{8}&&-\,d_3^{0}(l,i,j)\,d_3^{0}((\widetilde{li}),(\widetilde{ij}),k)\,B_1^{\gamma,0}(1,(\widetilde{k\widetilde{(ij})}),\widetilde{(\widetilde{li})(\widetilde{ij})})\,J_2^{(2)}(\lbrace p\rbrace_{2}) \nonumber\\
 \ph{9}&&-\,f_3^{0}(k,j,i)\,d_3^{0}(l,(\widetilde{kj}),(\widetilde{ji}))\,B_1^{\gamma,0}(1,\widetilde{(\widetilde{kj})(\widetilde{ji})},(\widetilde{l\widetilde{(kj})}))\,J_2^{(2)}(\lbrace p\rbrace_{2}) \nonumber\\
\snip \ph{10}&&+\,D_4^{0,a}(l,k,j,i)\,B_1^{\gamma,0}(1,(\widetilde{kji}),(\widetilde{lkj}))\,J_2^{(2)}(\lbrace p\rbrace_{2}) \nonumber\\
 \ph{11}&&-\,d_3^{0}(l,k,j)\,d_3^{0}((\widetilde{lk}),(\widetilde{kj}),i)\,B_1^{\gamma,0}(1,(\widetilde{i\widetilde{(kj})}),\widetilde{(\widetilde{lk})(\widetilde{kj})})\,J_2^{(2)}(\lbrace p\rbrace_{2}) \nonumber\\
 \ph{12}&&-\,f_3^{0}(i,j,k)\,d_3^{0}(l,(\widetilde{ij}),(\widetilde{jk}))\,B_1^{\gamma,0}(1,\widetilde{(\widetilde{ij})(\widetilde{jk})},(\widetilde{l\widetilde{(ij})}))\,J_2^{(2)}(\lbrace p\rbrace_{2}) \nonumber\\
\snip \ph{13}&&+\,D_4^0(1,i,j,k)\,B_1^{\gamma,0}(\overline{1},(\widetilde{ijk}),l)\,J_2^{(2)}(\lbrace p\rbrace_{2}) \nonumber\\
 \ph{14}&&-\,d_3^{0}(1,i,j)\,D_3^{0}(\overline{1},k,(\widetilde{ij}))\,B_1^{\gamma,0}(\overline{\overline{1}},(\widetilde{(\widetilde{ij})k}),l)\,J_2^{(2)}(\lbrace p\rbrace_{2}) \nonumber\\
 \ph{15}&&-\,d_3^{0}(1,k,j)\,D_3^{0}(\overline{1},i,(\widetilde{kj}))\,B_1^{\gamma,0}(\overline{\overline{1}},(\widetilde{(\widetilde{kj})i}),l)\,J_2^{(2)}(\lbrace p\rbrace_{3}) \nonumber\\
 \ph{16}&&-\,f_3^{0}(i,j,k)\,D_3^{0}(1,(\widetilde{ij}),(\widetilde{kj}))\,B_1^{\gamma,0}(\overline{1},\widetilde{(\widetilde{ij})(\widetilde{kj})},l)\,J_2^{(2)}(\lbrace p\rbrace_{2}) \nonumber\\
\snip \ph{17}&&+\,D_4^{0,c}(l,k,i,j)\,B_1^{\gamma,0}(1,(\widetilde{jik}),(\widetilde{lki}))\,J_2^{(2)}(\lbrace p\rbrace_{2}) \nonumber\\
\snip \ph{18}&&+\,D_4^{0,c}(l,i,k,j)\,B_1^{\gamma,0}(1,(\widetilde{jki}),(\widetilde{lik}))\,J_2^{(2)}(\lbrace p\rbrace_{2}) \nonumber\\
 \ph{19}&&-\,d_3^{0}(l,k,j)\,d_3^{0}((\widetilde{lk}),i,(\widetilde{kj}))\,B_1^{\gamma,0}(1,(\widetilde{i\widetilde{(kj})}),(\widetilde{i\widetilde{(lk})}))\,J_2^{(2)}(\lbrace p\rbrace_{2}) \nonumber\\
 \ph{20}&&-\,d_3^{0}(l,i,j)\,d_3^{0}((\widetilde{li}),k,(\widetilde{ij}))\,B_1^{\gamma,0}(1,(\widetilde{k\widetilde{(ij})}),(\widetilde{k\widetilde{(li})}))\,J_2^{(2)}(\lbrace p\rbrace_{2}) \nonumber\\
\snip \ph{21}&&-\,\tilde{A}_4^0(1,i,k,l)\,B_1^{\gamma,0}(\overline{1},j,(\widetilde{lki}))\,J_2^{(2)}(\lbrace p\rbrace_{2}) \nonumber\\
 \ph{22}&&+\,A_3^{0}(1,i,l)\,A_3^{0}(\overline{1},k,(\widetilde{li}))\,B_1^{\gamma,0}(\overline{\overline{1}},j,(\widetilde{k\widetilde{(li})}))\,J_2^{(2)}(\lbrace p\rbrace_{2}) \nonumber\\
 \ph{23}&&+\,A_3^{0}(1,k,l)\,A_3^{0}(\overline{1},i,(\widetilde{lk}))\,B_1^{\gamma,0}(\overline{\overline{1}},j,(\widetilde{i\widetilde{(lk})}))\,J_2^{(2)}(\lbrace p\rbrace_{2}) \nonumber\\
 \ph{24}&&+\frac{1}{2}\,\,d_3^{0}(1,i,j)\,d_3^{0}(\overline{1},k,(\widetilde{ij}))\,B_1^{\gamma,0}(\overline{1},(\widetilde{k\widetilde{(ij})}),l)\,J_2^{(2)}(\lbrace p\rbrace_{2}) \nonumber\\
 \ph{25}&&-\frac{1}{2}\,\,d_3^{0}(l,i,j)\,d_3^{0}(1,k,(\widetilde{ij}))\,B_1^{\gamma,0}(\overline{1},(\widetilde{k\widetilde{(ij})}),(\widetilde{li}))\,J_2^{(2)}(\lbrace p\rbrace_{2}) \nonumber\\
 \ph{26}&&-\frac{1}{2}\,\,A_3^{0}(1,i,l)\,d_3^{0}(\overline{1},k,j)\,B_1^{\gamma,0}(\overline{\overline{1}},(\widetilde{kj}),(\widetilde{li}))\,J_2^{(2)}(\lbrace p\rbrace_{2}) \nonumber\\
 \ph{27}&&-\frac{1}{2}\,\bigg[      +S^{FF}_{1i(\widetilde{ij})}       -S^{FF}_{\overline{1}i(\widetilde{k\widetilde{(ij})})}       -S^{FF}_{(\widetilde{li})i(\widetilde{ij})}       +S^{FF}_{(\widetilde{li})i(\widetilde{k\widetilde{(ij})})}       -S^{FF}_{1i(\widetilde{li})}       +S^{FF}_{\overline{1}i(\widetilde{li})}  \bigg]\nonumber\\
 &&\times d_3^{0}(1,k,(\widetilde{ij}))\,B_1^{\gamma,0}(\overline{1},(\widetilde{k\widetilde{(ij})}),(\widetilde{li}))\,J_2^{(2)}(\lbrace p\rbrace_{2}) \nonumber\\
 \ph{28}&&+\frac{1}{2}\,\,d_3^{0}(1,k,j)\,d_3^{0}(\overline{1},i,(\widetilde{kj}))\,B_1^{\gamma,0}(\overline{1},(\widetilde{i\widetilde{(kj})}),l)\,J_2^{(2)}(\lbrace p\rbrace_{2}) \nonumber\\
 \ph{29}&&-\frac{1}{2}\,\,d_3^{0}(l,k,j)\,d_3^{0}(1,i,(\widetilde{kj}))\,B_1^{\gamma,0}(\overline{1},(\widetilde{i\widetilde{(kj})}),(\widetilde{lk}))\,J_2^{(2)}(\lbrace p\rbrace_{2}) \nonumber\\
 \ph{30}&&-\frac{1}{2}\,\,A_3^{0}(1,k,l)\,d_3^{0}(\overline{1},i,j)\,B_1^{\gamma,0}(\overline{\overline{1}},(\widetilde{ij}),(\widetilde{lk}))\,J_2^{(2)}(\lbrace p\rbrace_{2}) \nonumber\\
 \ph{31}&&-\frac{1}{2}\,\bigg[      +S^{FF}_{1k(\widetilde{jk})}       -S^{FF}_{\overline{1}k(\widetilde{i\widetilde{(jk})})}       -S^{FF}_{(\widetilde{lk})k(\widetilde{jk})}       +S^{FF}_{(\widetilde{lk})k(\widetilde{i\widetilde{(jk})})}       -S^{FF}_{1k(\widetilde{lk})}       +S^{FF}_{\overline{1}k(\widetilde{lk})}  \bigg]\nonumber\\
 &&\times d_3^{0}(1,i,(\widetilde{jk}))\,B_1^{\gamma,0}(\overline{1},(\widetilde{i\widetilde{(jk})}),(\widetilde{lk}))\,J_2^{(2)}(\lbrace p\rbrace_{2}) \nonumber\\
 \ph{32}&&+\frac{1}{2}\,\,d_3^{0}(l,i,j)\,d_3^{0}((\widetilde{li}),k,(\widetilde{ij}))\,B_1^{\gamma,0}(1,(\widetilde{k\widetilde{(ij})}),(\widetilde{(\widetilde{li})k}))\,J_2^{(2)}(\lbrace p\rbrace_{2}) \nonumber\\
 \ph{33}&&-\frac{1}{2}\,\,d_3^{0}(1,i,j)\,d_3^{0}(l,k,(\widetilde{ij}))\,B_1^{\gamma,0}(\overline{1},(\widetilde{k\widetilde{(ij})}),(\widetilde{lk}))\,J_2^{(2)}(\lbrace p\rbrace_{2}) \nonumber\\
 \ph{34}&&-\frac{1}{2}\,\,A_3^{0}(1,i,l)\,d_3^{0}((\widetilde{li}),k,j)\,B_1^{\gamma,0}(\overline{1},(\widetilde{kj}),(\widetilde{k\widetilde{(li})}))\,J_2^{(2)}(\lbrace p\rbrace_{2}) \nonumber\\
 \ph{35}&&-\frac{1}{2}\,\bigg[      +S^{FF}_{(\widetilde{ij})i(\widetilde{li})}       -S^{FF}_{(\widetilde{k\widetilde{(ij})})i(\widetilde{k\widetilde{(li})})}       -S^{FF}_{1i(\widetilde{ij})}       +S^{FF}_{1i(\widetilde{k\widetilde{(ij})})}       -S^{FF}_{1i(\widetilde{li})}       +S^{FF}_{1i(\widetilde{k\widetilde{(li})})}  \bigg]\nonumber\\
 &&\times d_3^{0}((\widetilde{li}),k,(\widetilde{ij}))\,B_1^{\gamma,0}(1,(\widetilde{k\widetilde{(ij})}),(\widetilde{k\widetilde{(li})}))\,J_2^{(2)}(\lbrace p\rbrace_{2}) \nonumber\\
 \ph{36}&&+\frac{1}{2}\,\,d_3^{0}(l,k,j)\,d_3^{0}((\widetilde{lk}),i,(\widetilde{kj}))\,B_1^{\gamma,0}(1,(\widetilde{i\widetilde{(kj})}),(\widetilde{(\widetilde{lk})i}))\,J_2^{(2)}(\lbrace p\rbrace_{2}) \nonumber\\
 \ph{37}&&-\frac{1}{2}\,\,d_3^{0}(1,k,j)\,d_3^{0}(l,i,(\widetilde{kj}))\,B_1^{\gamma,0}(\overline{1},(\widetilde{i\widetilde{(kj})}),(\widetilde{li}))\,J_2^{(2)}(\lbrace p\rbrace_{2}) \nonumber\\
 \ph{38}&&-\frac{1}{2}\,\,A_3^{0}(1,k,l)\,d_3^{0}((\widetilde{lk}),i,j)\,B_1^{\gamma,0}(\overline{1},(\widetilde{ij}),(\widetilde{i\widetilde{(lk})}))\,J_2^{(2)}(\lbrace p\rbrace_{2}) \nonumber\\
 \ph{39}&&-\frac{1}{2}\,\bigg[      +S^{FF}_{(\widetilde{kj})k(\widetilde{lk})}       -S^{FF}_{(\widetilde{i\widetilde{(kj})})k(\widetilde{i\widetilde{(lk})})}       -S^{FF}_{1k(\widetilde{kj})}       +S^{FF}_{1k(\widetilde{i\widetilde{(kj})})}       -S^{FF}_{1k(\widetilde{lk})}       +S^{FF}_{1k(\widetilde{i\widetilde{(lk})})}  \bigg]\nonumber\\
 &&\times d_3^{0}((\widetilde{lk}),i,(\widetilde{kj}))\,B_1^{\gamma,0}(1,(\widetilde{i\widetilde{(kj})}),(\widetilde{i\widetilde{(lk})}))\,J_2^{(2)}(\lbrace p\rbrace_{2}) \nonumber\\
 \ph{40}&&-\frac{1}{2}\,\,A_3^{0}(1,i,l)\,A_3^{0}(\overline{1},k,(\widetilde{li}))\,B_1^{\gamma,0}(\overline{\overline{1}},j,(\widetilde{k\widetilde{(li})}))\,J_2^{(2)}(\lbrace p\rbrace_{2}) \nonumber\\
 \ph{41}&&+\frac{1}{2}\,\,d_3^{0}(1,i,j)\,A_3^{0}(\overline{1},k,l)\,B_1^{\gamma,0}(\overline{\overline{1}},(\widetilde{ij}),(\widetilde{lk}))\,J_2^{(2)}(\lbrace p\rbrace_{2}) \nonumber\\
 \ph{42}&&+\frac{1}{2}\,\,d_3^{0}(l,i,j)\,A_3^{0}(1,k,(\widetilde{li}))\,B_1^{\gamma,0}(\overline{1},(\widetilde{ij}),(\widetilde{(\widetilde{li})k}))\,J_2^{(2)}(\lbrace p\rbrace_{2}) \nonumber\\
 \ph{43}&&-\frac{1}{2}\,\bigg[      -S^{FF}_{1i(\widetilde{li})}       +S^{FF}_{\overline{1}i(\widetilde{k\widetilde{(li})})}       +S^{FF}_{(\widetilde{li})i(\widetilde{ij})}       -S^{FF}_{(\widetilde{k\widetilde{(li})})i(\widetilde{ij})}       +S^{FF}_{1i(\widetilde{ij})}       -S^{FF}_{\overline{1}i(\widetilde{ij})}  \bigg]\nonumber\\
 &&\times A_3^{0}(1,k,(\widetilde{li}))\,B_1^{\gamma,0}(\overline{1},(\widetilde{ij}),(\widetilde{k\widetilde{(li})}))\,J_2^{(2)}(\lbrace p\rbrace_{2}) \nonumber\\
 \ph{44}&&-\frac{1}{2}\,\,A_3^{0}(1,k,l)\,A_3^{0}(\overline{1},i,(\widetilde{lk}))\,B_1^{\gamma,0}(\overline{\overline{1}},j,(\widetilde{i\widetilde{(lk})}))\,J_2^{(2)}(\lbrace p\rbrace_{2}) \nonumber\\
 \ph{45}&&+\frac{1}{2}\,\,d_3^{0}(1,k,j)\,A_3^{0}(\overline{1},i,l)\,B_1^{\gamma,0}(\overline{\overline{1}},(\widetilde{kj}),(\widetilde{li}))\,J_2^{(2)}(\lbrace p\rbrace_{2}) \nonumber\\
 \ph{46}&&+\frac{1}{2}\,\,d_3^{0}(l,k,j)\,A_3^{0}(1,i,(\widetilde{lk}))\,B_1^{\gamma,0}(\overline{1},(\widetilde{kj}),(\widetilde{(\widetilde{lk})i}))\,J_2^{(2)}(\lbrace p\rbrace_{2}) \nonumber\\
 \ph{47}&&-\frac{1}{2}\,\bigg[      -S^{FF}_{1k(\widetilde{lk})}       +S^{FF}_{\overline{1}k(\widetilde{i\widetilde{(lk})})}       +S^{FF}_{(\widetilde{lk})k(\widetilde{kj})}       -S^{FF}_{(\widetilde{i\widetilde{(lk})})k(\widetilde{kj})}       +S^{FF}_{1k(\widetilde{kj})}       -S^{FF}_{\overline{1}k(\widetilde{kj})}  \bigg]\nonumber\\
 &&\times A_3^{0}(1,i,(\widetilde{lk}))\,B_1^{\gamma,0}(\overline{1},(\widetilde{kj}),(\widetilde{i\widetilde{(lk})}))\,J_2^{(2)}(\lbrace p\rbrace_{2}) .
\end{eqnarray}

\addtocontents{toc}{\protect\setcounter{tocdepth}{2}}
\subsubsection{B-type $\mathcal{O}(N^0)$ contribution}
\addtocontents{toc}{\protect\setcounter{tocdepth}{3}}

The $\tilde{B}_{3}^{\gamma,0}$ matrix element contributes at subleading colour, $1/N^2$ relative to leading colour.
The interferences of colour-ordered amplitudes which form this function result in one of the gluons
behaving as if it were an abelian gauge boson, colour connected only to the quark line.
The two remaining gluons are colour connected to the quark line
but also to each other. The abelian-like gluon is always the first gluon argument in the function $\tilde{B}_{3}^{\gamma,0}$. The subtracted
contribution to the cross section is given by
\ba
\sum_{P(i,j,k)}\tilde{B}_{3}^{\gamma,0}(\hat{1},i,j,k,l)-\sum_{P(i,j,k)}\tilde{B}_{3}^{\gamma,0,S}(\hat{1},i,j,k,l).
\ea
The abelian-like nature of the gluon $i$ (first argument) in the subtraction term can be seen by the fact that there are
no divergent collinear limits for this gluon with other gluons in the subtraction term,
\begin{eqnarray}
\lefteqn{{\tilde{B}_{3}^{\gamma,0,S}(\hat{1},i,j,k,l) =}} \nonumber \\
 \ph{1}&&+A_3^{0}(1,i,l)\,B_2^{\gamma,0}(\overline{1},j,k,(\widetilde{li}))\,J_2^{(3)}(\lbrace p\rbrace_{3}) \nonumber\\
 \ph{2}&&+d_3^{0}(1,j,k)\,\tilde{B}_2^{\gamma,0}(\overline{1},i,(\widetilde{jk}),l)\,J_2^{(3)}(\lbrace p\rbrace_{3}) \nonumber\\
 \ph{3}&&+d_3^{0}(l,k,j)\,\tilde{B}_2^{\gamma,0}(1,i,(\widetilde{jk}),(\widetilde{lk}))\,J_2^{(3)}(\lbrace p\rbrace_{3}) \nonumber\\
\snip \ph{4}&&+\,A_4^0(1,j,k,l)\,B_1^{\gamma,0}(\overline{1},i,(\widetilde{jkl}))\,J_2^{(2)}(\lbrace p\rbrace_{2}) \nonumber\\
 \ph{5}&&-\,d_3^{0}(1,j,k)\,A_3^{0}(\overline{1},(\widetilde{jk}),l)\,B_1^{\gamma,0}(\overline{\overline{1}},i,(\widetilde{l\widetilde{(jk})}))\,J_2^{(2)}(\lbrace p\rbrace_{2}) \nonumber\\
 \ph{6}&&-\,d_3^{0}(l,k,j)\,A_3^{0}(1,(\widetilde{jk}),(\widetilde{lk}))\,B_1^{\gamma,0}(\overline{1},i,\widetilde{(\widetilde{jk})(\widetilde{lk})})\,J_2^{(2)}(\lbrace p\rbrace_{2}) \nonumber\\
\snip \ph{7}&&+\,\tilde{A}_4^0(1,i,j,l)\,B_1^{\gamma,0}(\overline{1},k,(\widetilde{ijl}))\,J_2^{(2)}(\lbrace p\rbrace_{2}) \nonumber\\
 \ph{8}&&-\,A_3^{0}(1,i,l)\,A_3^{0}(\overline{1},j,(\widetilde{li}))\,B_1^{\gamma,0}(\overline{\overline{1}},k,(\widetilde{j\widetilde{(li})}))\,J_2^{(2)}(\lbrace p\rbrace_{3}) \nonumber\\
 \ph{9}&&-\,A_3^{0}(1,j,l)\,A_3^{0}(\overline{1},i,(\widetilde{lj}))\,B_1^{\gamma,0}(\overline{\overline{1}},k,(\widetilde{i\widetilde{(lj})}))\,J_2^{(2)}(\lbrace p\rbrace_{3}) \nonumber\\
 \ph{10}&&+\,A_3^{0}(1,j,l)\,A_3^{0}(\overline{1},i,(\widetilde{lj}))\,B_1^{\gamma,0}(\overline{\overline{1}},k,(\widetilde{i\widetilde{(lj})}))\,J_2^{(2)}(\lbrace p\rbrace_{2}) \nonumber\\
 \ph{11}&&-\,d_3^{0}(1,j,k)\,A_3^{0}(\overline{1},i,l)\,B_1^{\gamma,0}(\overline{\overline{1}},(\widetilde{jk}),(\widetilde{li}))\,J_2^{(2)}(\lbrace p\rbrace_{2}) \nonumber\\
 \ph{12}&&-\,d_3^{0}(l,k,j)\,A_3^{0}(1,i,(\widetilde{lk}))\,B_1^{\gamma,0}(\overline{1},(\widetilde{kj}),(\widetilde{i\widetilde{(lk})}))\,J_2^{(2)}(\lbrace p\rbrace_{2}) \nonumber\\
 \ph{13}&&-\bigg[      +S^{FF}_{1j(\widetilde{lj})}       -S^{FF}_{\overline{1}j(\widetilde{i\widetilde{(lj})})}       -S^{FF}_{1j(\widetilde{kj})}       +S^{FF}_{\overline{1}j(\widetilde{kj})}       -S^{FF}_{(\widetilde{lj})j(\widetilde{kj})}       +S^{FF}_{(\widetilde{i\widetilde{(lj})})j(\widetilde{kj})}  \bigg]\nonumber\\
 &&\times A_3^{0}(1,i,(\widetilde{lj}))\,B_1^{\gamma,0}(\overline{1},(\widetilde{kj}),(\widetilde{i\widetilde{(lj})}))\,J_2^{(2)}(\lbrace p\rbrace_{2}) .
\end{eqnarray}

\addtocontents{toc}{\protect\setcounter{tocdepth}{2}}
\subsubsection{B-type $\mathcal{O}(N^{-2})$ contribution}
\addtocontents{toc}{\protect\setcounter{tocdepth}{3}}
The $\tilde{\tilde{B}}_{3}^{\gamma,0}$ matrix element contributes to two subleading colour factors, as can be seen in Tab.~\ref{tab:channelsq}.
This function is symmetrised over all gluon momenta and therefore behaves as if all three gluons are abelian. This restricts the unresolved limits
of the final-state gluons to QED-like factorization. The subtracted cross section is given by
\ba
 \tilde{\tilde{B}}_{3}^{\gamma,0}(\hat{1},3,4,5,2)- \tilde{\tilde{B}}_{3}^{\gamma,0,S}(\hat{1},3,4,5,2) ,
\ea
where
\begin{eqnarray}
\lefteqn{{ \tilde{\tilde{B}}_{3}^{\gamma,0,S}(\hat{1},i,j,k,l)=}}\nonumber \\
 \ph{1}&&+A_3^{0}(1,i,l)\,\tilde{B}_2^{\gamma,0}(\overline{1},j,k,(\widetilde{li}))\,J_2^{(3)}(\lbrace p\rbrace_{3}) \nonumber\\
 \ph{2}&&+A_3^{0}(1,j,l)\,\tilde{B}_2^{\gamma,0}(\overline{1},i,k,(\widetilde{lj}))\,J_2^{(3)}(\lbrace p\rbrace_{3}) \nonumber\\
 \ph{3}&&+A_3^{0}(1,k,l)\,\tilde{B}_2^{\gamma,0}(\overline{1},i,j,(\widetilde{lk}))\,J_2^{(3)}(\lbrace p\rbrace_{3}) \nonumber\\
\snip \ph{4}&&+\,\tilde{A}_4^0(1,i,j,l)\,B_1^{\gamma,0}(\overline{1},k,(\widetilde{lij}))\,J_2^{(2)}(\lbrace p\rbrace_{2}) \nonumber\\
 \ph{5}&&-\,A_3^{0}(1,i,l)\,A_3^{0}(\overline{1},j,(\widetilde{li}))\,B_1^{\gamma,0}(\overline{\overline{1}},k,(\widetilde{j\widetilde{(li})}))\,J_2^{(2)}(\lbrace p\rbrace_{2}) \nonumber\\
 \ph{6}&&-\,A_3^{0}(1,j,l)\,A_3^{0}(\overline{1},i,(\widetilde{lj}))\,B_1^{\gamma,0}(\overline{\overline{1}},k,(\widetilde{i\widetilde{(lj})}))\,J_2^{(2)}(\lbrace p\rbrace_{2}) \nonumber\\
\snip \ph{7}&&+\,\tilde{A}_4^0(1,i,k,l)\,B_1^{\gamma,0}(\overline{1},j,(\widetilde{lik}))\,J_2^{(2)}(\lbrace p\rbrace_{2}) \nonumber\\
 \ph{8}&&-\,A_3^{0}(1,i,l)\,A_3^{0}(\overline{1},k,(\widetilde{li}))\,B_1^{\gamma,0}(\overline{\overline{1}},j,(\widetilde{k\widetilde{(li})}))\,J_2^{(2)}(\lbrace p\rbrace_{2}) \nonumber\\
 \ph{9}&&-\,A_3^{0}(1,k,l)\,A_3^{0}(\overline{1},i,(\widetilde{lk}))\,B_1^{\gamma,0}(\overline{\overline{1}},j,(\widetilde{i\widetilde{(lk})}))\,J_2^{(2)}(\lbrace p\rbrace_{2}) \nonumber\\
\snip \ph{10}&&+\,\tilde{A}_4^0(1,j,k,l)\,B_1^{\gamma,0}(\overline{1},i,(\widetilde{ljk}))\,J_2^{(2)}(\lbrace p\rbrace_{2}) \nonumber\\
 \ph{11}&&-\,A_3^{0}(1,j,l)\,A_3^{0}(\overline{1},k,(\widetilde{lj}))\,B_1^{\gamma,0}(\overline{\overline{1}},i,(\widetilde{k\widetilde{(lj})}))\,J_2^{(2)}(\lbrace p\rbrace_{2}) \nonumber\\
 \ph{12}&&-\,A_3^{0}(1,k,l)\,A_3^{0}(\overline{1},j,(\widetilde{lk}))\,B_1^{\gamma,0}(\overline{\overline{1}},i,(\widetilde{j\widetilde{(lk})}))\,J_2^{(2)}(\lbrace p\rbrace_{2}) .
\end{eqnarray}

\addtocontents{toc}{\protect\setcounter{tocdepth}{2}}
\subsubsection{C-type $\mathcal{O}(N_FN^1)$ contribution}
\addtocontents{toc}{\protect\setcounter{tocdepth}{3}}
\label{sec:nnloRRC}

The four-quark--one-gluon matrix element contains contributions from identical and non-identical
quark flavours for the two quark strings. As was the case at NLO in Secs.~\ref{sec:qnloRC}--\ref{sec:qnloRD}
the full four-quark contribution can be separated into a term derived from the non-identical flavour matrix elements,
given by the $C_{1}^{\gamma,0}$ functions, and a remainder of interference terms given by the $D_{1}^{\gamma,0}$ functions.

The $C_{1}^{\gamma,0}$ terms include both colour orderings and the vector boson coupling to both quark lines. The subtracted
contribution to the cross section is given by
\ba
\frac{1}{2}\sum_{P(3,4)} \left[ C_{1}^{\gamma,0}(\hat{1},5;4,3;2)-C_{1}^{\gamma,0,S}(\hat{1},5;4,3;2) \right],
\ea
where
\begin{eqnarray}
\lefteqn{{C_{1}^{\gamma,0,S}(\hat{1},i;j,k;l) =}} \nonumber \\
 \ph{1}&&+A_3^{0}(1,i,j)\,C_0^{\gamma,0}(\overline{1};(\widetilde{ji}),k;l)\,J_2^{(3)}(\lbrace p\rbrace_{3}) \nonumber\\
 \ph{2}&&+A_3^{0}(k,i,l)\,C_0^{\gamma,0}(1;j,(\widetilde{ki});(\widetilde{li}))\,J_2^{(3)}(\lbrace p\rbrace_{3}) \nonumber\\
 \ph{3}&&+E_3^{0}(l,j,k)\,B_{2,q}^{\gamma,0}(1,(\widetilde{jk}),i,(\widetilde{lj}))\,J_2^{(3)}(\lbrace p\rbrace_{3}) \nonumber\\
 \ph{4}&&+E_3^{0}(l,j,k)\,B_{2,q}^{\gamma,0}(1,i,(\widetilde{jk}),(\widetilde{lj}))\,J_2^{(3)}(\lbrace p\rbrace_{3}) \nonumber\\
\snip \ph{5}&&+\,E_4^{0,a}(l,j,k,i)\,B_{1,q}^{\gamma,0}(1,(\widetilde{ikj}),(\widetilde{ljk}))\,J_2^{(2)}(\lbrace p\rbrace_{2}) \nonumber\\
\snip \ph{6}&&+\,E_4^{0,b}(l,i,k,j)\,B_{1,q}^{\gamma,0}(1,(\widetilde{jki}),(\widetilde{lik}))\,J_2^{(2)}(\lbrace p\rbrace_{2}) \nonumber\\
 \ph{7}&&-\,E_3^{0}(l,j,k)\,d_3^{0}((\widetilde{lj}),i,(\widetilde{jk}))\,B_{1,q}^{\gamma,0}(1,(\widetilde{(\widetilde{jk})i}),(\widetilde{(\widetilde{lj})i}))\,J_2^{(2)}(\lbrace p\rbrace_{2}) \nonumber\\
 \ph{8}&&-\,E_3^{0}(l,j,k)\,d_3^{0}((\widetilde{lj}),(\widetilde{jk}),i)\,B_{1,q}^{\gamma,0}(1,(\widetilde{i\widetilde{(kj})}),\widetilde{(\widetilde{lj})(\widetilde{jk})})\,J_2^{(2)}(\lbrace p\rbrace_{2}) \nonumber\\
 \ph{9}&&-\,A_3^{0}(k,i,l)\,E_3^{0}((\widetilde{li}),(\widetilde{ki}),j)\,B_{1,q}^{\gamma,0}(1,(\widetilde{j\widetilde{(ki})}),\widetilde{(\widetilde{ki})(\widetilde{li})})\,J_2^{(2)}(\lbrace p\rbrace_{2}) \nonumber\\
\snip \ph{10}&&+\,E_4^0(1,k,j,i)\,B_{1,q}^{\gamma,0}(\overline{1},(\widetilde{ijk}),l)\,J_2^{(2)}(\lbrace p\rbrace_{2}) \nonumber\\
 \ph{11}&&-\,E_3^{0}(l,j,k)\,D_3^{0}(1,(\widetilde{kj}),i)\,B_{1,q}^{\gamma,0}(\overline{1},(\widetilde{i\widetilde{(kj})}),(\widetilde{lj}))\,J_2^{(2)}(\lbrace p\rbrace_{2}) \nonumber\\
 \ph{12}&&-\,A_3^{0}(1,i,j)\,E_3^{0}(\overline{1},k,(\widetilde{ji}))\,B_{1,q}^{\gamma,0}(\overline{\overline{1}},(\widetilde{k\widetilde{(ji})}),l)\,J_2^{(2)}(\lbrace p\rbrace_{2}) \nonumber\\
 \ph{13}&&-E_{3,q^\prime\to g}^{0}(k,1,l)\,B_{2,Q}^{\gamma,0}((\widetilde{kl}),\overline{1},i,j)\,J_2^{(3)}(\lbrace p\rbrace_{3}) \nonumber\\
 \ph{14}&&-E_{3,q^\prime\to g}^{0}(k,1,l)\,B_{2,Q}^{\gamma,0}((\widetilde{kl}),i,\overline{1},j)\,J_2^{(3)}(\lbrace p\rbrace_{3}) \nonumber\\
\snip \ph{15}&&-\,E_4^0(k,l,1,i)\,B_{1,Q}^{\gamma,0}((\widetilde{kli}),\overline{1},j)\,J_2^{(2)}(\lbrace p\rbrace_{2}) \nonumber\\
 \ph{16}&&+\,E_{3,q^\prime\to g}^{0}(k,1,l)\,D_3^{0}((\widetilde{kl}),\overline{1},i)\,B_{1,Q}^{\gamma,0}((\widetilde{(\widetilde{kl})i}),\overline{\overline{1}},j)\,J_2^{(2)}(\lbrace p\rbrace_{2}) \nonumber\\
\snip \ph{17}&&-\,E_4^0(k,1,l,i)\,B_{1,Q}^{\gamma,0}((\widetilde{kli}),\overline{1},j)\,J_2^{(2)}(\lbrace p\rbrace_{2}) \nonumber\\
 \ph{18}&&+\,E_{3,q^\prime\to g}^{0}(k,l,1)\,D_3^{0}((\widetilde{kl}),\overline{1},i)\,B_{1,Q}^{\gamma,0}((\widetilde{(\widetilde{lk})i}),\overline{\overline{1}},j)\,J_2^{(2)}(\lbrace p\rbrace_{2}) \nonumber\\
 \ph{19}&&+\,A_3^{0}(l,i,k)\,E_{3,q^\prime\to g}^{0}((\widetilde{ki}),1,(\widetilde{li}))\,B_{1,Q}^{\gamma,0}(\widetilde{(\widetilde{ki})(\widetilde{li})},\overline{1},j)\,J_2^{(2)}(\lbrace p\rbrace_{2}) \nonumber\\
\snip \ph{20}&&+\,B_4^0(k,1,l,j)\,B_{1,Q}^{\gamma,0}(i,\overline{1},(\widetilde{jlk}))\,J_2^{(2)}(\lbrace p\rbrace_{2}) \nonumber\\
 \ph{21}&&-2\,E_{3,q^\prime\to g}^{0}(k,1,l)\,a_{3,g\to q}^{0}((\widetilde{kl}),\overline{1},j)\,B_{1,Q}^{\gamma,0}(i,\overline{\overline{1}},(\widetilde{j\widetilde{(kl})}))\,J_2^{(2)}(\lbrace p\rbrace_{2}) \nonumber\\
\snip \ph{22}&&+\,B_4^0(k,1,l,j)\,\bar{B}_{1,Q}^{\gamma,0}((\widetilde{jlk}),i,\overline{1})\,J_2^{(2)}(\lbrace p\rbrace_{2}) \nonumber\\
 \ph{23}&&-\,E_{3,q^\prime\to g}^{0}(k,l,1)\,a_{3,g\to q}^{0}(j,\overline{1},(\widetilde{kl}))\,B_{1,Q}^{\gamma,0}((\widetilde{j\widetilde{(kl})}),i,\overline{\overline{1}})\,J_2^{(2)}(\lbrace p\rbrace_{2}) \nonumber\\
 \ph{24}&&-\,E_{3,q^\prime\to g}^{0}(k,1,l)\,a_{3,g\to q}^{0}((\widetilde{kl}),\overline{1},j)\,B_{1,Q}^{\gamma,0}(\overline{\overline{1}},i,(\widetilde{j\widetilde{(kl})}))\,J_2^{(2)}(\lbrace p\rbrace_{2}) \nonumber\\
 \ph{25}&&+\,A_{3,q}^{0}(1,i,j)\,E_{3,q^\prime\to g}^{0}(k,\overline{1},l)\,B_{1,Q}^{\gamma,0}((\widetilde{kl}),\overline{\overline{1}},(\widetilde{ij}))\,J_2^{(2)}(\lbrace p\rbrace_{2}) \nonumber\\
 \ph{26}&&+\,A_3^{0}(k,i,j)\,E_{3,q^\prime\to g}^{0}((\widetilde{ki}),1,l)\,B_{1,Q}^{\gamma,0}((\widetilde{(\widetilde{ki})l}),\overline{1},(\widetilde{ji}))\,J_2^{(2)}(\lbrace p\rbrace_{2}) \nonumber\\
 \ph{27}&&-\,E_{3,q^\prime\to g}^{0}(k,1,l)\,A_3^{0}((\widetilde{kl}),i,j)\,B_{1,Q}^{\gamma,0}((\widetilde{(\widetilde{kl})i}),\overline{1},(\widetilde{ij}))\,J_2^{(2)}(\lbrace p\rbrace_{2}) \nonumber\\
 \ph{28}&&-\bigg[      +S^{FF}_{\overline{1}i\widetilde{(\widetilde{ki})(\widetilde{li})}}       +S^{FF}_{(\widetilde{ki})ij}       -S^{FF}_{\widetilde{(\widetilde{ki})(\widetilde{li})}ij}       -S^{FF}_{1i(\widetilde{ki})}       -S^{FF}_{ji\overline{1}}       +S^{FF}_{1ij}  \bigg]\nonumber\\
 &&\times E_{3,q^\prime\to g}^{0}((\widetilde{ki}),1,(\widetilde{li}))\,B_{1,Q}^{\gamma,0}(\widetilde{(\widetilde{ki})(\widetilde{li})},\overline{1},j)\,J_2^{(2)}(\lbrace p\rbrace_{2}) .
\end{eqnarray}

In this subtraction term we once again retain information on the flavour of the quark line in the reduced matrix element and employ a symmetrisation
over the quark and antiquark for some reduced matrix elements indicated by a $\bar{B}$.

\addtocontents{toc}{\protect\setcounter{tocdepth}{2}}
\subsubsection{C-type $\mathcal{O}(N_F N^{-1})$ contribution}
\addtocontents{toc}{\protect\setcounter{tocdepth}{3}}
\label{sec:nnloRRCt}

The subleading-colour contribution to four-quark--one-gluon scattering can also be split into a term derived from non-identical
flavour matrix elements, $\tilde{C}_{1}^{\gamma,0}$ terms,  and a left-over set of interferences, $\tilde{D}_{1}^{\gamma,0}$ terms.
The $\tilde{C}_{1}^{\gamma,0}$ terms contain all colour orderings and vector boson couplings to both quark lines. The subtracted contribution
to the cross section is given by
\ba
\frac{1}{2}\sum_{P(3,4)}\left[\tilde{C}_{1}^{\gamma,0}(\hat{1},5;4,3;2)-\tilde{C}_{1}^{\gamma,0,S}(\hat{1},5;4,3;2)\right],
\ea
where
\begin{eqnarray}
\lefteqn{{\tilde{C}_{1}^{\gamma,0,S}(\hat{1},i;j,k;l)=}} \nonumber \\
 \ph{1}&&+A_3^{0}(1,i,l)\,C_0^{\gamma,0}(\overline{1};j,k;(\widetilde{il}))\,J_2^{(3)}(\lbrace p\rbrace_{3}) \nonumber\\
 \ph{2}&&+A_3^{0}(j,i,k)\,C_0^{\gamma,0}(1;(\widetilde{ji}),(\widetilde{ki});l)\,J_2^{(3)}(\lbrace p\rbrace_{3}) \nonumber\\
 \ph{3}&&-2A_3^{0}(1,i,k)\,C_0^{\gamma,0}(\overline{1};j,(\widetilde{ki});l)\,J_2^{(3)}(\lbrace p\rbrace_{3}) \nonumber\\
 \ph{4}&&-2A_3^{0}(j,i,l)\,C_0^{\gamma,0}(1;(\widetilde{ji}),k;(\widetilde{li}))\,J_2^{(3)}(\lbrace p\rbrace_{3}) \nonumber\\
 \ph{5}&&+2A_3^{0}(1,i,j)\,C_0^{\gamma,0}(\overline{1};(\widetilde{ji}),k;l)\,J_2^{(3)}(\lbrace p\rbrace_{3}) \nonumber\\
 \ph{6}&&+2A_3^{0}(l,i,k)\,C_0^{\gamma,0}(1,j,(\widetilde{ki}),(\widetilde{li}))\,J_2^{(3)}(\lbrace p\rbrace_{3}) \nonumber\\
 \ph{7}&&+E_3^{0}(l,j,k)\,\tilde{B}_{2,q}^{\gamma,0}(1,i,(\widetilde{kj}),(\widetilde{lj}))\,J_2^{(3)}(\lbrace p\rbrace_{3}) \nonumber\\
\snip \ph{8}&&+\,B_4^0(1,j,k,l)\,B_{1,q}^{\gamma,0}(\overline{1},i,(\widetilde{jkl}))\,J_2^{(2)}(\lbrace p\rbrace_{2}) \nonumber\\
 \ph{9}&&-\,E_3^{0}(l,j,k)\,A_3^{0}(1,(\widetilde{kj}),(\widetilde{lj}))\,B_{1,q}^{\gamma,0}(\overline{1},i,\widetilde{(\widetilde{kj})(\widetilde{lj})})\,J_2^{(2)}(\lbrace p\rbrace_{2}) \nonumber\\
\snip \ph{10}&&+\frac{1}{2}\,\,\tilde{E}_4^0(1,j,i,k)\,B_{1,q}^{\gamma,0}(\overline{1},(\widetilde{ijk}),l)\,J_2^{(2)}(\lbrace p\rbrace_{2}) \nonumber\\
 \ph{11}&&-\frac{1}{2}\,\,A_3^{0}(j,i,k)\,E_{3,q}^{0}(1,(\widetilde{ji}),(\widetilde{ki}))\,B_{1,q}^{\gamma,0}(\overline{1},\widetilde{(\widetilde{ji})(\widetilde{ki})},l)\,J_2^{(2)}(\lbrace p\rbrace_{2}) \nonumber\\
\snip \ph{12}&&+\frac{1}{2}\,\,\tilde{E}_4^0(l,j,i,k)\,B_{1,q}^{\gamma,0}(1,(\widetilde{kij}),(\widetilde{lji}))\,J_2^{(2)}(\lbrace p\rbrace_{2}) \nonumber\\
 \ph{13}&&-\frac{1}{2}\,\,A_3^{0}(j,i,k)\,E_3^{0}(l,(\widetilde{ji}),(\widetilde{ki}))\,B_{1,q}^{\gamma,0}(1,\widetilde{(\widetilde{ji})(\widetilde{ki})},(\widetilde{l\widetilde{(ji})}))\,J_2^{(2)}(\lbrace p\rbrace_{2}) \nonumber\\
 \ph{14}&&-\,E_3^{0}(l,j,k)\,A_3^{0}(1,i,(\widetilde{lj}))\,B_{1,q}^{\gamma,0}(\overline{1},(\widetilde{jk}),(\widetilde{i\widetilde{(lj})}))\,J_2^{(2)}(\lbrace p\rbrace_{2}) \nonumber\\
 \ph{15}&&+\,A_3^{0}(1,i,k)\,E_3^{0}(\overline{1},(\widetilde{ik}),j)\,B_{1,q}^{\gamma,0}(\overline{\overline{1}},(\widetilde{j\widetilde{(ik})}),l)\,J_2^{(2)}(\lbrace p\rbrace_{2}) \nonumber\\
 \ph{16}&&+\,A_3^{0}(j,i,l)\,E_3^{0}(1,k,(\widetilde{ji}))\,B_{1,q}^{\gamma,0}(\overline{1},(\widetilde{k\widetilde{(ji})}),(\widetilde{li}))\,J_2^{(2)}(\lbrace p\rbrace_{2}) \nonumber\\
 \ph{17}&&-\,A_3^{0}(1,i,j)\,E_3^{0}(\overline{1},k,(\widetilde{ji}))\,B_{1,q}^{\gamma,0}(\overline{\overline{1}},(\widetilde{k\widetilde{(ji})}),l)\,J_2^{(2)}(\lbrace p\rbrace_{2}) \nonumber\\
 \ph{18}&&-\,A_3^{0}(l,i,k)\,E_3^{0}(1,(\widetilde{ki}),j)\,B_{1,q}^{\gamma,0}(\overline{1},(\widetilde{(\widetilde{ki})j}),(\widetilde{li}))\,J_2^{(2)}(\lbrace p\rbrace_{2}) \nonumber\\
 \ph{19}&&-\bigg[      +S^{FF}_{1i(\widetilde{ki})}       +S^{FF}_{(\widetilde{ji})il}       -S^{FF}_{1i(\widetilde{ji})}       -S^{FF}_{li(\widetilde{ik})}  \bigg]\nonumber\\
 &&\times E_3^{0}(1,(\widetilde{ki}),(\widetilde{ji}))\,B_{1,q}^{\gamma,0}(\overline{1},\widetilde{(\widetilde{ki})(\widetilde{ji})},l)\,J_2^{(2)}(\lbrace p\rbrace_{2}) \nonumber\\
 \ph{20}&&+\,A_3^{0}(1,i,k)\,E_3^{0}(l,(\widetilde{ik}),j)\,B_{1,q}^{\gamma,0}(\overline{1},(\widetilde{j\widetilde{(ik})}),(\widetilde{l\widetilde{(ik})}))\,J_2^{(2)}(\lbrace p\rbrace_{2}) \nonumber\\
 \ph{21}&&+\,A_3^{0}(j,i,l)\,E_3^{0}((\widetilde{li}),k,(\widetilde{ji}))\,B_{1,q}^{\gamma,0}(1,(\widetilde{k\widetilde{(ji})}),(\widetilde{(\widetilde{li})k}))\,J_2^{(2)}(\lbrace p\rbrace_{2}) \nonumber\\
 \ph{22}&&-\,A_3^{0}(1,i,j)\,E_3^{0}(l,k,(\widetilde{ji}))\,B_{1,q}^{\gamma,0}(\overline{1},(\widetilde{k\widetilde{(ji})}),(\widetilde{lk}))\,J_2^{(2)}(\lbrace p\rbrace_{2}) \nonumber\\
 \ph{23}&&-\,A_3^{0}(l,i,k)\,E_3^{0}((\widetilde{li}),(\widetilde{ki}),j)\,B_{1,q}^{\gamma,0}(1,(\widetilde{(\widetilde{ki})j}),\widetilde{(\widetilde{li})(\widetilde{ki})})\,J_2^{(2)}(\lbrace p\rbrace_{2}) \nonumber\\
 \ph{24}&&-\bigg[      +S^{FF}_{1i(\widetilde{ki})}       +S^{FF}_{(\widetilde{ji})il}       -S^{FF}_{1i(\widetilde{ji})}       -S^{FF}_{li(\widetilde{ik})}  \bigg]\nonumber\\
 &&\times E_3^{0}(l,(\widetilde{ki}),(\widetilde{ji}))\,B_{1,q}^{\gamma,0}(1,\widetilde{(\widetilde{ki})(\widetilde{ji})},(\widetilde{l\widetilde{(ki})}))\,J_2^{(2)}(\lbrace p\rbrace_{2}) \nonumber\\
 \ph{25}&&-\frac{1}{2}\,E_{3,q^\prime\to g}^{0}(j,1,l)\,\tilde{\bar{B}}_{2,Q}^{\gamma,0}(k,\overline{1},i,(\widetilde{jl}))\,J_2^{(3)}(\lbrace p\rbrace_{3}) \nonumber\\
 \ph{26}&&-\frac{1}{2}\,E_{3,q^\prime\to g}^{0}(k,1,l)\,\tilde{\bar{B}}_{2,Q}^{\gamma,0}((\widetilde{kl}),i,\overline{1},j)\,J_2^{(3)}(\lbrace p\rbrace_{3}) \nonumber\\
\snip \ph{27}&&+\,B_4^0(j,1,l,k)\,\bar{B}_{1,Q}^{\gamma,0}(\overline{1},i,(\widetilde{jlk}))\,J_2^{(2)}(\lbrace p\rbrace_{2}) \nonumber\\
 \ph{28}&&-\frac{1}{2}\,\,E_{3,q^\prime\to g}^{0}(j,1,l)\,A_3^{0}((\widetilde{jl}),\overline{1},k)\,\bar{B}_{1,Q}^{\gamma,0}(\overline{\overline{1}},i,(\widetilde{(\widetilde{jl})k}))\,J_2^{(2)}(\lbrace p\rbrace_{2}) \nonumber\\
 \ph{29}&&-\frac{1}{2}\,\,E_{3,q^\prime\to g}^{0}(k,1,l)\,A_3^{0}((\widetilde{kl}),\overline{1},j)\,\bar{B}_{1,Q}^{\gamma,0}(\overline{\overline{1}},i,(\widetilde{(\widetilde{kl})j}))\,J_2^{(2)}(\lbrace p\rbrace_{2}) \nonumber\\
\snip \ph{30}&&-\frac{1}{2}\,\,\tilde{E}_4^0(j,1,i,l)\,B_{1,Q}^{\gamma,0}(k,\overline{1},(\widetilde{jil}))\,J_2^{(2)}(\lbrace p\rbrace_{2}) \nonumber\\
 \ph{31}&&+\frac{1}{2}\,\,A_3^{0}(1,i,l)\,E_{3,q^\prime\to g}^{0}(j,\overline{1},(\widetilde{il}))\,B_{1,Q}^{\gamma,0}(k,\overline{\overline{1}},(\widetilde{j\widetilde{(il})}))\,J_2^{(2)}(\lbrace p\rbrace_{2}) \nonumber\\
\snip \ph{32}&&-\frac{1}{2}\,\,\tilde{E}_4^0(k,1,i,l)\,B_{1,Q}^{\gamma,0}((\widetilde{kil}),\overline{1},j)\,J_2^{(2)}(\lbrace p\rbrace_{2}) \nonumber\\
 \ph{33}&&+\frac{1}{2}\,\,A_3^{0}(1,i,l)\,E_{3,q^\prime\to g}^{0}(k,\overline{1},(\widetilde{il}))\,B_{1,Q}^{\gamma,0}((\widetilde{k\widetilde{(il})}),\overline{\overline{1}},j)\,J_2^{(2)}(\lbrace p\rbrace_{2}) \nonumber\\
 \ph{34}&&+\frac{1}{2}\,\,E_{3,q^\prime\to g}^{0}(j,1,l)\,A_3^{0}((\widetilde{jl}),i,k)\,B_{1,Q}^{\gamma,0}((\widetilde{ki}),\overline{1},(\widetilde{(\widetilde{jl})i}))\,J_2^{(2)}(\lbrace p\rbrace_{2}) \nonumber\\
 \ph{35}&&+\frac{1}{2}\,\,E_{3,q^\prime\to g}^{0}(k,1,l)\,A_3^{0}((\widetilde{kl}),i,j)\,B_{1,Q}^{\gamma,0}((\widetilde{(\widetilde{kl})i}),\overline{1},(\widetilde{ij}))\,J_2^{(2)}(\lbrace p\rbrace_{2}) \nonumber\\
 \ph{36}&&-\,A_3^{0}(1,i,k)\,E_3^{0}((\widetilde{ik}),\overline{1},l)\,B_{1,Q}^{\gamma,0}((\widetilde{l\widetilde{(ik})}),\overline{\overline{1}},j)\,J_2^{(2)}(\lbrace p\rbrace_{2}) \nonumber\\
 \ph{37}&&-\,A_3^{0}(j,i,l)\,E_3^{0}(k,1,(\widetilde{li}))\,B_{1,Q}^{\gamma,0}((\widetilde{k\widetilde{(li})}),\overline{1},(\widetilde{ji}))\,J_2^{(2)}(\lbrace p\rbrace_{2}) \nonumber\\
 \ph{38}&&+\,A_3^{0}(1,i,j)\,E_3^{0}(k,\overline{1},l)\,B_{1,Q}^{\gamma,0}((\widetilde{lk}),\overline{\overline{1}},(\widetilde{ji}))\,J_2^{(2)}(\lbrace p\rbrace_{2}) \nonumber\\
 \ph{39}&&+\,A_3^{0}(l,i,k)\,E_3^{0}((\widetilde{ki}),1,(\widetilde{li}))\,B_{1,Q}^{\gamma,0}(\widetilde{(\widetilde{ki})(\widetilde{li})},\overline{1},j)\,J_2^{(2)}(\lbrace p\rbrace_{2}) \nonumber\\
 \ph{40}&&+\bigg[      +S^{FF}_{1i(\widetilde{ik})}       +S^{FF}_{ji(\widetilde{li})}       -S^{FF}_{1ij}       -S^{FF}_{(\widetilde{li})i(\widetilde{ik})}  \bigg]\nonumber\\
 &&\times E_3^{0}((\widetilde{ik}),1,(\widetilde{li}))\,B_{1,Q}^{\gamma,0}(\widetilde{(\widetilde{ik})(\widetilde{li})},\overline{1},j)\,J_2^{(2)}(\lbrace p\rbrace_{2}) \nonumber\\
 \ph{41}&&-\,A_3^{0}(1,i,k)\,E_3^{0}(j,\overline{1},l)\,B_{1,Q}^{\gamma,0}((\widetilde{ki}),\overline{\overline{1}},(\widetilde{lj}))\,J_2^{(2)}(\lbrace p\rbrace_{2}) \nonumber\\
 \ph{42}&&-\,A_3^{0}(j,i,l)\,E_3^{0}((\widetilde{ji}),1,(\widetilde{li}))\,B_{1,Q}^{\gamma,0}(k,\overline{1},\widetilde{(\widetilde{ji})(\widetilde{li})})\,J_2^{(2)}(\lbrace p\rbrace_{2}) \nonumber\\
 \ph{43}&&+\,A_3^{0}(1,i,j)\,E_3^{0}((\widetilde{ji}),\overline{1},l)\,B_{1,Q}^{\gamma,0}(k,\overline{\overline{1}},(\widetilde{l\widetilde{(ji})}))\,J_2^{(2)}(\lbrace p\rbrace_{2}) \nonumber\\
 \ph{44}&&+\,A_3^{0}(l,i,k)\,E_3^{0}(j,1,(\widetilde{li}))\,B_{1,Q}^{\gamma,0}((\widetilde{ki}),\overline{1},(\widetilde{j\widetilde{(li})}))\,J_2^{(2)}(\lbrace p\rbrace_{2}) \nonumber\\
 \ph{45}&&-\bigg[      +S^{FF}_{1i(\widetilde{ij})}       +S^{FF}_{ki(\widetilde{li})}       -S^{FF}_{1ik}       -S^{FF}_{(\widetilde{li})i(\widetilde{ij})}  \bigg]\nonumber\\
 &&\times E_3^{0}((\widetilde{ij}),1,(\widetilde{li}))\,B_{1,Q}^{\gamma,0}(k,\overline{1},\widetilde{(\widetilde{ij})(\widetilde{li})})\,J_2^{(2)}(\lbrace p\rbrace_{2}) .
\end{eqnarray}

\addtocontents{toc}{\protect\setcounter{tocdepth}{2}}
\subsubsection{D-type $\mathcal{O}(N^0)$ contribution}
\addtocontents{toc}{\protect\setcounter{tocdepth}{3}}
\label{sec:nnloRRD}

The interferences of quark orderings arising from the identical flavour contribution to four-quark--one-gluon
scattering gives the subtracted contribution to the cross section,
\ba
D_{1}^{\gamma,0}(\hat{1},5,2;3,4)-D_{1}^{\gamma,0,S}(\hat{1},5,2;3,4),
\ea
where
\begin{eqnarray}
\lefteqn{{D_{1}^{\gamma,0,S}(\hat{1},i,j;k,l) =}} \nonumber \\
 \ph{1}&&+A_3^{0}(1,i,k)\,D_0^{\gamma,0}(\overline{1},j;(\widetilde{ik}),l)\,J_2^{(3)}(\lbrace p\rbrace_{3}) \nonumber\\
 \ph{2}&&+A_3^{0}(j,i,l)\,D_0^{\gamma,0}(1,(\widetilde{ji});k,(\widetilde{li}))\,J_2^{(3)}(\lbrace p\rbrace_{3}) \nonumber\\
 \snip \ph{3}&&+2\,C_4^0(1,k,j,l)\,B_{1,q}^{\gamma,0}(\overline{1},i,(\widetilde{jkl}))\,J_2^{(2)}(\lbrace p\rbrace_{2}) \nonumber\\
 \ph{4}&&+2\,C_4^0(k,1,j,l)\,B_{1,q}^{\gamma,0}((\widetilde{jkl}),i,\overline{1})\,J_2^{(2)}(\lbrace p\rbrace_{2}) \nonumber\\
 \ph{5}&&+2\,C_4^0(l,j,k,1)\,B_{1,q}^{\gamma,0}(\overline{1},i,(\widetilde{jkl}))\,J_2^{(2)}(\lbrace p\rbrace_{2}) \nonumber\\
 \ph{6}&&+2\,C_4^0(j,l,k,1)\,B_{1,q}^{\gamma,0}(\overline{1},i,(\widetilde{jkl}))\,J_2^{(2)}(\lbrace p\rbrace_{2}) .
\end{eqnarray}

\addtocontents{toc}{\protect\setcounter{tocdepth}{2}}
\subsubsection{D-type $\mathcal{O}(N^{-2})$ contribution}
\addtocontents{toc}{\protect\setcounter{tocdepth}{3}}
\label{sec:nnloRRDt}

The interferences of quark orderings also contribute to the subleading colour and require subtraction,
\ba
\tilde{D}_{1}^{\gamma,0}(\hat{1},5,2;3,4)-\tilde{D}_{1}^{\gamma,0,S}(\hat{1},5,2;3,4),
\ea
where the subtraction term is given by
\begin{eqnarray}
\lefteqn{{\tilde{D}_{1}^{\gamma,0,S}(\hat{1},i,j;k,l) =}}\nonumber \\
 \ph{1}&&-A_3^{0}(1,i,k)\,D_0^{\gamma,0}(\overline{1},j;(\widetilde{ik}),l)\,J_2^{(3)}(\lbrace p\rbrace_{3}) \nonumber\\
 \ph{2}&&+A_3^{0}(1,i,l)\,D_0^{\gamma,0}(\overline{1},j;k,(\widetilde{il}))\,J_2^{(3)}(\lbrace p\rbrace_{3}) \nonumber\\
 \ph{3}&&+A_3^{0}(1,i,j)\,D_0^{\gamma,0}(\overline{1},(\widetilde{ij});k,l)\,J_2^{(3)}(\lbrace p\rbrace_{3}) \nonumber\\
 \ph{4}&&+A_3^{0}(j,i,k)\,D_0^{\gamma,0}(1,(\widetilde{ji});(\widetilde{ki}),l)\,J_2^{(3)}(\lbrace p\rbrace_{3}) \nonumber\\
 \ph{5}&&-A_3^{0}(j,i,l)\,D_0^{\gamma,0}(1,(\widetilde{ji});k,(\widetilde{li}))\,J_2^{(3)}(\lbrace p\rbrace_{3}) \nonumber\\
 \ph{6}&&+A_3^{0}(k,i,l)\,D_0^{\gamma,0}(1,j;(\widetilde{ki}),(\widetilde{li}))\,J_2^{(3)}(\lbrace p\rbrace_{3}) \nonumber\\
\snip \ph{7}&&+2\,C_4^0(1,k,j,l)\,B_{1,q}^{\gamma,0}(\overline{1},i,(\widetilde{jkl}))\,J_2^{(2)}(\lbrace p\rbrace_{2}) \nonumber\\
 \ph{8}&&+2\,C_4^0(k,1,j,l)\,B_{1,q}^{\gamma,0}((\widetilde{jkl}),i,\overline{1})\,J_2^{(2)}(\lbrace p\rbrace_{2}) \nonumber\\
 \ph{9}&&+2\,C_4^0(l,j,k,1)\,B_{1,q}^{\gamma,0}(\overline{1},i,(\widetilde{jkl}))\,J_2^{(2)}(\lbrace p\rbrace_{2}) \nonumber\\
 \ph{10}&&+2\,C_4^0(j,l,k,1)\,B_{1,q}^{\gamma,0}(\overline{1},i,(\widetilde{jkl}))\,J_2^{(2)}(\lbrace p\rbrace_{2}) .
\end{eqnarray}

\subsection*{Real-virtual}
\label{app:nnlosubRVq}

The real-virtual contributions to the quark channel receive contributions from the two-quark--two-gluon
and four-quark one-loop matrix elements, where the four-quark matrix elements include identical and non-identical
flavoured quarks. Each matrix element may be decomposed into its colour-stripped functions, as set out in
Tab.~\ref{tab:channelsq}.

\addtocontents{toc}{\protect\setcounter{tocdepth}{2}}
\subsubsection{B-type $\mathcal{O}(N^2)$ contribution}
\addtocontents{toc}{\protect\setcounter{tocdepth}{3}}

The leading-colour contribution is given by the $B_{2}^{\gamma,1}$ function, which is then summed
over both gluon permutations. The subtracted contribution to the cross section is given by
\ba
\sum_{P(i,j)} B_{2}^{\gamma,1}(\hat{1},i,j,2)-B_{2}^{\gamma,1,T}(\hat{1},i,j,2) ,
\ea
where the subtraction term for a single ordering is given by
\begin{eqnarray}
\lefteqn{{B_{2}^{\gamma,1,T}(\hat{1},i,j,k) =}} \nonumber\\
 \ph{1}&&-\bigg[ 
 +J_{2,QG}^{1,IF}(s_{1i})
 +J_{2,QG}^{1,FF}(s_{kj})
 +J_{2,GG}^{1,FF}(s_{ij})\bigg]\,B_2^{\gamma,0}(1,i,j,k)\,J_2^{(3)}(\lbrace p\rbrace_{3})\nonumber\\
 \ph{2}&&-\bigg[ 
 +J_{2,QG}^{1,IF}(s_{1j})
 +J_{2,QG}^{1,FF}(s_{ki})
 +J_{2,GG}^{1,FF}(s_{ji})\bigg]\,B_2^{\gamma,0}(1,j,i,k)\,J_2^{(3)}(\lbrace p\rbrace_{3})\nonumber\\
  \ph{3}&&+\,D_{3,q}^{0}(1,i,j)\,\bigg[\,B_1^{\gamma,1}(\overline{1},(\widetilde{ij}),k)\,\delta(1-x_1)\,\delta(1-x_2)\nonumber\\
  &&+\,\bigg( 
 +J_{2,QG}^{1,IF}(s_{\overline{1}(\widetilde{ij})})
 +J_{2,QG}^{1,FF}(s_{k(\widetilde{ij})})\bigg)\,B_1^{\gamma,0}(\overline{1},(\widetilde{ij}),k) \bigg]\,J_2^{(2)}(\lbrace p\rbrace_{2})\nonumber\\
  \ph{4}&&+\,d_3^{0}(k,j,i)\,\bigg[\,B_1^{\gamma,1}(1,(\widetilde{ij}),(\widetilde{kj}))\,\delta(1-x_1)\,\delta(1-x_2)\nonumber\\
  &&+\,\bigg( 
 +J_{2,QG}^{1,IF}(s_{1(\widetilde{ij})})
 +J_{2,QG}^{1,FF}(s_{(\widetilde{kj})(\widetilde{ij})})\bigg)\,B_1^{\gamma,0}(1,(\widetilde{ij}),(\widetilde{kj})) \bigg]\,J_2^{(2)}(\lbrace p\rbrace_{2})\nonumber\\
  \ph{5}&&+\,d_3^{0}(k,i,j)\,\bigg[\,B_1^{\gamma,1}(1,(\widetilde{ij}),(\widetilde{ki}))\,\delta(1-x_1)\,\delta(1-x_2)\nonumber\\
  &&+\,\bigg( 
 +J_{2,QG}^{1,FF}(s_{(\widetilde{ki})(\widetilde{ij})})
 +J_{2,QG}^{1,IF}(s_{1(\widetilde{ij})})\bigg)\,B_1^{\gamma,0}(1,(\widetilde{ij}),(\widetilde{ki})) \bigg]\,J_2^{(2)}(\lbrace p\rbrace_{2})\nonumber\\
  \ph{6}&&+\bigg[\,d_{3}^{1}(k,i,j)\,\delta(1-x_1)\,\delta(1-x_2)
  \nonumber\\&&\,  
   +\bigg( 
 +J_{2,QG}^{1,FF}(s_{kj})
 +J_{2,GG}^{1,FF}(s_{ij})
 -2J_{2,QG}^{1,FF}(s_{(\widetilde{ki})(\widetilde{ij})})
 +J_{2,QG}^{1,FF}(s_{ki})\bigg)\,d_3^{0}(k,i,j)\bigg]\nonumber\\
 &&\times B_1^{\gamma,0}(1,(\widetilde{ij}),(\widetilde{ki}))\,J_2^{(2)}(\lbrace p\rbrace_{2})\nonumber\\
  \ph{7}&&+\bigg[\,d_{3}^{1}(k,j,i)\,\delta(1-x_1)\,\delta(1-x_2)
  \nonumber\\&&\,  
   +\bigg( 
 +J_{2,QG}^{1,FF}(s_{kj})
 +J_{2,GG}^{1,FF}(s_{ij})
 -2J_{2,QG}^{1,FF}(s_{(\widetilde{kj})(\widetilde{ij})})
 +J_{2,QG}^{1,FF}(s_{ki})\bigg)\,d_3^{0}(k,j,i)\bigg]\nonumber\\
 &&\times B_1^{\gamma,0}(1,(\widetilde{ij}),(\widetilde{kj}))\,J_2^{(2)}(\lbrace p\rbrace_{2})\nonumber\\
  \ph{8}&&+\bigg[\,D_{3,q}^{1}(1,i,j)\,\delta(1-x_1)\,\delta(1-x_2)
  \nonumber\\&&\,  
   +\bigg( 
 +J_{2,QG}^{1,IF}(s_{1i})
 +J_{2,GG}^{1,FF}(s_{ij})
 -2J_{2,QG}^{1,IF}(s_{\overline{1}(\widetilde{ij})})
 +J_{2,QG}^{1,IF}(s_{1j})\bigg)\,D_3^{0}(1,i,j)\bigg]\nonumber\\
 &&\times B_1^{\gamma,0}(\overline{1},(\widetilde{ij}),k)\,J_2^{(2)}(\lbrace p\rbrace_{2})\nonumber\\
  \ph{9}&&-\bigg[\,\tilde{A}_{3,q}^{1}(1,i,k)\,\delta(1-x_1)\,\delta(1-x_2)
  \nonumber\\&&\,  
   +\bigg( 
 +J_{2,QQ}^{1,IF}(s_{1k})
 -J_{2,QQ}^{1,IF}(s_{\overline{1}(\widetilde{ik})})\bigg)\,A_3^{0}(1,i,k)\bigg]\,B_1^{\gamma,0}(\overline{1},j,(\widetilde{ki}))\,J_2^{(2)}(\lbrace p\rbrace_{2})\nonumber\\
  \ph{10}&&-\bigg[\,\tilde{A}_{3,q}^{1}(1,j,k)\,\delta(1-x_1)\,\delta(1-x_2)
  \nonumber\\&&\,  
   +\bigg( 
 +J_{2,QQ}^{1,IF}(s_{1k})
 -J_{2,QQ}^{1,IF}(s_{\overline{1}(\widetilde{jk})})\bigg)\,A_3^{0}(1,j,k)\bigg]\,B_1^{\gamma,0}(\overline{1},i,(\widetilde{kj}))\,J_2^{(2)}(\lbrace p\rbrace_{2})\nonumber\\
 \ph{11}&&+\frac{1}{2}\,\bigg[ +J_{2,QQ}^{1,IF}(s_{1k})
 -J_{2,QQ}^{1,IF}(s_{\overline{1}k})
 -J_{2,QG}^{1,IF}(s_{1j})\nonumber\\
 &&\phantom{+\frac{1}{2}\,}
 +J_{2,QG}^{1,IF}(s_{\overline{1}(\widetilde{ij})})
 +J_{2,QG}^{1,FF}(s_{kj})
 -J_{2,QG}^{1,FF}(s_{k(\widetilde{ij})})
 \nonumber\\
 &&\phantom{+\frac{1}{2}\,} +\bigg(  +{\cal S}^{FF}(s_{1j},s_{kj},x_{1j,kj})
 -{\cal S}^{FF}(s_{\overline{1}(\widetilde{ij})},s_{kj},x_{\overline{1}(\widetilde{ij}),kj})
 -{\cal S}^{FF}(s_{kj},s_{kj},1)\nonumber\\
 &&\phantom{+\frac{1}{2}\,}
 +{\cal S}^{FF}(s_{k(\widetilde{ij})},s_{kj},x_{k(\widetilde{ij}),kj})
 -{\cal S}^{FF}(s_{1k},s_{kj},x_{1k,kj})
 +{\cal S}^{FF}(s_{\overline{1}k},s_{kj},x_{\overline{1}k,kj})
 \bigg)\bigg]\nonumber\\
 &&\times d_3^{0}(1,i,j)\,B_1^{\gamma,0}(\overline{1},(\widetilde{ij}),k)\,J_2^{(2)}(\lbrace p\rbrace_{2}) \nonumber\\
 \ph{12}&&+\frac{1}{2}\,\bigg[ +J_{2,QQ}^{1,IF}(s_{1k})
 -J_{2,QQ}^{1,IF}(s_{\overline{1}k})
 -J_{2,QG}^{1,IF}(s_{1i})\nonumber\\
 &&\phantom{+\frac{1}{2}\,}
 +J_{2,QG}^{1,IF}(s_{\overline{1}(\widetilde{ji})})
 +J_{2,QG}^{1,FF}(s_{ki})
 -J_{2,QG}^{1,FF}(s_{k(\widetilde{ji})})
 \nonumber\\&&\phantom{+\frac{1}{2}\,}+\bigg(  +{\cal S}^{FF}(s_{1i},s_{ki},x_{1i,ki})
 -{\cal S}^{FF}(s_{\overline{1}(\widetilde{ji})},s_{ki},x_{\overline{1}(\widetilde{ji}),ki})
 -{\cal S}^{FF}(s_{ki},s_{ki},1)\nonumber\\
 &&\phantom{+\frac{1}{2}\,}
 +{\cal S}^{FF}(s_{k(\widetilde{ji})},s_{ki},x_{k(\widetilde{ji}),ki})
 -{\cal S}^{FF}(s_{1k},s_{ki},x_{1k,ki})
 +{\cal S}^{FF}(s_{\overline{1}k},s_{ki},x_{\overline{1}k,ki})
 \bigg)\bigg]\nonumber\\
 &&\times d_3^{0}(1,j,i)\,B_1^{\gamma,0}(\overline{1},(\widetilde{ij}),k)\,J_2^{(2)}(\lbrace p\rbrace_{2}) \nonumber\\
 \ph{13}&&+\frac{1}{2}\,\bigg[ +J_{2,QQ}^{1,IF}(s_{1k})
 -J_{2,QQ}^{1,IF}(s_{1(\widetilde{ki})})
 +J_{2,QG}^{1,FF}(s_{(\widetilde{ki})(\widetilde{ij})})\nonumber\\
 &&\phantom{+\frac{1}{2}\,}
 +J_{2,QG}^{1,IF}(s_{1j})
 -J_{2,QG}^{1,IF}(s_{1(\widetilde{ij})})
 -J_{2,QG}^{1,FF}(s_{kj})
 \nonumber\\&&\phantom{+\frac{1}{2}\,}+\bigg(  +{\cal S}^{FF}(s_{kj},s_{kj},1)
 -{\cal S}^{FF}(s_{(\widetilde{ki})(\widetilde{ij})},s_{kj},x_{(\widetilde{ki})(\widetilde{ij}),kj})
 -{\cal S}^{FF}(s_{1j},s_{kj},x_{1j,kj})\nonumber\\
 &&\phantom{+\frac{1}{2}\,}
 +{\cal S}^{FF}(s_{1(\widetilde{ij})},s_{kj},x_{1(\widetilde{ij}),kj})
 -{\cal S}^{FF}(s_{1k},s_{kj},x_{1k,kj})
 +{\cal S}^{FF}(s_{1(\widetilde{ki})},s_{kj},x_{1(\widetilde{ki}),kj})
 \bigg)\bigg]\nonumber\\
 &&\times d_3^{0}(k,i,j)\,B_1^{\gamma,0}(1,(\widetilde{ij}),(\widetilde{ki}))\,J_2^{(2)}(\lbrace p\rbrace_{2}) \nonumber\\
 \ph{14}&&+\frac{1}{2}\,\bigg[ +J_{2,QQ}^{1,IF}(s_{1k})
 -J_{2,QQ}^{1,IF}(s_{1(\widetilde{kj})})
 -J_{2,QG}^{1,FF}(s_{ki})\nonumber\\
 &&\phantom{+\frac{1}{2}\,}
 +J_{2,QG}^{1,FF}(s_{(\widetilde{kj})(\widetilde{ji})})
 +J_{2,QG}^{1,IF}(s_{1i})
 -J_{2,QG}^{1,IF}(s_{1(\widetilde{ji})})
 \nonumber\\&&\phantom{+\frac{1}{2}\,}+\bigg(  +{\cal S}^{FF}(s_{ki},s_{ki},1)
 -{\cal S}^{FF}(s_{(\widetilde{kj})(\widetilde{ji})},s_{ki},x_{(\widetilde{kj})(\widetilde{ji}),ki})
 -{\cal S}^{FF}(s_{1i},s_{ki},x_{1i,ki})\nonumber\\
 &&\phantom{+\frac{1}{2}\,}
 +{\cal S}^{FF}(s_{1(\widetilde{ji})},s_{ki},x_{1(\widetilde{ji}),ki})
 -{\cal S}^{FF}(s_{1k},s_{ki},x_{1k,ki})
 +{\cal S}^{FF}(s_{1(\widetilde{kj})},s_{ki},x_{1(\widetilde{kj}),ki})
 \bigg)\bigg]\nonumber\\
 &&\times d_3^{0}(k,j,i)\,B_1^{\gamma,0}(1,(\widetilde{ji}),(\widetilde{kj}))\,J_2^{(2)}(\lbrace p\rbrace_{2}) \nonumber\\
 \ph{15}&&+\frac{1}{2}\,\bigg[ +J_{2,QQ}^{1,IF}(s_{1k})
 -J_{2,QQ}^{1,IF}(s_{\overline{1}(\widetilde{ki})})
 -J_{2,QG}^{1,IF}(s_{1j})\nonumber\\
 &&\phantom{+\frac{1}{2}\,}
 +J_{2,QG}^{1,IF}(s_{\overline{1}j})
 -J_{2,QG}^{1,FF}(s_{kj})
 +J_{2,QG}^{1,FF}(s_{(\widetilde{ki})j})
 \nonumber\\&&\phantom{+\frac{1}{2}\,}+\bigg(  -{\cal S}^{FF}(s_{1k},s_{kj},x_{1k,kj})
 +{\cal S}^{FF}(s_{\overline{1}(\widetilde{ki})},s_{kj},x_{\overline{1}(\widetilde{ki}),kj})
 +{\cal S}^{FF}(s_{1j},s_{kj},x_{1j,kj})\nonumber\\
 &&\phantom{+\frac{1}{2}\,}
 -{\cal S}^{FF}(s_{\overline{1}j},s_{kj},x_{\overline{1}j,kj})
 \nonumber\\&&\phantom{+\frac{1}{2}\,} +{\cal S}^{FF}(s_{kj},s_{kj},1)
 -{\cal S}^{FF}(s_{(\widetilde{ki})j},s_{kj},x_{(\widetilde{ki})j,kj})
 \bigg)\bigg]\nonumber\\
 &&\times A_3^{0}(1,i,k)\,B_1^{\gamma,0}(\overline{1},j,(\widetilde{ki}))\,J_2^{(2)}(\lbrace p\rbrace_{2}) \nonumber\\
 \ph{16}&&+\frac{1}{2}\,\bigg[ +J_{2,QQ}^{1,IF}(s_{1k})
 -J_{2,QQ}^{1,IF}(s_{\overline{1}(\widetilde{kj})})
 -J_{2,QG}^{1,IF}(s_{1i})\nonumber\\
 &&\phantom{+\frac{1}{2}\,}
 +J_{2,QG}^{1,IF}(s_{\overline{1}i})
 -J_{2,QG}^{1,FF}(s_{ki})
 +J_{2,QG}^{1,FF}(s_{(\widetilde{kj})i})
 \nonumber\\&&\phantom{+\frac{1}{2}\,} +\bigg(  -{\cal S}^{FF}(s_{1k},s_{ki},x_{1k,ki})
 +{\cal S}^{FF}(s_{\overline{1}(\widetilde{kj})},s_{ki},x_{\overline{1}(\widetilde{kj}),ki})
 +{\cal S}^{FF}(s_{1i},s_{ki},x_{1i,ki})\nonumber\\
 &&\phantom{+\frac{1}{2}\,}
 -{\cal S}^{FF}(s_{\overline{1}i},s_{ki},x_{\overline{1}i,ki})
 +{\cal S}^{FF}(s_{ki},s_{ki},1)
 -{\cal S}^{FF}(s_{(\widetilde{kj})i},s_{ki},x_{(\widetilde{kj})i,ki})
 \bigg)\bigg]\nonumber\\
 &&\times A_3^{0}(1,j,k)\,B_1^{\gamma,0}(\overline{1},i,(\widetilde{kj}))\,J_2^{(2)}(\lbrace p\rbrace_{2}) .
\end{eqnarray}

\addtocontents{toc}{\protect\setcounter{tocdepth}{2}}
\subsubsection{B-type $\mathcal{O}(N^0)$ contribution}
\addtocontents{toc}{\protect\setcounter{tocdepth}{3}}

The subleading-colour matrix element contains a mixture of colour connection structures for both the
virtual gluon loop and the external gluons. This leads to a set of subtraction terms which are colour connected
only to the quark line, and a set which are also colour connected to gluons. The subtracted contribution to the cross section is given by
\ba
\sum_{P(i,j)}\tilde{B}_{2}^{\gamma,1}(\hat{1},i,j,2)-\tilde{B}_{2}^{\gamma,1,T}(\hat{1},i,j,2) ,
\ea
where
\begin{eqnarray}
\lefteqn{{\tilde{B}_{2}^{\gamma,1,T}(\hat{1},i,j,k) =}} \nonumber\\
  \ph{1}&&-J_{2,QQ}^{1,IF}(s_{1k})\,B_2^{\gamma,0}(1,i,j,k)\,J_2^{(3)}(\lbrace p\rbrace_{3})\nonumber\\
  \ph{2}&&-J_{2,QQ}^{1,IF}(s_{1k})\,B_2^{\gamma,0}(1,j,i,k)\,J_2^{(3)}(\lbrace p\rbrace_{3})\nonumber\\
 \ph{3}&&-\bigg[ 
 +J_{2,QG}^{1,IF}(s_{1j})
 +J_{2,QG}^{1,FF}(s_{kj})
 +J_{2,QG}^{1,IF}(s_{1i})
 +J_{2,QG}^{1,FF}(s_{ki})\bigg]\,\tilde{B}_2^{\gamma,0}(1,i,j,k)\,J_2^{(3)}(\lbrace p\rbrace_{3})\nonumber\\
  \ph{4}&&+J_{2,QQ}^{1,IF}(s_{1k})\,\tilde{B}_2^{\gamma,0}(1,j,i,k)\,J_2^{(3)}(\lbrace p\rbrace_{3})\nonumber\\
  \ph{5}&&+\,A_3^{0}(1,i,k)\,\bigg[\,B_1^{\gamma,1}(\overline{1},j,(\widetilde{ki}))\,\delta(1-x_1)\,\delta(1-x_2)\nonumber\\
  &&+\,\bigg( 
 +J_{2,QG}^{1,IF}(s_{\overline{1}j})
 +J_{2,QG}^{1,FF}(s_{(\widetilde{ki})j})\bigg)\,B_1^{\gamma,0}(\overline{1},j,(\widetilde{ki})) \bigg]\,J_2^{(2)}(\lbrace p\rbrace_{2})\nonumber\\
  \ph{6}&&+\,A_3^{0}(1,j,k)\,\bigg[\,B_1^{\gamma,1}(\overline{1},i,(\widetilde{kj}))\,\delta(1-x_1)\,\delta(1-x_2)\nonumber\\
  &&+\,\bigg( 
 +J_{2,QG}^{1,IF}(s_{\overline{1}i})
 +J_{2,QG}^{1,FF}(s_{(\widetilde{kj})i})\bigg)\,B_1^{\gamma,0}(\overline{1},i,(\widetilde{kj})) \bigg]\,J_2^{(2)}(\lbrace p\rbrace_{2})\nonumber\\
  \ph{7}&&+\,d_3^{0}(k,i,j)\,\bigg[\,\tilde{B}_1^{\gamma,1}(1,(\widetilde{ij}),(\widetilde{ki}))\,\delta(1-x_1)\,\delta(1-x_2)\nonumber\\
 &&\phantom{\bigg[}+J_{2,QQ}^{1,IF}(s_{1(\widetilde{ki})})\,B_1^{\gamma,0}(1,(\widetilde{ij}),(\widetilde{ki})) \bigg]\,J_2^{(2)}(\lbrace p\rbrace_{2})\nonumber\\
  \ph{8}&&+\,d_3^{0}(k,j,i)\,\bigg[\,\tilde{B}_1^{\gamma,1}(1,(\widetilde{ji}),(\widetilde{kj}))\,\delta(1-x_1)\,\delta(1-x_2)\nonumber\\
 &&\phantom{\bigg[}+J_{2,QQ}^{1,IF}(s_{1(\widetilde{kj})})\,B_1^{\gamma,0}(1,(\widetilde{ji}),(\widetilde{kj})) \bigg]\,J_2^{(2)}(\lbrace p\rbrace_{2})\nonumber\\
  \ph{9}&&+\,D_{3,q}^{0}(1,i,j)\,\bigg[\,\tilde{B}_1^{\gamma,1}(\overline{1},(\widetilde{ij}),k)\,\delta(1-x_1)\,\delta(1-x_2)\nonumber\\
 &&\phantom{\bigg[}+J_{2,QQ}^{1,IF}(s_{\overline{1}k})\,B_1^{\gamma,0}(\overline{1},(\widetilde{ij}),k) \bigg]\,J_2^{(2)}(\lbrace p\rbrace_{2})\nonumber\\
  \ph{10}&&+\bigg[\,A_{3,q}^{1}(1,i,k)\,\delta(1-x_1)\,\delta(1-x_2)
  \nonumber\\&&\,  
   +\bigg( 
 -J_{2,QQ}^{1,IF}(s_{\overline{1}(\widetilde{ki})})
 +J_{2,QG}^{1,IF}(s_{1i})
 +J_{2,QG}^{1,FF}(s_{ki})\bigg)\,A_3^{0}(1,i,k)\bigg]\,B_1^{\gamma,0}(\overline{1},j,(\widetilde{ki}))\,J_2^{(2)}(\lbrace p\rbrace_{2})\nonumber\\
  \ph{11}&&+\bigg[\,A_{3,q}^{1}(1,j,k)\,\delta(1-x_1)\,\delta(1-x_2)
  \nonumber\\&&\,  
   +\bigg( 
 -J_{2,QQ}^{1,IF}(s_{\overline{1}(\widetilde{kj})})
 +J_{2,QG}^{1,IF}(s_{1j})
 +J_{2,QG}^{1,FF}(s_{kj})\bigg)\,A_3^{0}(1,j,k)\bigg]\,B_1^{\gamma,0}(\overline{1},i,(\widetilde{kj}))\,J_2^{(2)}(\lbrace p\rbrace_{2})\nonumber\\
  \ph{12}&&+\bigg[\,\tilde{A}_{3,q}^{1}(1,i,k)\,\delta(1-x_1)\,\delta(1-x_2)
  \nonumber\\&&\,  
   +\bigg( 
 +J_{2,QQ}^{1,IF}(s_{1k})
 -J_{2,QQ}^{1,IF}(s_{\overline{1}(\widetilde{ki})})\bigg)\,A_3^{0}(1,i,k)\bigg]\,B_1^{\gamma,0}(\overline{1},j,(\widetilde{ki}))\,J_2^{(2)}(\lbrace p\rbrace_{2})\nonumber\\
  \ph{13}&&+\bigg[\,\tilde{A}_{3,q}^{1}(1,j,k)\,\delta(1-x_1)\,\delta(1-x_2)
  \nonumber\\&&\,  
   +\bigg( 
 -J_{2,QQ}^{1,IF}(s_{\overline{1}(\widetilde{kj})})
 +J_{2,QQ}^{1,IF}(s_{1k})\bigg)\,A_3^{0}(1,j,k)\bigg]\,B_1^{\gamma,0}(\overline{1},i,(\widetilde{kj}))\,J_2^{(2)}(\lbrace p\rbrace_{2})\nonumber\\
 \ph{14}&&+\bigg[ +J_{2,QG}^{1,IF}(s_{1j})
 +J_{2,QG}^{1,FF}(s_{kj})
 -J_{2,QQ}^{1,IF}(s_{1k})\nonumber\\
 &&\phantom{+}
 -J_{2,QG}^{1,IF}(s_{\overline{1}j})
 -J_{2,QG}^{1,FF}(s_{(\widetilde{ki})j})
 +J_{2,QQ}^{1,IF}(s_{\overline{1}(\widetilde{ki})})
 \nonumber\\&&\phantom{+} +\bigg(  +{\cal S}^{FF}(s_{1k},s_{kj},x_{1k,kj})
 -{\cal S}^{FF}(s_{\overline{1}(\widetilde{ki})},s_{kj},x_{\overline{1}(\widetilde{ki}),kj})
 -{\cal S}^{FF}(s_{1j},s_{kj},x_{1j,kj})\nonumber\\
 &&\phantom{+}
 +{\cal S}^{FF}(s_{\overline{1}j},s_{kj},x_{\overline{1}j,kj})
 \phantom{+} -{\cal S}^{FF}(s_{kj},s_{kj},1)
 +{\cal S}^{FF}(s_{(\widetilde{ki})j},s_{kj},x_{(\widetilde{ki})j,kj})
 \bigg)\bigg]\nonumber\\
 &&\times A_3^{0}(1,i,k)\,B_1^{\gamma,0}(\overline{1},j,(\widetilde{ki}))\,J_2^{(2)}(\lbrace p\rbrace_{2}) \nonumber\\
 \ph{15}&&+\bigg[ +J_{2,QG}^{1,IF}(s_{1i})
 -J_{2,QG}^{1,IF}(s_{\overline{1}i})
 +J_{2,QG}^{1,FF}(s_{ki})\nonumber\\
 &&\phantom{+}
 -J_{2,QG}^{1,FF}(s_{(\widetilde{kj})i})
 -J_{2,QQ}^{1,IF}(s_{1k})
 +J_{2,QQ}^{1,IF}(s_{\overline{1}(\widetilde{kj})})
 \nonumber\\&&\phantom{+}+\bigg(  +{\cal S}^{FF}(s_{1k},s_{ki},x_{1k,ki})
 -{\cal S}^{FF}(s_{\overline{1}(\widetilde{kj})},s_{ki},x_{\overline{1}(\widetilde{kj}),ki})
 -{\cal S}^{FF}(s_{1i},s_{ki},x_{1i,ki})\nonumber\\
 &&\phantom{+}
 +{\cal S}^{FF}(s_{\overline{1}i},s_{ki},x_{\overline{1}i,ki})
 \phantom{+} -{\cal S}^{FF}(s_{ki},s_{ki},1)
 +{\cal S}^{FF}(s_{(\widetilde{kj})i},s_{ki},x_{(\widetilde{kj})i,ki})
 \bigg)\bigg]\nonumber\\
 &&\times A_3^{0}(1,j,k)\,B_1^{\gamma,0}(\overline{1},i,(\widetilde{kj}))\,J_2^{(2)}(\lbrace p\rbrace_{2}) .
\end{eqnarray}

\addtocontents{toc}{\protect\setcounter{tocdepth}{2}}
\subsubsection{B-type $\mathcal{O}(N^{-2})$ contribution}
\addtocontents{toc}{\protect\setcounter{tocdepth}{3}}

The most subleading-colour contribution contains only abelian-like colour connections and therefore only contains
quark-antiquark antennae and quark-antiquark integrated real radiation functions. The subtracted contribution to the cross section is given
by
\ba
\tilde{\tilde{B}}_{2}^{\gamma,1}(\hat{1},3,4,2)-\tilde{\tilde{B}}_{2}^{\gamma,1,T}(\hat{1},3,4,2),
\ea
where
\begin{eqnarray}
\lefteqn{{\tilde{\tilde{B}}_{2}^{\gamma,1,T}(\hat{1},i,j,k) =}} \nonumber\\
  \ph{1}&&-J_{2,QQ}^{1,IF}(s_{1k})\,\tilde{B}_2^{\gamma,0}(1,i,j,k)\,J_2^{(3)}(\lbrace p\rbrace_{3})\nonumber\\
  \ph{2}&&+\,A_3^{0}(1,i,k)\,\bigg[\,\tilde{B}_1^{\gamma,1}(\overline{1},j,(\widetilde{ki}))\,\delta(1-x_1)\,\delta(1-x_2)\nonumber\\
 &&\phantom{\bigg[}+J_{2,QQ}^{1,IF}(s_{\overline{1}(\widetilde{ki})})\,B_1^{\gamma,0}(\overline{1},j,(\widetilde{ki})) \bigg]\,J_2^{(2)}(\lbrace p\rbrace_{2})\nonumber\\
  \ph{3}&&+\,A_3^{0}(1,j,k)\,\bigg[\,\tilde{B}_1^{\gamma,1}(\overline{1},i,(\widetilde{kj}))\,\delta(1-x_1)\,\delta(1-x_2)\nonumber\\
 &&\phantom{\bigg[}+J_{2,QQ}^{1,IF}(s_{\overline{1}(\widetilde{kj})})\,B_1^{\gamma,0}(\overline{1},i,(\widetilde{kj})) \bigg]\,J_2^{(2)}(\lbrace p\rbrace_{2})\nonumber\\
  \ph{4}&&+\bigg[\,\tilde{A}_{3,q}^{1}(1,i,k)\,\delta(1-x_1)\,\delta(1-x_2)
  \nonumber\\&&\,  
   +\bigg( 
 +J_{2,QQ}^{1,IF}(s_{1k})
 -J_{2,QQ}^{1,IF}(s_{\overline{1}(\widetilde{ki})})\bigg)\,A_3^{0}(1,i,k)\bigg]\,B_1^{\gamma,0}(\overline{1},j,(\widetilde{ki}))\,J_2^{(2)}(\lbrace p\rbrace_{2})\nonumber\\
  \ph{5}&&+\bigg[\,\tilde{A}_{3,q}^{1}(1,j,k)\,\delta(1-x_1)\,\delta(1-x_2)
  \nonumber\\&&\,  
   +\bigg( 
 +J_{2,QQ}^{1,IF}(s_{1k})
 -J_{2,QQ}^{1,IF}(s_{\overline{1}(\widetilde{kj})})\bigg)\,A_3^{0}(1,j,k)\bigg]\,B_1^{\gamma,0}(\overline{1},i,(\widetilde{kj}))\,J_2^{(2)}(\lbrace p\rbrace_{2}).
\end{eqnarray}

\addtocontents{toc}{\protect\setcounter{tocdepth}{2}}
\subsubsection{B-type $\mathcal{O}(N_F N^1)$ contribution}
\addtocontents{toc}{\protect\setcounter{tocdepth}{3}}

The two-quark--two-gluon matrix element with a closed quark loop contributes to two colour factors. The leading contribution is given by
the $\hat{B}_{2}^{\gamma,1}(\hat{1},i,j,2)$ function. The poles and unresolved limits of this matrix element are removed by the
subtraction term $\hat{B}_{2}^{\gamma,1,T}(\hat{1},i,j,2)$, and both of these functions are summed over the permutations of final-state gluons.

As was the case at NLO, there is also a contribution to the subtracted cross section from the integrated IC subtraction terms in Sec.~\ref{sec:nnloRRC}.
These are combined with the IC mass factorization terms to generate a finite subtraction term with no unresolved limits, $\hat{B}_{2,q\to g}^{\gamma,1,T}$.
The total subtracted contribution to the cross section is given by
\ba
\sum_{P(i,j)}\bigg[\hat{B}_{2}^{\gamma,1}(\hat{1},i,j,2)-\hat{B}_{2}^{\gamma,1,T}(\hat{1},i,j,2)\bigg]-\frac{1}{2}\sum_{P(4,2)}\hat{B}_{2,q\to g}^{\gamma,1,T}(\hat{1},3,4,2) ,
\ea
where
\begin{eqnarray}
\lefteqn{{\hat{B}_{2}^{\gamma,1,T}(\hat{1},i,j,k) =}} \nonumber\\
 \ph{1}&&-\bigg[ 
 +2\hat{J}_{2,QG}^{1,FF}(s_{ki})
 +2\hat{J}_{2,QG}^{1,FF}(s_{kj})\bigg]\,B_2^{\gamma,0}(1,i,j,k)\,J_2^{(3)}(\lbrace p\rbrace_{3})\nonumber\\
  \ph{2}&&+\,d_3^{0}(1,i,j)\,\bigg[\,\hat{B}_1^{\gamma,1}(\overline{1},(\widetilde{ij}),k)\,\delta(1-x_1)\,\delta(1-x_2)\nonumber\\
  &&+\,\bigg( 
 +\hat{J}_{2,QG}^{1,IF}(s_{\overline{1}(\widetilde{ij})})
 +\hat{J}_{2,QG}^{1,FF}(s_{k(\widetilde{ij})})\bigg)\,B_1^{\gamma,0}(\overline{1},(\widetilde{ij}),k) \bigg]\,J_2^{(2)}(\lbrace p\rbrace_{2})\nonumber\\
  \ph{3}&&+\,d_3^{0}(k,j,i)\,\bigg[\,\hat{B}_1^{\gamma,1}(1,(\widetilde{ij}),(\widetilde{kj}))\,\delta(1-x_1)\,\delta(1-x_2)\nonumber\\
  &&+\,\bigg( 
 +\hat{J}_{2,QG}^{1,IF}(s_{1(\widetilde{ij})})
 +\hat{J}_{2,QG}^{1,FF}(s_{(\widetilde{kj})(\widetilde{ij})})\bigg)\,B_1^{\gamma,0}(1,(\widetilde{ij}),(\widetilde{kj})) \bigg]\,J_2^{(2)}(\lbrace p\rbrace_{2})\nonumber\\
  \ph{4}&&+\bigg[\,\hat{d}_{3}^{1}(1,i,j)\,\delta(1-x_1)\,\delta(1-x_2)
  \nonumber\\&&\,  
   +\bigg( 
 +4\hat{J}_{2,QG}^{1,IF}(s_{1i})
 -\hat{J}_{2,QG}^{1,IF}(s_{\overline{1}(\widetilde{ij})})
 -\hat{J}_{2,QG}^{1,FF}(s_{k(\widetilde{ij})})\bigg)\,d_3^{0}(1,i,j)\bigg]\,B_1^{\gamma,0}(\overline{1},(\widetilde{ji}),k)\,J_2^{(2)}(\lbrace p\rbrace_{2})\nonumber\\
  \ph{5}&&+\bigg[\,\hat{d}_{3}^{1}(k,j,i)\,\delta(1-x_1)\,\delta(1-x_2)
  \nonumber\\&&\,  
   +\bigg( 
 +4\hat{J}_{2,QG}^{1,FF}(s_{kj})
 -\hat{J}_{2,QG}^{1,FF}(s_{(\widetilde{kj})(\widetilde{ji})})
 -\hat{J}_{2,QG}^{1,IF}(s_{1(\widetilde{ij})})\bigg)\,d_3^{0}(k,j,i)\bigg]\,B_1^{\gamma,0}(1,(\widetilde{ji}),(\widetilde{kj}))\,J_2^{(2)}(\lbrace p\rbrace_{2})\nonumber\\
 \ph{6}&&+\bigg[ +2\hat{J}_{2,QG}^{1,FF}(s_{kj})
 -4\hat{J}_{2,QG}^{1,IF}(s_{1i})
 +2\hat{J}_{2,QG}^{1,FF}(s_{ki})
\bigg]  d_3^{0}(1,i,j)\,B_1^{\gamma,0}(\overline{1},(\widetilde{ji}),k)\,J_2^{(2)}(\lbrace p\rbrace_{2})  \nonumber\\
 \ph{7}&&+\bigg[ -2\hat{J}_{2,QG}^{1,FF}(s_{kj})
 +2\hat{J}_{2,QG}^{1,FF}(s_{ki})
\bigg]  d_3^{0}(k,j,i)\,B_1^{\gamma,0}(1,(\widetilde{ji}),(\widetilde{kj}))\,J_2^{(2)}(\lbrace p\rbrace_{2})\,,  \nonumber\\
\end{eqnarray}

\begin{eqnarray}
\lefteqn{{\hat{B}_{0,q\to g}^{\gamma,1,T}(\hat{1},i,j,k) =}} \nonumber\\
  \ph{1}&&-2J_{2,GQ,q^\prime \to g}^{1,IF}(s_{1k})\,B_2^{\gamma,0}(k,1,i,j)\,J_2^{(3)}(\lbrace p\rbrace_{3})\nonumber\\
  \ph{2}&&-2J_{2,GQ,q^\prime \to g}^{1,IF}(s_{1k})\,B_2^{\gamma,0}(k,i,1,j)\,J_2^{(3)}(\lbrace p\rbrace_{3})\nonumber\\
 \ph{3}&& +4J_{2,GQ,q^\prime \to g}^{1,IF}(s_{1k})
\,D_3^{0}(k,1,i)\,B_1^{\gamma,0}((\widetilde{ki}),\overline{1},j)\,J_2^{(2)}(\lbrace p\rbrace_{2})  \nonumber\\
 \ph{4}&&+\bigg[ -2J_{2,GQ,q^\prime \to g}^{1,IF}(s_{1k})
 +2J_{2,GQ,q^\prime \to g}^{1,IF}(s_{(\widetilde{ki})1})
\bigg]  A_3^{0}(j,i,k)\,B_1^{\gamma,0}((\widetilde{ki}),1,(\widetilde{ji}))\,J_2^{(2)}(\lbrace p\rbrace_{2})  \nonumber\\
 \ph{5}&& -4J_{2,GQ,q^\prime \to g}^{1,IF}(s_{1k})
\,a_{3,g\to q}^{0}(k,1,j)\,B_1^{\gamma,0}(i,\overline{1},(\widetilde{kj}))\,J_2^{(2)}(\lbrace p\rbrace_{2})  \nonumber\\
 \ph{6}&& +2J_{2,GQ,q^\prime \to g}^{1,IF}(s_{\overline{1}k})
\,d_{3,g}^{0}(j,i,1)\,B_1^{\gamma,0}(k,\overline{1},(\widetilde{ji}))\,J_2^{(2)}(\lbrace p\rbrace_{2})  \nonumber\\
 \ph{7}&& -2J_{2,GQ,q^\prime \to g}^{1,IF}(s_{\overline{1}(\widetilde{ki})})
\,d_{3,g}^{0}(k,i,1)\,B_1^{\gamma,0}((\widetilde{ki}),\overline{1},j)\,J_2^{(2)}(\lbrace p\rbrace_{2})  \nonumber\\
 \ph{8}&& -2J_{2,GQ,q^\prime \to g}^{1,IF}(s_{1k})
\,a_{3,g\to q}^{0}(k,1,j)\,B_1^{\gamma,0}(\overline{1},i,(\widetilde{jk}))\,J_2^{(2)}(\lbrace p\rbrace_{2})  \nonumber\\
 \ph{9}&& -2J_{2,GQ,q^\prime \to g}^{1,IF}(s_{1k})
\,a_{3,g\to q}^{0}(j,1,k)\,B_1^{\gamma,0}((\widetilde{jk}),i,\overline{1})\,J_2^{(2)}(\lbrace p\rbrace_{2})\,.  \nonumber\\
\end{eqnarray}

\addtocontents{toc}{\protect\setcounter{tocdepth}{2}}
\subsubsection{B-type $\mathcal{O}(N_F N^{-1})$ contribution}
\addtocontents{toc}{\protect\setcounter{tocdepth}{3}}

The subleading-colour contribution to the two-quark--two-gluon matrix element with a closed quark loop, $\hat{\tilde{B}}_{2}^{\gamma,1}$,
has a subtraction terms which cancels all poles and all divergences, $\hat{\tilde{B}}_{2}^{\gamma,1,T}$, and a finite subtraction term which is
generated from the IC subtraction terms in Sec.~\ref{sec:nnloRRCt} and IC mass factorization counterterms, $\hat{\tilde{B}}_{2,q\to g}^{\gamma,1,T}$.
The total subtracted contribution to the cross section is given by
\ba
\hat{\tilde{B}}_{2}^{\gamma,1}(\hat{1},3,4,2)-\hat{\tilde{B}}_{2}^{\gamma,1,T}(\hat{1},3,4,2)-\hat{\tilde{B}}_{2,q\to g}^{\gamma,1,T}(\hat{1},3,4,2) ,
\ea
where
\begin{eqnarray}
\lefteqn{{\hat{\tilde{B}}_{2}^{\gamma,1,T}(\hat{1},i,j,k) =}} \nonumber\\
  \ph{1}&&-4\hat{J}_{2,QG}^{1,FF}(s_{ki})\,\tilde{B}_2^{\gamma,0}(1,i,j,k)\,J_2^{(3)}(\lbrace p\rbrace_{3})\nonumber\\
  \ph{2}&&+2\,A_3^{0}(1,i,k)\,\bigg[\,\hat{B}_1^{\gamma,1}(\overline{1},j,(\widetilde{ik}))\,\delta(1-x_1)\,\delta(1-x_2)\nonumber\\
 &&\phantom{\bigg[}+2\hat{J}_{2,QG}^{1,FF}(s_{(\widetilde{ki})j})\,B_1^{\gamma,0}(\overline{1},j,(\widetilde{ik})) \bigg]\,J_2^{(2)}(\lbrace p\rbrace_{2})\nonumber\\
  \ph{3}&&+2\bigg[\,\hat{A}_{3}^{1}(1,i,k)\,\delta(1-x_1)\,\delta(1-x_2)
 +2\hat{J}_{2,QG}^{1,FF}(s_{ki})\,A_3^{0}(1,i,k)\bigg]\,B_1^{\gamma,0}(\overline{1},j,(\widetilde{ik}))\,J_2^{(2)}(\lbrace p\rbrace_{2})\nonumber\\
 \ph{4}&&+\bigg[ +4\hat{J}_{2,QG}^{1,FF}(s_{kj})
 -4\hat{J}_{2,QG}^{1,FF}(s_{(\widetilde{ki})j})
\bigg]  A_3^{0}(1,i,k)\,B_1^{\gamma,0}(\overline{1},j,(\widetilde{ik}))\,J_2^{(2)}(\lbrace p\rbrace_{2})  ,
\end{eqnarray}

\begin{eqnarray}
\lefteqn{{\hat{\tilde{B}}_{0,q\to g}^{\gamma,1,T}(\hat{1},i,j,k) =}} \nonumber\\
 \ph{1}&&-\bigg[ 
 +J_{2,GQ,q^\prime \to g}^{1,IF}(s_{1k})
 +J_{2,GQ,q^\prime \to g}^{1,IF}(s_{1j})\bigg]\,\tilde{B}_2^{\gamma,0}(k,1,i,j)\,J_2^{(3)}(\lbrace p\rbrace_{3})\nonumber\\
 \ph{2}&&+\bigg[ +J_{2,GQ,q^\prime \to g}^{1,IF}(s_{1k})
 +J_{2,GQ,q^\prime \to g}^{1,IF}(s_{1j})
\bigg]  A_3^{0}(k,i,j)\,B_1^{\gamma,0}((\widetilde{ki}),1,(\widetilde{ij}))\,J_2^{(2)}(\lbrace p\rbrace_{2})  \nonumber\\
 \ph{3}&&+\bigg[ -J_{2,GQ,q^\prime \to g}^{1,IF}(s_{1k})
 -J_{2,GQ,q^\prime \to g}^{1,IF}(s_{1j})
\bigg]  a_{3,g\to q}^{0}(k,1,j)\,B_1^{\gamma,0}(\overline{1},i,(\widetilde{kj}))\,J_2^{(2)}(\lbrace p\rbrace_{2})  \nonumber\\
 \ph{4}&&+\bigg[ -J_{2,GQ,q^\prime \to g}^{1,IF}(s_{1k})
 -J_{2,GQ,q^\prime \to g}^{1,IF}(s_{1j})
\bigg]  a_{3,g\to q}^{0}(j,1,k)\,B_1^{\gamma,0}((\widetilde{kj}),i,\overline{1})\,J_2^{(2)}(\lbrace p\rbrace_{2}) . \nonumber\\
\end{eqnarray}

\addtocontents{toc}{\protect\setcounter{tocdepth}{2}}
\subsubsection{C-type $\mathcal{O}(N_FN^{1})$ contribution}
\addtocontents{toc}{\protect\setcounter{tocdepth}{3}}

The quark channel receives contributions from the four-quark one-loop scattering processes with identical and non-identical
flavour quarks. The full four-quark scattering process can be decomposed into a term calculated from the non-identical flavour
matrix element and the remaining interference terms, analogous to Secs.~\ref{sec:qnloRC}--\ref{sec:qnloRD} for the tree-level
four-quark matrix elements. The corresponding one-loop matrix elements have an additional colour decomposition coming from the
virtual loop.

The leading colour contribution to the cross section is given by
\ba
\sum_{P(3,4)}\left[C_{0}^{\gamma,1}(\hat{1};4,3;2)-C_{0}^{\gamma,1,T}(\hat{1};4,3;2)\right],
\ea
where
\begin{eqnarray}
\lefteqn{{C_{0}^{\gamma,1,T}(\hat{1};j,k;i) =}} \nonumber\\
 \ph{1}&&-\bigg[ 
 +J_{2,QQ}^{1,IF}(s_{1j})
 +J_{2,QQ}^{1,FF}(s_{ik})\bigg]\,C_0^{\gamma,0}(1;j,k;i)\,J_2^{(3)}(\lbrace p\rbrace_{3})\nonumber\\
  \ph{2}&&+\frac{1}{2}\,\,E_3^{0}(i,j,k)\,\bigg[\,B_{1,q}^{\gamma,1}(1,(\widetilde{jk}),(\widetilde{ij}))\,\delta(1-x_1)\,\delta(1-x_2)\nonumber\\
  &&+\,\bigg( 
 +J_{2,QG}^{1,FF}(s_{(\widetilde{ij})(\widetilde{jk})})
 +J_{2,QG}^{1,IF}(s_{1(\widetilde{jk})})\bigg)\,B_{1,q}^{\gamma,0}(1,(\widetilde{jk}),(\widetilde{ij})) \bigg]\,J_2^{(2)}(\lbrace p\rbrace_{2})\nonumber\\
  \ph{3}&&+\frac{1}{2}\,\,E_{3,q}^{0}(1,j,k)\,\bigg[\,B_{1,q}^{\gamma,1}(\overline{1},(\widetilde{jk}),i)\,\delta(1-x_1)\,\delta(1-x_2)\nonumber\\
  &&+\,\bigg( 
 +J_{2,QG}^{1,FF}(s_{i(\widetilde{jk})})
 +J_{2,QG}^{1,IF}(s_{\overline{1}(\widetilde{jk})})\bigg)\,B_{1,q}^{\gamma,0}(\overline{1},(\widetilde{jk}),i) \bigg]\,J_2^{(2)}(\lbrace p\rbrace_{2})\nonumber\\
  \ph{4}&&+\frac{1}{2}\,\bigg[\,E_{3}^{1}(i,j,k)\,\delta(1-x_1)\,\delta(1-x_2)
  \nonumber\\&&\,  
   +\bigg( 
 +J_{2,QQ}^{1,FF}(s_{ij})
 +J_{2,QQ}^{1,FF}(s_{ik})
 -2J_{2,QG}^{1,FF}(s_{(\widetilde{ij})(\widetilde{jk})})\bigg)\,E_3^{0}(i,j,k)\bigg]\,B_{1,q}^{\gamma,0}(1,(\widetilde{jk}),(\widetilde{ij}))\,J_2^{(2)}(\lbrace p\rbrace_{2})\nonumber\\
  \ph{5}&&+\frac{1}{2}\,\bigg[\,E_{3}^{1}(1,j,k)\,\delta(1-x_1)\,\delta(1-x_2)
  \nonumber\\&&\,  
   +\bigg( 
 +J_{2,QQ}^{1,IF}(s_{1j})
 +J_{2,QQ}^{1,IF}(s_{1k})
 -2J_{2,QG}^{1,IF}(s_{\overline{1}(\widetilde{jk})})\bigg)\,E_3^{0}(1,j,k)\bigg]\,B_{1,q}^{\gamma,0}(\overline{1},(\widetilde{jk}),i)\,J_2^{(2)}(\lbrace p\rbrace_{2})\nonumber\\
  \ph{6}&&-\,E_{3,q^\prime\to g}^{0}(k,1,i)\,\bigg[\,B_{1,Q}^{\gamma,1}((\widetilde{ik}),\overline{1},j)\,\delta(1-x_1)\,\delta(1-x_2)\nonumber\\
  &&+\,\bigg( 
 +J_{2,GQ}^{1,IF}(s_{\overline{1}(\widetilde{ik})})
 +J_{2,GQ}^{1,IF}(s_{\overline{1}j})\nonumber\\
 &&\phantom{+}
 -J_{2,QG,g \to q}^{1,IF}(s_{\overline{1}(\widetilde{ik})})
 -J_{2,QG,g \to q}^{1,IF}(s_{\overline{1}j})\bigg)\,B_{1,Q}^{\gamma,0}((\widetilde{ik}),\overline{1},j) \bigg]\,J_2^{(2)}(\lbrace p\rbrace_{2})\nonumber\\
  \ph{7}&&-\bigg[\,E_{3,q^\prime}^{1}(k,1,i)\,\delta(1-x_1)\,\delta(1-x_2)
  \nonumber\\&&\,  
   +\bigg( 
 +J_{2,QQ}^{1,IF}(s_{1k})
 +J_{2,QQ}^{1,FF}(s_{ki})\nonumber\\
 &&\phantom{-}
 -2J_{2,GQ}^{1,IF}(s_{\overline{1}(\widetilde{ki})})
 +2J_{2,QG,g \to q}^{1,IF}(s_{\overline{1}(\widetilde{ki})})\bigg)\,E_{3,q^\prime\to g}^{0}(k,1,i)\bigg]\,B_{1,Q}^{\gamma,0}((\widetilde{ki}),\overline{1},j)\,J_2^{(2)}(\lbrace p\rbrace_{2})\nonumber\\
 \ph{8}&&-\bigg[ +J_{2,QQ}^{1,IF}(s_{1j})
 -J_{2,QQ}^{1,IF}(s_{1k})
 +J_{2,QQ}^{1,FF}(s_{kj})
 -J_{2,QQ}^{1,FF}(s_{j(\widetilde{ki})})\nonumber\\
 &&\phantom{-}
 +J_{2,GQ}^{1,IF}(s_{\overline{1}(\widetilde{ki})})
 -J_{2,GQ}^{1,IF}(s_{\overline{1}j})
 -J_{2,QG,g \to q}^{1,IF}(s_{\overline{1}(\widetilde{ki})})
 +J_{2,QG,g \to q}^{1,IF}(s_{\overline{1}j})\nonumber\\
 &&\phantom{-}
 +\bigg(  -{\cal S}^{FF}(s_{1j},s_{ki},x_{1j,ki})
 +{\cal S}^{FF}(s_{1k},s_{ki},x_{1k,ki})
 -{\cal S}^{FF}(s_{kj},s_{ki},x_{kj,ki})\nonumber\\
 &&\phantom{-}
 +{\cal S}^{FF}(s_{j(\widetilde{ki})},s_{ki},x_{j(\widetilde{ki}),ki})
-{\cal S}^{FF}(s_{\overline{1}(\widetilde{ki})},s_{ki},x_{\overline{1}(\widetilde{ki}),ki})
 +{\cal S}^{FF}(s_{\overline{1}j},s_{ki},x_{\overline{1}j,ki})
 \bigg)\bigg]\nonumber\\
 &&\times E_{3,q^\prime\to g}^{0}(k,1,i)\,B_{1,Q}^{\gamma,0}((\widetilde{ki}),\overline{1},j)\,J_2^{(2)}(\lbrace p\rbrace_{2}) .
\end{eqnarray}

\addtocontents{toc}{\protect\setcounter{tocdepth}{2}}
\subsubsection{C-type $\mathcal{O}(N_FN^{-1})$ contribution}
\addtocontents{toc}{\protect\setcounter{tocdepth}{3}}

The subleading-colour term constructed from non-identical flavour four-quark matrix elements has the contribution to the
cross section given by
\ba
\tilde{C}_{0}^{\gamma,1}(\hat{1};4,3;2)-\tilde{C}_{0}^{\gamma,1,T}(\hat{1};4,3;2),
\ea
where
\begin{eqnarray}
\lefteqn{{\tilde{C}_{0}^{\gamma,1,T}(\hat{1};j,k;i) =}} \nonumber\\
 \ph{1}&&+\bigg[ 
 +2J_{2,QQ}^{1,IF}(s_{1k})
 -2J_{2,QQ}^{1,FF}(s_{ki})
 -J_{2,QQ}^{1,IF}(s_{1i})\nonumber\\
 &&
 -J_{2,QQ}^{1,FF}(s_{jk})
 +2J_{2,QQ}^{1,FF}(s_{ij})
 -2J_{2,QQ}^{1,IF}(s_{1j})\bigg]\,C_0^{\gamma,0}(1;j,k;i)\,J_2^{(3)}(\lbrace p\rbrace_{3})\nonumber\\
  \ph{2}&&+\frac{1}{2}\,\bigg[\,\tilde{E}_{3}^{1}(i,j,k)\,\delta(1-x_1)\,\delta(1-x_2)
 +J_{2,QQ}^{1,FF}(s_{jk})\,E_3^{0}(i,j,k)\bigg]\nonumber\\
 &&\times B_{1,q}^{\gamma,0}(1,(\widetilde{jk}),(\widetilde{ij}))\,J_2^{(2)}(\lbrace p\rbrace_{2})\nonumber\\
  \ph{3}&&+\frac{1}{2}\,\bigg[\,\tilde{E}_{3,q}^{1}(1,j,k)\,\delta(1-x_1)\,\delta(1-x_2)
 +J_{2,QQ}^{1,FF}(s_{jk})\,E_{3,q}^{0}(1,j,k)\bigg]\nonumber\\
 &&\times B_{1,q}^{\gamma,0}(\overline{1},(\widetilde{jk}),i)\,J_2^{(2)}(\lbrace p\rbrace_{2})\nonumber\\
  \ph{4}&&+\frac{1}{2}\,\,E_3^{0}(i,j,k)\,\bigg[\,\tilde{B}_{1,q}^{\gamma,1}(1,(\widetilde{jk}),(\widetilde{ij}))\,\delta(1-x_1)\,\delta(1-x_2)\nonumber\\
 &&\phantom{\bigg[}+J_{2,QQ}^{1,IF}(s_{1(\widetilde{ij})})\,B_{1,q}^{\gamma,0}(1,(\widetilde{jk}),(\widetilde{ij})) \bigg]\,J_2^{(2)}(\lbrace p\rbrace_{2})\nonumber\\
  \ph{5}&&+\frac{1}{2}\,\,E_{3,q}^{0}(1,j,k)\,\bigg[\,\tilde{B}_{1,q}^{\gamma,1}(\overline{1},(\widetilde{jk}),i)\,\delta(1-x_1)\,\delta(1-x_2)\nonumber\\
 &&\phantom{\bigg[}+J_{2,QQ}^{1,IF}(s_{\overline{1}i})\,B_{1,q}^{\gamma,0}(\overline{1},(\widetilde{jk}),i) \bigg]\,J_2^{(2)}(\lbrace p\rbrace_{2})\nonumber\\
 \ph{6}&&-\bigg[ +J_{2,QQ}^{1,IF}(s_{1k})
 +J_{2,QQ}^{1,FF}(s_{ij})
 -J_{2,QQ}^{1,IF}(s_{1j})
 -J_{2,QQ}^{1,FF}(s_{ki})
 \nonumber\\ &&+\bigg(  -{\cal S}^{FF}(s_{1k},s_{kj},x_{1k,kj})
 -{\cal S}^{FF}(s_{ij},s_{kj},x_{ij,kj})
 +{\cal S}^{FF}(s_{1j},s_{kj},x_{1j,kj})\nonumber\\
 &&
 +{\cal S}^{FF}(s_{ki},s_{kj},x_{ki,kj})
 \bigg)\bigg]\,E_3^{0}(i,j,k)\,B_{1,q}^{\gamma,0}(1,(\widetilde{jk}),(\widetilde{ij}))\,J_2^{(2)}(\lbrace p\rbrace_{2}) \nonumber\\
\ph{7}&&-\bigg[ +J_{2,QQ}^{1,IF}(s_{1k})
 +J_{2,QQ}^{1,FF}(s_{ij})
 -J_{2,QQ}^{1,IF}(s_{1j})
 -J_{2,QQ}^{1,FF}(s_{ki})
 \nonumber\\ &&+\bigg(  -{\cal S}^{FF}(s_{1k},s_{kj},x_{1k,kj})
 -{\cal S}^{FF}(s_{ij},s_{kj},x_{ij,kj})
 +{\cal S}^{FF}(s_{1j},s_{kj},x_{1j,kj})\nonumber\\
 &&
 +{\cal S}^{FF}(s_{ki},s_{kj},x_{ki,kj})
 \bigg)\bigg]\,E_3^{0}(1,j,k)\,B_{1,q}^{\gamma,0}(\bar{1},(\widetilde{jk}),i)\,J_2^{(2)}(\lbrace p\rbrace_{2}) \nonumber\\ 
 \ph{8}&&-\frac{1}{2}\,\,E_{3,q^\prime\to g}^{0}(j,1,i)\,\bigg[\,\tilde{B}_{1,Q}^{\gamma,1}(k,\overline{1},(\widetilde{ij}))\,\delta(1-x_1)\,\delta(1-x_2)\nonumber\\
 &&\phantom{\bigg[}+J_{2,QQ}^{1,FF}(s_{k(\widetilde{ij})})\,B_{1,Q}^{\gamma,0}(k,\overline{1},(\widetilde{ij})) \bigg]\,J_2^{(2)}(\lbrace p\rbrace_{2})\nonumber\\
  \ph{9}&&-\frac{1}{2}\,\,E_{3,q^\prime\to g}^{0}(k,1,i)\,\bigg[\,\tilde{B}_{1,Q}^{\gamma,1}((\widetilde{ik}),\overline{1},j)\,\delta(1-x_1)\,\delta(1-x_2)\nonumber\\
 &&\phantom{\bigg[}+J_{2,QQ}^{1,FF}(s_{j(\widetilde{ik})})\,B_{1,Q}^{\gamma,0}((\widetilde{ik}),\overline{1},j) \bigg]\,J_2^{(2)}(\lbrace p\rbrace_{2})\nonumber\\
  \ph{10}&&-\frac{1}{2}\,\bigg[\,\tilde{E}_{3,q^\prime}^{1}(j,1,i)\,\delta(1-x_1)\,\delta(1-x_2)
 +J_{2,QQ}^{1,IF}(s_{1i})\,E_{3,q^\prime\to g}^{0}(j,1,i)\bigg]\nonumber\\
 &&\times B_{1,Q}^{\gamma,0}(k,\overline{1},(\widetilde{ij}))\,J_2^{(2)}(\lbrace p\rbrace_{2})\nonumber\\
  \ph{11}&&-\frac{1}{2}\,\bigg[\,\tilde{E}_{3,q^\prime}^{1}(k,1,i)\,\delta(1-x_1)\,\delta(1-x_2)
 +J_{2,QQ}^{1,IF}(s_{1i})\,E_{3,q^\prime\to g}^{0}(k,1,i)\bigg]\nonumber\\
 &&\times B_{1,Q}^{\gamma,0}((\widetilde{ik}),\overline{1},j)\,J_2^{(2)}(\lbrace p\rbrace_{2})\nonumber\\
 \ph{12}&&+\bigg[ +J_{2,QQ}^{1,IF}(s_{1k})
 +J_{2,QQ}^{1,FF}(s_{ij})
 -J_{2,QQ}^{1,IF}(s_{1j})
 -J_{2,QQ}^{1,FF}(s_{ki})
 \nonumber\\ &&+\bigg(  -{\cal S}^{FF}(s_{1k},s_{jk},x_{1k,jk})
 -{\cal S}^{FF}(s_{ij},s_{jk},x_{ij,jk})
 +{\cal S}^{FF}(s_{1j},s_{jk},x_{1j,jk})\nonumber\\
 &&
 +{\cal S}^{FF}(s_{ki},s_{jk},x_{ki,jk})
 \bigg)\bigg]\,E_3^{0}(j,1,i)\,B_{1,Q}^{\gamma,0}(k,\overline{1},(\widetilde{ij}))\,J_2^{(2)}(\lbrace p\rbrace_{2})
\nonumber\\
 \ph{13}&&+\bigg[ +J_{2,QQ}^{1,IF}(s_{1j})
 +J_{2,QQ}^{1,FF}(s_{ik})
 -J_{2,QQ}^{1,IF}(s_{1k})
 -J_{2,QQ}^{1,FF}(s_{ji})
 \nonumber\\ &&+\bigg(  -{\cal S}^{FF}(s_{1j},s_{kj},x_{1j,kj})
 -{\cal S}^{FF}(s_{ik},s_{kj},x_{ik,kj})
 +{\cal S}^{FF}(s_{1k},s_{kj},x_{1k,kj})\nonumber\\
 &&
 +{\cal S}^{FF}(s_{ji},s_{kj},x_{ji,kj})
 \bigg)\bigg]\,E_3^{0}(k,1,i)\,B_{1,Q}^{\gamma,0}(j,\overline{1},(\widetilde{ik}))\,J_2^{(2)}(\lbrace p\rbrace_{2})
 .
\end{eqnarray}

\addtocontents{toc}{\protect\setcounter{tocdepth}{2}}
\subsubsection{C-type $\mathcal{O}(N^2_F)$ contribution}
\addtocontents{toc}{\protect\setcounter{tocdepth}{3}}

The contributions with four quarks and a closed quark loop can be written purely in terms of the
non-identical flavour matrix elements with a closed quark loop, $\hat{C}_{0}^{\gamma,1}$. The subtracted contribution to the cross section is given by
\ba
\hat{C}_{0}^{\gamma,1}(\hat{1};4,3;2)-\hat{C}_{0}^{\gamma,1,T}(\hat{1};4,3;2),
\ea
where
\begin{eqnarray}
\lefteqn{{\hat{C}_{0}^{\gamma,1,T}(\hat{1};j,k;i) =}} \nonumber\\
  \ph{1}&&+\frac{1}{2}\,\,E_3^{0}(i,j,k)\,\hat{B}_{1,q}^{\gamma,1}(1,(\widetilde{jk}),(\widetilde{ij}))\,\delta(1-x_1)\,\delta(1-x_2)\nonumber\\
  \ph{2}&&+\frac{1}{2}\,\,E_{3,q}^{0}(1,j,k)\,\hat{B}_{1,q}^{\gamma,1}(\overline{1},(\widetilde{jk}),i)\,\delta(1-x_1)\,\delta(1-x_2)\nonumber\\
  \ph{3}&&+\frac{1}{2}\,\,\hat{E}_{3}^{1}(i,j,k)\,B_{1,q}^{\gamma,0}(1,(\widetilde{jk}),(\widetilde{ij}))\,\delta(1-x_1)\,\delta(1-x_2)\nonumber\\
  \ph{4}&&+\frac{1}{2}\,\,\hat{E}_{3,q}^{1}(1,j,k)\,B_{1,q}^{\gamma,0}(\overline{1},(\widetilde{jk}),i)\,\delta(1-x_1)\,\delta(1-x_2)\nonumber\\
  \ph{5}&&-\,E_{3,q^\prime\to g}^{0}(j,1,i)\,\hat{B}_{1,Q}^{\gamma,1}(k,\overline{1},(\widetilde{ij}))\,\delta(1-x_1)\,\delta(1-x_2)\nonumber\\
  \ph{6}&&-\,\hat{E}_{3,q^\prime}^{1}(j,1,i)\,B_{1,Q}^{\gamma,0}(k,\overline{1},(\widetilde{ij}))\,\delta(1-x_1)\,\delta(1-x_2).
\end{eqnarray}

\addtocontents{toc}{\protect\setcounter{tocdepth}{2}}
\subsubsection{D-type $\mathcal{O}(N^{0})$ contribution}
\addtocontents{toc}{\protect\setcounter{tocdepth}{3}}

The remaining four-quark one-loop interferences between different quark orderings also contribute to two colour factors. The
leading-colour term has the subtracted contribution to the cross section given by
\ba
{D}_{0}^{\gamma,1}(\hat{1},2;3,4)-{D}_{0}^{\gamma,1,T}(\hat{1},2;3,4).
\ea
The matrix element contains explicit poles but no single unresolved divergences for the same reason that the tree-level
${D}_{0}^{\gamma,0}$ was finite in all limits. The subtraction term is then given by
\begin{eqnarray}
\lefteqn{{D_{0}^{\gamma,1,T}(\hat{1},j;k,i) =}} \nonumber\\
 \ph{1}&&+\bigg[ 
 +J_{2,QQ}^{1,IF}(s_{1k})
 +J_{2,QQ}^{1,FF}(s_{ji})\bigg]\,D_0^{\gamma,0}(1,j,k,i)\,J_2^{(3)}(\lbrace p\rbrace_{3}).
\end{eqnarray}

\addtocontents{toc}{\protect\setcounter{tocdepth}{2}}
\subsubsection{D-type $\mathcal{O}(N^{-2})$ contribution}
\addtocontents{toc}{\protect\setcounter{tocdepth}{3}}

The subleading-colour contribution to the identical-quark interferences similarly has no single-unresolved limits and requires only
the poles to be removed. The contribution to the cross section is given by
\ba
\tilde{D}_{0}^{\gamma,1}(\hat{1},2;3,4)-\tilde{D}_{0}^{\gamma,1,T}(\hat{1},2;3,4),
\ea
where
\begin{eqnarray}
\lefteqn{{\tilde{D}_{0}^{\gamma,1,T}(\hat{1},j;k,i) =}} \nonumber\\
 \ph{1}&&+\bigg[ 
 +J_{2,QQ}^{1,IF}(s_{1k})
 -J_{2,QQ}^{1,IF}(s_{1i})
 -J_{2,QQ}^{1,IF}(s_{1j})\nonumber\\
 &&
 -J_{2,QQ}^{1,FF}(s_{jk})
 +J_{2,QQ}^{1,FF}(s_{ji})
 -J_{2,QQ}^{1,FF}(s_{ki})\bigg]\,D_0^{\gamma,0}(1,j,k,i)\,J_2^{(3)}(\lbrace p\rbrace_{3}).
\end{eqnarray}

\subsection{Gluon-initiated subtraction terms}

The gluon channel receives contributions from the sub-processes: $g\gamma\to q\bar{q}gg$ and $g\gamma\to q\bar{q}q\bar{q}$ at tree level,
$g\gamma\to q\bar{q}g$ at one-loop and $g\gamma\to q\bar{q}$ at two loops. Each of these contributions can be separated into
their respective colour factors which are listed in Tab.~\ref{tab:channelsg}.
\label{app:nnlog}

\subsection*{Double real}
\label{app:nnlosubRRg}

The double-real correction to the gluon channel contains two-quark--three-gluon, and four-quark--one-gluon
 matrix elements with one of the gluons crossed into the initial-state.

\addtocontents{toc}{\protect\setcounter{tocdepth}{2}}
\subsubsection{B-type $\mathcal{O}(N^{2})$ contribution}
\addtocontents{toc}{\protect\setcounter{tocdepth}{3}}

The leading-colour contribution is given by the $B_{3}^{\gamma,0}$ function, summed over all gluon permutations,
where one of the gluons is in the initial-state. The subtraction term is not summed over any orderings but is constructed
to cancel against the full set of matrix elements due to the intricate singularity structure of the subtraction term,
\ba
\sum_{P(\hat{1},i,j)}\bar{B}_{3}^{\gamma,0}(4,\hat{1},i,j,5)-B_{3}^{\gamma,0,S}(4,\hat{1},2,3,5),
\ea
where the permutation sum $P(\hat{1},i,j)$ runs over the set $\{\hat{1},2,3\}$ and the subtraction term is given by
\begin{eqnarray}
\lefteqn{{B_{3}^{\gamma,0,S}(k,\hat{1},i,j,l) =}} \nonumber \\
 \ph{1}&&+f_{3,g}^{0}(1,j,i)\,\bar{B}_2^{\gamma,0}(k,(\widetilde{ji}),\overline{1},l)\,J_2^{(3)}(\lbrace p\rbrace_{3}) \nonumber\\
 \ph{2}&&+f_{3,g}^{0}(1,i,j)\,\bar{B}_2^{\gamma,0}(k,(\widetilde{ji}),\overline{1},l)\,J_2^{(3)}(\lbrace p\rbrace_{3}) \nonumber\\
 \ph{3}&&+f_{3,g}^{0}(1,j,i)\,\bar{B}_2^{\gamma,0}(k,\overline{1},(\widetilde{ji}),l)\,J_2^{(3)}(\lbrace p\rbrace_{3}) \nonumber\\
 \ph{4}&&+f_{3,g}^{0}(1,i,j)\,\bar{B}_2^{\gamma,0}(k,\overline{1},(\widetilde{ji}),l)\,J_2^{(3)}(\lbrace p\rbrace_{3}) \nonumber\\
 \ph{5}&&+d_3^{0}(l,j,i)\,\bar{B}_2^{\gamma,0}(k,1,(\widetilde{ji}),(\widetilde{lj}))\,J_2^{(3)}(\lbrace p\rbrace_{3}) \nonumber\\
 \ph{6}&&+d_3^{0}(k,j,i)\,\bar{B}_2^{\gamma,0}((\widetilde{kj}),(\widetilde{ji}),1,l)\,J_2^{(3)}(\lbrace p\rbrace_{3}) \nonumber\\
 \ph{7}&&+d_3^{0}(l,i,j)\,\bar{B}_2^{\gamma,0}(k,1,(\widetilde{ij}),(\widetilde{li}))\,J_2^{(3)}(\lbrace p\rbrace_{3}) \nonumber\\
 \ph{8}&&+d_3^{0}(k,i,j)\,\bar{B}_2^{\gamma,0}((\widetilde{ki}),(\widetilde{ij}),1,l)\,J_2^{(3)}(\lbrace p\rbrace_{3}) \nonumber\\
 \ph{9}&&+d_{3,g}^{0}(k,i,1)\,\bar{B}_2^{\gamma,0}((\widetilde{ki}),\overline{1},j,l)\,J_2^{(3)}(\lbrace p\rbrace_{3}) \nonumber\\
 \ph{10}&&+d_{3,g}^{0}(k,j,1)\,\bar{B}_2^{\gamma,0}((\widetilde{kj}),\overline{1},i,l)\,J_2^{(3)}(\lbrace p\rbrace_{3}) \nonumber\\
 \ph{11}&&+d_{3,g}^{0}(l,i,1)\,\bar{B}_2^{\gamma,0}(k,j,\overline{1},(\widetilde{li}))\,J_2^{(3)}(\lbrace p\rbrace_{3}) \nonumber\\
 \ph{12}&&+d_{3,g}^{0}(l,j,1)\,\bar{B}_2^{\gamma,0}(k,i,\overline{1},(\widetilde{lj}))\,J_2^{(3)}(\lbrace p\rbrace_{3}) \nonumber\\
\snip \ph{13}&&+\,D_4^0(k,i,j,1)\,{\bar{B}}_1^{\gamma,0}((\widetilde{kij}),\overline{1},l)\,J_2^{(2)}(\lbrace p\rbrace_{2}) \nonumber\\
 \ph{14}&&-\,d_3^{0}(k,i,j)\,D_3^{0}((\widetilde{ki}),(\widetilde{ij}),1)\,{\bar{B}}_1^{\gamma,0}(\widetilde{(\widetilde{ki})(\widetilde{ij})},\overline{1},l)\,J_2^{(2)}(\lbrace p\rbrace_{2}) \nonumber\\
 \ph{15}&&-\,f_{3,g}^{0}(1,j,i)\,D_3^{0}(k,(\widetilde{ij}),\overline{1})\,{\bar{B}}_1^{\gamma,0}((\widetilde{k\widetilde{(ij})}),\overline{\overline{1}},l)\,J_2^{(2)}(\lbrace p\rbrace_{2}) \nonumber\\
\snip \ph{16}&&+\,D_4^0(k,1,i,j)\,{\bar{B}}_1^{\gamma,0}((\widetilde{kij}),\overline{1},l)\,J_2^{(2)}(\lbrace p\rbrace_{2}) \nonumber\\
 \ph{17}&&-\,d_3^{0}(k,j,i)\,D_3^{0}((\widetilde{kj}),(\widetilde{ji}),1)\,{\bar{B}}_1^{\gamma,0}(\widetilde{(\widetilde{kj})(\widetilde{ji})},\overline{1},l)\,J_2^{(2)}(\lbrace p\rbrace_{2}) \nonumber\\
 \ph{18}&&-\,f_{3,g}^{0}(1,i,j)\,D_3^{0}(k,(\widetilde{ij}),\overline{1})\,{\bar{B}}_1^{\gamma,0}((\widetilde{k\widetilde{(ij})}),\overline{\overline{1}},l)\,J_2^{(2)}(\lbrace p\rbrace_{2}) \nonumber\\
\snip \ph{19}&&+\,D_4^0(l,i,j,1)\,{\bar{B}}_1^{\gamma,0}(k,\overline{1},(\widetilde{lij}))\,J_2^{(2)}(\lbrace p\rbrace_{2}) \nonumber\\
 \ph{20}&&-\,d_3^{0}(l,i,j)\,D_3^{0}((\widetilde{li}),(\widetilde{ij}),1)\,{\bar{B}}_1^{\gamma,0}(k,\overline{1},\widetilde{(\widetilde{li})(\widetilde{ij})})\,J_2^{(2)}(\lbrace p\rbrace_{2}) \nonumber\\
 \ph{21}&&-\,f_{3,g}^{0}(1,j,i)\,D_3^{0}(l,(\widetilde{ij}),\overline{1})\,{\bar{B}}_1^{\gamma,0}(k,\overline{\overline{1}},(\widetilde{l\widetilde{(ij})}))\,J_2^{(2)}(\lbrace p\rbrace_{2}) \nonumber\\
\snip \ph{22}&&+\,D_4^0(l,1,i,j)\,{\bar{B}}_1^{\gamma,0}(k,\overline{1},(\widetilde{lij}))\,J_2^{(2)}(\lbrace p\rbrace_{2}) \nonumber\\
 \ph{23}&&-\,d_3^{0}(l,j,i)\,D_3^{0}((\widetilde{lj}),(\widetilde{ji}),1)\,{\bar{B}}_1^{\gamma,0}(k,\overline{1},\widetilde{(\widetilde{lj})(\widetilde{ji})})\,J_2^{(2)}(\lbrace p\rbrace_{2}) \nonumber\\
 \ph{24}&&-\,f_{3,g}^{0}(1,i,j)\,D_3^{0}(l,(\widetilde{ij}),\overline{1})\,{\bar{B}}_1^{\gamma,0}(k,\overline{\overline{1}},(\widetilde{l\widetilde{(ij})}))\,J_2^{(2)}(\lbrace p\rbrace_{2}) \nonumber\\
 \ph{25}&&-2\,A_3^{0}(k,1,l)\,D_3^{0}(\overline{1},i,j)\,{\bar{B}}_1^{\gamma,0}((\widetilde{kl}),\overline{\overline{1}},(\widetilde{ij}))\,J_2^{(2)}(\lbrace p\rbrace_{2}) \nonumber\\
\snip \ph{26}&&+\,D_4^0(k,i,1,j)\,{\bar{B}}_1^{\gamma,0}((\widetilde{kij}),\overline{1},l)\,J_2^{(2)}(\lbrace p\rbrace_{2}) \nonumber\\
 \ph{27}&&-\,d_{3,g}^{0}(k,i,1)\,D_3^{0}((\widetilde{ki}),\overline{1},j)\,{\bar{B}}_1^{\gamma,0}((\widetilde{j\widetilde{(ki})}),\overline{\overline{1}},l)\,J_2^{(2)}(\lbrace p\rbrace_{2}) \nonumber\\
 \ph{28}&&-\,d_{3,g}^{0}(k,j,1)\,D_3^{0}((\widetilde{kj}),\overline{1},i)\,{\bar{B}}_1^{\gamma,0}((\widetilde{i\widetilde{(kj})}),\overline{\overline{1}},l)\,J_2^{(2)}(\lbrace p\rbrace_{2}) \nonumber\\
\snip \ph{29}&&+\,D_4^0(l,i,1,j)\,{\bar{B}}_1^{\gamma,0}(k,\overline{1},(\widetilde{lij}))\,J_2^{(2)}(\lbrace p\rbrace_{2}) \nonumber\\
 \ph{30}&&-\,d_{3,g}^{0}(l,i,1)\,D_3^{0}((\widetilde{li}),\overline{1},j)\,{\bar{B}}_1^{\gamma,0}(k,\overline{\overline{1}},(\widetilde{j\widetilde{(li})}))\,J_2^{(2)}(\lbrace p\rbrace_{2}) \nonumber\\
 \ph{31}&&-\,d_{3,g}^{0}(l,j,1)\,D_3^{0}((\widetilde{lj}),\overline{1},i)\,{\bar{B}}_1^{\gamma,0}(k,\overline{\overline{1}},(\widetilde{i\widetilde{(lj})}))\,J_2^{(2)}(\lbrace p\rbrace_{2}) \nonumber\\
\snip \ph{32}&&-\,A_4^0(k,1,i,l)\,{\bar{B}}_1^{\gamma,0}((\widetilde{kil}),\overline{1},j)\,J_2^{(2)}(\lbrace p\rbrace_{2}) \nonumber\\
\snip \ph{33}&&-\,A_4^0(k,i,1,l)\,{\bar{B}}_1^{\gamma,0}((\widetilde{kil}),\overline{1},j)\,J_2^{(2)}(\lbrace p\rbrace_{2}) \nonumber\\
 \ph{34}&&+\,d_{3,g}^{0}(k,i,1)\,A_3^{0}((\widetilde{ki}),\overline{1},l)\,{\bar{B}}_1^{\gamma,0}((\widetilde{l\widetilde{(ki})}),\overline{\overline{1}},j)\,J_2^{(2)}(\lbrace p\rbrace_{2}) \nonumber\\
 \ph{35}&&+\,d_{3,g}^{0}(l,i,1)\,A_3^{0}((\widetilde{li}),\overline{1},k)\,{\bar{B}}_1^{\gamma,0}((\widetilde{k\widetilde{(li})}),\overline{\overline{1}},j)\,J_2^{(2)}(\lbrace p\rbrace_{2}) \nonumber\\
 \ph{36}&&+\,A_3^{0}(k,1,l)\,A_3^{0}(\overline{1},i,(\widetilde{kl}))\,{\bar{B}}_1^{\gamma,0}((\widetilde{i\widetilde{(kl})}),\overline{\overline{1}},j)\,J_2^{(2)}(\lbrace p\rbrace_{2}) \nonumber\\
\snip \ph{37}&&-\,A_4^0(k,1,j,l)\,{\bar{B}}_1^{\gamma,0}((\widetilde{kjl}),\overline{1},i)\,J_2^{(2)}(\lbrace p\rbrace_{2}) \nonumber\\
\snip \ph{38}&&-\,A_4^0(k,j,1,l)\,{\bar{B}}_1^{\gamma,0}((\widetilde{kjl}),\overline{1},i)\,J_2^{(2)}(\lbrace p\rbrace_{2}) \nonumber\\
 \ph{39}&&+\,d_{3,g}^{0}(k,j,1)\,A_3^{0}((\widetilde{kj}),\overline{1},l)\,{\bar{B}}_1^{\gamma,0}((\widetilde{l\widetilde{(kj})}),\overline{\overline{1}},i)\,J_2^{(2)}(\lbrace p\rbrace_{2}) \nonumber\\
 \ph{40}&&+\,d_{3,g}^{0}(l,j,1)\,A_3^{0}((\widetilde{lj}),\overline{1},k)\,{\bar{B}}_1^{\gamma,0}((\widetilde{k\widetilde{(lj})}),\overline{\overline{1}},i)\,J_2^{(2)}(\lbrace p\rbrace_{2}) \nonumber\\
 \ph{41}&&+\,A_3^{0}(k,1,l)\,A_3^{0}(\overline{1},j,(\widetilde{kl}))\,{\bar{B}}_1^{\gamma,0}((\widetilde{j\widetilde{(kl})}),\overline{\overline{1}},i)\,J_2^{(2)}(\lbrace p\rbrace_{2}) \nonumber\\
\snip \ph{42}&&-\,\tilde{A}_4^0(k,j,1,l)\,{\bar{B}}_1^{\gamma,0}((\widetilde{kjl}),\overline{1},i)\,J_2^{(2)}(\lbrace p\rbrace_{2}) \nonumber\\
 \ph{43}&&+\,A_3^{0}(k,j,l)\,A_3^{0}((\widetilde{kj}),1,(\widetilde{jl}))\,{\bar{B}}_1^{\gamma,0}(\widetilde{(\widetilde{kj})(\widetilde{jl})},\overline{1},i)\,J_2^{(2)}(\lbrace p\rbrace_{2}) \nonumber\\
 \ph{44}&&+\,A_3^{0}(k,1,l)\,A_3^{0}(\overline{1},j,(\widetilde{kl}))\,{\bar{B}}_1^{\gamma,0}((\widetilde{j\widetilde{(kl})}),\overline{\overline{1}},i)\,J_2^{(2)}(\lbrace p\rbrace_{2}) \nonumber\\
\snip \ph{45}&&-\,\tilde{A}_4^0(k,i,1,l)\,{\bar{B}}_1^{\gamma,0}((\widetilde{kil}),\overline{1},j)\,J_2^{(2)}(\lbrace p\rbrace_{2}) \nonumber\\
 \ph{46}&&+\,A_3^{0}(k,i,l)\,A_3^{0}((\widetilde{ki}),1,(\widetilde{il}))\,{\bar{B}}_1^{\gamma,0}(\widetilde{(\widetilde{ki})(\widetilde{il})},\overline{1},j)\,J_2^{(2)}(\lbrace p\rbrace_{2}) \nonumber\\
 \ph{47}&&+\,A_3^{0}(k,1,l)\,A_3^{0}(\overline{1},i,(\widetilde{kl}))\,{\bar{B}}_1^{\gamma,0}((\widetilde{i\widetilde{(kl})}),\overline{\overline{1}},j)\,J_2^{(2)}(\lbrace p\rbrace_{2}) \nonumber\\
\snip \ph{48}&&-\,\tilde{A}_{4}^{0,a}(k,i,j,l)\,{\bar{B}}_1^{\gamma,0}((\widetilde{kij}),1,(\widetilde{lji}))\,J_2^{(2)}(\lbrace p\rbrace_{2}) \nonumber\\
\snip \ph{49}&&-\,\tilde{A}_{4}^{0,a}(k,j,i,l)\,{\bar{B}}_1^{\gamma,0}((\widetilde{kji}),1,(\widetilde{lij}))\,J_2^{(2)}(\lbrace p\rbrace_{2}) \nonumber\\
 \ph{50}&&+\,A_3^{0}(k,i,l)\,A_3^{0}((\widetilde{ki}),j,(\widetilde{il}))\,{\bar{B}}_1^{\gamma,0}((\widetilde{(\widetilde{ki})j}),1,(\widetilde{j\widetilde{(il})}))\,J_2^{(2)}(\lbrace p\rbrace_{2}) \nonumber\\
 \ph{51}&&+\,A_3^{0}(k,j,l)\,A_3^{0}((\widetilde{kj}),i,(\widetilde{jl}))\,{\bar{B}}_1^{\gamma,0}((\widetilde{(\widetilde{kj})i}),1,(\widetilde{i\widetilde{(jl})}))\,J_2^{(2)}(\lbrace p\rbrace_{2}) \nonumber\\
 \ph{52}&&+\,f_{3,g}^{0}(1,j,i)\,d_{3,g\to q}^{0}(k,\overline{1},(\widetilde{ij}))\,{\bar{B}}_1^{\gamma,0}((\widetilde{k\widetilde{(ij})}),\overline{\overline{1}},l)\,J_2^{(2)}(\lbrace p\rbrace_{2}) \nonumber\\
 \ph{53}&&-\,d_3^{0}(k,j,i)\,d_{3,g\to q}^{0}((\widetilde{kj}),1,(\widetilde{ji}))\,{\bar{B}}_1^{\gamma,0}(\widetilde{(\widetilde{kj})(\widetilde{ji})},\overline{1},l)\,J_2^{(2)}(\lbrace p\rbrace_{2}) \nonumber\\
 \ph{54}&&+\,A_3^{0}(k,j,l)\,d_{3,g\to q}^{0}((\widetilde{kj}),1,i)\,{\bar{B}}_1^{\gamma,0}((\widetilde{i\widetilde{(kj})}),\overline{1},(\widetilde{lj}))\,J_2^{(2)}(\lbrace p\rbrace_{2}) \nonumber\\
 \ph{55}&&-\,d_{3,g}^{0}(l,j,1)\,d_{3,g\to q}^{0}(k,\overline{1},i)\,{\bar{B}}_1^{\gamma,0}((\widetilde{ki}),\overline{\overline{1}},(\widetilde{lj}))\,J_2^{(2)}(\lbrace p\rbrace_{2}) \nonumber\\
 \ph{56}&&-\bigg[      +S^{FF}_{(\widetilde{ij})j1}       -S^{FF}_{(\widetilde{kj})j(\widetilde{ji})}       +S^{FF}_{(\widetilde{kj})jl}       -S^{FF}_{lj1}  \bigg]\nonumber\\
 &&\times d_{3,g\to q}^{0}((\widetilde{kj}),1,(\widetilde{ij}))\,{\bar{B}}_1^{\gamma,0}(\widetilde{(\widetilde{kj})(\widetilde{ij})},\overline{1},l)\,J_2^{(2)}(\lbrace p\rbrace_{2}) \nonumber\\
 \ph{57}&&+\,f_{3,g}^{0}(1,i,j)\,d_{3,g\to q}^{0}(k,\overline{1},(\widetilde{ji}))\,{\bar{B}}_1^{\gamma,0}((\widetilde{k\widetilde{(ji})}),\overline{\overline{1}},l)\,J_2^{(2)}(\lbrace p\rbrace_{2}) \nonumber\\
 \ph{58}&&-\,d_3^{0}(k,i,j)\,d_{3,g\to q}^{0}((\widetilde{ki}),1,(\widetilde{ij}))\,{\bar{B}}_1^{\gamma,0}(\widetilde{(\widetilde{ki})(\widetilde{ij})},\overline{1},l)\,J_2^{(2)}(\lbrace p\rbrace_{2}) \nonumber\\
 \ph{59}&&+\,A_3^{0}(k,i,l)\,d_{3,g\to q}^{0}((\widetilde{ki}),1,j)\,{\bar{B}}_1^{\gamma,0}((\widetilde{j\widetilde{(ki})}),\overline{1},(\widetilde{li}))\,J_2^{(2)}(\lbrace p\rbrace_{2}) \nonumber\\
 \ph{60}&&-\,d_{3,g}^{0}(l,i,1)\,d_{3,g\to q}^{0}(k,\overline{1},j)\,{\bar{B}}_1^{\gamma,0}((\widetilde{kj}),\overline{\overline{1}},(\widetilde{li}))\,J_2^{(2)}(\lbrace p\rbrace_{2}) \nonumber\\
 \ph{61}&&-\bigg[      +S^{FF}_{(\widetilde{ji})i1}       -S^{FF}_{(\widetilde{ki})i(\widetilde{ij})}       +S^{FF}_{(\widetilde{ki})il}       -S^{FF}_{li1}  \bigg]\nonumber\\
 &&\times d_{3,g\to q}^{0}((\widetilde{ki}),1,(\widetilde{ji}))\,{\bar{B}}_1^{\gamma,0}(\widetilde{(\widetilde{ki})(\widetilde{ji})},\overline{1},l)\,J_2^{(2)}(\lbrace p\rbrace_{2}) \nonumber\\
 \ph{62}&&+\,f_{3,g}^{0}(1,j,i)\,d_{3,g\to q}^{0}(l,\overline{1},(\widetilde{ij}))\,{\bar{B}}_1^{\gamma,0}(k,\overline{\overline{1}},(\widetilde{l\widetilde{(ij})}))\,J_2^{(2)}(\lbrace p\rbrace_{2}) \nonumber\\
 \ph{63}&&-\,d_3^{0}(l,j,i)\,d_{3,g\to q}^{0}((\widetilde{lj}),1,(\widetilde{ji}))\,{\bar{B}}_1^{\gamma,0}(k,\overline{1},\widetilde{(\widetilde{lj})(\widetilde{ji})})\,J_2^{(2)}(\lbrace p\rbrace_{2}) \nonumber\\
 \ph{64}&&+\,A_3^{0}(l,j,k)\,d_{3,g\to q}^{0}((\widetilde{lj}),1,i)\,{\bar{B}}_1^{\gamma,0}((\widetilde{kj}),\overline{1},(\widetilde{i\widetilde{(lj})}))\,J_2^{(2)}(\lbrace p\rbrace_{2}) \nonumber\\
 \ph{65}&&-\,d_{3,g}^{0}(k,j,1)\,d_{3,g\to q}^{0}(l,\overline{1},i)\,{\bar{B}}_1^{\gamma,0}((\widetilde{kj}),\overline{\overline{1}},(\widetilde{li}))\,J_2^{(2)}(\lbrace p\rbrace_{2}) \nonumber\\
 \ph{66}&&-\bigg[      +S^{FF}_{(\widetilde{ij})j1}       -S^{FF}_{(\widetilde{lj})j(\widetilde{ji})}       +S^{FF}_{(\widetilde{lj})jk}       -S^{FF}_{kj1}  \bigg]\nonumber\\
 &&\times d_{3,g\to q}^{0}((\widetilde{lj}),1,(\widetilde{ij}))\,{\bar{B}}_1^{\gamma,0}(k,\overline{1},\widetilde{(\widetilde{lj})(\widetilde{ij})})\,J_2^{(2)}(\lbrace p\rbrace_{2}) \nonumber\\
 \ph{67}&&+\,f_{3,g}^{0}(1,i,j)\,d_{3,g\to q}^{0}(l,\overline{1},(\widetilde{ji}))\,{\bar{B}}_1^{\gamma,0}(k,\overline{\overline{1}},(\widetilde{l\widetilde{(ji})}))\,J_2^{(2)}(\lbrace p\rbrace_{2}) \nonumber\\
 \ph{68}&&-\,d_3^{0}(l,i,j)\,d_{3,g\to q}^{0}((\widetilde{li}),1,(\widetilde{ij}))\,{\bar{B}}_1^{\gamma,0}(k,\overline{1},\widetilde{(\widetilde{li})(\widetilde{ij})})\,J_2^{(2)}(\lbrace p\rbrace_{2}) \nonumber\\
 \ph{69}&&+\,A_3^{0}(l,i,k)\,d_{3,g\to q}^{0}((\widetilde{li}),1,j)\,{\bar{B}}_1^{\gamma,0}((\widetilde{ki}),\overline{1},(\widetilde{j\widetilde{(li})}))\,J_2^{(2)}(\lbrace p\rbrace_{2}) \nonumber\\
 \ph{70}&&-\,d_{3,g}^{0}(k,i,1)\,d_{3,g\to q}^{0}(l,\overline{1},j)\,{\bar{B}}_1^{\gamma,0}((\widetilde{ki}),\overline{\overline{1}},(\widetilde{lj}))\,J_2^{(2)}(\lbrace p\rbrace_{2}) \nonumber\\
 \ph{71}&&-\bigg[      +S^{FF}_{(\widetilde{ji})i1}       -S^{FF}_{(\widetilde{li})i(\widetilde{ij})}       +S^{FF}_{(\widetilde{li})ik}       -S^{FF}_{ki1}  \bigg]\nonumber\\
 &&\times d_{3,g\to q}^{0}((\widetilde{li}),1,(\widetilde{ji}))\,{\bar{B}}_1^{\gamma,0}(k,\overline{1},\widetilde{(\widetilde{li})(\widetilde{ji})})\,J_2^{(2)}(\lbrace p\rbrace_{2}) \nonumber\\
 \ph{72}&&+2\,d_3^{0}(k,i,j)\,A_3^{0}((\widetilde{ki}),1,l)\,{\bar{B}}_1^{\gamma,0}((\widetilde{l\widetilde{(ki})}),\overline{1},(\widetilde{ij}))\,J_2^{(2)}(\lbrace p\rbrace_{2}) \nonumber\\
 \ph{73}&&+2\,d_3^{0}(k,j,i)\,A_3^{0}((\widetilde{kj}),1,l)\,{\bar{B}}_1^{\gamma,0}((\widetilde{l\widetilde{(kj})}),\overline{1},(\widetilde{ji}))\,J_2^{(2)}(\lbrace p\rbrace_{2}) \nonumber\\
 \ph{74}&&+2\,d_3^{0}(l,i,j)\,A_3^{0}((\widetilde{li}),1,k)\,{\bar{B}}_1^{\gamma,0}((\widetilde{k\widetilde{(li})}),\overline{1},(\widetilde{ij}))\,J_2^{(2)}(\lbrace p\rbrace_{2}) \nonumber\\
 \ph{75}&&+2\,d_3^{0}(l,j,i)\,A_3^{0}((\widetilde{lj}),1,k)\,{\bar{B}}_1^{\gamma,0}((\widetilde{k\widetilde{(lj})}),\overline{1},(\widetilde{ji}))\,J_2^{(2)}(\lbrace p\rbrace_{2}) \nonumber\\
 \ph{76}&&-2\,A_3^{0}(k,1,l)\,d_3^{0}((\widetilde{kl}),i,j)\,{\bar{B}}_1^{\gamma,0}((\widetilde{(\widetilde{kl})i}),\overline{1},(\widetilde{ij}))\,J_2^{(2)}(\lbrace p\rbrace_{2}) \nonumber\\
 \ph{77}&&-2\,A_3^{0}(k,1,l)\,d_3^{0}((\widetilde{kl}),j,i)\,{\bar{B}}_1^{\gamma,0}((\widetilde{(\widetilde{kl})j}),\overline{1},(\widetilde{ji}))\,J_2^{(2)}(\lbrace p\rbrace_{2}) \nonumber\\
 \ph{78}&&-2\,A_3^{0}(k,i,l)\,A_3^{0}((\widetilde{ki}),1,(\widetilde{il}))\,{\bar{B}}_1^{\gamma,0}(\widetilde{(\widetilde{ki})(\widetilde{il})},\overline{1},j)\,J_2^{(2)}(\lbrace p\rbrace_{2}) \nonumber\\
 \ph{79}&&-2\,A_3^{0}(k,j,l)\,A_3^{0}((\widetilde{kj}),1,(\widetilde{jl}))\,{\bar{B}}_1^{\gamma,0}(\widetilde{(\widetilde{kj})(\widetilde{jl})},\overline{1},i)\,J_2^{(2)}(\lbrace p\rbrace_{2}) \nonumber\\
 \ph{80}&&+2\bigg[      +S^{FF}_{\overline{1}ij}       -S^{FF}_{\overline{1}i\widetilde{(\widetilde{ki})(\widetilde{li})}}       -S^{FF}_{(\widetilde{ki})ij}       -S^{FF}_{(\widetilde{li})ij}       +S^{FF}_{\widetilde{(\widetilde{ki})(\widetilde{li})}ij}       +S^{FF}_{(\widetilde{ki})i(\widetilde{li})}  \bigg]\nonumber\\
 &&\times A_3^{0}((\widetilde{ki}),1,(\widetilde{li}))\,{\bar{B}}_1^{\gamma,0}(\widetilde{(\widetilde{ki})(\widetilde{li})},\overline{1},j)\,J_2^{(2)}(\lbrace p\rbrace_{2}) \nonumber\\
 \ph{81}&&+2\bigg[      +S^{FF}_{\overline{1}ji}       -S^{FF}_{\overline{1}j\widetilde{(\widetilde{kj})(\widetilde{lj})}}       -S^{FF}_{(\widetilde{kj})ji}       -S^{FF}_{(\widetilde{lj})ji}       +S^{FF}_{\widetilde{(\widetilde{kj})(\widetilde{lj})}ji}       +S^{FF}_{(\widetilde{kj})j(\widetilde{lj})}  \bigg]\nonumber\\
 &&\times A_3^{0}((\widetilde{kj}),1,(\widetilde{lj}))\,{\bar{B}}_1^{\gamma,0}(\widetilde{(\widetilde{kj})(\widetilde{lj})},\overline{1},i)\,J_2^{(2)}(\lbrace p\rbrace_{2}) \nonumber\\
 \ph{82}&&+\frac{1}{2}\,\,d_{3,g}^{0}(k,i,1)\,d_{3,g}^{0}((\widetilde{ki}),j,\overline{1})\,{\bar{B}}_1^{\gamma,0}((\widetilde{j\widetilde{(ki})}),\overline{\overline{1}},l)\,J_2^{(2)}(\lbrace p\rbrace_{2}) \nonumber\\
 \ph{83}&&-\frac{1}{2}\,\,d_{3,g}^{0}(l,i,1)\,d_{3,g}^{0}(k,j,\overline{1})\,{\bar{B}}_1^{\gamma,0}((\widetilde{kj}),\overline{\overline{1}},(\widetilde{li}))\,J_2^{(2)}(\lbrace p\rbrace_{2}) \nonumber\\
 \ph{84}&&-\frac{1}{2}\,\,A_3^{0}(k,i,l)\,d_{3,g}^{0}((\widetilde{ki}),j,1)\,{\bar{B}}_1^{\gamma,0}((\widetilde{j\widetilde{(ki})}),\overline{1},(\widetilde{il}))\,J_2^{(2)}(\lbrace p\rbrace_{2}) \nonumber\\
 \ph{85}&&-\frac{1}{2}\,\bigg[      +S^{FF}_{(\widetilde{ki})i1}       -S^{FF}_{(\widetilde{j\widetilde{(ki})})i\overline{1}}       -S^{FF}_{1i(\widetilde{li})}       +S^{FF}_{\overline{1}i(\widetilde{li})}       -S^{FF}_{(\widetilde{li})i(\widetilde{ki})}       +S^{FF}_{(\widetilde{li})i(\widetilde{j\widetilde{(ki})})}  \bigg]\nonumber\\
 &&\times d_{3,g}^{0}((\widetilde{ki}),j,1)\,{\bar{B}}_1^{\gamma,0}((\widetilde{j\widetilde{(ki})}),\overline{1},(\widetilde{li}))\,J_2^{(2)}(\lbrace p\rbrace_{2}) \nonumber\\
 \ph{86}&&+\frac{1}{2}\,\,d_{3,g}^{0}(k,j,1)\,d_{3,g}^{0}((\widetilde{kj}),i,\overline{1})\,{\bar{B}}_1^{\gamma,0}((\widetilde{i\widetilde{(kj})}),\overline{\overline{1}},l)\,J_2^{(2)}(\lbrace p\rbrace_{2}) \nonumber\\
 \ph{87}&&-\frac{1}{2}\,\,d_{3,g}^{0}(l,j,1)\,d_{3,g}^{0}(k,i,\overline{1})\,{\bar{B}}_1^{\gamma,0}((\widetilde{ki}),\overline{\overline{1}},(\widetilde{lj}))\,J_2^{(2)}(\lbrace p\rbrace_{2}) \nonumber\\
 \ph{88}&&-\frac{1}{2}\,\,A_3^{0}(k,j,l)\,d_{3,g}^{0}((\widetilde{kj}),i,1)\,{\bar{B}}_1^{\gamma,0}((\widetilde{i\widetilde{(kj})}),\overline{1},(\widetilde{jl}))\,J_2^{(2)}(\lbrace p\rbrace_{2}) \nonumber\\
 \ph{89}&&+\frac{1}{2}\,\bigg[      -S^{FF}_{(\widetilde{kj})j1}       +S^{FF}_{(\widetilde{i\widetilde{(kj})})j\overline{1}}       +S^{FF}_{1j(\widetilde{lj})}       -S^{FF}_{\overline{1}j(\widetilde{lj})}       +S^{FF}_{(\widetilde{lj})j(\widetilde{kj})}       -S^{FF}_{(\widetilde{lj})j(\widetilde{i\widetilde{(kj})})}  \bigg]\nonumber\\
 &&\times d_{3,g}^{0}((\widetilde{kj}),i,1)\,{\bar{B}}_1^{\gamma,0}((\widetilde{i\widetilde{(kj})}),\overline{1},(\widetilde{lj}))\,J_2^{(2)}(\lbrace p\rbrace_{2}) \nonumber\\
 \ph{90}&&+\frac{1}{2}\,\,d_{3,g}^{0}(l,i,1)\,d_{3,g}^{0}((\widetilde{li}),j,\overline{1})\,{\bar{B}}_1^{\gamma,0}(k,\overline{\overline{1}},(\widetilde{j\widetilde{(li})}))\,J_2^{(2)}(\lbrace p\rbrace_{2}) \nonumber\\
 \ph{91}&&-\frac{1}{2}\,\,d_{3,g}^{0}(k,i,1)\,d_{3,g}^{0}(l,j,\overline{1})\,{\bar{B}}_1^{\gamma,0}((\widetilde{ki}),\overline{\overline{1}},(\widetilde{lj}))\,J_2^{(2)}(\lbrace p\rbrace_{2}) \nonumber\\
 \ph{92}&&-\frac{1}{2}\,\,A_3^{0}(l,i,k)\,d_{3,g}^{0}((\widetilde{li}),j,1)\,{\bar{B}}_1^{\gamma,0}((\widetilde{ik}),\overline{1},(\widetilde{j\widetilde{(li})}))\,J_2^{(2)}(\lbrace p\rbrace_{2}) \nonumber\\
 \ph{93}&&+\frac{1}{2}\,\bigg[      -S^{FF}_{(\widetilde{li})i1}       +S^{FF}_{(\widetilde{j\widetilde{(li})})i\overline{1}}       +S^{FF}_{(\widetilde{ki})i1}       -S^{FF}_{(\widetilde{ki})i\overline{1}}       +S^{FF}_{(\widetilde{ki})i(\widetilde{li})}       -S^{FF}_{(\widetilde{ki})i(\widetilde{j\widetilde{(li})})}  \bigg]\nonumber\\
 &&\times d_{3,g}^{0}((\widetilde{li}),j,1)\,{\bar{B}}_1^{\gamma,0}((\widetilde{ki}),\overline{1},(\widetilde{j\widetilde{(li})}))\,J_2^{(2)}(\lbrace p\rbrace_{2}) \nonumber\\
 \ph{94}&&+\frac{1}{2}\,\,d_{3,g}^{0}(l,j,1)\,d_{3,g}^{0}((\widetilde{lj}),i,\overline{1})\,{\bar{B}}_1^{\gamma,0}(k,\overline{\overline{1}},(\widetilde{i\widetilde{(lj})}))\,J_2^{(2)}(\lbrace p\rbrace_{2}) \nonumber\\
 \ph{95}&&-\frac{1}{2}\,\,d_{3,g}^{0}(k,j,1)\,d_{3,g}^{0}(l,i,\overline{1})\,{\bar{B}}_1^{\gamma,0}((\widetilde{kj}),\overline{\overline{1}},(\widetilde{li}))\,J_2^{(2)}(\lbrace p\rbrace_{2}) \nonumber\\
 \ph{96}&&-\frac{1}{2}\,\,A_3^{0}(l,j,k)\,d_{3,g}^{0}((\widetilde{lj}),i,1)\,{\bar{B}}_1^{\gamma,0}((\widetilde{jk}),\overline{1},(\widetilde{i\widetilde{(lj})}))\,J_2^{(2)}(\lbrace p\rbrace_{2}) \nonumber\\
 \ph{97}&&+\frac{1}{2}\,\bigg[      -S^{FF}_{(\widetilde{lj})j1}       +S^{FF}_{(\widetilde{i\widetilde{(lj})})j\overline{1}}       +S^{FF}_{1j(\widetilde{kj})}       -S^{FF}_{\overline{1}j(\widetilde{kj})}       +S^{FF}_{(\widetilde{kj})j(\widetilde{lj})}       -S^{FF}_{(\widetilde{kj})j(\widetilde{i\widetilde{(lj})})}  \bigg]\nonumber\\
 &&\times d_{3,g}^{0}((\widetilde{lj}),i,1)\,{\bar{B}}_1^{\gamma,0}((\widetilde{kj}),\overline{1},(\widetilde{i\widetilde{(lj})}))\,J_2^{(2)}(\lbrace p\rbrace_{2}) \nonumber\\
 \ph{98}&&-\frac{1}{2}\,\,A_3^{0}(k,j,l)\,A_3^{0}((\widetilde{kj}),i,(\widetilde{lj}))\,{\bar{B}}_1^{\gamma,0}((\widetilde{i\widetilde{(kj})}),1,(\widetilde{i\widetilde{(lj})}))\,J_2^{(2)}(\lbrace p\rbrace_{2}) \nonumber\\
 \ph{99}&&+\frac{1}{2}\,\,d_{3,g}^{0}(k,j,1)\,A_3^{0}((\widetilde{kj}),i,l)\,{\bar{B}}_1^{\gamma,0}((\widetilde{i\widetilde{(kj})}),\overline{1},(\widetilde{il}))\,J_2^{(2)}(\lbrace p\rbrace_{2}) \nonumber\\
 \ph{100}&&+\frac{1}{2}\,\,d_{3,g}^{0}(l,j,1)\,A_3^{0}((\widetilde{lj}),i,k)\,{\bar{B}}_1^{\gamma,0}((\widetilde{ik}),\overline{1},(\widetilde{i\widetilde{(lj})}))\,J_2^{(2)}(\lbrace p\rbrace_{2}) \nonumber\\
 \ph{101}&&+\frac{1}{2}\,\bigg[      +S^{FF}_{(\widetilde{kj})j(\widetilde{lj})}       -S^{FF}_{(\widetilde{i\widetilde{(kj})})j(\widetilde{i\widetilde{(lj})})}       -S^{FF}_{(\widetilde{kj})j1}       +S^{FF}_{(\widetilde{i\widetilde{(kj})})j1}       -S^{FF}_{(\widetilde{lj})j1}       +S^{FF}_{(\widetilde{i\widetilde{(lj})})j1}  \bigg]\nonumber\\
 &&\times A_3^{0}((\widetilde{kj}),i,(\widetilde{lj}))\,{\bar{B}}_1^{\gamma,0}((\widetilde{i\widetilde{(kj})}),1,(\widetilde{i\widetilde{(lj})}))\,J_2^{(2)}(\lbrace p\rbrace_{2}) \nonumber\\
 \ph{102}&&-\frac{1}{2}\,\,A_3^{0}(k,i,l)\,A_3^{0}((\widetilde{ki}),j,(\widetilde{li}))\,{\bar{B}}_1^{\gamma,0}((\widetilde{j\widetilde{(ki})}),1,(\widetilde{j\widetilde{(li})}))\,J_2^{(2)}(\lbrace p\rbrace_{2}) \nonumber\\
 \ph{103}&&+\frac{1}{2}\,\,d_{3,g}^{0}(k,i,1)\,A_3^{0}((\widetilde{ki}),j,l)\,{\bar{B}}_1^{\gamma,0}((\widetilde{j\widetilde{(ki})}),\overline{1},(\widetilde{jl}))\,J_2^{(2)}(\lbrace p\rbrace_{2}) \nonumber\\
 \ph{104}&&+\frac{1}{2}\,\,d_{3,g}^{0}(l,i,1)\,A_3^{0}((\widetilde{li}),j,k)\,{\bar{B}}_1^{\gamma,0}((\widetilde{jk}),\overline{1},(\widetilde{j\widetilde{(li})}))\,J_2^{(2)}(\lbrace p\rbrace_{2}) \nonumber\\
 \ph{105}&&+\frac{1}{2}\,\bigg[      +S^{FF}_{(\widetilde{ki})i(\widetilde{li})}       -S^{FF}_{(\widetilde{j\widetilde{(ki})})i(\widetilde{j\widetilde{(li})})}       -S^{FF}_{(\widetilde{ki})i1}       +S^{FF}_{(\widetilde{j\widetilde{(ki})})i1}       -S^{FF}_{(\widetilde{li})i1}       +S^{FF}_{(\widetilde{j\widetilde{(li})})i1}  \bigg]\nonumber\\
 &&\times A_3^{0}((\widetilde{ki}),j,(\widetilde{li}))\,{\bar{B}}_1^{\gamma,0}((\widetilde{j\widetilde{(ki})}),1,(\widetilde{j\widetilde{(li})}))\,J_2^{(2)}(\lbrace p\rbrace_{2}) \nonumber\\
 \ph{106}&&-A_3^{0}(k,1,l)\,\bar{B}_2^{\gamma,0}(\overline{1},j,i,(\widetilde{kl}))\,J_2^{(3)}(\lbrace p\rbrace_{3}) \nonumber\\
 \ph{107}&&-A_3^{0}(k,1,l)\,\bar{B}_2^{\gamma,0}(\overline{1},i,j,(\widetilde{kl}))\,J_2^{(3)}(\lbrace p\rbrace_{3}) \nonumber\\
\snip \ph{108}&&-\,A_4^0(k,1,i,l)\,{\bar{B}}_1^{\gamma,0}(\overline{1},j,(\widetilde{kil}))\,J_2^{(2)}(\lbrace p\rbrace_{2}) \nonumber\\
\snip \ph{109}&&-\,A_4^0(k,i,1,l)\,{\bar{B}}_1^{\gamma,0}(\overline{1},j,(\widetilde{kil}))\,J_2^{(2)}(\lbrace p\rbrace_{2}) \nonumber\\
 \ph{110}&&+\,d_{3,g}^{0}(k,i,1)\,A_3^{0}((\widetilde{ki}),\overline{1},l)\,{\bar{B}}_1^{\gamma,0}(\overline{\overline{1}},j,(\widetilde{l\widetilde{(ki})}))\,J_2^{(2)}(\lbrace p\rbrace_{2}) \nonumber\\
 \ph{111}&&+\,d_{3,g}^{0}(l,i,1)\,A_3^{0}((\widetilde{li}),\overline{1},k)\,{\bar{B}}_1^{\gamma,0}(\overline{\overline{1}},j,(\widetilde{k\widetilde{(li})}))\,J_2^{(2)}(\lbrace p\rbrace_{2}) \nonumber\\
 \ph{112}&&+\,A_3^{0}(k,1,l)\,A_3^{0}(\overline{1},i,(\widetilde{kl}))\,{\bar{B}}_1^{\gamma,0}(\overline{\overline{1}},j,(\widetilde{i\widetilde{(kl})}))\,J_2^{(2)}(\lbrace p\rbrace_{2}) \nonumber\\
\snip \ph{113}&&-\,A_4^0(k,1,j,l)\,{\bar{B}}_1^{\gamma,0}(\overline{1},i,(\widetilde{kjl}))\,J_2^{(2)}(\lbrace p\rbrace_{2}) \nonumber\\
\snip \ph{114}&&-\,A_4^0(k,j,1,l)\,{\bar{B}}_1^{\gamma,0}(\overline{1},i,(\widetilde{kjl}))\,J_2^{(2)}(\lbrace p\rbrace_{2}) \nonumber\\
 \ph{115}&&+\,d_{3,g}^{0}(k,j,1)\,A_3^{0}((\widetilde{kj}),\overline{1},l)\,{\bar{B}}_1^{\gamma,0}(\overline{\overline{1}},i,(\widetilde{l\widetilde{(kj})}))\,J_2^{(2)}(\lbrace p\rbrace_{2}) \nonumber\\
 \ph{116}&&+\,d_{3,g}^{0}(l,j,1)\,A_3^{0}((\widetilde{lj}),\overline{1},k)\,{\bar{B}}_1^{\gamma,0}(\overline{\overline{1}},i,(\widetilde{k\widetilde{(lj})}))\,J_2^{(2)}(\lbrace p\rbrace_{2}) \nonumber\\
 \ph{117}&&+\,A_3^{0}(k,1,l)\,A_3^{0}(\overline{1},j,(\widetilde{kl}))\,{\bar{B}}_1^{\gamma,0}(\overline{\overline{1}},i,(\widetilde{j\widetilde{(kl})}))\,J_2^{(2)}(\lbrace p\rbrace_{2}) \nonumber\\
 \ph{118}&&+\,f_{3,g}^{0}(1,j,i)\,d_{3,g\to q}^{0}(k,\overline{1},(\widetilde{ij}))\,{\bar{B}}_1^{\gamma,0}(\overline{\overline{1}},(\widetilde{k\widetilde{(ij})}),l)\,J_2^{(2)}(\lbrace p\rbrace_{2}) \nonumber\\
 \ph{119}&&-\,d_3^{0}(k,j,i)\,d_{3,g\to q}^{0}((\widetilde{kj}),1,(\widetilde{ji}))\,{\bar{B}}_1^{\gamma,0}(\overline{1},\widetilde{(\widetilde{kj})(\widetilde{ji})},l)\,J_2^{(2)}(\lbrace p\rbrace_{2}) \nonumber\\
 \ph{120}&&+\,A_3^{0}(k,j,l)\,d_{3,g\to q}^{0}((\widetilde{kj}),1,i)\,{\bar{B}}_1^{\gamma,0}(\overline{1},(\widetilde{i\widetilde{(kj})}),(\widetilde{lj}))\,J_2^{(2)}(\lbrace p\rbrace_{2}) \nonumber\\
 \ph{121}&&-\,d_{3,g}^{0}(l,j,1)\,d_{3,g\to q}^{0}(k,\overline{1},i)\,{\bar{B}}_1^{\gamma,0}(\overline{\overline{1}},(\widetilde{ki}),(\widetilde{lj}))\,J_2^{(2)}(\lbrace p\rbrace_{2}) \nonumber\\
 \ph{122}&&-\bigg[      +S^{FF}_{(\widetilde{ij})j1}       -S^{FF}_{(\widetilde{kj})j(\widetilde{ji})}       +S^{FF}_{(\widetilde{kj})jl}       -S^{FF}_{lj1}  \bigg]\nonumber\\
 &&\times d_{3,g\to q}^{0}((\widetilde{kj}),1,(\widetilde{ij}))\,{\bar{B}}_1^{\gamma,0}(\overline{1},\widetilde{(\widetilde{kj})(\widetilde{ij})},l)\,J_2^{(2)}(\lbrace p\rbrace_{2}) \nonumber\\
 \ph{123}&&+\,f_{3,g}^{0}(1,i,j)\,d_{3,g\to q}^{0}(k,\overline{1},(\widetilde{ji}))\,{\bar{B}}_1^{\gamma,0}(\overline{\overline{1}},(\widetilde{k\widetilde{(ji})}),l)\,J_2^{(2)}(\lbrace p\rbrace_{2}) \nonumber\\
 \ph{124}&&-\,d_3^{0}(k,i,j)\,d_{3,g\to q}^{0}((\widetilde{ki}),1,(\widetilde{ij}))\,{\bar{B}}_1^{\gamma,0}(\overline{1},\widetilde{(\widetilde{ki})(\widetilde{ij})},l)\,J_2^{(2)}(\lbrace p\rbrace_{2}) \nonumber\\
 \ph{125}&&+\,A_3^{0}(k,i,l)\,d_{3,g\to q}^{0}((\widetilde{ki}),1,j)\,{\bar{B}}_1^{\gamma,0}(\overline{1},(\widetilde{j\widetilde{(ki})}),(\widetilde{li}))\,J_2^{(2)}(\lbrace p\rbrace_{2}) \nonumber\\
 \ph{126}&&-\,d_{3,g}^{0}(l,i,1)\,d_{3,g\to q}^{0}(k,\overline{1},j)\,{\bar{B}}_1^{\gamma,0}(\overline{\overline{1}},(\widetilde{kj}),(\widetilde{li}))\,J_2^{(2)}(\lbrace p\rbrace_{2}) \nonumber\\
 \ph{127}&&-\bigg[      +S^{FF}_{(\widetilde{ji})i1}       -S^{FF}_{(\widetilde{ki})i(\widetilde{ij})}       +S^{FF}_{(\widetilde{ki})il}       -S^{FF}_{li1}  \bigg]\nonumber\\
 &&\times d_{3,g\to q}^{0}((\widetilde{ki}),1,(\widetilde{ji}))\,{\bar{B}}_1^{\gamma,0}(\overline{1},\widetilde{(\widetilde{ki})(\widetilde{ji})},l)\,J_2^{(2)}(\lbrace p\rbrace_{2}) \nonumber\\
 \ph{128}&&+\,f_{3,g}^{0}(1,j,i)\,d_{3,g\to q}^{0}(l,\overline{1},(\widetilde{ij}))\,{\bar{B}}_1^{\gamma,0}(\overline{\overline{1}},(\widetilde{l\widetilde{(ij})}),k)\,J_2^{(2)}(\lbrace p\rbrace_{2}) \nonumber\\
 \ph{129}&&-\,d_3^{0}(l,j,i)\,d_{3,g\to q}^{0}((\widetilde{lj}),1,(\widetilde{ji}))\,{\bar{B}}_1^{\gamma,0}(\overline{1},\widetilde{(\widetilde{lj})(\widetilde{ji})},k)\,J_2^{(2)}(\lbrace p\rbrace_{2}) \nonumber\\
 \ph{130}&&+\,A_3^{0}(l,j,k)\,d_{3,g\to q}^{0}((\widetilde{lj}),1,i)\,{\bar{B}}_1^{\gamma,0}(\overline{1},(\widetilde{i\widetilde{(lj})}),(\widetilde{kj}))\,J_2^{(2)}(\lbrace p\rbrace_{2}) \nonumber\\
 \ph{131}&&-\,d_{3,g}^{0}(k,j,1)\,d_{3,g\to q}^{0}(l,\overline{1},i)\,{\bar{B}}_1^{\gamma,0}(\overline{\overline{1}},(\widetilde{li}),(\widetilde{kj}))\,J_2^{(2)}(\lbrace p\rbrace_{2}) \nonumber\\
 \ph{132}&&-\bigg[      +S^{FF}_{(\widetilde{ij})j1}       -S^{FF}_{(\widetilde{lj})j(\widetilde{ji})}       +S^{FF}_{(\widetilde{lj})jk}       -S^{FF}_{kj1}  \bigg]\nonumber\\
 &&\times d_{3,g\to q}^{0}((\widetilde{lj}),1,(\widetilde{ij}))\,{\bar{B}}_1^{\gamma,0}(\overline{1},\widetilde{(\widetilde{lj})(\widetilde{ij})},k)\,J_2^{(2)}(\lbrace p\rbrace_{2}) \nonumber\\
 \ph{133}&&+\,f_{3,g}^{0}(1,i,j)\,d_{3,g\to q}^{0}(l,\overline{1},(\widetilde{ji}))\,{\bar{B}}_1^{\gamma,0}(\overline{\overline{1}},(\widetilde{l\widetilde{(ji})}),k)\,J_2^{(2)}(\lbrace p\rbrace_{2}) \nonumber\\
 \ph{134}&&-\,d_3^{0}(l,i,j)\,d_{3,g\to q}^{0}((\widetilde{li}),1,(\widetilde{ij}))\,{\bar{B}}_1^{\gamma,0}(\overline{1},\widetilde{(\widetilde{li})(\widetilde{ij})},k)\,J_2^{(2)}(\lbrace p\rbrace_{2}) \nonumber\\
 \ph{135}&&+\,A_3^{0}(l,i,k)\,d_{3,g\to q}^{0}((\widetilde{li}),1,j)\,{\bar{B}}_1^{\gamma,0}(\overline{1},(\widetilde{j\widetilde{(li})}),(\widetilde{ki}))\,J_2^{(2)}(\lbrace p\rbrace_{2}) \nonumber\\
 \ph{136}&&-\,d_{3,g}^{0}(k,i,1)\,d_{3,g\to q}^{0}(l,\overline{1},j)\,{\bar{B}}_1^{\gamma,0}(\overline{\overline{1}},(\widetilde{lj}),(\widetilde{ki}))\,J_2^{(2)}(\lbrace p\rbrace_{2}) \nonumber\\
 \ph{137}&&-\bigg[      +S^{FF}_{(\widetilde{ji})i1}       -S^{FF}_{(\widetilde{li})i(\widetilde{ij})}       +S^{FF}_{(\widetilde{li})ik}       -S^{FF}_{ki1}  \bigg]\nonumber\\
 &&\times d_{3,g\to q}^{0}((\widetilde{li}),1,(\widetilde{ji}))\,{\bar{B}}_1^{\gamma,0}(\overline{1},\widetilde{(\widetilde{li})(\widetilde{ji})},k)\,J_2^{(2)}(\lbrace p\rbrace_{2}) \nonumber\\
 \ph{138}&&+\,d_3^{0}(k,i,j)\,A_3^{0}((\widetilde{ki}),1,l)\,{\bar{B}}_1^{\gamma,0}(\overline{1},(\widetilde{ji}),(\widetilde{l\widetilde{(ki})}))\,J_2^{(2)}(\lbrace p\rbrace_{2}) \nonumber\\
 \ph{139}&&+\,d_3^{0}(k,j,i)\,A_3^{0}((\widetilde{kj}),1,l)\,{\bar{B}}_1^{\gamma,0}(\overline{1},(\widetilde{ij}),(\widetilde{l\widetilde{(kj})}))\,J_2^{(2)}(\lbrace p\rbrace_{2}) \nonumber\\
 \ph{140}&&+\,d_3^{0}(l,i,j)\,A_3^{0}((\widetilde{li}),1,k)\,{\bar{B}}_1^{\gamma,0}(\overline{1},(\widetilde{ji}),(\widetilde{k\widetilde{(li})}))\,J_2^{(2)}(\lbrace p\rbrace_{2}) \nonumber\\
 \ph{141}&&+\,d_3^{0}(l,j,i)\,A_3^{0}((\widetilde{lj}),1,k)\,{\bar{B}}_1^{\gamma,0}(\overline{1},(\widetilde{ij}),(\widetilde{k\widetilde{(lj})}))\,J_2^{(2)}(\lbrace p\rbrace_{2}) \nonumber\\
 \ph{142}&&-\,A_3^{0}(k,j,l)\,A_3^{0}((\widetilde{kj}),1,(\widetilde{jl}))\,{\bar{B}}_1^{\gamma,0}(\overline{1},i,\widetilde{(\widetilde{kj})(\widetilde{jl})})\,J_2^{(2)}(\lbrace p\rbrace_{2}) \nonumber\\
 \ph{143}&&-\,A_3^{0}(k,i,l)\,A_3^{0}((\widetilde{ki}),1,(\widetilde{il}))\,{\bar{B}}_1^{\gamma,0}(\overline{1},j,\widetilde{(\widetilde{ki})(\widetilde{il})})\,J_2^{(2)}(\lbrace p\rbrace_{2}) \nonumber\\
 \ph{144}&&+\bigg[      +S^{FF}_{\overline{1}ij}       -S^{FF}_{\overline{1}i\widetilde{(\widetilde{ki})(\widetilde{li})}}       -S^{FF}_{(\widetilde{ki})ij}       -S^{FF}_{(\widetilde{li})ij}       +S^{FF}_{(\widetilde{ki})i(\widetilde{li})}       +S^{FF}_{ji\widetilde{(\widetilde{ki})(\widetilde{li})}}  \bigg]\nonumber\\
 &&\times A_3^{0}((\widetilde{ki}),1,(\widetilde{li}))\,{\bar{B}}_1^{\gamma,0}(\overline{1},j,\widetilde{(\widetilde{ki})(\widetilde{li})})\,J_2^{(2)}(\lbrace p\rbrace_{2}) \nonumber\\
 \ph{145}&&+\bigg[      +S^{FF}_{ij\widetilde{(\widetilde{kj})(\widetilde{lj})}}       +S^{FF}_{\overline{1}ji}       -S^{FF}_{\overline{1}j\widetilde{(\widetilde{kj})(\widetilde{lj})}}       -S^{FF}_{(\widetilde{kj})ji}       -S^{FF}_{(\widetilde{lj})ji}       +S^{FF}_{(\widetilde{kj})j(\widetilde{lj})}  \bigg]\nonumber\\
 &&\times A_3^{0}((\widetilde{kj}),1,(\widetilde{lj}))\,{\bar{B}}_1^{\gamma,0}(\overline{1},i,\widetilde{(\widetilde{kj})(\widetilde{lj})})\,J_2^{(2)}(\lbrace p\rbrace_{2}) .
\end{eqnarray}

\addtocontents{toc}{\protect\setcounter{tocdepth}{2}}
\subsubsection{B-type $\mathcal{O}(N^{0})$ contribution}
\addtocontents{toc}{\protect\setcounter{tocdepth}{3}}

The subleading-colour matrix element, $\tilde{B}_{3}^{\gamma,0}$, is summed over all gluon permutations, including the initial-state gluon.
The subtraction term is constructed in a way such that it is summed only over final-state gluon permutations.

The form of the IF collinear unresolved limits depend on whether the initial-state parton is abelian-like or not and so the subtraction term is
constructed to reflect these possibilities. The total subtracted contribution to the cross section
is given by
\ba
\sum_{P(\hat{1},i,j)}\tilde{\bar{B}}_{3}^{\gamma,0}(3,\hat{1},i,j,4)-\sum_{P(i,j)}\tilde{B}_{3}^{\gamma,0,S}(3,\hat{1},i,j,4),
\ea
where
\begin{eqnarray}
\lefteqn{{\tilde{B}_{3}^{\gamma,0,S}(k,\hat{1},i,j,l) =}} \nonumber \\
 \ph{1}&&+A_3^{0}(k,i,l)\,\bar{B}_2^{\gamma,0}((\widetilde{ki}),j,1,(\widetilde{li}))\,J_2^{(3)}(\lbrace p\rbrace_{3}) \nonumber\\
 \ph{2}&&+A_3^{0}(k,i,l)\,\bar{B}_2^{\gamma,0}((\widetilde{ki}),1,j,(\widetilde{li}))\,J_2^{(3)}(\lbrace p\rbrace_{3}) \nonumber\\
 \ph{3}&&+d_{3,g}^{0}(k,j,1)\,\tilde{\bar{B}}_2^{\gamma,0}((\widetilde{kj}),i,\overline{1},l)\,J_2^{(3)}(\lbrace p\rbrace_{3}) \nonumber\\
 \ph{4}&&+d_{3,g}^{0}(l,j,1)\,\tilde{\bar{B}}_2^{\gamma,0}(k,i,\overline{1},(\widetilde{lj}))\,J_2^{(3)}(\lbrace p\rbrace_{3}) \nonumber\\
 \ph{5}&&+d_3^{0}(k,i,j)\,\tilde{\bar{B}}_2^{\gamma,0}((\widetilde{ki}),1,(\widetilde{ij}),l)\,J_2^{(3)}(\lbrace p\rbrace_{2}) \nonumber\\
 \ph{6}&&+d_3^{0}(l,j,i)\,\tilde{\bar{B}}_2^{\gamma,0}(k,1,(\widetilde{ij}),(\widetilde{lj}))\,J_2^{(3)}(\lbrace p\rbrace_{2}) \nonumber\\
\snip \ph{7}&&+\,A_4^0(k,i,j,l)\,{\bar{B}}_1^{\gamma,0}((\widetilde{kij}),1,(\widetilde{lji}))\,J_2^{(2)}(\lbrace p\rbrace_{2}) \nonumber\\
 \ph{8}&&-\,d_3^{0}(k,i,j)\,A_3^{0}((\widetilde{ki}),(\widetilde{ji}),l)\,{\bar{B}}_1^{\gamma,0}(\widetilde{(\widetilde{ki})(\widetilde{ji})},1,(\widetilde{l\widetilde{(ji})}))\,J_2^{(2)}(\lbrace p\rbrace_{3}) \nonumber\\
 \ph{9}&&-\,d_3^{0}(l,j,i)\,A_3^{0}(k,(\widetilde{ji}),(\widetilde{lj}))\,{\bar{B}}_1^{\gamma,0}((\widetilde{k\widetilde{(ji})}),1,\widetilde{(\widetilde{ji})(\widetilde{lj})})\,J_2^{(2)}(\lbrace p\rbrace_{2}) \nonumber\\
\snip \ph{10}&&+\,\tilde{A}_{4}^{0,a}(k,i,j,l)\,{\bar{B}}_1^{\gamma,0}((\widetilde{kij}),1,(\widetilde{lji}))\,J_2^{(2)}(\lbrace p\rbrace_{2}) \nonumber\\
\snip \ph{11}&&+\,\tilde{A}_{4}^{0,a}(k,j,i,l)\,{\bar{B}}_1^{\gamma,0}((\widetilde{kji}),1,(\widetilde{lij}))\,J_2^{(2)}(\lbrace p\rbrace_{2}) \nonumber\\
 \ph{12}&&-\,A_3^{0}(k,i,l)\,A_3^{0}((\widetilde{ki}),j,(\widetilde{li}))\,{\bar{B}}_1^{\gamma,0}((\widetilde{(\widetilde{ki})j}),1,(\widetilde{j\widetilde{(li})}))\,J_2^{(2)}(\lbrace p\rbrace_{2}) \nonumber\\
 \ph{13}&&-\,A_3^{0}(k,j,l)\,A_3^{0}((\widetilde{kj}),i,(\widetilde{lj}))\,{\bar{B}}_1^{\gamma,0}((\widetilde{(\widetilde{kj})i}),1,(\widetilde{i\widetilde{(lj})}))\,J_2^{(2)}(\lbrace p\rbrace_{2}) \nonumber\\
 \ph{14}&&+\,A_3^{0}(k,j,l)\,A_3^{0}((\widetilde{kj}),i,(\widetilde{lj}))\,{\bar{B}}_1^{\gamma,0}((\widetilde{(\widetilde{kj})i}),1,(\widetilde{i\widetilde{(lj})}))\,J_2^{(2)}(\lbrace p\rbrace_{2}) \nonumber\\
 \ph{15}&&-\,d_{3,g}^{0}(k,j,1)\,A_3^{0}((\widetilde{kj}),i,l)\,{\bar{B}}_1^{\gamma,0}((\widetilde{(\widetilde{kj})i}),\overline{1},(\widetilde{li}))\,J_2^{(2)}(\lbrace p\rbrace_{2}) \nonumber\\
 \ph{16}&&-\,d_{3,g}^{0}(l,j,1)\,A_3^{0}(k,i,(\widetilde{lj}))\,{\bar{B}}_1^{\gamma,0}((\widetilde{ki}),\overline{1},(\widetilde{(\widetilde{lj})i}))\,J_2^{(2)}(\lbrace p\rbrace_{2}) \nonumber\\
 \ph{17}&&-\bigg[      +S^{FF}_{(\widetilde{kj})j(\widetilde{lj})}       -S^{FF}_{(\widetilde{i\widetilde{(kj})})j(\widetilde{i\widetilde{(lj})})}       -S^{FF}_{(\widetilde{kj})j1}       +S^{FF}_{(\widetilde{i\widetilde{(kj})})j1}       -S^{FF}_{(\widetilde{lj})j1}       +S^{FF}_{(\widetilde{i\widetilde{(lj})})j1}  \bigg]\nonumber\\
 &&\times A_3^{0}((\widetilde{kj}),i,(\widetilde{lj}))\,{\bar{B}}_1^{\gamma,0}((\widetilde{i\widetilde{(kj})}),1,(\widetilde{i\widetilde{(lj})}))\,J_2^{(2)}(\lbrace p\rbrace_{2}) \nonumber\\
 \ph{18}&&-A_3^{0}(k,1,l)\,\tilde{\bar{B}}_2^{\gamma,0}(\overline{1},i,j,(\widetilde{kl}))\,J_2^{(3)}(\lbrace p\rbrace_{3}) \nonumber\\
 \ph{19}&&-A_3^{0}(k,1,l)\,\bar{B}_2^{\gamma,0}(\overline{1},i,j,(\widetilde{kl}))\,J_2^{(3)}(\lbrace p\rbrace_{3}) \nonumber\\
\snip \ph{20}&&-\,\tilde{A}_4^0(k,1,i,l)\,{\bar{B}}_1^{\gamma,0}(\overline{1},j,(\widetilde{kli}))\,J_2^{(2)}(\lbrace p\rbrace_{2}) \nonumber\\
 \ph{21}&&+\,A_3^{0}(k,1,l)\,A_3^{0}(\overline{1},i,(\widetilde{kl}))\,{\bar{B}}_1^{\gamma,0}(\overline{\overline{1}},j,(\widetilde{i\widetilde{(kl})}))\,J_2^{(2)}(\lbrace p\rbrace_{2}) \nonumber\\
 \ph{22}&&+\,A_3^{0}(k,i,l)\,A_3^{0}((\widetilde{ki}),1,(\widetilde{li}))\,{\bar{B}}_1^{\gamma,0}(\overline{1},j,\widetilde{(\widetilde{ki})(\widetilde{li})})\,J_2^{(2)}(\lbrace p\rbrace_{2}) \nonumber\\
\snip \ph{23}&&-\,\tilde{A}_4^0(k,1,j,l)\,{\bar{B}}_1^{\gamma,0}(\overline{1},i,(\widetilde{klj}))\,J_2^{(2)}(\lbrace p\rbrace_{2}) \nonumber\\
 \ph{24}&&+\,A_3^{0}(k,1,l)\,A_3^{0}(\overline{1},j,(\widetilde{kl}))\,{\bar{B}}_1^{\gamma,0}(\overline{\overline{1}},i,(\widetilde{j\widetilde{(kl})}))\,J_2^{(2)}(\lbrace p\rbrace_{2}) \nonumber\\
 \ph{25}&&+\,A_3^{0}(k,j,l)\,A_3^{0}((\widetilde{kj}),1,(\widetilde{lj}))\,{\bar{B}}_1^{\gamma,0}(\overline{1},i,\widetilde{(\widetilde{kj})(\widetilde{lj})})\,J_2^{(2)}(\lbrace p\rbrace_{2}) \nonumber\\
\snip \ph{26}&&-\,A_4^0(k,j,1,l)\,{\bar{B}}_1^{\gamma,0}(\overline{1},i,(\widetilde{kjl}))\,J_2^{(2)}(\lbrace p\rbrace_{2}) \nonumber\\
 \ph{27}&&+\,d_{3,g}^{0}(k,j,1)\,A_3^{0}((\widetilde{kj}),\overline{1},l)\,{\bar{B}}_1^{\gamma,0}(\overline{\overline{1}},i,(\widetilde{l\widetilde{(kj})}))\,J_2^{(2)}(\lbrace p\rbrace_{2}) \nonumber\\
\snip \ph{28}&&-\,A_4^0(k,1,j,l)\,{\bar{B}}_1^{\gamma,0}(\overline{1},i,(\widetilde{kjl}))\,J_2^{(2)}(\lbrace p\rbrace_{2}) \nonumber\\
 \ph{29}&&+\,d_{3,g}^{0}(l,j,1)\,A_3^{0}((\widetilde{lj}),\overline{1},k)\,{\bar{B}}_1^{\gamma,0}(\overline{\overline{1}},i,(\widetilde{k\widetilde{(lj})}))\,J_2^{(2)}(\lbrace p\rbrace_{2}) \nonumber\\
 \ph{30}&&+\,A_3^{0}(k,1,l)\,A_3^{0}(\overline{1},j,(\widetilde{kl}))\,{\bar{B}}_1^{\gamma,0}(\overline{\overline{1}},i,(\widetilde{j\widetilde{(kl})}))\,J_2^{(2)}(\lbrace p\rbrace_{2}) \nonumber\\
 \ph{31}&&-\,A_3^{0}(k,i,l)\,A_3^{0}((\widetilde{ki}),1,(\widetilde{li}))\,{\bar{B}}_1^{\gamma,0}(\overline{1},j,\widetilde{(\widetilde{ki})(\widetilde{li})})\,J_2^{(2)}(\lbrace p\rbrace_{2}) \nonumber\\
 \ph{32}&&+\,d_3^{0}(k,i,j)\,A_3^{0}((\widetilde{ki}),1,l)\,{\bar{B}}_1^{\gamma,0}(\overline{1},(\widetilde{ij}),(\widetilde{(\widetilde{ki})l}))\,J_2^{(2)}(\lbrace p\rbrace_{2}) \nonumber\\
 \ph{33}&&+\,d_3^{0}(l,i,j)\,A_3^{0}((\widetilde{li}),1,k)\,{\bar{B}}_1^{\gamma,0}(\overline{1},(\widetilde{ij}),(\widetilde{(\widetilde{li})k}))\,J_2^{(2)}(\lbrace p\rbrace_{2}) \nonumber\\
 \ph{34}&&+\bigg[      +S^{FF}_{\overline{1}ij}       -S^{FF}_{\overline{1}i\widetilde{(\widetilde{ki})(\widetilde{li})}}       -S^{FF}_{(\widetilde{ki})ij}       -S^{FF}_{(\widetilde{li})ij}       +S^{FF}_{(\widetilde{ki})i(\widetilde{li})}       +S^{FF}_{ji\widetilde{(\widetilde{ki})(\widetilde{li})}}  \bigg]\nonumber\\
 &&\times A_3^{0}((\widetilde{ki}),1,(\widetilde{li}))\,{\bar{B}}_1^{\gamma,0}(\overline{1},j,\widetilde{(\widetilde{li})(\widetilde{ki})})\,J_2^{(2)}(\lbrace p\rbrace_{2}) .
\end{eqnarray}

\addtocontents{toc}{\protect\setcounter{tocdepth}{2}}
\subsubsection{B-type $\mathcal{O}(N^{-2})$ contribution}
\addtocontents{toc}{\protect\setcounter{tocdepth}{3}}

The most subleading-colour two-quark--three-gluon matrix element contains only QED-like unresolved limits and is not summed
over any gluon permutations as it is completely symmetric in all gluon arguments. Its contribution to the
cross section is given by
\ba
\tilde{\tilde{\bar{B}}}_{3}^{\gamma,0}(3,\hat{1},2,5,4)-\tilde{\tilde{B}}_{3}^{\gamma,0,S}(3,\hat{1},2,5,4),
\ea

\begin{eqnarray}
\lefteqn{{\tilde{\tilde{B}}_{3}^{\gamma,0,S}(k,\hat{1},i,j,l) =}}\nonumber \\
 \ph{1}&&+A_3^{0}(k,i,l)\,\tilde{\bar{B}}_2^{\gamma,0}((\widetilde{ki}),1,j,(\widetilde{li}))\,J_2^{(3)}(\lbrace p\rbrace_{3}) \nonumber\\
 \ph{2}&&+A_3^{0}(k,j,l)\,\tilde{\bar{B}}_2^{\gamma,0}((\widetilde{kj}),1,i,(\widetilde{lj}))\,J_2^{(3)}(\lbrace p\rbrace_{3}) \nonumber\\
\snip \ph{3}&&+\,\tilde{A}_{4}^{0,a}(k,i,j,l)\,{\bar{B}}_1^{\gamma,0}((\widetilde{kij}),1,(\widetilde{lji}))\,J_2^{(2)}(\lbrace p\rbrace_{2}) \nonumber\\
\snip \ph{4}&&+\,\tilde{A}_{4}^{0,a}(k,j,i,l)\,{\bar{B}}_1^{\gamma,0}((\widetilde{kji}),1,(\widetilde{lij}))\,J_2^{(2)}(\lbrace p\rbrace_{2}) \nonumber\\
 \ph{5}&&-\,A_3^{0}(k,i,l)\,A_3^{0}((\widetilde{ki}),j,(\widetilde{li}))\,{\bar{B}}_1^{\gamma,0}((\widetilde{j\widetilde{(ki})}),1,(\widetilde{j\widetilde{(li})}))\,J_2^{(2)}(\lbrace p\rbrace_{2}) \nonumber\\
 \ph{6}&&-\,A_3^{0}(k,j,l)\,A_3^{0}((\widetilde{kj}),i,(\widetilde{lj}))\,{\bar{B}}_1^{\gamma,0}((\widetilde{i\widetilde{(kj})}),1,(\widetilde{i\widetilde{(lj})}))\,J_2^{(2)}(\lbrace p\rbrace_{2}) \nonumber\\
 \ph{7}&&-A_3^{0}(k,1,l)\,\tilde{\bar{B}}_2^{\gamma,0}(\overline{1},i,j,(\widetilde{kl}))\,J_2^{(3)}(\lbrace p\rbrace_{3}) \nonumber\\
\snip \ph{8}&&-\,\tilde{A}_4^0(k,1,i,l)\,{\bar{B}}_1^{\gamma,0}(\overline{1},j,(\widetilde{kil}))\,J_2^{(2)}(\lbrace p\rbrace_{2}) \nonumber\\
 \ph{9}&&+\,A_3^{0}(k,1,l)\,A_3^{0}(\overline{1},i,(\widetilde{kl}))\,{\bar{B}}_1^{\gamma,0}(\overline{\overline{1}},j,(\widetilde{i\widetilde{(kl})}))\,J_2^{(2)}(\lbrace p\rbrace_{2}) \nonumber\\
 \ph{10}&&+\,A_3^{0}(k,i,l)\,A_3^{0}((\widetilde{ki}),1,(\widetilde{li}))\,{\bar{B}}_1^{\gamma,0}(\overline{1},j,[(\widetilde{ki}), (\widetilde{li})])\,J_2^{(2)}(\lbrace p\rbrace_{2}) \nonumber\\
\snip \ph{11}&&-\,\tilde{A}_4^0(k,1,j,l)\,{\bar{B}}_1^{\gamma,0}(\overline{1},i,(\widetilde{kjl}))\,J_2^{(2)}(\lbrace p\rbrace_{2}) \nonumber\\
 \ph{12}&&+\,A_3^{0}(k,1,l)\,A_3^{0}(\overline{1},j,(\widetilde{kl}))\,{\bar{B}}_1^{\gamma,0}(\overline{\overline{1}},i,(\widetilde{j\widetilde{(kl})}))\,J_2^{(2)}(\lbrace p\rbrace_{2}) \nonumber\\
 \ph{13}&&+\,A_3^{0}(k,j,l)\,A_3^{0}((\widetilde{kj}),1,(\widetilde{lj}))\,{\bar{B}}_1^{\gamma,0}(\overline{1},i,[(\widetilde{kj}), (\widetilde{lj})])\,J_2^{(2)}(\lbrace p\rbrace_{2}) .
\end{eqnarray}

\addtocontents{toc}{\protect\setcounter{tocdepth}{2}}
\subsubsection{C-type $\mathcal{O}(N_FN^{1})$ contribution}
\addtocontents{toc}{\protect\setcounter{tocdepth}{3}}
\label{sec:nnloRRCg}

The four-quark--one-gluon matrix element has contributions from identical and non-identical
flavour quark lines. As was the case in Secs.~\ref{sec:nnloRRC}--\ref{sec:nnloRRD} for the quark channel,
the gluon-initiated matrix elements can be decomposed into $C_{1}^{\gamma,0}$ terms and $D_{1}^{\gamma,0}$,
where each term can be further decomposed into leading colour, subleading colour and closed quark loop terms.

The leading-colour contribution to the cross section is given by
\ba
C_{1}^{\gamma,0}(3,\hat{1};5,4;2)-C_{1}^{\gamma,0,S}(3,\hat{1};5,4;2),
\ea
where
\begin{eqnarray}
\lefteqn{{C_{1}^{\gamma,0,S}(k,\hat{1};i,j;l) =}} \nonumber \\
 \ph{1}&&+E_3^{0}(l,i,j)\,B_{2,q}^{\gamma,0}(k,(\widetilde{ji}),1,(\widetilde{li}))\,J_2^{(3)}(\lbrace p\rbrace_{3}) \nonumber\\
 \ph{2}&&+E_3^{0}(k,j,i)\,B_{2,q}^{\gamma,0}((\widetilde{kj}),1,(\widetilde{ji}),l)\,J_2^{(3)}(\lbrace p\rbrace_{3}) \nonumber\\
 \ph{3}&&-a_{3,g\to q}^{0}(k,1,l)\,C_0^{\gamma,0}(\overline{1};i,j;(\widetilde{kl}))\,J_2^{(3)}(\lbrace p\rbrace_{3}) \nonumber\\
 \ph{4}&&-a_{3,g\to q}^{0}(l,1,k)\,C_0^{\gamma,0}((\widetilde{kl});i,j;\overline{1})\,J_2^{(3)}(\lbrace p\rbrace_{3}) \nonumber\\
 \ph{5}&&-a_{3,g\to q}^{0}(i,1,j)\,C_0^{\gamma,0}(k;\overline{1},(\widetilde{ij});l)\,J_2^{(3)}(\lbrace p\rbrace_{3}) \nonumber\\
 \ph{6}&&-a_{3,g\to q}^{0}(j,1,i)\,C_0^{\gamma,0}(k;(\widetilde{ij}),\overline{1};l)\,J_2^{(3)}(\lbrace p\rbrace_{3}) \nonumber\\
 \ph{7}&&+\,E_3^{0}(k,j,i)\,a_{3,g\to q}^{0}((\widetilde{kj}),1,l)\,B_{1,q}^{\gamma,0}(\overline{1},(\widetilde{ij}),(\widetilde{(\widetilde{kj})l}))\,J_2^{(2)}(\lbrace p\rbrace_{2}) \nonumber\\
 \ph{8}&&+\,E_3^{0}(l,i,j)\,a_{3,g\to q}^{0}((\widetilde{li}),1,k)\,B_{1,q}^{\gamma,0}((\widetilde{(\widetilde{li})k}),(\widetilde{ij}),\overline{1})\,J_2^{(2)}(\lbrace p\rbrace_{2}) \nonumber\\
\snip \ph{9}&&+\,E_4^0(l,i,j,1)\,B_{1,q}^{\gamma,0}(k,\overline{1},(\widetilde{lij}))\,J_2^{(2)}(\lbrace p\rbrace_{2}) \nonumber\\
\snip \ph{10}&&+\,E_4^0(k,j,i,1)\,B_{1,q}^{\gamma,0}((\widetilde{kji}),\overline{1},l)\,J_2^{(2)}(\lbrace p\rbrace_{2}) \nonumber\\
 \ph{11}&&-\,a_{3,g\to q}^{0}(l,1,k)\,E_3^{0}(\overline{1},j,i)\,B_{1,q}^{\gamma,0}((\widetilde{kl}),\overline{\overline{1}},(\widetilde{ij}))\,J_2^{(2)}(\lbrace p\rbrace_{2}) \nonumber\\
 \ph{12}&&-\,a_{3,g\to q}^{0}(k,1,l)\,E_3^{0}(\overline{1},j,i)\,B_{1,q}^{\gamma,0}((\widetilde{ij}),\overline{\overline{1}},(\widetilde{kl}))\,J_2^{(2)}(\lbrace p\rbrace_{2}) \nonumber\\
 \ph{13}&&+\,E_3^{0}(k,j,i)\,a_{3,g\to q}^{0}((\widetilde{kj}),1,l)\,B_{1,q}^{\gamma,0}((\widetilde{ij}),\overline{1},(\widetilde{(\widetilde{kj})l}))\,J_2^{(2)}(\lbrace p\rbrace_{2}) \nonumber\\
 \ph{14}&&+\,E_3^{0}(l,i,j)\,a_{3,g\to q}^{0}((\widetilde{li}),1,k)\,B_{1,q}^{\gamma,0}((\widetilde{(\widetilde{li})k}),\overline{1},(\widetilde{ij}))\,J_2^{(2)}(\lbrace p\rbrace_{2}) \nonumber\\
 \ph{15}&&-\,a_{3,g\to q}^{0}(i,1,j)\,E_3^{0}(k,(\widetilde{ij}),\overline{1})\,B_{1,q}^{\gamma,0}((\widetilde{k\widetilde{(ij})}),\overline{\overline{1}},l)\,J_2^{(2)}(\lbrace p\rbrace_{2}) \nonumber\\
 \ph{16}&&-\,a_{3,g\to q}^{0}(j,1,i)\,E_3^{0}(l,(\widetilde{ij}),\overline{1})\,B_{1,q}^{\gamma,0}(k,\overline{\overline{1}},(\widetilde{l\widetilde{(ij})}))\,J_2^{(2)}(\lbrace p\rbrace_{2}) \nonumber\\
 \ph{17}&&-\,E_3^{0}(k,j,i)\,d_{3,g\to q}^{0}((\widetilde{kj}),1,(\widetilde{ij}))\,B_{1,q}^{\gamma,0}(\widetilde{(\widetilde{kj})(\widetilde{ij})},\overline{1},l)\,J_2^{(2)}(\lbrace p\rbrace_{2}) \nonumber\\
 \ph{18}&&-\,E_3^{0}(l,i,j)\,d_{3,g}^{0}(k,(\widetilde{ij}),1)\,B_{1,q}^{\gamma,0}((\widetilde{k\widetilde{(ij})}),\overline{1},(\widetilde{li}))\,J_2^{(2)}(\lbrace p\rbrace_{2}) \nonumber\\
 \ph{19}&&-\,E_3^{0}(l,i,j)\,d_{3,g\to q}^{0}((\widetilde{li}),1,(\widetilde{ij}))\,B_{1,q}^{\gamma,0}(k,\overline{1},\widetilde{(\widetilde{li})(\widetilde{ij})})\,J_2^{(2)}(\lbrace p\rbrace_{2}) \nonumber\\
 \ph{20}&&-\,E_3^{0}(k,j,i)\,d_{3,g}^{0}(l,(\widetilde{ij}),1)\,B_{1,q}^{\gamma,0}((\widetilde{kj}),\overline{1},(\widetilde{l\widetilde{(ij})}))\,J_2^{(2)}(\lbrace p\rbrace_{2}) \nonumber\\
 \ph{21}&&+E_3^{0}(j,k,l)\,B_{2,Q}^{\gamma,0}((\widetilde{jk}),1,(\widetilde{lk}),i)\,J_2^{(3)}(\lbrace p\rbrace_{3}) \nonumber\\
 \ph{22}&&+E_3^{0}(i,l,k)\,B_{2,Q}^{\gamma,0}(j,(\widetilde{lk}),1,(\widetilde{il}))\,J_2^{(3)}(\lbrace p\rbrace_{3}) \nonumber\\
 \ph{23}&&+\,E_3^{0}(i,l,k)\,a_{3,g\to q}^{0}((\widetilde{il}),1,j)\,B_{1,Q}^{\gamma,0}((\widetilde{(\widetilde{il})j}),(\widetilde{kl}),\overline{1})\,J_2^{(2)}(\lbrace p\rbrace_{2}) \nonumber\\
 \ph{24}&&+\,E_3^{0}(j,k,l)\,a_{3,g\to q}^{0}((\widetilde{jk}),1,i)\,B_{1,Q}^{\gamma,0}(\overline{1},(\widetilde{kl}),(\widetilde{(\widetilde{jk})i}))\,J_2^{(2)}(\lbrace p\rbrace_{2}) \nonumber\\
\snip \ph{25}&&+\,E_4^0(i,l,k,1)\,B_{1,Q}^{\gamma,0}(j,\overline{1},(\widetilde{kli}))\,J_2^{(2)}(\lbrace p\rbrace_{2}) \nonumber\\
\snip \ph{26}&&+\,E_4^0(j,k,l,1)\,B_{1,Q}^{\gamma,0}((\widetilde{lkj}),\overline{1},i)\,J_2^{(2)}(\lbrace p\rbrace_{2}) \nonumber\\
 \ph{27}&&-\,a_{3,g\to q}^{0}(i,1,j)\,E_3^{0}(\overline{1},k,l)\,B_{1,Q}^{\gamma,0}((\widetilde{ij}),\overline{\overline{1}},(\widetilde{kl}))\,J_2^{(2)}(\lbrace p\rbrace_{2}) \nonumber\\
 \ph{28}&&-\,a_{3,g\to q}^{0}(j,1,i)\,E_3^{0}(\overline{1},k,l)\,B_{1,Q}^{\gamma,0}((\widetilde{kl}),\overline{\overline{1}},(\widetilde{ij}))\,J_2^{(2)}(\lbrace p\rbrace_{2}) \nonumber\\
 \ph{29}&&+\,E_3^{0}(i,l,k)\,a_{3,g\to q}^{0}((\widetilde{il}),1,j)\,B_{1,Q}^{\gamma,0}((\widetilde{(\widetilde{il})j}),\overline{1},(\widetilde{kl}))\,J_2^{(2)}(\lbrace p\rbrace_{2}) \nonumber\\
 \ph{30}&&+\,E_3^{0}(j,k,l)\,a_{3,g\to q}^{0}((\widetilde{jk}),1,i)\,B_{1,Q}^{\gamma,0}((\widetilde{kl}),\overline{1},(\widetilde{(\widetilde{jk})i}))\,J_2^{(2)}(\lbrace p\rbrace_{2}) \nonumber\\
 \ph{31}&&-\,a_{3,g\to q}^{0}(k,1,l)\,E_3^{0}(i,(\widetilde{kl}),\overline{1})\,B_{1,Q}^{\gamma,0}(j,\overline{\overline{1}},(\widetilde{i\widetilde{(kl})}))\,J_2^{(2)}(\lbrace p\rbrace_{2}) \nonumber\\
 \ph{32}&&-\,a_{3,g\to q}^{0}(l,1,k)\,E_3^{0}(j,(\widetilde{kl}),\overline{1})\,B_{1,Q}^{\gamma,0}((\widetilde{j\widetilde{(kl})}),\overline{\overline{1}},i)\,J_2^{(2)}(\lbrace p\rbrace_{2}) \nonumber\\
 \ph{33}&&-\,E_3^{0}(i,l,k)\,d_{3,g\to q}^{0}((\widetilde{il}),1,(\widetilde{kl}))\,B_{1,Q}^{\gamma,0}(j,\overline{1},\widetilde{(\widetilde{il})(\widetilde{kl})})\,J_2^{(2)}(\lbrace p\rbrace_{2}) \nonumber\\
 \ph{34}&&-\,E_3^{0}(j,k,l)\,d_{3,g}^{0}(i,(\widetilde{kl}),1)\,B_{1,Q}^{\gamma,0}((\widetilde{jk}),\overline{1},(\widetilde{i\widetilde{(kl})}))\,J_2^{(2)}(\lbrace p\rbrace_{2}) \nonumber\\
 \ph{35}&&-\,E_3^{0}(j,k,l)\,d_{3,g\to q}^{0}((\widetilde{jk}),1,(\widetilde{kl}))\,B_{1,Q}^{\gamma,0}(\widetilde{(\widetilde{jk})(\widetilde{kl})},\overline{1},i)\,J_2^{(2)}(\lbrace p\rbrace_{2}) \nonumber\\
 \ph{36}&&-\,E_3^{0}(i,l,k)\,d_{3,g}^{0}(j,(\widetilde{kl}),1)\,B_{1,Q}^{\gamma,0}((\widetilde{j\widetilde{(kl})}),\overline{1},(\widetilde{il}))\,J_2^{(2)}(\lbrace p\rbrace_{2}) .
\end{eqnarray}

\addtocontents{toc}{\protect\setcounter{tocdepth}{2}}
\subsubsection{C-type $\mathcal{O}(N_FN^{-1})$ contribution}
\addtocontents{toc}{\protect\setcounter{tocdepth}{3}}
\label{sec:nnloRRCtg}

The subleading-colour matrix element constructed from squared non-identical quark amplitudes contains both colour orderings
and diagrams where the vector boson couples to either quark line. The contribution to the cross section is given by
\ba
\tilde{C}_{1}^{\gamma,0}(3,\hat{1};5,4;2)-\tilde{C}_{1}^{\gamma,0,S}(3,\hat{1};5,4;2),
\ea
where
\begin{eqnarray}
\lefteqn{{\tilde{C}_{1}^{\gamma,0,S}(k,\hat{1};i,j;l) =}} \nonumber\\
 \ph{1}&&+\frac{1}{2}\,E_3^{0}(k,i,j)\,\tilde{B}_{2,q}^{\gamma,0}((\widetilde{ki}),1,(\widetilde{ij}),l)\,J_2^{(3)}(\lbrace p\rbrace_{3}) \nonumber\\
 \ph{2}&&+\frac{1}{2}\,E_3^{0}(l,i,j)\,\tilde{B}_{2,q}^{\gamma,0}(k,(\widetilde{ij}),1,(\widetilde{li}))\,J_2^{(3)}(\lbrace p\rbrace_{3}) \nonumber\\
 \ph{3}&&-a_{3,g\to q}^{0}(i,1,j)\,C_0^{\gamma,0}(k;\overline{1},(\widetilde{ij});l)\,J_2^{(3)}(\lbrace p\rbrace_{3}) \nonumber\\
 \ph{4}&&-a_{3,g\to q}^{0}(k,1,l)\,C_0^{\gamma,0}(\overline{1};i,j;(\widetilde{kl}))\,J_2^{(3)}(\lbrace p\rbrace_{3}) \nonumber\\
 \ph{5}&&-a_{3,g\to q}^{0}(j,1,i)\,C_0^{\gamma,0}(k;(\widetilde{ij}),\overline{1};l)\,J_2^{(3)}(\lbrace p\rbrace_{3}) \nonumber\\
 \ph{6}&&-a_{3,g\to q}^{0}(l,1,k)\,C_0^{\gamma,0}((\widetilde{kl});i,j;\overline{1})\,J_2^{(3)}(\lbrace p\rbrace_{3}) \nonumber\\
\snip \ph{7}&&+\,B_4^0(k,i,j,l)\,B_{1,q}^{\gamma,0}((\widetilde{kij}),1,(\widetilde{lji}))\,J_2^{(2)}(\lbrace p\rbrace_{2}) \nonumber\\
 \ph{8}&&-\frac{1}{2}\,\,E_3^{0}(k,i,j)\,A_3^{0}((\widetilde{ki}),(\widetilde{ij}),l)\,B_{1,q}^{\gamma,0}(\widetilde{(\widetilde{ki})(\widetilde{ij})},1,(\widetilde{l\widetilde{(ij})}))\,J_2^{(2)}(\lbrace p\rbrace_{2}) \nonumber\\
 \ph{9}&&-\frac{1}{2}\,\,E_3^{0}(l,i,j)\,A_3^{0}((\widetilde{li}),(\widetilde{ij}),k)\,B_{1,q}^{\gamma,0}((\widetilde{k\widetilde{(ij})}),1,\widetilde{(\widetilde{li})(\widetilde{ij})})\,J_2^{(2)}(\lbrace p\rbrace_{2}) \nonumber\\
\snip \ph{10}&&+\frac{1}{2}\,\,\tilde{E}_4^0(k,i,1,j)\,B_{1,q}^{\gamma,0}((\widetilde{kij}),\overline{1},l)\,J_2^{(2)}(\lbrace p\rbrace_{2}) \nonumber\\
 \ph{11}&&-\frac{1}{2}\,\,A_3^{0}(i,1,j)\,E_{3,q^\prime\to g}^{0}(k,\overline{1},(\widetilde{ij}))\,B_{1,q}^{\gamma,0}((\widetilde{k\widetilde{(ij})}),\overline{\overline{1}},l)\,J_2^{(2)}(\lbrace p\rbrace_{2}) \nonumber\\
\snip \ph{12}&&+\frac{1}{2}\,\,\tilde{E}_4^0(l,i,1,j)\,B_{1,q}^{\gamma,0}(k,\overline{1},(\widetilde{lij}))\,J_2^{(2)}(\lbrace p\rbrace_{2}) \nonumber\\
 \ph{13}&&-\frac{1}{2}\,\,A_3^{0}(i,1,j)\,E_{3,q^\prime\to g}^{0}(l,\overline{1},(\widetilde{ij}))\,B_{1,q}^{\gamma,0}(k,\overline{\overline{1}},(\widetilde{l\widetilde{(ij})}))\,J_2^{(2)}(\lbrace p\rbrace_{2}) \nonumber\\
 \ph{14}&&+\frac{1}{2}\,\,E_3^{0}(k,i,j)\,a_{3,g\to q}^{0}((\widetilde{ki}),1,l)\,B_{1,q}^{\gamma,0}(\overline{1},(\widetilde{ij}),(\widetilde{l\widetilde{(ki})}))\,J_2^{(2)}(\lbrace p\rbrace_{2}) \nonumber\\
 \ph{15}&&+\frac{1}{2}\,\,E_3^{0}(k,i,j)\,a_{3,g\to q}^{0}(l,1,(\widetilde{ki}))\,B_{1,q}^{\gamma,0}((\widetilde{l\widetilde{(ki})}),(\widetilde{ij}),\overline{1})\,J_2^{(2)}(\lbrace p\rbrace_{2}) \nonumber\\
 \ph{16}&&+\frac{1}{2}\,\,E_3^{0}(l,i,j)\,a_{3,g\to q}^{0}(k,1,(\widetilde{li}))\,B_{1,q}^{\gamma,0}(\overline{1},(\widetilde{ij}),(\widetilde{k\widetilde{(li})}))\,J_2^{(2)}(\lbrace p\rbrace_{2}) \nonumber\\
 \ph{17}&&+\frac{1}{2}\,\,E_3^{0}(l,i,j)\,a_{3,g\to q}^{0}((\widetilde{li}),1,k)\,B_{1,q}^{\gamma,0}((\widetilde{k\widetilde{(li})}),(\widetilde{ij}),\overline{1})\,J_2^{(2)}(\lbrace p\rbrace_{2}) \nonumber\\
 \ph{18}&&+\frac{1}{2}\,E_3^{0}(i,k,l)\,\tilde{B}_{2,Q}^{\gamma,0}(j,(\widetilde{kl}),1,(\widetilde{ik}))\,J_2^{(3)}(\lbrace p\rbrace_{3}) \nonumber\\
 \ph{19}&&+\frac{1}{2}\,E_3^{0}(j,k,l)\,\tilde{B}_{2,Q}^{\gamma,0}((\widetilde{jk}),1,(\widetilde{kl}),i)\,J_2^{(3)}(\lbrace p\rbrace_{3}) \nonumber\\
\snip \ph{20}&&+\,B_4^0(i,k,l,j)\,B_{1,Q}^{\gamma,0}((\widetilde{jlk}),1,(\widetilde{ikl}))\,J_2^{(2)}(\lbrace p\rbrace_{2}) \nonumber\\
 \ph{21}&&-\frac{1}{2}\,\,E_3^{0}(i,k,l)\,A_3^{0}((\widetilde{ik}),(\widetilde{kl}),j)\,B_{1,Q}^{\gamma,0}((\widetilde{j\widetilde{(kl})}),1,\widetilde{(\widetilde{ik})(\widetilde{kl})})\,J_2^{(2)}(\lbrace p\rbrace_{2}) \nonumber\\
 \ph{22}&&-\frac{1}{2}\,\,E_3^{0}(j,k,l)\,A_3^{0}((\widetilde{jk}),(\widetilde{kl}),i)\,B_{1,Q}^{\gamma,0}(\widetilde{(\widetilde{jk})(\widetilde{kl})},1,(\widetilde{i\widetilde{(kl})}))\,J_2^{(2)}(\lbrace p\rbrace_{2}) \nonumber\\
\snip \ph{23}&&+\frac{1}{2}\,\,\tilde{E}_4^0(i,k,1,l)\,B_{1,Q}^{\gamma,0}(j,\overline{1},(\widetilde{ikl}))\,J_2^{(2)}(\lbrace p\rbrace_{2}) \nonumber\\
 \ph{24}&&-\frac{1}{2}\,\,A_3^{0}(k,1,l)\,E_{3,q^\prime\to g}^{0}(i,\overline{1},(\widetilde{kl}))\,B_{1,Q}^{\gamma,0}(j,\overline{\overline{1}},(\widetilde{i\widetilde{(kl})}))\,J_2^{(2)}(\lbrace p\rbrace_{2}) \nonumber\\
\snip \ph{25}&&+\frac{1}{2}\,\,\tilde{E}_4^0(j,k,1,l)\,B_{1,Q}^{\gamma,0}((\widetilde{jkl}),\overline{1},i)\,J_2^{(2)}(\lbrace p\rbrace_{2}) \nonumber\\
 \ph{26}&&-\frac{1}{2}\,\,A_3^{0}(k,1,l)\,E_{3,q^\prime\to g}^{0}(j,\overline{1},(\widetilde{kl}))\,B_{1,Q}^{\gamma,0}((\widetilde{j\widetilde{(kl})}),\overline{\overline{1}},i)\,J_2^{(2)}(\lbrace p\rbrace_{2}) \nonumber\\
 \ph{27}&&+\frac{1}{2}\,\,E_3^{0}(i,k,l)\,a_{3,g\to q}^{0}((\widetilde{ik}),1,j)\,B_{1,Q}^{\gamma,0}((\widetilde{j\widetilde{(ik})}),(\widetilde{kl}),\overline{1})\,J_2^{(2)}(\lbrace p\rbrace_{2}) \nonumber\\
 \ph{28}&&+\frac{1}{2}\,\,E_3^{0}(i,k,l)\,a_{3,g\to q}^{0}(j,1,(\widetilde{ik}))\,B_{1,Q}^{\gamma,0}(\overline{1},(\widetilde{kl}),(\widetilde{j\widetilde{(ik})}))\,J_2^{(2)}(\lbrace p\rbrace_{2}) \nonumber\\
 \ph{29}&&+\frac{1}{2}\,\,E_3^{0}(j,k,l)\,a_{3,g\to q}^{0}((\widetilde{jk}),1,i)\,B_{1,Q}^{\gamma,0}(\overline{1},(\widetilde{kl}),(\widetilde{i\widetilde{(jk})}))\,J_2^{(2)}(\lbrace p\rbrace_{2}) \nonumber\\
 \ph{30}&&+\frac{1}{2}\,\,E_3^{0}(j,k,l)\,a_{3,g\to q}^{0}(i,1,(\widetilde{jk}))\,B_{1,Q}^{\gamma,0}((\widetilde{i\widetilde{(jk})}),(\widetilde{kl}),\overline{1})\,J_2^{(2)}(\lbrace p\rbrace_{2}) .
\end{eqnarray}

\addtocontents{toc}{\protect\setcounter{tocdepth}{2}}
\subsubsection{D-type $\mathcal{O}(N^{0})$ contribution}
\addtocontents{toc}{\protect\setcounter{tocdepth}{3}}
\label{sec:nnloRRDg}

The leading-colour term coming from interferences of quark orderings is given by the $D_{1}^{\gamma,0}$ function,
which generates the contributions to the cross section,
\ba
D_{1}^{\gamma,0}(3,\hat{1},2;4,5)-D_{1}^{\gamma,0,S}(3,\hat{1},2;4,5),
\ea
where
\begin{eqnarray}
\lefteqn{{{D}_{1}^{\gamma,0,S}(k,\hat{1},j;i,l) =}} \nonumber\\
 \ph{1}&&-A_3^{0}(i,1,k)\,D_0^{\gamma,0}(\overline{1},j;(\widetilde{ik}),l)\,J_2^{(3)}(\lbrace p\rbrace_{3}) \nonumber\\
 \ph{2}&&-A_3^{0}(j,1,l)\,D_0^{\gamma,0}(i,\overline{1};k,(\widetilde{jl}))\,J_2^{(3)}(\lbrace p\rbrace_{3}) \nonumber\\
\snip \ph{3}&&+2\,C_4^0(i,k,j,l)\,B_{1,q}^{\gamma,0}((\widetilde{ikj}),1,(\widetilde{kjl}))\,J_2^{(2)}(\lbrace p\rbrace_{2}) \nonumber\\
 \ph{4}&&+2\,C_4^0(k,i,j,l)\,B_{1,q}^{\gamma,0}((\widetilde{kij}),1,(\widetilde{ijl}))\,J_2^{(2)}(\lbrace p\rbrace_{2}) \nonumber\\
 \ph{5}&&+2\,C_4^0(l,j,i,k)\,B_{1,q}^{\gamma,0}((\widetilde{jik}),1,(\widetilde{lji}))\,J_2^{(2)}(\lbrace p\rbrace_{2}) \nonumber\\
 \ph{6}&&+2\,C_4^0(j,l,i,k)\,B_{1,q}^{\gamma,0}((\widetilde{lik}),1,(\widetilde{jli}))\,J_2^{(2)}(\lbrace p\rbrace_{2}) 
\end{eqnarray}

\addtocontents{toc}{\protect\setcounter{tocdepth}{2}}
\subsubsection{D-type $\mathcal{O}(N^{-2})$ contribution}
\addtocontents{toc}{\protect\setcounter{tocdepth}{3}}
\label{sec:nnloRRDtg}

The subleading-colour matrix element $\tilde{D}_{1}^{\gamma,0}$ has the same infrared structure as the ${D}_{1}^{\gamma,0}$
function discussed above and so their subtraction terms are identical,
\ba
\tilde{D}_{1}^{\gamma,0}(3,\hat{1},2;4,5)-\tilde{D}_{1}^{\gamma,0,S}(3,\hat{1},2;4,5),
\ea
where
\begin{eqnarray}
\lefteqn{{{\tilde{D}}_{1}^{\gamma,0,S}(k,\hat{1},j;i,l) =}} \nonumber\\
 \ph{1}&&-A_3^{0}(i,1,k)\,D_0^{\gamma,0}(\overline{1},j;(\widetilde{ik}),l)\,J_2^{(3)}(\lbrace p\rbrace_{3}) \nonumber\\
 \ph{2}&&-A_3^{0}(j,1,l)\,D_0^{\gamma,0}(i,\overline{1};k,(\widetilde{jl}))\,J_2^{(3)}(\lbrace p\rbrace_{3}) \nonumber\\
\snip \ph{3}&&+2\,C_4^0(i,k,j,l)\,B_{1,q}^{\gamma,0}((\widetilde{ikj}),1,(\widetilde{kjl}))\,J_2^{(2)}(\lbrace p\rbrace_{2}) \nonumber\\
 \ph{4}&&+2\,C_4^0(k,i,j,l)\,B_{1,q}^{\gamma,0}((\widetilde{kij}),1,(\widetilde{ijl}))\,J_2^{(2)}(\lbrace p\rbrace_{2}) \nonumber\\
 \ph{5}&&+2\,C_4^0(l,j,i,k)\,B_{1,q}^{\gamma,0}((\widetilde{jik}),1,(\widetilde{lji}))\,J_2^{(2)}(\lbrace p\rbrace_{2}) \nonumber\\
 \ph{6}&&+2\,C_4^0(j,l,i,k)\,B_{1,q}^{\gamma,0}((\widetilde{lik}),1,(\widetilde{jli}))\,J_2^{(2)}(\lbrace p\rbrace_{2}) .
\end{eqnarray}

\subsection*{Real-virtual}
\label{app:nnlosubRVg}

The gluon-induced real-virtual corrections have at most three partons in the final state and
therefore only the two-quark--two-gluon one-loop matrix elements are necessary. These matrix elements have a colour decomposition
given in Tab.~\ref{tab:channelsg} and we present the subtraction terms for the colour-stripped matrix elements in
this section.

\addtocontents{toc}{\protect\setcounter{tocdepth}{2}}
\subsubsection{B-type $\mathcal{O}(N^{2})$ contribution}
\addtocontents{toc}{\protect\setcounter{tocdepth}{3}}

The leading-colour matrix element is given by the function $B_{2}^{\gamma,1}$, which is summed
over both gluon orderings. The subtraction term $B_{2}^{\gamma,1,T}$, is constructed to cancel against the full set of orderings and
is thus not summed over any permutations of gluons. The IC terms from the RR and the IC mass factorization kernels combine to form a finite
subtraction term, $B_{2,g\to q}^{\gamma,1,T}$. The total contribution to the cross section is given by
\ba
\sum_{P(\hat{1},2)}\bar{B}_{2}^{\gamma,1}(3,\hat{1},2,4)-B_{2}^{\gamma,1,T}(3,\hat{1},2,4)-B_{2,g\to q}^{\gamma,1,T}(3,\hat{1},2,4) ,
\ea

\begin{eqnarray}
\lefteqn{{B_{2}^{\gamma,1,T}(j,\hat{1},i,k) =}} \nonumber\\
 \ph{1}&&-\bigg[ 
 +J_{2,GG}^{1,IF}(s_{1i})
 +J_{2,QG}^{1,FF}(s_{ki})
 +J_{2,GQ}^{1,IF}(s_{1j})\bigg]\,\bar{B}_2^{\gamma,0}(j,1,i,k)\,J_2^{(3)}(\lbrace p\rbrace_{3})\nonumber\\
 \ph{2}&&-\bigg[ 
 +J_{2,GQ}^{1,IF}(s_{1k})
 +J_{2,QG}^{1,FF}(s_{ji})
 +J_{2,GG}^{1,IF}(s_{1i})\bigg]\,\bar{B}_2^{\gamma,0}(j,i,1,k)\,J_2^{(3)}(\lbrace p\rbrace_{3})\nonumber\\
  \ph{3}&&+\,d_{3,g}^{0}(j,i,1)\,\bigg[\,\bar{B}_1^{\gamma,1}((\widetilde{ji}),\overline{1},k)\,\delta(1-x_1)\,\delta(1-x_2)\nonumber\\
  &&+\,\bigg( 
 +J_{2,GQ}^{1,IF}(s_{\overline{1}(\widetilde{ji})})
 +J_{2,GQ}^{1,IF}(s_{\overline{1}k})\bigg)\,\bar{B}_1^{\gamma,0}((\widetilde{ji}),\overline{1},k) \bigg]\,J_2^{(2)}(\lbrace p\rbrace_{2})\nonumber\\
  \ph{4}&&+\,d_{3,g}^{0}(k,i,1)\,\bigg[\,\bar{B}_1^{\gamma,1}(j,\overline{1},(\widetilde{ki}))\,\delta(1-x_1)\,\delta(1-x_2)\nonumber\\
  &&+\,\bigg( 
 +J_{2,GQ}^{1,IF}(s_{\overline{1}(\widetilde{ki})})
 +J_{2,GQ}^{1,IF}(s_{\overline{1}j})\bigg)\,\bar{B}_1^{\gamma,0}(j,\overline{1},(\widetilde{ki})) \bigg]\,J_2^{(2)}(\lbrace p\rbrace_{2})\nonumber\\
  \ph{5}&&+\bigg[\,d_{3,g}^{1}(j,i,1)\,\delta(1-x_1)\,\delta(1-x_2)
  \nonumber\\&&\,  
   +\bigg( 
 -2J_{2,GQ}^{1,IF}(s_{\overline{1}(\widetilde{ji})})
 +J_{2,GQ}^{1,IF}(s_{1j})
 +J_{2,QG}^{1,FF}(s_{ji})
 +J_{2,GG}^{1,IF}(s_{1i})\bigg)\,d_{3,g}^{0}(j,i,1)\bigg]\nonumber\\
 &&\times \bar{B}_1^{\gamma,0}((\widetilde{ji}),\overline{1},k)\,J_2^{(2)}(\lbrace p\rbrace_{2})\nonumber\\
  \ph{6}&&+\bigg[\,d_{3,g\to q}^{1}(j,1,i)\,\delta(1-x_1)\,\delta(1-x_2)
  \nonumber\\&&\,  
   +\bigg( 
 -2J_{2,GQ}^{1,IF}(s_{\overline{1}(\widetilde{ji})})
 +J_{2,GQ}^{1,IF}(s_{1j})
 +J_{2,QG}^{1,FF}(s_{ji})
 +J_{2,GG}^{1,IF}(s_{1i})\bigg)\,d_{3,g\to q}^{0}(j,1,i)\bigg]\nonumber\\
 &&\times \bar{B}_1^{\gamma,0}((\widetilde{ji}),\overline{1},k)\,J_2^{(2)}(\lbrace p\rbrace_{2})\nonumber\\
  \ph{7}&&+\bigg[\,d_{3,g}^{1}(k,i,1)\,\delta(1-x_1)\,\delta(1-x_2)
  \nonumber\\&&\,  
   +\bigg( 
 +J_{2,GG}^{1,IF}(s_{1i})
 +J_{2,GQ}^{1,IF}(s_{1k})
 +J_{2,QG}^{1,FF}(s_{ki})
 -2J_{2,GQ}^{1,IF}(s_{\overline{1}(\widetilde{ki})})\bigg)\,d_{3,g}^{0}(k,i,1)\bigg]\nonumber\\
 &&\times \bar{B}_1^{\gamma,0}(j,\overline{1},(\widetilde{ki}))\,J_2^{(2)}(\lbrace p\rbrace_{2})\nonumber\\
  \ph{8}&&+\bigg[\,d_{3,g\to q}^{1}(k,1,i)\,\delta(1-x_1)\,\delta(1-x_2)
  \nonumber\\&&\,  
   +\bigg( 
 +J_{2,GG}^{1,IF}(s_{1i})
 +J_{2,GQ}^{1,IF}(s_{1k})
 +J_{2,QG}^{1,FF}(s_{ki})
 -2J_{2,GQ}^{1,IF}(s_{\overline{1}(\widetilde{ki})})\bigg)\,d_{3,g\to q}^{0}(k,1,i)\bigg]\nonumber\\
 &&\times \bar{B}_1^{\gamma,0}(j,\overline{1},(\widetilde{ki}))\,J_2^{(2)}(\lbrace p\rbrace_{2})\nonumber\\
  \ph{9}&&-\bigg[\,\tilde{A}_{3,g}^{1}(j,1,k)\,\delta(1-x_1)\,\delta(1-x_2)
  \nonumber\\&&\,  
   +\bigg( 
 +J_{2,QQ}^{1,FF}(s_{jk})
 -J_{2,QQ}^{1,IF}(s_{\overline{1}(\widetilde{jk})})\bigg)\,A_{3,g\to q}^{0}(j,1,k)\bigg]\,\bar{B}_1^{\gamma,0}((\widetilde{jk}),\overline{1},i)\,J_2^{(2)}(\lbrace p\rbrace_{2})\nonumber\\
  \ph{10}&&-\bigg[\,\tilde{A}_{3}^{1}(j,i,k)\,\delta(1-x_1)\,\delta(1-x_2)
  \nonumber\\&&\,  
   +\bigg( 
 +J_{2,QQ}^{1,FF}(s_{jk})
 -J_{2,QQ}^{1,FF}(s_{(\widetilde{ji})(\widetilde{ki})})\bigg)\,A_3^{0}(j,i,k)\bigg]\,\bar{B}_1^{\gamma,0}((\widetilde{ji}),1,(\widetilde{ki}))\,J_2^{(2)}(\lbrace p\rbrace_{2})\nonumber\\
  \ph{11}&&-\bigg[\,A_{3,g}^{1}(j,1,k)\,\delta(1-x_1)\,\delta(1-x_2)
  \nonumber\\&&\,  
   +\bigg( 
 -J_{2,QQ}^{1,IF}(s_{\overline{1}(\widetilde{jk})})
 +J_{2,GQ}^{1,IF}(s_{1k})
 +J_{2,GQ}^{1,IF}(s_{1j})\bigg)\,A_{3,g\to q}^{0}(j,1,k)\bigg]\nonumber\\
 &&\times \bar{B}_1^{\gamma,0}((\widetilde{jk}),\overline{1},i)\,J_2^{(2)}(\lbrace p\rbrace_{2})\nonumber\\
 \ph{12}&&-\bigg[ +J_{2,QQ}^{1,FF}(s_{jk})
 +J_{2,GG}^{1,IF}(s_{1i})
 -J_{2,QG}^{1,FF}(s_{ji})
 -J_{2,GQ}^{1,IF}(s_{1k})
 \nonumber\\ &&+\bigg(  -{\cal S}^{FF}(s_{1i},s_{ji},x_{1i,ji})
 +{\cal S}^{FF}(s_{ji},s_{ji},1)
 -{\cal S}^{FF}(s_{jk},s_{ji},x_{jk,ji})\nonumber\\
 &&
 +{\cal S}^{FF}(s_{1k},s_{ji},x_{1k,ji})
 \bigg)\bigg]\,d_{3,g\to q}^{0}(j,1,i)\,\bar{B}_1^{\gamma,0}((\widetilde{ji}),\overline{1},k)\,J_2^{(2)}(\lbrace p\rbrace_{2}) \nonumber\\
 \ph{13}&&-\bigg[ +J_{2,GG}^{1,IF}(s_{1i})
 -J_{2,QG}^{1,FF}(s_{ki})
 +J_{2,QQ}^{1,FF}(s_{kj})
 -J_{2,GQ}^{1,IF}(s_{1j})
 \nonumber\\ &&+\bigg(  -{\cal S}^{FF}(s_{1i},s_{ki},x_{1i,ki})
 +{\cal S}^{FF}(s_{ki},s_{ki},1)
 -{\cal S}^{FF}(s_{kj},s_{ki},x_{kj,ki})\nonumber\\
 &&
 +{\cal S}^{FF}(s_{1j},s_{ki},x_{1j,ki})
 \bigg)\bigg]\,d_{3,g\to q}^{0}(k,1,i)\,\bar{B}_1^{\gamma,0}(j,\overline{1},(\widetilde{ki}))\,J_2^{(2)}(\lbrace p\rbrace_{2}) \nonumber\\
 \ph{14}&&+2\bigg[ +J_{2,QQ}^{1,FF}(s_{jk})
 -J_{2,QQ}^{1,IF}(s_{\overline{1}(\widetilde{jk})})
 +J_{2,QG}^{1,FF}(s_{(\widetilde{jk})i})\nonumber\\
 &&\phantom{+2}
 +J_{2,QG}^{1,IF}(s_{\overline{1}i})
 -J_{2,QG}^{1,FF}(s_{ji})
 -J_{2,QG}^{1,FF}(s_{ki})\nonumber\\
 &&\phantom{+2}
 +\bigg(  +{\cal S}^{FF}(s_{ji},s_{jk},x_{ji,jk})
 +{\cal S}^{FF}(s_{ki},s_{jk},x_{ki,jk})
 -{\cal S}^{FF}(s_{jk},s_{jk},1)\nonumber\\
 &&\phantom{+2}
 -{\cal S}^{FF}(s_{(\widetilde{jk})i},s_{jk},x_{(\widetilde{jk})i,jk})
 -{\cal S}^{FF}(s_{\overline{1}i},s_{jk},x_{\overline{1}i,jk})
 +{\cal S}^{FF}(s_{\overline{1}(\widetilde{jk})},s_{jk},x_{\overline{1}(\widetilde{jk}),jk})
 \bigg)\bigg]\nonumber\\
 &&\times A_{3,g\to q}^{0}(j,1,k)\,\bar{B}_1^{\gamma,0}((\widetilde{jk}),\overline{1},i)\,J_2^{(2)}(\lbrace p\rbrace_{2}) \nonumber\\
 \ph{15}&&-\frac{1}{2}\,\bigg[ +J_{2,GQ}^{1,IF}(s_{1j})
 -J_{2,GQ}^{1,IF}(s_{\overline{1}(\widetilde{ji})})
 -J_{2,GQ}^{1,IF}(s_{1k})\nonumber\\
 &&\phantom{+2}
 +J_{2,GQ}^{1,IF}(s_{\overline{1}k})
 -J_{2,QQ}^{1,FF}(s_{jk})
 +J_{2,QQ}^{1,FF}(s_{(\widetilde{ji})k})\nonumber\\
 &&\phantom{+2}
+\bigg(  -{\cal S}^{FF}(s_{1j},s_{jk},x_{1j,jk})
 +{\cal S}^{FF}(s_{\overline{1}(\widetilde{ji})},s_{jk},x_{\overline{1}(\widetilde{ji}),jk})
 +{\cal S}^{FF}(s_{1k},s_{jk},x_{1k,jk})\nonumber\\
 &&\phantom{+2}
 -{\cal S}^{FF}(s_{\overline{1}k},s_{jk},x_{\overline{1}k,jk})
 +{\cal S}^{FF}(s_{jk},s_{jk},1)
 -{\cal S}^{FF}(s_{(\widetilde{ji})k},s_{jk},x_{(\widetilde{ji})k,jk})
 \bigg)\bigg]\nonumber\\
 &&\times d_{3,g}^{0}(j,i,1)\,\bar{B}_1^{\gamma,0}((\widetilde{ji}),\overline{1},k)\,J_2^{(2)}(\lbrace p\rbrace_{2}) \nonumber\\
 \ph{16}&&-\frac{1}{2}\,\bigg[ +J_{2,GQ}^{1,IF}(s_{1k})
 -J_{2,GQ}^{1,IF}(s_{\overline{1}(\widetilde{ki})})
 -J_{2,GQ}^{1,IF}(s_{1j})\nonumber\\
 &&\phantom{+2}
 +J_{2,GQ}^{1,IF}(s_{\overline{1}j})
 -J_{2,QQ}^{1,FF}(s_{kj})
 +J_{2,QQ}^{1,FF}(s_{(\widetilde{ki})j})\nonumber\\
 &&\phantom{+2}
 +\bigg(  -{\cal S}^{FF}(s_{1k},s_{kj},x_{1k,kj})
 +{\cal S}^{FF}(s_{\overline{1}(\widetilde{ki})},s_{kj},x_{\overline{1}(\widetilde{ki}),kj})
 +{\cal S}^{FF}(s_{1j},s_{kj},x_{1j,kj})\nonumber\\
 &&\phantom{+2}
 -{\cal S}^{FF}(s_{\overline{1}j},s_{kj},x_{\overline{1}j,kj})
 +{\cal S}^{FF}(s_{kj},s_{kj},1)
 -{\cal S}^{FF}(s_{(\widetilde{ki})j},s_{kj},x_{(\widetilde{ki})j,kj})
 \bigg)\bigg]\nonumber\\
 &&\times d_{3,g}^{0}(k,i,1)\,\bar{B}_1^{\gamma,0}(j,\overline{1},(\widetilde{ki}))\,J_2^{(2)}(\lbrace p\rbrace_{2}) \nonumber\\
 \ph{17}&&+\frac{1}{2}\,\bigg[ +J_{2,QQ}^{1,FF}(s_{jk})
 -J_{2,QQ}^{1,FF}(s_{(\widetilde{ji})(\widetilde{ki})})
 -J_{2,GQ}^{1,IF}(s_{1j})\nonumber\\
 &&\phantom{+2}
 +J_{2,GQ}^{1,IF}(s_{1(\widetilde{ji})})
 -J_{2,GQ}^{1,IF}(s_{1k})
 +J_{2,GQ}^{1,IF}(s_{1(\widetilde{ki})})\nonumber\\
 &&\phantom{+2}
+\bigg(  -{\cal S}^{FF}(s_{jk},s_{jk},1)
 +{\cal S}^{FF}(s_{(\widetilde{ji})(\widetilde{ki})},s_{jk},x_{(\widetilde{ji})(\widetilde{ki}),jk})
 +{\cal S}^{FF}(s_{1j},s_{jk},x_{1j,jk})\nonumber\\
 &&\phantom{+2}
 -{\cal S}^{FF}(s_{1(\widetilde{ji})},s_{jk},x_{1(\widetilde{ji}),jk})
 +{\cal S}^{FF}(s_{1k},s_{jk},x_{1k,jk})
 -{\cal S}^{FF}(s_{1(\widetilde{ki})},s_{jk},x_{1(\widetilde{ki}),jk})
 \bigg)\bigg]\nonumber\\
 &&\times A_3^{0}(j,i,k)\,\bar{B}_1^{\gamma,0}((\widetilde{ji}),1,(\widetilde{ki}))\,J_2^{(2)}(\lbrace p\rbrace_{2}) \nonumber\\
 \ph{18}&&+\bigg[ -2J_{2,QQ,g \to q}^{1,IF}(s_{1k})
 +2J_{2,QQ,g \to q}^{1,IF}(s_{\overline{1}k})
\bigg]  D_3^{0}(1,i,j)\,\bar{B}_1^{\gamma,0}((\widetilde{ij}),\overline{1},k)\,J_2^{(2)}(\lbrace p\rbrace_{2})  \nonumber\\
 \ph{19}&&+\bigg[ +2J_{2,QQ,g \to q}^{1,IF}(s_{1k})
 -2J_{2,QQ,g \to q}^{1,IF}(s_{\overline{1}(\widetilde{ki})})
\bigg]  A_3^{0}(1,i,k)\,\bar{B}_1^{\gamma,0}(j,\overline{1},(\widetilde{ik}))\,J_2^{(2)}(\lbrace p\rbrace_{2})  \nonumber\\
 \ph{20}&&+\bigg[ +2J_{2,QQ,g \to q}^{1,IF}(s_{1k})
 -2J_{2,QQ,g \to q}^{1,IF}(s_{\overline{1}(\widetilde{kj})})
\bigg]  A_3^{0}(1,j,k)\,\bar{B}_1^{\gamma,0}((\widetilde{jk}),\overline{1},i)\,J_2^{(2)}(\lbrace p\rbrace_{2})  \nonumber\\
 \ph{21}&&+\bigg[ -2J_{2,QQ,g \to q}^{1,IF}(s_{1k})
 +2J_{2,QQ,g \to q}^{1,IF}(s_{1(\widetilde{ki})})
\bigg]  d_3^{0}(k,i,j)\,\bar{B}_1^{\gamma,0}((\widetilde{ij}),1,(\widetilde{ki}))\,J_2^{(2)}(\lbrace p\rbrace_{2})  \nonumber\\
 \ph{22}&&+\bigg[ -2J_{2,QQ,g \to q}^{1,IF}(s_{1k})
 +2J_{2,QQ,g \to q}^{1,IF}(s_{1(\widetilde{kj})})
\bigg]  d_3^{0}(k,j,i)\,\bar{B}_1^{\gamma,0}((\widetilde{ji}),1,(\widetilde{kj}))\,J_2^{(2)}(\lbrace p\rbrace_{2})  \nonumber\\
 \ph{23}&&+\bigg[ -2J_{2,QG}^{1,IF}(s_{\overline{1}(\widetilde{ki})})
 +2J_{2,GQ}^{1,IF}(s_{\overline{1}(\widetilde{ki})})
\bigg]  d_{3,g\to q}^{0}(k,1,i)\,\bar{B}_1^{\gamma,0}(j,\overline{1},(\widetilde{ki}))\,J_2^{(2)}(\lbrace p\rbrace_{2})  \nonumber\\
 \ph{24}&&+\bigg[ -2J_{2,QG}^{1,IF}(s_{\overline{1}(\widetilde{ji})})
 +2J_{2,GQ}^{1,IF}(s_{\overline{1}(\widetilde{ji})})
\bigg]  d_{3,g\to q}^{0}(j,1,i)\,\bar{B}_1^{\gamma,0}((\widetilde{ji}),\overline{1},k)\,J_2^{(2)}(\lbrace p\rbrace_{2})  \nonumber\\
  \ph{27}&&-\,A_3^{0}(j,1,k)\,\bigg[\,\bar{B}_{1}^{\gamma,1}(\overline{1},i,(\widetilde{jk}))\,\delta(1-x_1)\,\delta(1-x_2)\nonumber\\
  &&+\,\bigg( 
 +J_{2,QG}^{1,IF}(s_{\overline{1}i})
 +J_{2,QG}^{1,FF}(s_{(\widetilde{jk})i})\bigg)\,{\bar{B}}_1^{\gamma,0}(\overline{1},i,(\widetilde{jk})) \bigg]\,J_2^{(2)}(\lbrace p\rbrace_{2})\nonumber\\
  \ph{28}&&-\bigg[\,A_{3,g}^{1}(j,1,k)\,\delta(1-x_1)\,\delta(1-x_2)
  \nonumber\\&&\,  
   +\bigg( 
 -J_{2,QQ}^{1,IF}(s_{\overline{1}(\widetilde{jk})})
 +J_{2,GQ}^{1,IF}(s_{1k})
 +J_{2,GQ}^{1,IF}(s_{1j})\bigg)\,A_3^{0}(j,1,k)\bigg]\,{\bar{B}}_1^{\gamma,0}(\overline{1},i,(\widetilde{kj}))\,J_2^{(2)}(\lbrace p\rbrace_{2})\nonumber\\
 \ph{29}&&-\bigg[ +J_{2,QQ}^{1,FF}(s_{jk})
 -J_{2,GQ}^{1,IF}(s_{1k})
 -J_{2,QG}^{1,FF}(s_{ji})
 +J_{2,GG}^{1,IF}(s_{1i})
 \nonumber\\ &&+\bigg(  -{\cal S}^{FF}(s_{1i},s_{ji},x_{1i,ji})
 +{\cal S}^{FF}(s_{ji},s_{ji},1)
 -{\cal S}^{FF}(s_{jk},s_{ji},x_{jk,ji})\nonumber\\
 &&\phantom{+2}
 +{\cal S}^{FF}(s_{1k},s_{ji},x_{1k,ji})
 \bigg)\bigg]\,d_{3,g\to q}^{0}(j,1,i)\,{\bar{B}}_1^{\gamma,0}(\overline{1},(\widetilde{ji}),k)\,J_2^{(2)}(\lbrace p\rbrace_{2}) \nonumber\\
 \ph{30}&&-\bigg[ +J_{2,GG}^{1,IF}(s_{1i})
 +J_{2,QQ}^{1,FF}(s_{kj})
 -J_{2,GQ}^{1,IF}(s_{1j})
 -J_{2,QG}^{1,FF}(s_{ki})
 \nonumber\\ &&+\bigg(  -{\cal S}^{FF}(s_{1i},s_{ki},x_{1i,ki})
 +{\cal S}^{FF}(s_{ki},s_{ki},1)
 -{\cal S}^{FF}(s_{kj},s_{ki},x_{kj,ki})\nonumber\\
 &&\phantom{+2}
 +{\cal S}^{FF}(s_{1j},s_{ki},x_{1j,ki})
 \bigg)\bigg]\, d_{3,g\to q}^{0}(k,1,i)\,{\bar{B}}_1^{\gamma,0}(\overline{1},(\widetilde{ki}),j)\,J_2^{(2)}(\lbrace p\rbrace_{2}) \nonumber\\
 \ph{31}&&+\bigg[ +J_{2,QQ}^{1,FF}(s_{jk})
 -J_{2,QG}^{1,FF}(s_{ji})
 -J_{2,QG}^{1,FF}(s_{ki})\nonumber\\
 &&\phantom{+2}
 -J_{2,QQ}^{1,IF}(s_{\overline{1}(\widetilde{jk})})
 +J_{2,QG}^{1,FF}(s_{(\widetilde{jk})i})
 +J_{2,QG}^{1,IF}(s_{\overline{1}i})
 \nonumber\\&&\phantom{+} +\bigg(  +{\cal S}^{FF}(s_{ji},s_{jk},x_{ji,jk})
 +{\cal S}^{FF}(s_{ki},s_{jk},x_{ki,jk})
 -{\cal S}^{FF}(s_{jk},s_{jk},1)\nonumber\\
 &&\phantom{+2}
 -{\cal S}^{FF}(s_{(\widetilde{jk})i},s_{jk},x_{(\widetilde{jk})i,jk})
 -{\cal S}^{FF}(s_{\overline{1}i},s_{jk},x_{\overline{1}i,jk})
 +{\cal S}^{FF}(s_{\overline{1}(\widetilde{jk})},s_{jk},x_{\overline{1}(\widetilde{jk}),jk})
 \bigg)\bigg]\nonumber\\
 &&\times A_3^{0}(j,1,k)\,{\bar{B}}_1^{\gamma,0}(\overline{1},i,(\widetilde{jk}))\,J_2^{(2)}(\lbrace p\rbrace_{2}) ,
\end{eqnarray}

\begin{eqnarray}
\lefteqn{{B_{2,g\to q}^{\gamma,1,T}(j,\hat{1},i,k) =}} \nonumber\\
  \ph{25}&&-J_{2,QQ,g \to q}^{1,IF}(s_{1k})\,\bar{B}_2^{\gamma,0}(1,i,j,k)\,J_2^{(3)}(\lbrace p\rbrace_{3})\nonumber\\
  \ph{26}&&-J_{2,QQ,g \to q}^{1,IF}(s_{1k})\,\bar{B}_2^{\gamma,0}(1,j,i,k)\,J_2^{(3)}(\lbrace p\rbrace_{3})\nonumber\\
 \ph{32}&&+\bigg[ +J_{2,QQ,g \to q}^{1,IF}(s_{1k})
 -J_{2,QQ,g \to q}^{1,IF}(s_{\overline{1}(\widetilde{ki})})
\bigg]  A_3^{0}(1,i,k)\,{\bar{B}}_1^{\gamma,0}(\overline{1},j,(\widetilde{ik}))\,J_2^{(2)}(\lbrace p\rbrace_{2})  \nonumber\\
 \ph{33}&&+\bigg[ +J_{2,QQ,g \to q}^{1,IF}(s_{1k})
 -J_{2,QQ,g \to q}^{1,IF}(s_{\overline{1}(\widetilde{kj})})
\bigg]  A_3^{0}(1,j,k)\,{\bar{B}}_1^{\gamma,0}(\overline{1},i,(\widetilde{jk}))\,J_2^{(2)}(\lbrace p\rbrace_{2})  \nonumber\\
 \ph{34}&& +J_{2,QQ,g \to q}^{1,IF}(s_{1(\widetilde{ki})})
\,d_3^{0}(k,i,j)\,\bar{B}_1^{\gamma,0}(1,(\widetilde{ij}),(\widetilde{ki}))\,J_2^{(2)}(\lbrace p\rbrace_{2})  \nonumber\\
 \ph{35}&& +J_{2,QQ,g \to q}^{1,IF}(s_{\overline{1}k})
\,d_3^{0}(1,j,i)\,{\bar{B}}_1^{\gamma,0}(\overline{1},(\widetilde{ij}),k)\,J_2^{(2)}(\lbrace p\rbrace_{2})  \nonumber\\
 \ph{36}&& +J_{2,QQ,g \to q}^{1,IF}(s_{1(\widetilde{kj})})
\,d_3^{0}(k,j,i)\,{\bar{B}}_1^{\gamma,0}(1,(\widetilde{ji}),(\widetilde{kj}))\,J_2^{(2)}(\lbrace p\rbrace_{2})  \nonumber\\
 \ph{37}&& +J_{2,QQ,g \to q}^{1,IF}(s_{\overline{1}k})
\,d_3^{0}(1,i,j)\,{\bar{B}}_1^{\gamma,0}(\overline{1},(\widetilde{ji}),k)\,J_2^{(2)}(\lbrace p\rbrace_{2})  .
\end{eqnarray}

\addtocontents{toc}{\protect\setcounter{tocdepth}{2}}
\subsubsection{B-type $\mathcal{O}(N^{0})$ contribution}
\addtocontents{toc}{\protect\setcounter{tocdepth}{3}}

The subleading-colour matrix element $\tilde{B}_{2}^{\gamma,1}(3,\hat{1},2,4)$ has its poles and divergences regulated
by the subtraction term $\tilde{B}_{2}^{\gamma,1,T}(3,\hat{1},2,4)$ and we also construct the finite IC subtraction term
which contains integrated IC real subtraction term and IC mass factorization counterterms. The total contribution to the cross
section is given by
\ba
\sum_{P(\hat{1},2)}\tilde{\bar{B}}_{2}^{\gamma,1}(3,\hat{1},2,4)-\tilde{B}_{2}^{\gamma,1,T}(3,\hat{1},2,4)-\tilde{B}_{2,g\to q}^{\gamma,1,T}(3,\hat{1},2,4) ,
\ea
where
\begin{eqnarray}
\lefteqn{{\tilde{B}_{2}^{\gamma,1,T}(j,\hat{1},i,k) =}} \nonumber\\
  \ph{1}&&-J_{2,QQ}^{1,FF}(s_{jk})\,\bar{B}_2^{\gamma,0}(j,1,i,k)\,J_2^{(3)}(\lbrace p\rbrace_{3})\nonumber\\
  \ph{2}&&-J_{2,QQ}^{1,FF}(s_{jk})\,\bar{B}_2^{\gamma,0}(j,i,1,k)\,J_2^{(3)}(\lbrace p\rbrace_{3})\nonumber\\
 \ph{3}&&-\bigg[ 
 +J_{2,GQ}^{1,IF}(s_{1k})
 +J_{2,GQ}^{1,IF}(s_{1j})\bigg]\,\tilde{\bar{B}}_2^{\gamma,0}(j,1,i,k)\,J_2^{(3)}(\lbrace p\rbrace_{3})\nonumber\\
 \ph{4}&&-\bigg[ 
 +J_{2,QG}^{1,FF}(s_{ji})
 +J_{2,QG}^{1,FF}(s_{ki})\bigg]\,\tilde{\bar{B}}_2^{\gamma,0}(j,1,i,k)\,J_2^{(3)}(\lbrace p\rbrace_{3})\nonumber\\
  \ph{5}&&+J_{2,QQ}^{1,FF}(s_{jk})\,\tilde{\bar{B}}_2^{\gamma,0}(j,1,i,k)\,J_2^{(3)}(\lbrace p\rbrace_{3})\nonumber\\
  \ph{6}&&+\,A_3^{0}(j,i,k)\,\bigg[\,\bar{B}_1^{\gamma,1}((\widetilde{ji}),1,(\widetilde{ki}))\,\delta(1-x_1)\,\delta(1-x_2)\nonumber\\
  &&+\,\bigg( 
 +J_{2,GQ}^{1,IF}(s_{1(\widetilde{ji})})
 +J_{2,GQ}^{1,IF}(s_{1(\widetilde{ki})})\bigg)\,\bar{B}_1^{\gamma,0}((\widetilde{ji}),1,(\widetilde{ki})) \bigg]\,J_2^{(2)}(\lbrace p\rbrace_{2})\nonumber\\
  \ph{7}&&+\,d_{3,g}^{0}(j,i,1)\,\bigg[\,\tilde{\bar{B}}_1^{\gamma,1}((\widetilde{ji}),\overline{1},k)\,\delta(1-x_1)\,\delta(1-x_2)\nonumber\\
 &&\phantom{\bigg[}+J_{2,QQ}^{1,FF}(s_{(\widetilde{ji})k})\,\bar{B}_1^{\gamma,0}((\widetilde{ji}),\overline{1},k) \bigg]\,J_2^{(2)}(\lbrace p\rbrace_{2})\nonumber\\
  \ph{8}&&+\,d_{3,g}^{0}(k,i,1)\,\bigg[\,\tilde{\bar{B}}_1^{\gamma,1}(j,\overline{1},(\widetilde{ki}))\,\delta(1-x_1)\,\delta(1-x_2)\nonumber\\
 &&\phantom{\bigg[}+J_{2,QQ}^{1,FF}(s_{(\widetilde{ki})j})\,\bar{B}_1^{\gamma,0}(j,\overline{1},(\widetilde{ki})) \bigg]\,J_2^{(2)}(\lbrace p\rbrace_{2})\nonumber\\
  \ph{9}&&+\bigg[\,A_{3}^{1}(j,i,k)\,\delta(1-x_1)\,\delta(1-x_2)
  \nonumber\\&&\,  
   +\bigg( 
 +J_{2,QG}^{1,FF}(s_{ji})
 +J_{2,QG}^{1,FF}(s_{ki})
 -J_{2,QQ}^{1,FF}(s_{(\widetilde{ki})(\widetilde{ji})})\bigg)\,A_3^{0}(j,i,k)\bigg]\nonumber\\
 &&\times \bar{B}_1^{\gamma,0}((\widetilde{ji}),1,(\widetilde{ki}))\,J_2^{(2)}(\lbrace p\rbrace_{2})\nonumber\\
  \ph{10}&&+\bigg[\,\tilde{A}_{3}^{1}(j,i,k)\,\delta(1-x_1)\,\delta(1-x_2)
  \nonumber\\&&\,  
   +\bigg( 
 +J_{2,QQ}^{1,FF}(s_{jk})
 -J_{2,QQ}^{1,FF}(s_{(\widetilde{ki})(\widetilde{ji})})\bigg)\,A_3^{0}(j,i,k)\bigg]\,\bar{B}_1^{\gamma,0}((\widetilde{ji}),1,(\widetilde{ki}))\,J_2^{(2)}(\lbrace p\rbrace_{2})\nonumber\\
 \ph{11}&&-\bigg[ +J_{2,QQ}^{1,FF}(s_{jk})
 -J_{2,QQ}^{1,FF}(s_{(\widetilde{ji})(\widetilde{ki})})
 -J_{2,GQ}^{1,IF}(s_{1j})\nonumber\\
 &&\phantom{-}
 +J_{2,GQ}^{1,IF}(s_{1(\widetilde{ji})})
 -J_{2,GQ}^{1,IF}(s_{1k})
 +J_{2,GQ}^{1,IF}(s_{1(\widetilde{ki})})
 \nonumber\\&&+\bigg(  -{\cal S}^{FF}(s_{jk},s_{jk},1)
 +{\cal S}^{FF}(s_{(\widetilde{ji})(\widetilde{ki})},s_{jk},x_{(\widetilde{ji})(\widetilde{ki}),jk})
 +{\cal S}^{FF}(s_{1j},s_{jk},x_{1j,jk})\nonumber\\
 &&\phantom{-}
 -{\cal S}^{FF}(s_{1(\widetilde{ij})},s_{jk},x_{1(\widetilde{ij}),jk})
 +{\cal S}^{FF}(s_{1k},s_{jk},x_{1k,jk})
 -{\cal S}^{FF}(s_{1(\widetilde{ki})},s_{jk},x_{1(\widetilde{ki}),jk})
 \bigg)\bigg]\nonumber\\
 &&\times A_3^{0}(j,i,k)\,\bar{B}_1^{\gamma,0}((\widetilde{ji}),1,(\widetilde{ki}))\,J_2^{(2)}(\lbrace p\rbrace_{2}) \nonumber\\
   \ph{15}&&-\,A_3^{0}(j,1,k)\,\bigg[\,\bar{B}_{1}^{\gamma,1}(\overline{1},i,(\widetilde{jk}))\,\delta(1-x_1)\,\delta(1-x_2)\nonumber\\
  &&+\,\bigg( 
 +J_{2,QG}^{1,IF}(s_{\overline{1}i})
 +J_{2,QG}^{1,FF}(s_{(\widetilde{jk})i})\bigg)\,{\bar{B}}_1^{\gamma,0}(\overline{1},i,(\widetilde{jk})) \bigg]\,J_2^{(2)}(\lbrace p\rbrace_{2})\nonumber\\
  \ph{16}&&-\,A_3^{0}(j,1,k)\,\bigg[\,\tilde{\bar{B}}_{1}^{\gamma,1}(\overline{1},i,(\widetilde{jk}))\,\delta(1-x_1)\,\delta(1-x_2)\nonumber\\
 &&\phantom{\bigg[}+J_{2,QQ}^{1,IF}(s_{\overline{1}(\widetilde{jk})})\,{\bar{B}}_1^{\gamma,0}(\overline{1},i,(\widetilde{jk})) \bigg]\,J_2^{(2)}(\lbrace p\rbrace_{2})\nonumber\\
  \ph{17}&&-\bigg[\,A_{3,g}^{1}(j,1,k)\,\delta(1-x_1)\,\delta(1-x_2)
  \nonumber\\&&\,  
   +\bigg( 
 -J_{2,QQ}^{1,IF}(s_{\overline{1}(\widetilde{jk})})
 +J_{2,GQ}^{1,IF}(s_{1j})
 +J_{2,GQ}^{1,IF}(s_{1k})\bigg)\,A_3^{0}(j,1,k)\bigg]\nonumber\\
 &&\times{\bar{B}}_1^{\gamma,0}(\overline{1},i,(\widetilde{jk}))\,J_2^{(2)}(\lbrace p\rbrace_{2})\nonumber\\
  \ph{18}&&-\bigg[\,\tilde{A}_{3,g}^{1}(j,1,k)\,\delta(1-x_1)\,\delta(1-x_2)
  \nonumber\\&&\,  
   +\bigg( 
 -J_{2,QQ}^{1,IF}(s_{\overline{1}(\widetilde{jk})})
 +J_{2,QQ}^{1,FF}(s_{jk})\bigg)\,A_3^{0}(j,1,k)\bigg]\,{\bar{B}}_1^{\gamma,0}(\overline{1},i,(\widetilde{jk}))\,J_2^{(2)}(\lbrace p\rbrace_{2})\nonumber\\
 \ph{19}&&-\bigg[ +J_{2,QQ}^{1,IF}(s_{(\widetilde{jk})\overline{1}})
 +J_{2,QG}^{1,FF}(s_{ji})
 +J_{2,QG}^{1,FF}(s_{ki})\nonumber\\
 &&\phantom{-}
 -J_{2,QG}^{1,IF}(s_{\overline{1}i})
 -J_{2,QG}^{1,FF}(s_{(\widetilde{jk})i})
 -J_{2,QQ}^{1,FF}(s_{jk})
 \nonumber\\&&\phantom{-} +\bigg(  -{\cal S}^{FF}(s_{(\widetilde{jk})\overline{1}},s_{jk},x_{(\widetilde{jk})\overline{1},jk})
 -{\cal S}^{FF}(s_{ij},s_{jk},x_{ij,jk})
 -{\cal S}^{FF}(s_{ki},s_{jk},x_{ki,jk})\nonumber\\
 &&\phantom{-}
 +{\cal S}^{FF}(s_{(\widetilde{jk})i},s_{jk},x_{(\widetilde{jk})i,jk})
 +{\cal S}^{FF}(s_{\overline{1}i},s_{jk},x_{\overline{1}i,jk})
 +{\cal S}^{FF}(s_{jk},s_{jk},1)
 \bigg)\bigg]\nonumber\\
 &&\times A_3^{0}(j,1,k)\,{\bar{B}}_1^{\gamma,0}(\overline{1},i,(\widetilde{jk}))\,J_2^{(2)}(\lbrace p\rbrace_{2}) ,
\end{eqnarray}

\begin{eqnarray}
\lefteqn{{\tilde{B}_{2,g\to q}^{\gamma,1,T}(j,\hat{1},i,k) =}} \nonumber\\
  \ph{12}&&-J_{2,QQ,g \to q}^{1,IF}(s_{1k})\,\bar{B}_2^{\gamma,0}(1,i,j,k)\,J_2^{(3)}(\lbrace p\rbrace_{3})\nonumber\\
  \ph{13}&&-J_{2,QQ,g \to q}^{1,IF}(s_{1k})\,\bar{B}_2^{\gamma,0}(1,j,i,k)\,J_2^{(3)}(\lbrace p\rbrace_{3})\nonumber\\
  \ph{14}&&-J_{2,QQ,g \to q}^{1,IF}(s_{1k})\,\tilde{\bar{B}}_2^{\gamma,0}(1,i,j,k)\,J_2^{(3)}(\lbrace p\rbrace_{3})\nonumber\\
 \ph{20}&& +J_{2,QQ,g \to q}^{1,IF}(s_{k1})
\,A_3^{0}(1,i,k)\,{\bar{B}}_1^{\gamma,0}(\overline{1},j,(\widetilde{ik}))\,J_2^{(2)}(\lbrace p\rbrace_{2})  \nonumber\\
 \ph{21}&& +J_{2,QQ,g \to q}^{1,IF}(s_{k1})
\,A_3^{0}(1,j,k)\,{\bar{B}}_1^{\gamma,0}(\overline{1},i,(\widetilde{jk}))\,J_2^{(2)}(\lbrace p\rbrace_{2})  \nonumber\\
 \ph{22}&&+\bigg[ +J_{2,QQ,g \to q}^{1,IF}(s_{1k})
 -J_{2,QQ,g \to q}^{1,IF}(s_{\overline{1}(\widetilde{ki})})
\bigg]  A_3^{0}(1,i,k)\,{\bar{B}}_1^{\gamma,0}(\overline{1},j,(\widetilde{ik}))\,J_2^{(2)}(\lbrace p\rbrace_{2})  \nonumber\\
 \ph{23}&&+\bigg[ -J_{2,QQ,g \to q}^{1,IF}(s_{\overline{1}(\widetilde{kj})})
 +J_{2,QQ,g \to q}^{1,IF}(s_{1k})
\bigg]  A_3^{0}(1,j,k)\,{\bar{B}}_1^{\gamma,0}(\overline{1},i,(\widetilde{jk}))\,J_2^{(2)}(\lbrace p\rbrace_{2})  \nonumber\\
 \ph{24}&& +J_{2,QQ,g \to q}^{1,IF}(s_{1(\widetilde{ki})})
\,d_3^{0}(k,i,j)\,{\bar{B}}_1^{\gamma,0}(1,(\widetilde{ij}),(\widetilde{ki}))\,J_2^{(2)}(\lbrace p\rbrace_{2})  \nonumber\\
 \ph{25}&& +J_{2,QQ,g \to q}^{1,IF}(s_{\overline{1}k})
\,d_3^{0}(1,j,i)\,{\bar{B}}_1^{\gamma,0}(\overline{1},(\widetilde{ij}),k)\,J_2^{(2)}(\lbrace p\rbrace_{2})  \nonumber\\
 \ph{26}&& +J_{2,QQ,g \to q}^{1,IF}(s_{1(\widetilde{kj})})
\,d_3^{0}(k,j,i)\,{\bar{B}}_1^{\gamma,0}(1,(\widetilde{ji}),(\widetilde{kj}))\,J_2^{(2)}(\lbrace p\rbrace_{2})  \nonumber\\
 \ph{27}&& +J_{2,QQ,g \to q}^{1,IF}(s_{\overline{1}k})
\,d_3^{0}(1,i,j)\,{\bar{B}}_1^{\gamma,0}(\overline{1},(\widetilde{ji}),k)\,J_2^{(2)}(\lbrace p\rbrace_{2})\nonumber\\
  \ph{1}&&-\frac{1}{2}J_{2,QQ,g \to q}^{1,IF}(s_{1k})\,D_0^{\gamma,0}(1,j;k,i)\,J_2^{(3)}(\lbrace p\rbrace_{3})\nonumber\\
  \ph{2}&&-\frac{1}{2}J_{2,QQ,g \to q}^{1,IF}(s_{1k})\,D_0^{\gamma,0}(i,1;j,k)\,J_2^{(3)}(\lbrace p\rbrace_{3}).
\end{eqnarray}

We note that the last two lines of $\tilde{B}_{2,g\to q}^{\gamma,1,T}$ come from the integrated version of the IC terms in the
$\tilde{D}_{0}^{\gamma,0,S}$ subtraction term because this contributes to the same colour factor as the $\tilde{B}_{2}^{\gamma,1}$
matrix element. These two lines are therefore finite on their own.

\addtocontents{toc}{\protect\setcounter{tocdepth}{2}}
\subsubsection{B-type $\mathcal{O}(N^{-2})$ contribution}
\addtocontents{toc}{\protect\setcounter{tocdepth}{3}}

The subleading-colour two-quark--two-gluon matrix element $\tilde{\tilde{B}}_{2}^{\gamma,1}$ has its poles and divergences
removed by the subtraction term $\tilde{\tilde{B}}_{2}^{\gamma,1,S}$. We also construct the finite IC subtraction term, $\tilde{\tilde{B}}_{2,g\to q}^{\gamma,1,T}$.
The total subtracted contribution to the cross section is given by
\ba
\tilde{\tilde{\bar{B}}}_{2}^{\gamma,1}(3,\hat{1},2,4)-\tilde{\tilde{B}}_{2}^{\gamma,1,T}(3,\hat{1},2,4)-\tilde{\tilde{B}}_{2,g\to q}^{\gamma,1,T}(3,\hat{1},2,4),
\ea
where
\begin{eqnarray}
\lefteqn{{\tilde{\tilde{B}}_{2}^{\gamma,1,T}(j,\hat{1},i,k) =}} \nonumber\\
  \ph{1}&&-J_{2,QQ}^{1,FF}(s_{jk})\,\tilde{\bar{B}}_2^{\gamma,0}(j,1,i,k)\,J_2^{(3)}(\lbrace p\rbrace_{3})\nonumber\\
  \ph{2}&&+\,A_3^{0}(j,i,k)\,\bigg[\,\tilde{\bar{B}}_1^{\gamma,1}((\widetilde{ji}),1,(\widetilde{ki}))\,\delta(1-x_1)\,\delta(1-x_2)\nonumber\\
 &&\phantom{\bigg[}+J_{2,QQ}^{1,FF}(s_{(\widetilde{ji})(\widetilde{ki})})\,\bar{B}_1^{\gamma,0}((\widetilde{ji}),1,(\widetilde{ki})) \bigg]\,J_2^{(2)}(\lbrace p\rbrace_{2})\nonumber\\
  \ph{3}&&+\bigg[\,\tilde{A}_{3}^{1}(j,i,k)\,\delta(1-x_1)\,\delta(1-x_2)
  \nonumber\\&&\,  
   +\bigg( 
 +J_{2,QQ}^{1,FF}(s_{jk})
 -J_{2,QQ}^{1,FF}(s_{(\widetilde{ji})(\widetilde{ki})})\bigg)\,A_3^{0}(j,i,k)\bigg]\,\bar{B}_1^{\gamma,0}((\widetilde{ji}),1,(\widetilde{ki}))\,J_2^{(2)}(\lbrace p\rbrace_{2})\nonumber\\
  \ph{5}&&-\,A_3^{0}(j,1,k)\,\bigg[\,\tilde{\bar{B}}_{1}^{\gamma,1}(\overline{1},i,(\widetilde{jk}))\,\delta(1-x_1)\,\delta(1-x_2)\nonumber\\
 &&\phantom{\bigg[}+J_{2,QQ}^{1,IF}(s_{\overline{1}(\widetilde{jk})})\,{\bar{B}}_1^{\gamma,0}(\overline{1},i,(\widetilde{jk})) \bigg]\,J_2^{(2)}(\lbrace p\rbrace_{2})\nonumber\\
  \ph{6}&&-\bigg[\,\tilde{A}_{3,g}^{1}(j,1,k)\,\delta(1-x_1)\,\delta(1-x_2)
  \nonumber\\&&\,  
   +\bigg( 
 +J_{2,QQ}^{1,FF}(s_{jk})
 -J_{2,QQ}^{1,IF}(s_{\overline{1}(\widetilde{jk})})\bigg)\,A_3^{0}(j,1,k)\bigg]\,{\bar{B}}_1^{\gamma,0}(\overline{1},i,(\widetilde{jk}))\,J_2^{(2)}(\lbrace p\rbrace_{2}),
\end{eqnarray}

\begin{eqnarray}
\lefteqn{{\tilde{\tilde{B}}_{2,g\to q}^{\gamma,1,T}(j,\hat{1},i,k) =}} \nonumber\\
  \ph{4}&&-J_{2,QQ,g \to q}^{1,IF}(s_{1k})\,\tilde{\bar{B}}_2^{\gamma,0}(1,i,j,k)\,J_2^{(3)}(\lbrace p\rbrace_{3})\nonumber\\
 \ph{7}&& +J_{2,QQ,g \to q}^{1,IF}(s_{1k})
\,A_{3,q}^{0}(1,i,k)\,{\bar{B}}_1^{\gamma,0}(\overline{1},j,(\widetilde{ki}))\,J_2^{(2)}(\lbrace p\rbrace_{2})  \nonumber\\
 \ph{8}&& +J_{2,QQ,g \to q}^{1,IF}(s_{1k})
\,A_{3,q}^{0}(1,j,k)\,{\bar{B}}_1^{\gamma,0}(\overline{1},i,(\widetilde{kj}))\,J_2^{(2)}(\lbrace p\rbrace_{2}) \nonumber\\
  \ph{1}&&-\frac{1}{2}J_{2,QQ,g \to q}^{1,IF}(s_{1k})\,D_0^{\gamma,0}(1,j;k,i)\,J_2^{(3)}(\lbrace p\rbrace_{3})\nonumber\\
  \ph{2}&&-\frac{1}{2}J_{2,QQ,g \to q}^{1,IF}(s_{1j})\,D_0^{\gamma,0}(i,1;k,j)\,J_2^{(3)}(\lbrace p\rbrace_{3}).
\end{eqnarray}

As in the section above, we note that the last two lines of the IC subtraction term are the integrated version of $\tilde{D}_{1}^{\gamma,0,S}$, combined with
the appropriate IC mass factorization counterterms to render it finite.
They are included here, as these terms come with the same colour factor as the $\tilde{\tilde{B}}_{2}^{\gamma,1}$
matrix element and are finite on their own.

\addtocontents{toc}{\protect\setcounter{tocdepth}{2}}
\subsubsection{B-type $\mathcal{O}(N_FN^{1})$ contribution}
\addtocontents{toc}{\protect\setcounter{tocdepth}{3}}

The matrix elements containing a closed quark loop contribute to two colour factors in Tab.~\ref{tab:channelsg}. The matrix
element, $\hat{B}_{2}^{\gamma,1}$, is summed over both gluon orderings but the subtraction term, $\hat{B}_{2}^{\gamma,1,T}$, is constructed to cancel the poles
and divergences of the full matrix element and so is not summed over any permutations. We also construct
a finite IC subtraction term, $\hat{B}_{2,g\to q}^{\gamma,1,T}$. The leading-colour contribution to
the cross section is given by
\ba
\sum_{P(\hat{1},2)}\hat{\bar{B}}_{2}^{\gamma,1}(3,\hat{1},2,4)-\hat{B}_{2}^{\gamma,1,T}(3,\hat{1},2,4)-B_{2,g\to q}^{\gamma,1,T}(3,\hat{1},2,4) ,
\ea
where
\begin{eqnarray}
\lefteqn{{\hat{B}_{2}^{\gamma,1,T}(j,\hat{1},i,k) =}} \nonumber\\
 \ph{1}&&-\bigg[ 
 +2\hat{J}_{2,GQ}^{1,IF}(s_{ji})
 +2\hat{J}_{2,QG}^{1,FF}(s_{ji})\bigg]\,\bar{B}_2^{\gamma,0}(j,1,i,k)\,J_2^{(3)}(\lbrace p\rbrace_{3})\nonumber\\
 \ph{2}&&-\bigg[ 
 +2\hat{J}_{2,GQ}^{1,IF}(s_{ji})
 +2\hat{J}_{2,QG}^{1,FF}(s_{ki})\bigg]\,\bar{B}_2^{\gamma,0}(j,i,1,k)\,J_2^{(3)}(\lbrace p\rbrace_{3})\nonumber\\
  \ph{3}&&+\,D_{3,g}^{0}(j,i,1)\,\bigg[\,\hat{\bar{B}}_{1}^{\gamma,1}((\widetilde{ji}),\overline{1},k)\,\delta(1-x_1)\,\delta(1-x_2)\nonumber\\
 &&\phantom{\bigg[}+2\hat{J}_{2,GQ}^{1,IF}(s_{k\overline{1}})\,{\bar{B}}_1^{\gamma,0}((\widetilde{ji}),\overline{1},k) \bigg]\,J_2^{(2)}(\lbrace p\rbrace_{2})\nonumber\\
  \ph{4}&&+\,D_{3,g}^{0}(k,i,1)\,\bigg[\,\hat{\bar{B}}_{1}^{\gamma,1}(j,\overline{1},(\widetilde{ki}))\,\delta(1-x_1)\,\delta(1-x_2)\nonumber\\
 &&\phantom{\bigg[}+2\hat{J}_{2,GQ}^{1,IF}(s_{j\overline{1}})\,{\bar{B}}_1^{\gamma,0}(j,\overline{1},(\widetilde{ki})) \bigg]\,J_2^{(2)}(\lbrace p\rbrace_{2})\nonumber\\
  \ph{5}&&-\,A_3^{0}(k,1,j)\,\bigg[\,\hat{\bar{B}}_{1}^{\gamma,1}(\overline{1},i,(\widetilde{jk}))\,\delta(1-x_1)\,\delta(1-x_2)\nonumber\\
 &&\phantom{\bigg[}+2\hat{J}_{2,GQ}^{1,IF}(s_{(\widetilde{jk})i})\,{\bar{B}}_1^{\gamma,0}(\overline{1},i,(\widetilde{jk})) \bigg]\,J_2^{(2)}(\lbrace p\rbrace_{2})\nonumber\\
  \ph{6}&&-\,A_3^{0}(k,1,j)\,\bigg[\,\hat{\bar{B}}_{1}^{\gamma,1}((\widetilde{jk}),\overline{1},i)\,\delta(1-x_1)\,\delta(1-x_2)\nonumber\\
 &&\phantom{\bigg[}+2\hat{J}_{2,GQ}^{1,IF}(s_{(\widetilde{jk})\overline{1}})\,{\bar{B}}_1^{\gamma,0}((\widetilde{jk}),\overline{1},i) \bigg]\,J_2^{(2)}(\lbrace p\rbrace_{2})\nonumber\\
  \ph{7}&&-\bigg[\,\hat{A}_{3,g}^{1}(j,1,k)\,\delta(1-x_1)\,\delta(1-x_2)
 +2\hat{J}_{2,QG}^{1,FF}(s_{jk})\,A_{3,g\to q}^{0}(j,1,k)\bigg]\nonumber\\
 &&\times{\bar{B}}_1^{\gamma,0}(\overline{1},i,(\widetilde{kj}))\,J_2^{(2)}(\lbrace p\rbrace_{2})\nonumber\\
 \ph{8}&&+\bigg[ +2\hat{J}_{2,QG}^{1,FF}(s_{jk})
 -2\hat{J}_{2,QG}^{1,FF}(s_{ki})
\bigg]  a_{3,g\to q}^{0}(k,1,j)\,{\bar{B}}_1^{\gamma,0}(\overline{1},i,(\widetilde{kj}))\,J_2^{(2)}(\lbrace p\rbrace_{2})  \nonumber\\
 \ph{9}&&+\bigg[ -2\hat{J}_{2,QG}^{1,FF}(s_{ji})
 +2\hat{J}_{2,QG}^{1,FF}(s_{jk})
\bigg]  a_{3,g\to q}^{0}(j,1,k)\,{\bar{B}}_1^{\gamma,0}(\overline{1},i,(\widetilde{kj}))\,J_2^{(2)}(\lbrace p\rbrace_{2})  \nonumber\\
  \ph{10}&&-\bigg[\,\hat{A}_{3,g}^{1}(j,1,k)\,\delta(1-x_1)\,\delta(1-x_2)
 +2\hat{J}_{2,QG}^{1,FF}(s_{jk})\,A_{3,g\to q}^{0}(j,1,k)\bigg]\nonumber\\
 &&\times{\bar{B}}_1^{\gamma,0}((\widetilde{kj}),\overline{1},i)\,J_2^{(2)}(\lbrace p\rbrace_{2})\nonumber\\
 \ph{11}&&+\bigg[ +2\hat{J}_{2,QG}^{1,FF}(s_{jk})
 -2\hat{J}_{2,QG}^{1,FF}(s_{ki})
\bigg]  a_{3,g\to q}^{0}(k,1,j)\,{\bar{B}}_1^{\gamma,0}((\widetilde{jk}),\overline{1},i)\,J_2^{(2)}(\lbrace p\rbrace_{2})  \nonumber\\
 \ph{12}&&+\bigg[ -2\hat{J}_{2,QG}^{1,FF}(s_{ji})
 +2\hat{J}_{2,QG}^{1,FF}(s_{jk})
\bigg]  a_{3,g\to q}^{0}(j,1,k)\,{\bar{B}}_1^{\gamma,0}((\widetilde{jk}),\overline{1},i)\,J_2^{(2)}(\lbrace p\rbrace_{2})  \nonumber\\
  \ph{13}&&+\bigg[\,\hat{D}_{3,g}^{1}(j,i,1)\,\delta(1-x_1)\,\delta(1-x_2)
 +2\hat{J}_{2,QG}^{1,IF}(s_{1j})\,D_{3,g}^{0}(j,i,1)\bigg]\nonumber\\
 &&\times{\bar{B}}_1^{\gamma,0}((\widetilde{ji}),\overline{1},k)\,J_2^{(2)}(\lbrace p\rbrace_{2})\nonumber\\
 \ph{14}&&+\bigg[ +2\hat{J}_{2,QG}^{1,FF}(s_{ki})
 -2\hat{J}_{2,QG}^{1,IF}(s_{1j})
\bigg]  d_{3,g}^{0}(j,i,1)\,{\bar{B}}_1^{\gamma,0}((\widetilde{ji}),\overline{1},k)\,J_2^{(2)}(\lbrace p\rbrace_{2})  \nonumber\\
 \ph{15}&& +2\hat{J}_{2,QG}^{1,FF}(s_{ki})
\,d_{3,g\to q}^{0}(k,1,i)\,{\bar{B}}_1^{\gamma,0}(j,\overline{1},(\widetilde{ki}))\,J_2^{(2)}(\lbrace p\rbrace_{2})  \nonumber\\
 \ph{16}&& -2\hat{J}_{2,QG}^{1,IF}(s_{1j})
\,d_{3,g\to q}^{0}(j,1,i)\,{\bar{B}}_1^{\gamma,0}((\widetilde{ji}),\overline{1},k)\,J_2^{(2)}(\lbrace p\rbrace_{2})  \nonumber\\
  \ph{17}&&+\bigg[\,\hat{D}_{3,g}^{1}(k,i,1)\,\delta(1-x_1)\,\delta(1-x_2)
 +2\hat{J}_{2,QG}^{1,IF}(s_{1k})\,D_{3,g}^{0}(k,i,1)\bigg]\nonumber\\
 &&\times{\bar{B}}_1^{\gamma,0}(j,\overline{1},(\widetilde{ki}))\,J_2^{(2)}(\lbrace p\rbrace_{2})\nonumber\\
 \ph{18}&&+\bigg[ +2\hat{J}_{2,QG}^{1,FF}(s_{ji})
 -2\hat{J}_{2,QG}^{1,IF}(s_{1k})
\bigg]  d_{3,g}^{0}(k,i,1)\,{\bar{B}}_1^{\gamma,0}(j,\overline{1},(\widetilde{ki}))\,J_2^{(2)}(\lbrace p\rbrace_{2})  \nonumber\\
 \ph{19}&& +2\hat{J}_{2,QG}^{1,FF}(s_{ji})
\,d_{3,g\to q}^{0}(j,1,i)\,{\bar{B}}_1^{\gamma,0}((\widetilde{ji}),\overline{1},k)\,J_2^{(2)}(\lbrace p\rbrace_{2})  \nonumber\\
 \ph{20}&& -2\hat{J}_{2,QG}^{1,IF}(s_{1k})
\,d_{3,g\to q}^{0}(k,1,i)\,{\bar{B}}_1^{\gamma,0}(j,\overline{1},(\widetilde{ki}))\,J_2^{(2)}(\lbrace p\rbrace_{2})  .
\end{eqnarray}

\begin{eqnarray}
\lefteqn{{\hat{B}_{2,g\to q}^{\gamma,1,T}(j,\hat{1},k,i) =}} \nonumber\\
  \ph{1}&&-J_{2,QQ,g \to q}^{1,IF}(s_{1i})\,C_0^{\gamma,0}(1;j,k;i)\,J_2^{(3)}(\lbrace p\rbrace_{3})\nonumber\\
  \ph{2}&&-J_{2,QQ,g \to q}^{1,IF}(s_{1i})\,C_0^{\gamma,0}(i;j,k;1)\,J_2^{(3)}(\lbrace p\rbrace_{3})\nonumber\\
 \ph{3}&& +J_{2,QQ,g \to q}^{1,IF}(s_{1(\widetilde{ij})})
\,E_3^{0}(i,j,k)\,{\bar{B}}_{1,q}^{\gamma,0}(1,(\widetilde{jk}),(\widetilde{ij}))\,J_2^{(2)}(\lbrace p\rbrace_{2})  \nonumber\\
 \ph{4}&& +J_{2,QQ,g \to q}^{1,IF}(s_{\overline{1}i})
\,E_{3,q}^{0}(1,j,k)\,{\bar{B}}_{1,q}^{\gamma,0}(\overline{1},(\widetilde{jk}),i)\,J_2^{(2)}(\lbrace p\rbrace_{2})  \nonumber\\
 \ph{5}&&+\bigg[ -2J_{2,QQ,g \to q}^{1,IF}(s_{1i})
 +2J_{2,QQ,g \to q}^{1,IF}(s_{\overline{1}i})
\bigg]  E_{3,q}^{0}(1,j,k)\,\bar{B}_{1,q}^{\gamma,0}((\widetilde{jk}),\overline{1},i)\,J_2^{(2)}(\lbrace p\rbrace_{2})  \nonumber\\
 \ph{6}&& -J_{2,QQ,g \to q}^{1,IF}(s_{1i})
\,E_{3,q^\prime\to g}^{0}(j,1,i)\,\bar{B}_{1,Q}^{\gamma,0}(k,\overline{1},(\widetilde{ij}))\,J_2^{(2)}(\lbrace p\rbrace_{2})  \nonumber\\
 \ph{7}&& -J_{2,QQ,g \to q}^{1,IF}(s_{1i})
\,E_{3,q^\prime\to g}^{0}(k,1,i)\,\bar{B}_{1,Q}^{\gamma,0}((\widetilde{ik}),\overline{1},j)\,J_2^{(2)}(\lbrace p\rbrace_{2})  .
\end{eqnarray}

\addtocontents{toc}{\protect\setcounter{tocdepth}{2}}
\subsubsection{B-type $\mathcal{O}(N_FN^{-1})$ contribution}
\addtocontents{toc}{\protect\setcounter{tocdepth}{3}}

The subleading-colour contribution to the closed quark-loop matrix element is given by the $\hat{\tilde{B}}_{2}^{\gamma,1}$ function.
The poles and divergences of this function are removed by the subtraction term $\hat{\tilde{B}}_{2}^{\gamma,1,T}$. We also construct
the finite IC subtraction term, $\hat{\tilde{B}}_{2,g\to q}^{\gamma,1,T}$ which is derived by integrating the IC terms in Sec.~\ref{sec:nnloRRCtg}
and combining with the relevant IC mass factorization kernels. The total contribution to the cross section is given by
\ba
\hat{\tilde{\bar{B}}}_{2}^{\gamma,1}(3,\hat{1},2,4)-\hat{\tilde{B}}_{2}^{\gamma,1,T}(3,\hat{1},2,4)-\hat{\tilde{B}}_{2,g\to q}^{\gamma,1,T}(3,\hat{1},2,4) ,
\ea
where
\begin{eqnarray}
\lefteqn{{\hat{\tilde{B}}_{2}^{\gamma,1,T}(j,\hat{1},i,k) =}} \nonumber\\
 \ph{1}&&-\bigg[ 
 +\hat{J}_{2,QG}^{1,FF}(s_{ji})
 +\hat{J}_{2,QG}^{1,FF}(s_{ki})
 +2\hat{J}_{2,GQ}^{1,IF}(s_{jk})\bigg]\,\tilde{\bar{B}}_2^{\gamma,0}(j,1,i,k)\,J_2^{(3)}(\lbrace p\rbrace_{3})\nonumber\\
  \ph{2}&&-\,A_3^{0}(j,1,k)\,\bigg[\ \hat{\bar{B}}_{1}^{\gamma,1}(\overline{1},i,(\widetilde{jk}))\,\delta(1-x_1)\,\delta(1-x_2)\nonumber\\
 &&\phantom{\bigg[}+2\hat{J}_{2,QG}^{1,FF}(s_{(\widetilde{jk})i})\,{\bar{B}}_1^{\gamma,0}(\overline{1},i,(\widetilde{jk})) \bigg]\,J_2^{(2)}(\lbrace p\rbrace_{2})\nonumber\\
  \ph{3}&&+\,A_3^{0}(j,i,k)\,\bigg[\,\hat{\bar{B}}_{1}^{\gamma,1}((\widetilde{ji}),1,(\widetilde{ik}))\,\delta(1-x_1)\,\delta(1-x_2)\nonumber\\
 &&\phantom{\bigg[}+2\hat{J}_{2,GQ}^{1,IF}(s_{(\widetilde{ji})(\widetilde{ik})})\,{\bar{B}}_1^{\gamma,0}((\widetilde{ji}),1,(\widetilde{ik})) \bigg]\,J_2^{(2)}(\lbrace p\rbrace_{2})\nonumber\\
  \ph{4}&&+\bigg[\,\hat{A}_{3}^{1}(j,i,k)\,\delta(1-x_1)\,\delta(1-x_2)
  \nonumber\\&&\,  
   +\bigg( 
 +\hat{J}_{2,QG}^{1,FF}(s_{ji})
 +\hat{J}_{2,QG}^{1,FF}(s_{ki})\bigg)\,A_3^{0}(j,i,k)\bigg]\,{\bar{B}}_1^{\gamma,0}((\widetilde{ji}),1,(\widetilde{ik}))\,J_2^{(2)}(\lbrace p\rbrace_{2})\nonumber\\
  \ph{5}&&-\bigg[\,\hat{A}_{3,g}^{1}(j,1,k)\,\delta(1-x_1)\,\delta(1-x_2)
 +2\hat{J}_{2,GQ}^{1,IF}(s_{jk})\,A_3^{0}(j,1,k)\bigg]\,{\bar{B}}_1^{\gamma,0}(\overline{1},i,(\widetilde{jk}))\,J_2^{(2)}(\lbrace p\rbrace_{2})\nonumber\\
 \ph{6}&&+\bigg[ -\hat{J}_{2,QG}^{1,FF}(s_{ki})
 -\hat{J}_{2,QG}^{1,FF}(s_{ji})
 +2\hat{J}_{2,QG}^{1,FF}(s_{(\widetilde{jk})i})
\bigg]  A_3^{0}(j,1,k)\,{\bar{B}}_1^{\gamma,0}(\overline{1},i,(\widetilde{jk}))\,J_2^{(2)}(\lbrace p\rbrace_{2}) ,\nonumber\\
\end{eqnarray}

\begin{eqnarray}
\lefteqn{{\hat{\tilde{B}}_{2,g\to q}^{\gamma,1,T}(j,\hat{1},i,k) =}} \nonumber\\
  \ph{1}&&-J_{2,QQ,g \to q}^{1,IF}(s_{1i})\,C_0^{\gamma,0}(1;j,k;i)\,J_2^{(3)}(\lbrace p\rbrace_{3})\nonumber\\
  \ph{2}&&-J_{2,QQ,g \to q}^{1,IF}(s_{1i})\,C_0^{\gamma,0}(i;j,k;1)\,J_2^{(3)}(\lbrace p\rbrace_{3})\nonumber\\
 \ph{3}&& +J_{2,QQ,g \to q}^{1,IF}(s_{1(\widetilde{ij})})
\,E_3^{0}(i,j,k)\,{\bar{B}}_{1,q}^{\gamma,0}(1,(\widetilde{jk}),(\widetilde{ij}))\,J_2^{(2)}(\lbrace p\rbrace_{2})  \nonumber\\
 \ph{4}&& +J_{2,QQ,g \to q}^{1,IF}(s_{\overline{1}i})
\,E_3^{0}(1,j,k)\,{\bar{B}}_{1,q}^{\gamma,0}(\overline{1},(\widetilde{jk}),i)\,J_2^{(2)}(\lbrace p\rbrace_{2})  \nonumber\\
 \ph{5}&& -J_{2,QQ,g \to q}^{1,IF}(s_{1i})
\,E_{3,q^\prime\to g}^{0}(j,1,i)\,\bar{B}_{1,Q}^{\gamma,0}(k,\overline{1},(\widetilde{ji}))\,J_2^{(2)}(\lbrace p\rbrace_{2})  \nonumber\\
 \ph{6}&& -J_{2,QQ,g \to q}^{1,IF}(s_{1i})
\,E_{3,q^\prime\to g}^{0}(k,1,i)\,\bar{B}_{1,Q}^{\gamma,0}((\widetilde{ki}),\overline{1},j)\,J_2^{(2)}(\lbrace p\rbrace_{2})  .
\end{eqnarray}


\begin{thebibliography}{10}

\bibitem{disbook}
  R.~Devenish and A.~Cooper-Sarkar,
 {\it Deep inelastic scattering}, Oxford University Press (Oxford, 2004).


\bibitem{Newman:2013ada}
  P.~Newman, M.~Wing,
  Rev.\ Mod.\ Phys.\  {\bf 86} (2014)  1037
 [arXiv:1308.3368 [hep-ex]].

\bibitem{lo}
  K.H.~Streng, T.F.~Walsh and P.M.~Zerwas,
  Z.\ Phys.\ C {\bf 2} (1979) 237;
   R.~D.~Peccei and R.~R\"uckl,
  Nucl.\ Phys.\ B {\bf 162} (1980) 125;
 C.~Rumpf, G.~Kramer and J.~Willrodt,
  Z.\ Phys.\ C {\bf 7} (1981) 337.


\bibitem{h1a12}
  C.~Adloff {\it et al.} [H1 Collaboration],
  Eur.\ Phys.\ J.\ C {\bf 19} (2001) 289
  [hep-ex/0010054].

\bibitem{h1b1}
  C.~Adloff {\it et al.} [H1 Collaboration],
  Phys.\ Lett.\ B {\bf 542} (2002) 193
  [hep-ex/0206029].

\bibitem{h1c2}
 A.~Aktas {\it et al.} [H1 Collaboration],
  Eur.\ Phys.\ J.\ C {\bf 33} (2004) 477
  [hep-ex/0310019].




\bibitem{h1d12}
  A.~Aktas {\it et al.} [H1 Collaboration],
  Phys.\ Lett.\ B {\bf 653} (2007) 134
  [arXiv:0706.3722 [hep-ex]].

\bibitem{h1e12}
  F.~D.~Aaron {\it et al.} [H1 Collaboration],
  Eur.\ Phys.\ J.\ C {\bf 65} (2010) 363
  [arXiv:0904.3870 [hep-ex]].


\bibitem{h1highq2}
 V.~Andreev {\it et al.} [H1 Collaboration],
  Eur.\ Phys.\ J.\ C {\bf 75} (2015) 65
  [arXiv:1406.4709 [hep-ex]].

\bibitem{h1lowq2}
  V.~Andreev {\it et al.} [H1 Collaboration],
  arXiv:1611.03421 [hep-ex].


\bibitem{zeusa2}
  J.~Breitweg {\it et al.} [ZEUS Collaboration],
  Phys.\ Lett.\ B {\bf 507} (2001) 70
  [hep-ex/0102042].

\bibitem{zeusb1}
  S.~Chekanov {\it et al.} [ZEUS Collaboration],
  Phys.\ Lett.\ B {\bf 547} (2002) 164
  [hep-ex/0208037];
  Eur.\ Phys.\ J.\ C {\bf 44} (2005) 183
  [hep-ex/0502007].



\bibitem{zeusc12}
 S.~Chekanov {\it et al.} [ZEUS Collaboration],
  Nucl.\ Phys.\ B {\bf 765} (2007) 1
  [hep-ex/0608048].

\bibitem{zeusd1}
S.~Chekanov {\it et al.} [ZEUS Collaboration],
  Phys.\ Lett.\ B {\bf 649} (2007) 12
  [hep-ex/0701039];
 H.~Abramowicz {\it et al.} [ZEUS Collaboration],
  Phys.\ Lett.\ B {\bf 691} (2010) 127
  [arXiv:1003.2923 [hep-ex]].






\bibitem{zeus2j}
 H.~Abramowicz {\it et al.} [ZEUS Collaboration],
  Eur.\ Phys.\ J.\ C {\bf 70} (2010) 965
  [arXiv:1010.6167 [hep-ex]].


\bibitem{graudenz}
  D.~Graudenz,
  Phys.\ Rev.\ D {\bf 49} (1994) 3291
   [hep-ph/9307311];
  hep-ph/9710244.

\bibitem{mirkes}
 E.~Mirkes and D.~Zeppenfeld,
  Phys.\ Lett.\ B {\bf 380} (1996) 205
 [hep-ph/9511448].

\bibitem{jetvip}
 M.~Klasen, G.~Kramer and B.~P\"otter,
  Eur.\ Phys.\ J.\ C {\bf 1} (1998) 261
  [hep-ph/9703302];
  B.~P\"otter,
  Comput.\ Phys.\ Commun.\  {\bf 119} (1999) 45
  [hep-ph/9806437].



\bibitem{nagy}
  Z.~Nagy and Z.~Trocsanyi,
  Phys.\ Rev.\ Lett.\  {\bf 87} (2001) 082001
  [hep-ph/0104315].


\bibitem{laporta}
 S.~Laporta,
  Int.\ J.\ Mod.\ Phys.\ A {\bf 15} (2000) 5087
  [hep-ph/0102033].

\bibitem{gr}
  T.~Gehrmann and E.~Remiddi,
  Nucl.\ Phys.\ B {\bf 580} (2000) 485
  [hep-ph/9912329].


\bibitem{henn}
  J.~M.~Henn,
  Phys.\ Rev.\ Lett.\  {\bf 110} (2013) 251601
  [arXiv:1304.1806 [hep-th]].



 \bibitem{secdec}
  T.~Binoth and G.~Heinrich,
  Nucl.\ Phys.\ B {\bf 693} (2004) 134
 [hep-ph/0402265];
  C.~Anastasiou, K.~Melnikov and F.~Petriello,
  Phys.\ Rev.\ D {\bf 69} (2004) 076010
  [hep-ph/0311311].

 \bibitem{qtsub}
  S.~Catani and M.~Grazzini,
  Phys.\ Rev.\ Lett.\  {\bf 98} (2007) 222002
 [hep-ph/0703012].




 \bibitem{ourant}
  A.~Gehrmann-De Ridder, T.~Gehrmann and E.W.N.\ Glover,
  JHEP {\bf 0509} (2005) 056
 [hep-ph/0505111];
  Phys.\ Lett.\ B {\bf 612} (2005) 49
  [hep-ph/0502110];
  Phys.\ Lett.\ B {\bf 612} (2005) 36
  [hep-ph/0501291].

\bibitem{currie}
 J.~Currie, E.~W.~N.~Glover and S.~Wells,
  JHEP {\bf 1304} (2013) 066
  [arXiv:1301.4693 [hep-ph]].



 \bibitem{stripper}
  M.~Czakon,
  Phys.\ Lett.\ B {\bf 693} (2010) 259
 [arXiv:1005.0274];
   R.~Boughezal, K.~Melnikov and F.~Petriello,
  Phys.\ Rev.\ D {\bf 85} (2012) 034025
    [arXiv:1111.7041 [hep-ph]].


  \bibitem{njettiness}
  R.~Boughezal, C.~Focke, X.~Liu and F.~Petriello,
  Phys.\ Rev.\ Lett.\  {\bf 115} (2015)  062002
  [arXiv:1504.02131 [hep-ph]];
    R.~Boughezal, X.~Liu and F.~Petriello,
  Phys.\ Rev.\ D {\bf 91} (2015)  094035;
    J.~Gaunt, M.~Stahlhofen, F.~J.~Tackmann and J.~R.~Walsh,
  JHEP {\bf 1509} (2015) 058
  [arXiv:1505.04794 [hep-ph]].

\bibitem{trocsanyi}
  G.~Somogyi and Z.~Trocsanyi,
  JHEP {\bf 0808} (2008) 042
   [arXiv:0807.0509 [hep-ph]];
 V.~Del Duca, C.~Duhr, A.~Kardos, G.~Somogyi, Z.~Sz\"or, Z.~Trocsanyi and Z.~Tulipant,
  Phys.\ Rev.\ D {\bf 94} (2016)  074019
    [arXiv:1606.03453 [hep-ph]].

\bibitem{dynnlo}
 S.~Catani, L.~Cieri, G.~Ferrera, D.~de Florian and M.~Grazzini,
  Phys.\ Rev.\ Lett.\  {\bf 103} (2009) 082001
  [arXiv:0903.2120 [hep-ph]].

\bibitem{babisdy}
 K.~Melnikov and F.~Petriello,
  Phys.\ Rev.\ D {\bf 74} (2006) 114017
  [hep-ph/0609070].

\bibitem{babishiggs}
 C.~Anastasiou, K.~Melnikov and F.~Petriello,
  Nucl.\ Phys.\ B {\bf 724} (2005) 197
  [hep-ph/0501130].

\bibitem{hnnlo}
 M.~Grazzini,
  JHEP {\bf 0802} (2008) 043
  [arXiv:0801.3232 [hep-ph]].

\bibitem{our3j}
 A.~Gehrmann-De Ridder, T.~Gehrmann, E.W.N.~Glover and G.~Heinrich,
  JHEP {\bf 0711} (2007) 058
  [arXiv:0710.0346 [hep-ph]];
  Comput.\ Phys.\ Commun.\  {\bf 185} (2014) 3331
 [arXiv:1402.4140 [hep-ph]].


\bibitem{weinzierl3j}
 S.~Weinzierl,
  JHEP {\bf 0906} (2009) 041
  [arXiv:0904.1077 [hep-ph]].

\bibitem{twogamma}
  S.~Catani, L.~Cieri, D.~de Florian, G.~Ferrera and M.~Grazzini,
  Phys.\ Rev.\ Lett.\  {\bf 108} (2012) 072001
  [arXiv:1110.2375 [hep-ph]];
   J.~M.~Campbell, R.~K.~Ellis, Y.~Li and C.~Williams,
  JHEP {\bf 1607} (2016) 148
  [arXiv:1603.02663 [hep-ph]].


\bibitem{vh}
  G.~Ferrera, M.~Grazzini and F.~Tramontano,
  Phys.\ Rev.\ Lett.\  {\bf 107} (2011) 152003
  [arXiv:1107.1164 [hep-ph]].



\bibitem{vgamma}
 M.~Grazzini, S.~Kallweit and D.~Rathlev,
  JHEP {\bf 1507} (2015) 085
  [arXiv:1504.01330 [hep-ph]].

\bibitem{czakon1}
M.~Czakon, P.~Fiedler and A.~Mitov,
  Phys.\ Rev.\ Lett.\  {\bf 110} (2013) 252004
  [arXiv:1303.6254 [hep-ph]].

\bibitem{czakon2}
  M.~Czakon, D.~Heymes and A.~Mitov,
  Phys.\ Rev.\ Lett.\  {\bf 116} (2016) 082003
  [arXiv:1511.00549 [hep-ph]].






\bibitem{hjet}
  R.~Boughezal, F.~Caola, K.~Melnikov, F.~Petriello and M.~Schulze,
  Phys.\ Rev.\ Lett.\  {\bf 115} (2015) 082003
  [arXiv:1504.07922 [hep-ph]];
   F.~Caola, K.~Melnikov and M.~Schulze,
  Phys.\ Rev.\ D {\bf 92} (2015)  074032
  [arXiv:1508.02684 [hep-ph]].

\bibitem{ourhj}
 X.~Chen, J.~Cruz-Martinez, T.~Gehrmann, E.~W.~N.~Glover and M.~Jaquier,
  JHEP {\bf 1610} (2016) 066
  [arXiv:1607.08817 [hep-ph]].



\bibitem{wjet}
 R.~Boughezal, C.~Focke, X.~Liu and F.~Petriello,
  Phys.\ Rev.\ Lett.\  {\bf 115} (2015)  062002
  [arXiv:1504.02131 [hep-ph]].

\bibitem{ourzj}
 A.~Gehrmann-De Ridder, T.~Gehrmann, E.~W.~N.~Glover, A.~Huss and T.~A.~Morgan,
  Phys.\ Rev.\ Lett.\  {\bf 117} (2016)  022001
  [arXiv:1507.02850 [hep-ph]];
  JHEP {\bf 1607} (2016) 133
  [arXiv:1605.04295 [hep-ph]];
  JHEP {\bf 1611} (2016) 094
  [arXiv:1610.01843 [hep-ph]].

\bibitem{zjet}
 R.~Boughezal, J.~M.~Campbell, R.~K.~Ellis, C.~Focke, W.~T.~Giele, X.~Liu and F.~Petriello,
  Phys.\ Rev.\ Lett.\  {\bf 116} (2016)  152001
  [arXiv:1512.01291 [hep-ph]].

\bibitem{mcfmgam}
  J.~M.~Campbell, R.~K.~Ellis and C.~Williams,
  arXiv:1612.04333 [hep-ph].


\bibitem{zz}
  F.~Cascioli {\it et al.},
  Phys.\ Lett.\ B {\bf 735} (2014) 311
  [arXiv:1405.2219 [hep-ph]];
  M.~Grazzini, S.~Kallweit and D.~Rathlev,
  Phys.\ Lett.\ B {\bf 750} (2015) 407
  [arXiv:1507.06257 [hep-ph]].



\bibitem{ww}
  T.~Gehrmann, M.~Grazzini, S.~Kallweit, P.~Maierh\"{o}fer, A.~von Manteuffel, S.~Pozzorini, D.~Rathlev and L.~Tancredi,
  Phys.\ Rev.\ Lett.\  {\bf 113} (2014)   212001
  [arXiv:1408.5243 [hep-ph]];
  M.~Grazzini, S.~Kallweit, S.~Pozzorini, D.~Rathlev and M.~Wiesemann,
  JHEP {\bf 1608} (2016) 140
  [arXiv:1605.02716 [hep-ph]].



\bibitem{zw}
 M.~Grazzini, S.~Kallweit, D.~Rathlev and M.~Wiesemann,
  Phys.\ Lett.\ B {\bf 761} (2016) 179
  [arXiv:1604.08576 [hep-ph]].

\bibitem{abelof}
  G.~Abelof, R.~Boughezal, X.~Liu and F.~Petriello,
  Phys.\ Lett.\ B {\bf 763} (2016) 52
  [arXiv:1607.04921 [hep-ph]].



\bibitem{2jnew}
 J.~Currie, E.~W.~N.~Glover and J.~Pires,
  Phys.\ Rev.\ Lett.\  {\bf 118} (2017)  072002
  [arXiv:1611.01460 [hep-ph]];
 J.~Currie, A.~Gehrmann-De Ridder, T.~Gehrmann, E.~W.~N.~Glover, A.~Huss and J.~Pires,
  arXiv:1705.10271 [hep-ph].


\bibitem{mcfmnnlo}
 R.~Boughezal, J.~M.~Campbell, R.~K.~Ellis, C.~Focke, W.~Giele, X.~Liu, F.~Petriello and C.~Williams,
  Eur.\ Phys.\ J.\ C {\bf 77} (2017)  7
  [arXiv:1605.08011 [hep-ph]].

\bibitem{disprl}
 J.~Currie, T.~Gehrmann and J.~Niehues,
  Phys.\ Rev.\ Lett.\  {\bf 117} (2016) 042001
  [arXiv:1606.03991 [hep-ph]].




\bibitem{meRR}
  K.~Hagiwara and D.~Zeppenfeld,
  Nucl.\ Phys.\ B {\bf 313} (1989) 560;
  F.~A.~Berends, W.~T.~Giele and H.~Kuijf,
  Nucl.\ Phys.\ B {\bf 321} (1989) 39;
  N.~K.~Falck, D.~Graudenz and G.~Kramer,
  Nucl.\ Phys.\ B {\bf 328} (1989) 317.

\bibitem{meRV}
 E.~W.~N.~Glover and D.~J.~Miller,
  Phys.\ Lett.\ B {\bf 396} (1997) 257
 [hep-ph/9609474];
 Z.~Bern, L.~J.~Dixon, D.~A.~Kosower and S.~Weinzierl,
  Nucl.\ Phys.\ B {\bf 489} (1997) 3
 [hep-ph/9610370];
 J.~M.~Campbell, E.~W.~N.~Glover and D.~J.~Miller,
  Phys.\ Lett.\ B {\bf 409} (1997) 503
 [hep-ph/9706297];
  Z.~Bern, L.~J.~Dixon and D.~A.~Kosower,
  Nucl.\ Phys.\ B {\bf 513} (1998) 3
 [hep-ph/9708239].



\bibitem{meVV}
   L.W.~Garland, T.~Gehrmann, E.W.N.~Glover, A.~Koukoutsakis and E.~Remiddi,
  Nucl.\ Phys.\ B {\bf 627} (2002) 107
 [hep-ph/0112081];
  Nucl.\ Phys.\ B {\bf 642} (2002) 227
 [hep-ph/0206067];
 T.~Gehrmann and L.~Tancredi,
  JHEP {\bf 1202} (2012) 004
  [arXiv:1112.1531 [hep-ph]].




\bibitem{openloops}
F.~Cascioli, P.~Maierhofer and S.~Pozzorini,
  Phys.\ Rev.\ Lett.\  {\bf 108} (2012) 111601
  [arXiv:1111.5206 [hep-ph]].

\bibitem{blackhat}
 C.~F.~Berger, Z.~Bern, L.~J.~Dixon, F.~Febres Cordero, D.~Forde, H.~Ita, D.~A.~Kosower and D.~Maitre,
  Phys.\ Rev.\ D {\bf 78} (2008) 036003
  [arXiv:0803.4180 [hep-ph]].

\bibitem{gosam}
 G.~Cullen, N.~Greiner, G.~Heinrich, G.~Luisoni, P.~Mastrolia, G.~Ossola, T.~Reiter and F.~Tramontano,
  Eur.\ Phys.\ J.\ C {\bf 72} (2012) 1889
  [arXiv:1111.2034 [hep-ph]].

\bibitem{amcnlo}
  J.~Alwall {\it et al.},
  JHEP {\bf 1407} (2014) 079
  [arXiv:1405.0301 [hep-ph]].



\bibitem{gavin}
  G.~P.~Salam,
  Eur.\ Phys.\ J.\ C {\bf 67} (2010) 637
    [arXiv:0906.1833 [hep-ph]].

\bibitem{ant}
 D.~A.~Kosower,
  Phys.\ Rev.\ D {\bf 57} (1998) 5410
   [hep-ph/9710213];
  Phys.\ Rev.\ D {\bf 67} (2003) 116003
  [hep-ph/0212097];
  J.~M.~Campbell, M.~A.~Cullen and E.~W.~N.~Glover,
  Eur.\ Phys.\ J.\ C {\bf 9} (1999) 245
  [hep-ph/9809429].


\bibitem{nloant}
 A.~Daleo, T.~Gehrmann and D.~Maitre,
  JHEP {\bf 0704} (2007) 016
  [hep-ph/0612257].


\bibitem{daleo}
 A.~Daleo, A.~Gehrmann-De Ridder, T.~Gehrmann and G.~Luisoni,
  JHEP {\bf 1001} (2010) 118
  [arXiv:0912.0374 [hep-ph]].

\bibitem{monni}
 R.~Boughezal, A.~Gehrmann-De Ridder and M.~Ritzmann,
  JHEP {\bf 1102} (2011) 098
   [arXiv:1011.6631 [hep-ph]];
  A.~Gehrmann-De Ridder, T.~Gehrmann and M.~Ritzmann,
  JHEP {\bf 1210} (2012) 047
  [arXiv:1207.5779 [hep-ph]];
  T.~Gehrmann and P.~F.~Monni,
  JHEP {\bf 1112} (2011) 049
   [arXiv:1107.4037 [hep-ph]].

\bibitem{vegas}
  G.~P.~Lepage,
  J.\ Comput.\ Phys.\  {\bf 27} (1978) 192.


\bibitem{joao}
  E.W.N.~Glover and J.~Pires,
  JHEP {\bf 1006} (2010) 096
 [arXiv:1003.2824 [hep-ph]];
  A.~Gehrmann-De Ridder, E.W.N.~Glover and J.~Pires,
  JHEP {\bf 1202} (2012) 141
  [arXiv:1112.3613 [hep-ph]].





\bibitem{disme}
 T.~Gehrmann and E.~Remiddi,
  Nucl.\ Phys.\ B {\bf 640} (2002) 379
  [hep-ph/0207020];
 T.~Gehrmann and E.~W.~N.~Glover,
  Phys.\ Lett.\ B {\bf 676} (2009) 146
  [arXiv:0904.2665 [hep-ph]].


\bibitem{sherpa}
  T.~Gleisberg, S.~H\"oche, F.~Krauss, M.~Sch\"onherr, S.~Schumann, F.~Siegert and J.~Winter,
  JHEP {\bf 0902} (2009) 007
  [arXiv:0811.4622 [hep-ph]].

\bibitem{hoeche}
 T.~Carli, T.~Gehrmann and S.~H\"oche,
  Eur.\ Phys.\ J.\ C {\bf 67} (2010) 73
  [arXiv:0912.3715 [hep-ph]].

\bibitem{klasen}
 T.~Biek\"otter, M.~Klasen and G.~Kramer,
  Phys.\ Rev.\ D {\bf 92} (2015) 074037
  [arXiv:1508.07153 [hep-ph]].

\bibitem{sudakov}
  S.~Catani and B.R.~Webber,
  JHEP {\bf 9710} (1997) 005
[hep-ph/9710333].

\bibitem{czakonpdf}
  M.~Czakon, M.~L.~Mangano, A.~Mitov and J.~Rojo,
  JHEP {\bf 1307} (2013) 167
  [arXiv:1303.7215 [hep-ph]].

\bibitem{pdg}
 C.~Patrignani {\it et al.} [Particle Data Group],
  Chin.\ Phys.\ C {\bf 40} (2016) 100001.


\end{thebibliography}
\end{document}